# Comparative Results: Group Search Optimizer and Central Force Optimization

## Richard A. Formato[1]

**Abstract:** This note compares the performance of two multidimensional search and optimization algorithms: Group Search Optimizer and Central Force Optimization. GSO is a new state-of-the-art algorithm that has gained some notoriety, consequently providing an excellent yardstick for measuring the performance of other algorithms. CFO is a novel deterministic metaheuristic that has performed well against GSO in previous tests. The CFO implementation reported here includes architectural improvements in errant probe retrieval and decision space adaptation that result in even better performance. Detailed results are provided for the twenty-three function benchmark suite used to evaluate GSO. CFO performs better than or essentially as well as GSO on twenty functions and nearly as well on one of the remaining three.

*Ver. 3, 24 February 2010* (Fig. A2(b) replaced for improved visualization; minor typos corrected).

*Ver. 2, 22 February 2010* (Reset of decision space boundaries to initial values after time loop explicitly added in Fig. 1 because previously it was implied but explicitly included only in the source code listing in Appendix 3; Fig. A2(b) and discussion added to provide 3D IPD visualization; Reference [6] updated).

*14 February 2010*
*Saint Augustine, Florida*

**Comparative Results: Group Search Optimizer and Central Force Optimization**

**Richard A. Formato[1]**

**1. Introduction**

This note compares two multidimensional search and optimization algorithms: Group Search Optimizer (GSO) and Central Force Optimization (CFO) [1-11]. GSO is a new, state-of-the-art metaheuristic that has gained some notoriety [12,13]. It mimics animal foraging behavior using the strategies of "producing" (searching for food) and "scrounging" (joining resources discovered by other group members). GSO was tested against a suite of twenty-three benchmark functions with dimensionalities ranging from 2 to 30, and also on a limited basis for scalability to 300 dimensions. This report compares GSO and CFO using the 2-30D benchmark suite. CFO performs exceptionally well, and, unlike GSO which is fundamentally stochastic, returns precisely the same results for every run with the same setup parameters. CFO's further development lies in two areas: theory and architecture. On the theoretical front, there is reason to believe that the theory of Near Earth objects may be very useful [7-9], but this issue is not considered here. Rather, this note describes an improved architecture that was determined empirically. It is likely that most of CFO's future architectural improvements similarly will be empirical in nature, and this note suggests some directions for such work.

**2. CFO Implementation**

CFO pseudocode for the implementation used here appears in Fig. 1 [for completeness the metaheuristic is described in detail in Appendix 1]. CFO is quite different from inherently stochastic algorithms because it is completely *deterministic*. Its probe trajectories are computed with absolute precision at every step. Nevertheless, CFO benefits from some measure of *pseudorandomness* because more information is generated about decision space $\Omega$'s topology [11]. Three methods are employed toward this end: (1) variable IPD (initial probe distribution); (2) variable repositioning factor; and (3) decision space adaptation in which $\Omega$'s size is periodically reduced. Each of these methods is "pseudorandom" in nature because the numbers involved are completely arbitrary but precisely known, either by specification or calculation. Thus, each variable is known with absolute precision at every step (iteration) of the CFO run. This type of "randomness" is fundamentally different from a truly random process in which values are computed from probability distributions and consequently unknowable in advance.

The CFO algorithm comprises three loops as shown in Fig. 1. The time step ( $j$ ) loop is common to all CFO implementations, while the outer $\left( \frac{N_p}{N_d} \right)$ and $\gamma$ loops define the IPD as described in Appendix 1. The errant probe retrieval scheme in Fig. 1 has been modified from the one used in [11] because it was possible that probes would be positioned beyond $\Omega$'s new boundary after shrinking [14]. The new scheme addresses this outlier issue, but it still is possible that many probes could be placed directly on the new boundary. This contingency, however, seems to be less problematic because the probes remain inside $\Omega$'s initial boundary, and placing them on the new one is no different than shrinking $\Omega$ by a different amount. This view is supported by some numerical experiments that failed to reveal significant differences when probes were placed within the new boundary by some fraction of $\Omega$'s extent along the coordinate direction in question. Concerning adaptation of $\Omega$'s boundaries, note that at the start of each new $\left(\dfrac{N_p}{N_d}\right) \cdot \gamma$ run the decision space boundaries must be reset to their starting values before any shrinking is applied [step (h) in Fig. 1].

Two other changes have been made from the implementation in [11]: (1) when $\Omega$ is shrunk; and (2) the test for early termination. Instead of shrinking every 20th step, $\Omega$ now is re-sized every 10th step starting at step #20. The test for early termination is a difference between the average best fitness over 25 steps (instead of 50), including the current step, and the current best fitness of less than $10^{-6}$ (absolute value). These changes, along with the new errant probe retrieval scheme, have improved CFO's performance compared to GSO. The complete source code listing for this version of CFO (in Power Basic) appears in Appendix 3 and is available in electronic format upon request to the author.

---

**For** $N_p / N_d = 2$ to $\left(\dfrac{N_p}{N_d}\right)_{MAX}$ by $2$ :

**For** $\gamma = \gamma_{start}$ to $\gamma_{stop}$ by $\Delta\gamma$ :

   (a.1)        Compute initial probe distribution.
   (a.2)        Compute initial fitness matrix.
   (a.3)        Assign initial probe accelerations.
   (a.4)        Set initial $F_{rep}$ .

**For** $j = 0$ to $N_t$ (or earlier termination criterion – see text):

   (b)        Compute probe position vectors, $\vec{R}_j^{\,p}, 1 \le p \le N_p$ [eq.(A2)].

   (c)        Retrieve errant probes ($1 \le p \le N_p$):

             If $\vec{R}_j^{\,p} \cdot \hat{e}_i < x_i^{min}$ $\therefore$

             $\vec{R}_j^{\,p} \cdot \hat{e}_i = \max\{x_i^{min} + F_{rep}(\vec{R}_{j-1}^{\,p} \cdot \hat{e}_i - x_i^{min}), x_i^{min}\}$

             If $\vec{R}_j^{\,p} \cdot \hat{e}_i > x_i^{max}$ $\therefore$

             $\vec{R}_j^{\,p} \cdot \hat{e}_i = \min\{x_i^{max} - F_{rep}(x_i^{max} - \vec{R}_{j-1}^{\,p} \cdot \hat{e}_i), x_i^{max}\}$

   (d)        Compute fitness matrix for current probe distribution.

   (e)        Compute accelerations using current probe distribution and fitnesses [eq. (A1)].

   (f)        Increment $F_{rep}$ by $\Delta F_{rep}$: $\begin{cases} F_{rep} = F_{rep} + \Delta F_{rep} \\ \text{If } F_{rep} > 1 \therefore F_{rep} = \Delta F_{rep} \end{cases}$.

   (g)        If $j \ge 20$ and $j \, MOD \, 10 = 0$ $\therefore$ Shrink $\Omega$ around $\vec{R}_{best}$ (see text).

**Next** $j$

   (h)        Reset $\Omega$'s boundaries to their starting values before shrinking.

**Next** $\gamma$

**Next** $N_p / N_d$

Fig. 1. CFO Pseudocode.

## 3. GSO-CFO Comparison

CFO was tested against the same twenty-three function 2-30D benchmark suite used to evaluate GSO in [12]. In that paper, GSO was compared to two other algorithms, "PSO" and "GA". The benchmark functions, their decision spaces and characteristics, and the other algorithms are discussed in detail in [12]. PSO is a Particle Swarm implemented using "PSOt," a MATLAB-based toolbox that includes standard and variant PSO algorithms. GA is a Genetic Algorithm implemented using the GAOT toolbox (genetic algorithm optimization toolbox). Recommended default parameter values were used for PSOt and GA as described in [12]. Thus, while this note compares CFO and GSO directly, it indirectly compares CFO to PSO and GA as well.

Table 1 summarizes CFO's results using the same function numbering as [12]. Detailed CFO results for each test function appear in Appendix 2. $f_{max}$ is the known global *maximum* (note that the negative of each benchmark in [12] is used here because, unlike the other algorithms, CFO locates maxima, not minima). $< \cdot >$ denotes average value. Because GSO, PSO and GA are inherently stochastic, their performance can only be described statistically. The statistical data for those algorithms in Table 2 are reproduced from [12]. By contrast, CFO's results are repeatable over runs with the same parameters because CFO is completely deterministic; no statistical description is needed for CFO.

CFO's best fitness returned by the best run in the set of runs with variable $N_p/N_d$ and variable $\gamma$ is tabulated in Table 1. For each value of $N_p/N_d$, eleven runs were made with $0 \le \gamma \le 1$ in increments of $0.1$. For functions $f_{14} - f_{23}$, $2 \le N_p/N_d \le 14$ by 2 (77 runs total), while for $f_1 - f_{13}$, $2 \le N_p/N_d \le 6$ by 2 (33 runs total) [ $N_p/N_d$ reduced to avoid excessive runtimes]. Table 1 shows the $\gamma$ value corresponding to the best fitness, $\gamma_{best}$, and the corresponding best value of $N_p/N_d$. The number of function evaluations, $N_{eval}$, is tabulated for the single best run, and for the group of runs used to determine $\gamma_{best}$ and the best number of probes per axis. While $N_t$ was set to 1000 to permit complete evolution of the best fitness, in every case the run was terminated much sooner because the early termination criterion was met.

CFO returned the best fitness on all seven functions in the first group of high dimensionality unimodal functions ($f_1 - f_7$). In the second set of six high dimensionality multimodal functions with many local maxima ($f_8 - f_{13}$), CFO performed best on two ($f_9, f_{10}$) and essentially the same as the best other algorithm (GSO) on $f_8$. CFO performed nearly as well as GSO on $f_{11}$ (best fitnesses, GSO/CFO, -0.030792/-0.0441976). In the last group of ten multimodal functions with few local maxima ($f_{14} - f_{23}$), CFO returned the best fitness on six functions ($f_{15}$, $f_{16}$, $f_{20} - f_{23}$), essentially equal fitnesses on three ($f_{14,}$ $f_{17,}$ $f_{18}$), and very slightly worse fitness on one ($f_{19}$: GSO/CFO, 3.8628/3.86269). CFO's performance was not as good as GSO's on only two of the twenty three functions ($f_{12}$, $f_{13}$).

## 4. Conclusion

Even though CFO is not as highly developed as the other algorithms in [12], it performed very well against all of them. CFO performed better than or essentially as well as GSO, GA and PSO on twenty of twenty three test functions, and only slightly worse on one of the remaining three. By contrast, in [12] GSO returned the best performance on only fifteen test functions compared to PSO and GA. Thus, CFO's performance is considerably better than GSO's, which in turn performed better than PSO or GA. In the work reported here, CFO's performance was improved by making changes in the probe retrieval scheme and in decision space adaptation. Further improvements in these areas, as well as in the IPD, likely will result in even better performance.

## Table 1. CFO Comparative Results for 23 Benchmark Functions

( $N_d$ = Function Dimension, $f_{max}$ = Known Global Maximum)

| Test Function* | $N_d$ | $f_{max}$* | <Best Fitness>/ Other Algorithm | CFO Best Fitness | $\gamma_{best}$ | Best $N_p/N_d$ | $N_{eval}$ Best Run | Total |
|---|---|---|---|---|---|---|---|---|
| Unimodal Functions (other algorithms: average of 1000 runs) | | | | | | | | |
| $f_1$ | 30 | 0 | -3.6927x10⁻³⁷ / PSO | **0** | 0.5 | 2 | 2,760 | 405,780 |
| $f_2$ | 30 | 0 | -2.9168x10⁻²⁴ / PSO | **0** | 0.5 | 2 | 15,960 | 330,300 |
| $f_3$ | 30 | 0 | -1.1979x10⁻³ / PSO | **-3.857x10⁻⁵** | 0.5 | 2 | 4,380 | 859,740 |
| $f_4$ | 30 | 0 | -0.1078 / GSO | **0** | 0.5 | 2 | 17,340 | 178,140 |
| $f_5$ | 30 | 0 | -37.3582 / PSO | **-2.05081x10⁻³** | 0.1 | 4 | 10,200 | 477,540 |
| $f_6$ | 30 | 0 | -1.6000x10⁻² / GSO | **0** | 0.5 | 2 | 2,760 | 227,940 |
| $f_7$ | 30 | 0 | -9.9024x10⁻³ / PSO | **-2.3835x10⁻⁴** | 0.9 | 6 | 18,180 | 399,960 |
| Multimodal Functions, Many Local Maxima (other algorithms: avg 1000 runs) | | | | | | | | |
| $f_8$ | 30 | 12,569.5 | **12,569.4882** / GSO | **12,569.4866** | 0.5 | 4 | 7,440 | 326.820 |
| $f_9$ | 30 | 0 | -0.6509 / GA | **0** | 0.5 | 2 | 2,760 | 359,040 |
| $f_{10}$ | 30 | 0 | -2.6548x10⁻⁵ / PSO | **0** | 0.5 | 2 | 2,760 | 478,560 |
| $f_{11}$ | 30 | 0 | **-3.0792x10⁻²** / PSO | -4.41976x10⁻² | 0.1 | 6 | 35,280 | 266,700 |
| $f_{12}$ | 30 | 0 | **-2.7648x10⁻¹¹** / GSO | -2.067x10⁻⁵ | 0.5 | 2 | 2,160 | 233.280 |
| $f_{13}$ | 30 | 0 | **-4.6948x10⁻⁵** / GSO | -1.02803x10⁻³ | 0.7 | 4 | 10,920 | 317,280 |
| Multimodal Functions, Few Local Maxima (other algorithms: avg 50 runs) | | | | | | | | |
| $f_{14}$ | 2 | -1 | **-0.9980** / GSO | **-0.9980** | 0.2 | 12 | 2,784 | 79,136 |
| $f_{15}$ | 4 | -0.0003075 | -3.7713x10⁻⁴ / PSO | **-3.6196x10⁻⁴** | 0.5 | 12 | 7,008 | 171,216 |
| $f_{16}$ | 2 | 1.0316285 | 1.031628 / PSO | **1.03162821** | 0.5 | 12 | 3,624 | 74,832 |
| $f_{17}$ | 2 | -0.398 | **-0.3979** / GSO | **-0.39795354** | 0.6 | 4 | 416 | 73,444 |
| $f_{18}$ | 2 | -3 | **-3** / GSO | **-3.0000001** | 0.6 | 6 | 2,148 | 94,668 |
| $f_{19}$ | 3 | 3.86 | **3.8628** / GSO | 3.86268376 | 0.2 | 14 | 2,100 | 128,286 |
| $f_{20}$ | 6 | 3.32 | 3.2697 / GSO | **3.32157899** | 0.5 | 12 | 10,440 | 408,084 |
| $f_{21}$ | 4 | 10 | 7.5439 / PSO | **10.15319585** | 0.4 | 6 | 2,376 | 210,296 |
| $f_{22}$ | 4 | 10 | 8.3553 / PSO | **10.4029108** | 0.4 | 10 | 7,640 | 256,776 |
| $f_{23}$ | 4 | 10 | 8.9439 / PSO | **10.53633734** | 1.0 | 12 | 9,264 | 242,848 |

*Negative of the functions in [12] are computed by CFO because CFO searches for maxima instead of minima.

Note: best fitnesses highlighted in **red**; essentially equal fitnesses in **blue**.

**Appendix 1.  CFO Metaheuristic**

CFO searches for the global *maxima* of an *objective function* $f(x_1, x_2, ..., x_{N_d})$ defined on an $N_d$-dimensional decision space $\Omega$: $x_i^{\min} \leq x_i \leq x_i^{\max}$, $1 \leq i \leq N_d$. The $x_i$ are the *decision variables*, and $i$ the coordinate number. The term *fitness* refers to the value of $f(\vec{x})$ at point $\vec{x}$ in $\Omega$. There is no *a priori* information about the objective function's maxima, that is, $f(\vec{x})$'s topology or "landscape" is unknown.

CFO searches $\Omega$ by flying "probes" through the space at discrete "time" steps (iterations). Each probe's location is specified by its position vector computed from two *equations of motion* that analogize their real-world counterparts for material objects moving through physical space under the influence of gravity without energy dissipation.

Probe $p$'s position vector at step $j$ is $\vec{R}_j^p = \sum_{k=1}^{N_d} x_k^{p,j} \hat{e}_k$, where the $x_k^{p,j}$ are its coordinates and $\hat{e}_k$ the unit vector along the $x_k$-axis. The indices $p$, $1 \leq p \leq N_p$, and $j$, $0 \leq j \leq N_t$, respectively, are the probe number and iteration number, where $N_p$ and $N_t$ are the corresponding *total* number of probes and *total* number of time steps.

***Equations of Motion.*** In metaphorical "CFO space" each of the $N_p$ probes experiences an acceleration created by the "gravitational pull" of "masses" in $\Omega$. Probe $p$'s acceleration at step $j-1$ is given by

$$\vec{a}_{j-1}^p = G \sum_{\substack{k=1 \\ k \neq p}}^{N_p} U(M_{j-1}^k - M_{j-1}^p) \cdot (M_{j-1}^k - M_{j-1}^p)^\alpha \times \frac{(\vec{R}_{j-1}^k - \vec{R}_{j-1}^p)}{\left\| \vec{R}_{j-1}^k - \vec{R}_{j-1}^p \right\|^\beta}, \quad \text{(A1)}$$

which is the first of CFO's two equations of motion. In (A1), $M_{j-1}^p = f(x_1^{p,j-1}, x_2^{p,j-1}, ..., x_{N_d}^{p,j-1})$ is the objective function's fitness at probe $p$'s location at time step $j-1$. Each of the other probes at that step (iteration) has associated with it fitness $M_{j-1}^k$, $k = 1, ..., p-1, p+1, ..., N_p$. $G$ is CFO's *gravitational constant*, and $U(\cdot)$ is the Unit Step function, $U(z) = \begin{cases} 1, & z \geq 0 \\ 0, & otherwise \end{cases}$.

The acceleration $\vec{a}_{j-1}^p$ causes probe $p$ to move from position $\vec{R}_{j-1}^p$ at step $j-1$ to $\vec{R}_j^p$ at step $j$ according to the trajectory equation

$$\vec{R}_j^p = \vec{R}_{j-1}^p + \frac{1}{2} \vec{a}_{j-1}^p \Delta t^2, \quad j \geq 1 \quad , \quad \text{(A2)}$$

which is CFO's second equation of motion.

CFO's equations of motion, (A1) and (A2), combine to compute a new probe distribution at each time step using "masses" discovered by the probe distribution at the previous step. $\Delta t$ is the "time" interval between steps during which the acceleration is constant. Note that CFO's terminology has no significance beyond reflecting its kinematic roots, as does the factor ½ in (A2). The gravitational constant, $G$, and time increment, $\Delta t$, have direct analogues in Newton's equations of motion for real masses moving under real gravity through three-dimensional physical space. The CFO exponents $\alpha$ and $\beta$, by contrast, have no analogues in Nature. They provide added flexibility to the algorithm designer who, in metaphorical "CFO space," is free to change how "gravity" varies with distance, or mass, or both if doing so creates a more effective algorithm.

**CFO "Mass."** The concept of "mass" in CFO space is very important and quite different than it is in real space. Mass in the physical Universe is an inherent, immutable property of matter, whereas in CFO space it is a positive-definite *user-defined function* of the objective function's fitness [not (necessarily) the fitness itself]. For example, in equation (A1) mass is defined as $MASS_{CFO} = U(M_{j-1}^k - M_{j-1}^p) \cdot (M_{j-1}^k - M_{j-1}^p)^\alpha$ [difference in fitness values raised to the $\alpha$ power multiplied by the Unit Step]. A different function can be used if it results in better performance. In this specific implementation the Unit Step is a critical element because it prevents negative mass. Without the Unit Step CFO mass could be negative depending on which fitness is greater. But mass in the real Universe is positive, and consequently the force of gravity always attractive. By contrast, in metaphorical CFO space mass can be positive or negative depending on how it is defined, with very undesirable effects resulting from the wrong definition. Negative mass creates a *repulsive* gravitational force that flies probes away from maxima instead of toward them, thus defeating CFO's very purpose.

**Initial Probe Distribution.** Every CFO run begins with a *user-specified* Initial Probe Distribution (IPD) defined by two parameters: **(1)** the total number of probes used, $N_p$; and **(2)** where the probes are placed inside $\Omega$. The CFO implementation reported here employs a *pseudorandom* variable IPD comprising an orthogonal array of $N_p/N_d$ probes per axis deployed uniformly on "probe lines" parallel to the coordinate axes and intersecting at a point along $\Omega$'s principal diagonal. Pseudorandomness is defined as an arbitrary numerical sequence that is precisely known by specification or calculation. CFO's fundamentally deterministic nature is not altered by injecting pseudorandomness because at every step CFO's calculations are repeatable with absolute precision (see [11] for a discussion of why pseudorandomness is important in CFO).

Fig. A1 provides a two-dimensional (2D) example of this type of IPD (nine probes shown on each probe line, two overlapping, but any number may be used). The probe lines are parallel to the $x_1$ and $x_2$ axes intersecting at a point on $\Omega$'s principal diagonal marked by position vector $\vec{D} = \vec{X}_{\min} + \gamma(\vec{X}_{\max} - \vec{X}_{\min})$, where $\vec{X}_{\min} = \sum_{i=1}^{N_d} x_i^{\min} \hat{e}_i$ and $\vec{X}_{\max} = \sum_{i=1}^{N_d} x_i^{\max} \hat{e}_i$ are the diagonal's endpoint vectors. The parameter $0 \le \gamma \le 1$ determines where the probe lines intersect along the diagonal. Fig. A2(a) shows a typical 2D IPD for different values of $\gamma$, while Fig. A2(b) provides a 3D example for various $\gamma$ values. In Fig. A2(b) the principal diagonal is shown in red and the three orthogonal probe lines in blue. In this example each probe line contains six equally spaced probes. For certain values of $\gamma$ three probes overlap at the probe lines' intersection point on the principal diagonal. Of course, this IPD procedure is generalized to the $N_d$-dimensional decision space $\Omega$ to create $N_d$ probe lines parallel to the $N_d$ coordinate axes.

While Fig. A1 shows equal numbers of probes on each probe line, a different number of probes per axis can be used instead. For example, if equal probe spacing were desired in a decision space with unequal boundaries, or if overlapping probes were to be excluded in a symmetrical space, then unequal numbers could be used. Unequal numbers also might be appropriate if *a priori* knowledge of $\Omega$'s landscape, however obtained, suggests denser sampling in one region. While the variable $N_p / N_d$ IPD of Fig. A1 was used for the results reported here, any number of other variable IPDs could be used instead. The key idea is that the IPD must be pseudorandom in the sense of uncorrelated with the decision space landscape in order to provide better sampling of the landscape.

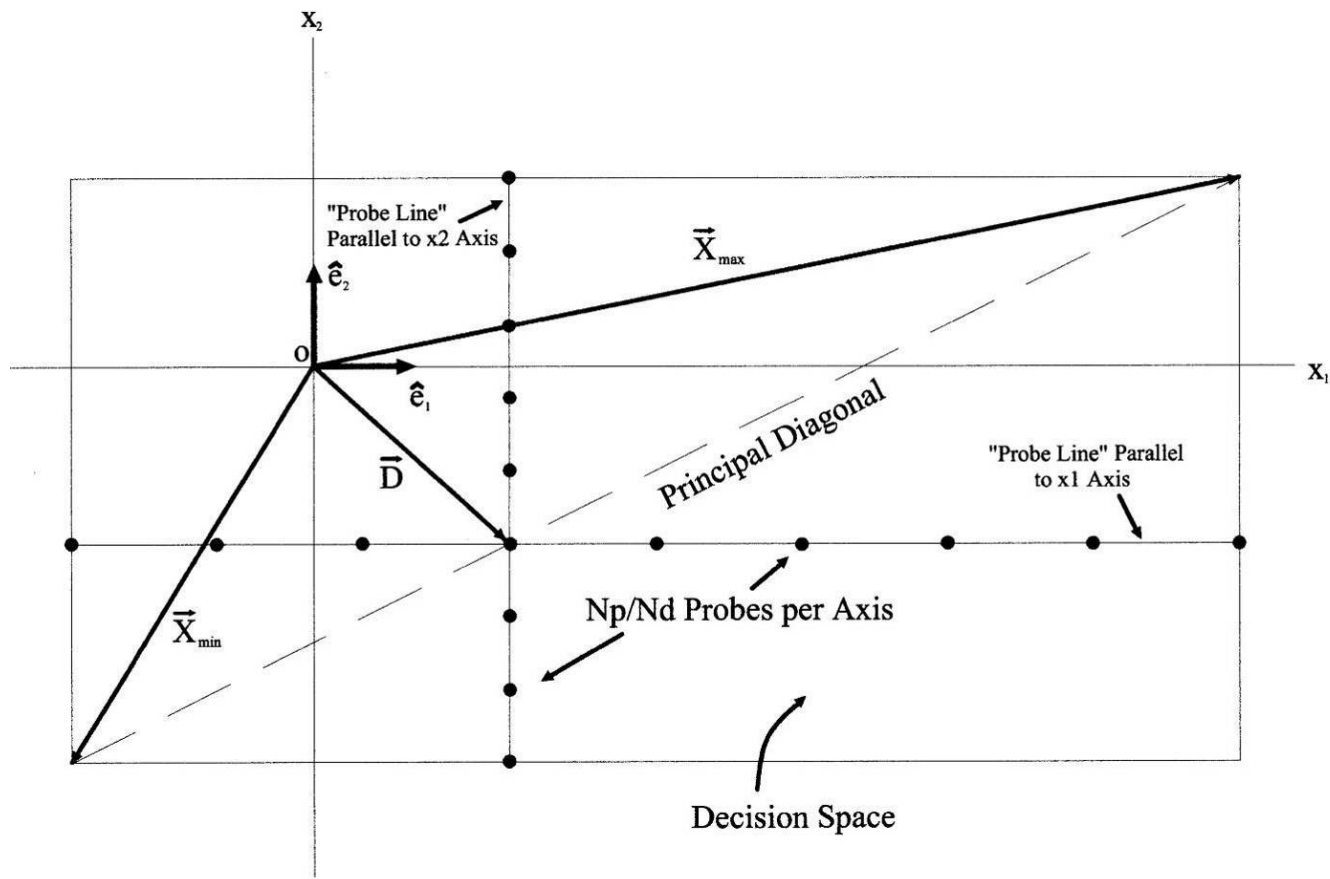

Fig. A1.  Variable 2D Initial Probe Distribution used for CFO Runs Reported in This Note.

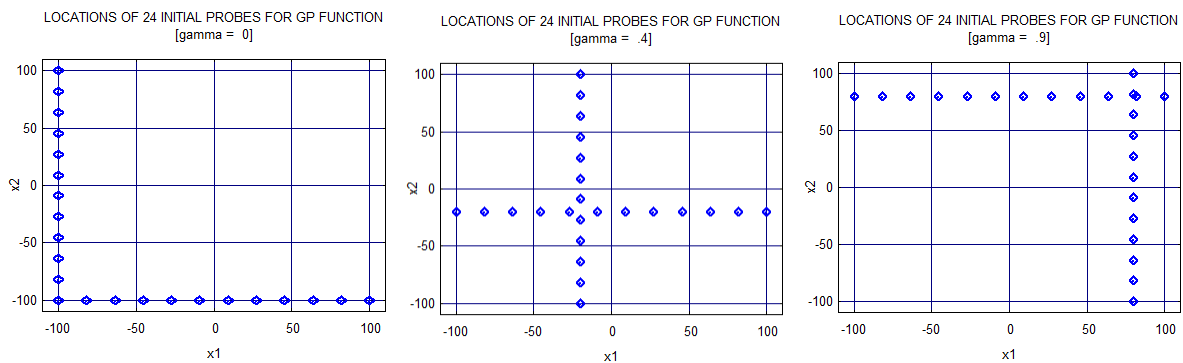

Fig. A2(a).  Typical 2D IPD's for Different Values of $\gamma$.

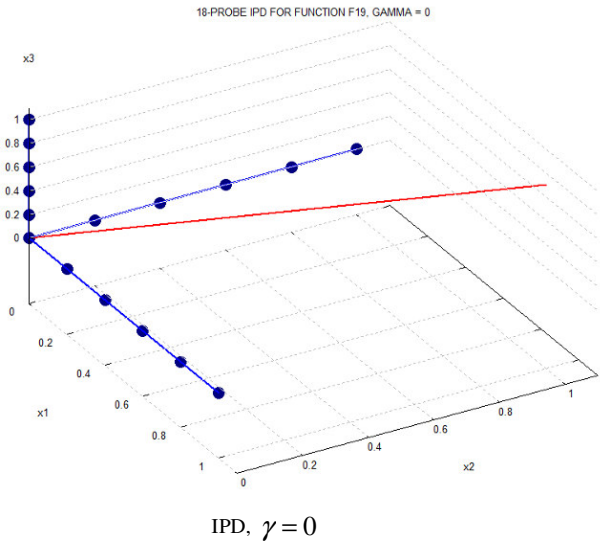

IPD, $\gamma = 0$

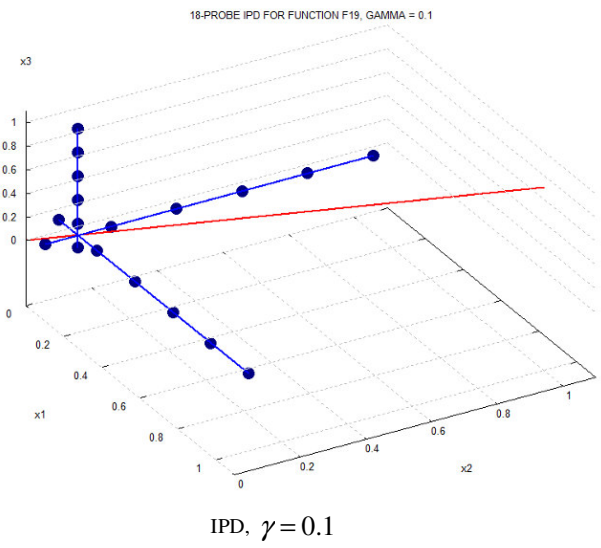

IPD, $\gamma = 0.1$

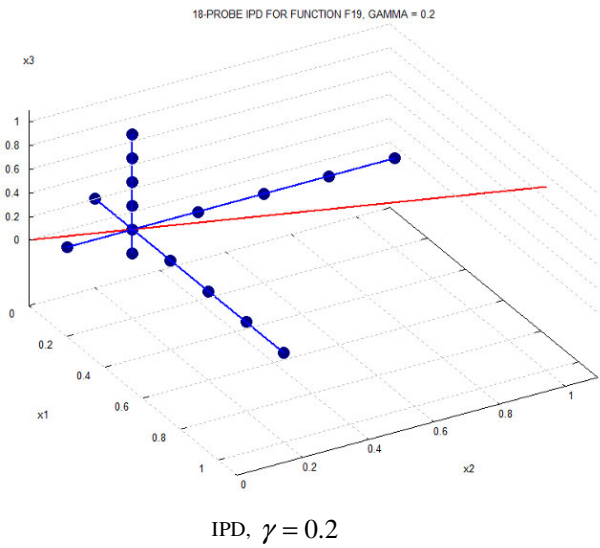

IPD, $\gamma = 0.2$

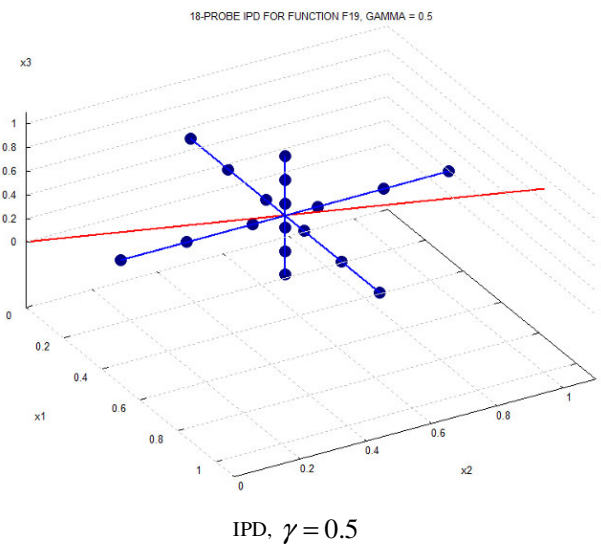

IPD, $\gamma = 0.5$

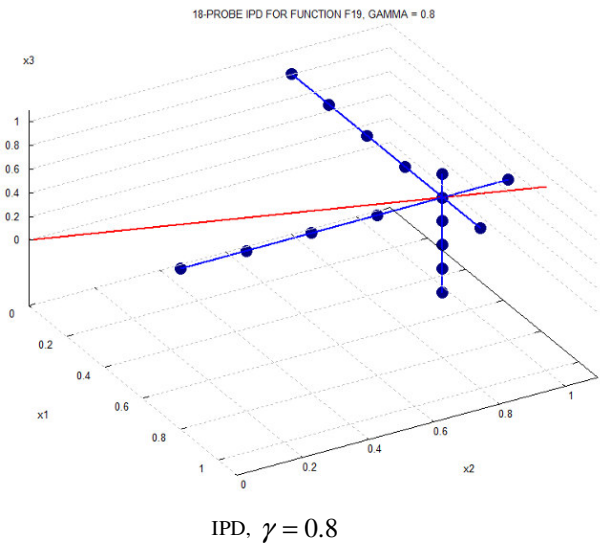

IPD, $\gamma = 0.8$

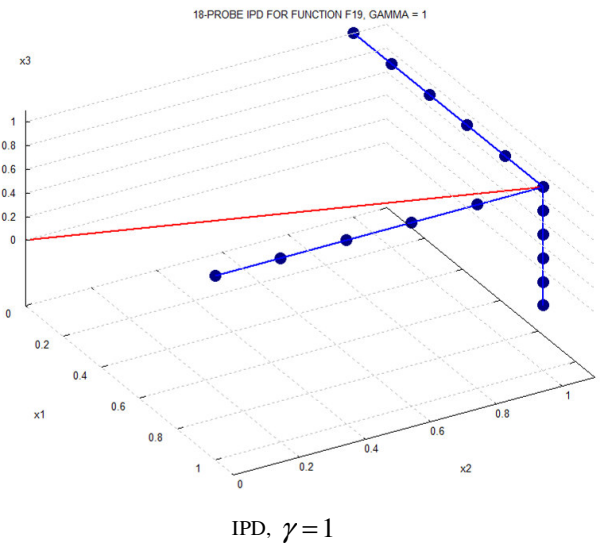

IPD, $\gamma = 1$

Fig. A2(b).  Typical Variable 3D IPD (Initial Probe Distribution) for the GSO $f_{19}$ Function (probes as filled circles).

***Errant Probes.*** At any iteration in a CFO run, it is possible that a probe's acceleration computed from equation (A1) may be too great to keep it inside $\Omega$. If any coordinate $x_i < x_i^{\min}$ or $x_i > x_i^{\max}$, the probe enters a region of *unfeasible* solutions that are not valid for the problem at hand. The question (which arises in many algorithms) is what to do with an errant probe. While many schemes are possible, a simple, empirically determined one is used here. On a coordinate-by-coordinate basis, probes flying outside $\Omega$ are placed a fraction $\Delta F_{rep} \leq F_{rep} \leq 1$ of the distance between the probe's starting coordinate and the corresponding boundary coordinate. $F_{rep}$ is the "repositioning factor" introduced in [2]. See step (c) in the pseudocode of Fig. 1 for details.

Variable $F_{rep}$ is another pseudorandom component of CFO. $F_{rep}$ starts at an arbitrary initial value which is then incremented by an arbitrary amount $\Delta F_{rep}$ at each iteration (subject to $\Delta F_{rep} \leq F_{rep} \leq 1$). This process is inherently pseudorandom because $F_{rep}$ is deterministic but arbitrary and uncorrelated with $\Omega$'s topology. Placing errant probes pseudorandomly throughout $\Omega$ as the run progresses generates more information about its topology. This particular scheme seems to work well across a wide range of objective functions. Of course, many other procedures could be used to set $F_{rep}$'s value instead, some no doubt better than others.

***Decision Space Adaptation:*** This CFO implementation includes adaptive reconfiguration of the decision space in order to improve convergence speed. This feature, too, is pseudorandom in nature because the manner in which $\Omega$'s boundaries are changed is arbitrary and uncorrelated with the landscape. Fig. A3 illustrates in 2D how $\Omega$'s size is adaptively reduced around $\vec{R}_{best}$, the location of the probe with best fitness throughout the run up to the current iteration. $\Omega$ is shrunk every $10^{th}$ step beginning at step 20. $\Omega$'s boundary coordinates are reduced by one-half the distance from the best probe's position to each boundary on a coordinate-by-coordinate basis, *i.e.*, $x_i^{\prime\min} = x_i^{\min} + \dfrac{\vec{R}_{best} \cdot \hat{e}_i - x_i^{\min}}{2}$ and $x_i^{\prime\max} = x_i^{\max} - \dfrac{x_i^{\max} - \vec{R}_{best} \cdot \hat{e}_i}{2}$, where the primed coordinate is $\Omega$'s new boundary, and the dot denotes vector inner product. For clarity in the diagram, Fig. A3 shows $\vec{R}_{best}$ as being fixed whereas generally it varies throughout a run. Changing $\Omega$'s boundary every ten steps instead of some other interval was chosen arbitrarily (another, probably better approach, might be a reactive adaptation based on performance measures such as convergence speed or fitness saturation).

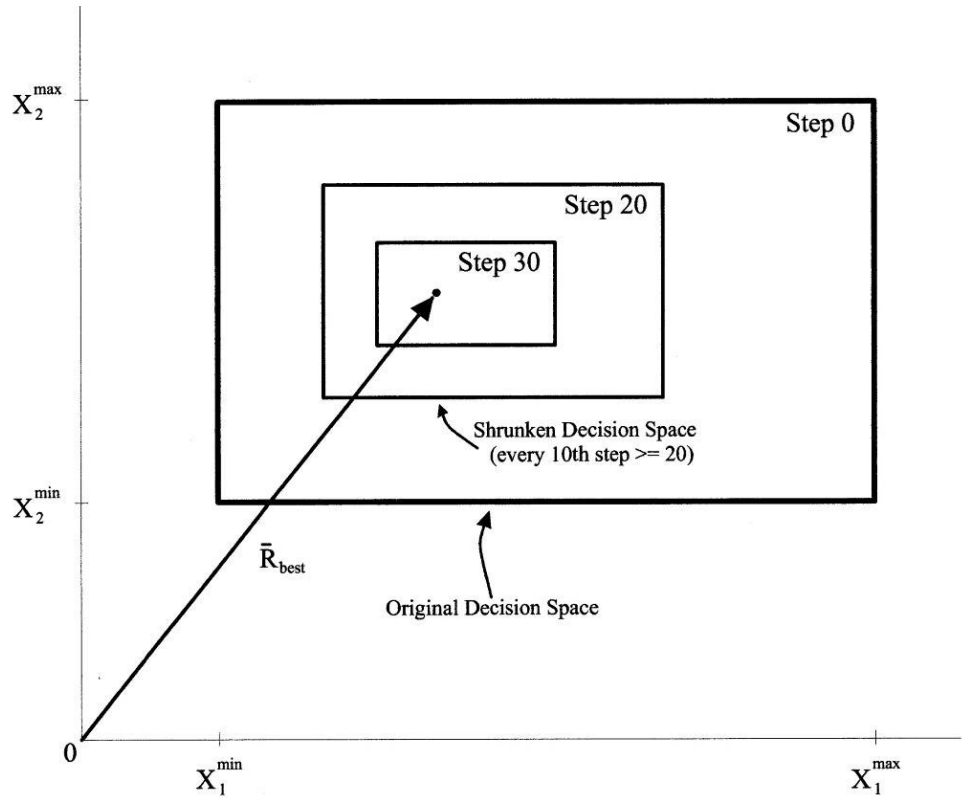

Fig. A3. Schematic 2D Decision Space Adaptation (constant $\vec{R}_{best}$ for illustration only)

*Ver. 3, 24 February 2010* (Fig. A2(b) replaced for improved visualization; minor typos corrected).

*Ver. 2, 22 February 2010* (Reset of decision space boundaries to initial values after time loop explicitly added in Fig. 1 because previously it was implied but explicitly included only in the source code listing in Appendix 3; Fig. A2(b) and discussion added to provide 3D IPD visualization; Reference [6] updated).

*14 February 2010*
*Saint Augustine, Florida*

[1]Richard A. Formato, JD, PhD
Registered Patent Attorney & Consulting Engineer
P.O. Box 1714, Harwich, MA  02645  USA
rf2@ieee.org

## Appendix 2. CFO Summary Data for the GSO Benchmark Suite [12]

In the plots below, $D_{avg}$ is the normalized average distance between the probe with the best fitness and all other probes at each time step, viz., $D_{avg} = \dfrac{1}{L \cdot (N_p - 1)} \sum_{p=1}^{N_p} \sqrt{\sum_{i=1}^{N_d} (x_i^{p,j} - x_i^{p*,j})^2}$ where $p*$ is the number of the probe with the best fitness, and $L = \sqrt{\sum_{i=1}^{N_d} (x_i^{\max} - x_i^{\min})^2}$ is the length of $\Omega$'s principal diagonal (see [7,11] for a discussion of $D_{avg}$). In the tables below: $N_{eval}$ is the number of function evaluations. "V" denotes that $F_{rep}$ was variable as discussed above. Note that the tabulated value is $F_{rep}$'s ending value when the run terminated (not the initial value). *Fitness* tabulates the best fitness returned during the run. The *Initial Probes* column shows the type of initial probe distribution, in this case probes UNIFORMly spaced along probe lines parallel to $\Omega$'s axes (notated "P-AXIS") as described above and shown in Figs. A1 and A2. Other column headings are self-explanatory. Note that, because CFO is deterministic, these data can be used to validate implementations based on the pseudocode in Fig. 1.

### F1

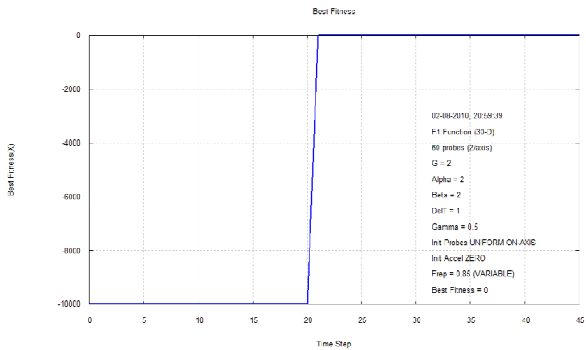

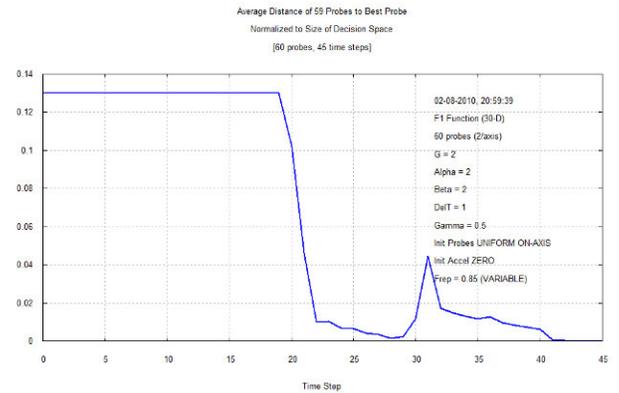

Run ID: 02-08-2010, 20:59:39 FUNCTION: F1

| Run # | Gamma | Nt | Nd | Np | G | DelT | Alpha | Beta | #Steps | Neval | Frep | Fitness | Initial Probes |
|-------|-------|------|----|-----|-----|------|-------|------|--------|-------|----------|----------------|----------------|
| 0 | 0.000 | 1000 | 30 | 60 | 2.0 | 1.0 | 2.00 | 2.00 | 0 | 0 | 0.50000V | -9999.00000000 | UNIFORM P-AXIS |
| 1 | 0.000 | 1000 | 30 | 60 | 2.0 | 1.0 | 2.00 | 2.00 | 67 | 4080 | 0.05000V | -23.95817193 | UNIFORM P-AXIS |
| 2 | 0.100 | 1000 | 30 | 60 | 2.0 | 1.0 | 2.00 | 2.00 | 143 | 8700 | 0.10000V | -83.88344902 | UNIFORM P-AXIS |
| 3 | 0.200 | 1000 | 30 | 60 | 2.0 | 1.0 | 2.00 | 2.00 | 69 | 4200 | 0.15000V | -192.03220245 | UNIFORM P-AXIS |
| 4 | 0.300 | 1000 | 30 | 60 | 2.0 | 1.0 | 2.00 | 2.00 | 70 | 4260 | 0.20000V | -0.37297992 | UNIFORM P-AXIS |
| 5 | 0.400 | 1000 | 30 | 60 | 2.0 | 1.0 | 2.00 | 2.00 | 165 | 9960 | 0.20000V | -451.28837349 | UNIFORM P-AXIS |
| 6 | 0.500 | 1000 | 30 | 60 | 2.0 | 1.0 | 2.00 | 2.00 | 45 | 2760 | 0.85000V | 0.00000000 | UNIFORM P-AXIS |
| 7 | 0.600 | 1000 | 30 | 60 | 2.0 | 1.0 | 2.00 | 2.00 | 141 | 8520 | 0.90000V | -623.06247227 | UNIFORM P-AXIS |
| 8 | 0.700 | 1000 | 30 | 60 | 2.0 | 1.0 | 2.00 | 2.00 | 50 | 3060 | 0.15000V | -566.66190400 | UNIFORM P-AXIS |
| 9 | 0.800 | 1000 | 30 | 60 | 2.0 | 1.0 | 2.00 | 2.00 | 79 | 4800 | 0.65000V | -749.64379305 | UNIFORM P-AXIS |
| 10 | 0.900 | 1000 | 30 | 60 | 2.0 | 1.0 | 2.00 | 2.00 | 169 | 10200 | 0.40000V | -73.57166861 | UNIFORM P-AXIS |
| 11 | 1.000 | 1000 | 30 | 60 | 2.0 | 1.0 | 2.00 | 2.00 | 65 | 3960 | 0.90000V | -421.87500000 | UNIFORM P-AXIS |
| 12 | 0.000 | 1000 | 30 | 120 | 2.0 | 1.0 | 2.00 | 2.00 | 91 | 11040 | 0.30000V | -0.05004136 | UNIFORM P-AXIS |
| 13 | 0.100 | 1000 | 30 | 120 | 2.0 | 1.0 | 2.00 | 2.00 | 72 | 8760 | 0.30000V | -0.01775796 | UNIFORM P-AXIS |
| 14 | 0.200 | 1000 | 30 | 120 | 2.0 | 1.0 | 2.00 | 2.00 | 92 | 11160 | 0.35000V | -0.00089833 | UNIFORM P-AXIS |
| 15 | 0.300 | 1000 | 30 | 120 | 2.0 | 1.0 | 2.00 | 2.00 | 165 | 19920 | 0.20000V | -0.12885020 | UNIFORM P-AXIS |
| 16 | 0.400 | 1000 | 30 | 120 | 2.0 | 1.0 | 2.00 | 2.00 | 108 | 13080 | 0.20000V | -0.00283524 | UNIFORM P-AXIS |
| 17 | 0.500 | 1000 | 30 | 120 | 2.0 | 1.0 | 2.00 | 2.00 | 217 | 26160 | 0.90000V | -0.00000219 | UNIFORM P-AXIS |
| 18 | 0.600 | 1000 | 30 | 120 | 2.0 | 1.0 | 2.00 | 2.00 | 108 | 13080 | 0.20000V | -0.00283524 | UNIFORM P-AXIS |
| 19 | 0.700 | 1000 | 30 | 120 | 2.0 | 1.0 | 2.00 | 2.00 | 80 | 9720 | 0.70000V | -0.08407900 | UNIFORM P-AXIS |
| 20 | 0.800 | 1000 | 30 | 120 | 2.0 | 1.0 | 2.00 | 2.00 | 85 | 10320 | 0.95000V | -0.00038449 | UNIFORM P-AXIS |
| 21 | 0.900 | 1000 | 30 | 120 | 2.0 | 1.0 | 2.00 | 2.00 | 57 | 6960 | 0.50000V | -0.04264091 | UNIFORM P-AXIS |
| 22 | 1.000 | 1000 | 30 | 120 | 2.0 | 1.0 | 2.00 | 2.00 | 105 | 12720 | 0.05000V | -0.00299090 | UNIFORM P-AXIS |
| 23 | 0.000 | 1000 | 30 | 180 | 2.0 | 1.0 | 2.00 | 2.00 | 59 | 10800 | 0.60000V | -0.08461177 | UNIFORM P-AXIS |
| 24 | 0.100 | 1000 | 30 | 180 | 2.0 | 1.0 | 2.00 | 2.00 | 91 | 16560 | 0.30000V | -0.07247290 | UNIFORM P-AXIS |
| 25 | 0.200 | 1000 | 30 | 180 | 2.0 | 1.0 | 2.00 | 2.00 | 53 | 9720 | 0.30000V | -0.33188771 | UNIFORM P-AXIS |
| 26 | 0.300 | 1000 | 30 | 180 | 2.0 | 1.0 | 2.00 | 2.00 | 175 | 31680 | 0.70000V | -0.08457834 | UNIFORM P-AXIS |
| 27 | 0.400 | 1000 | 30 | 180 | 2.0 | 1.0 | 2.00 | 2.00 | 72 | 13140 | 0.30000V | -0.08222241 | UNIFORM P-AXIS |
| 28 | 0.500 | 1000 | 30 | 180 | 2.0 | 1.0 | 2.00 | 2.00 | 108 | 19620 | 0.20000V | -0.00263043 | UNIFORM P-AXIS |
| 29 | 0.600 | 1000 | 30 | 180 | 2.0 | 1.0 | 2.00 | 2.00 | 205 | 37080 | 0.30000V | -0.00328457 | UNIFORM P-AXIS |
| 30 | 0.700 | 1000 | 30 | 180 | 2.0 | 1.0 | 2.00 | 2.00 | 125 | 22680 | 0.10000V | -0.06670581 | UNIFORM P-AXIS |
| 31 | 0.800 | 1000 | 30 | 180 | 2.0 | 1.0 | 2.00 | 2.00 | 53 | 9720 | 0.30000V | -0.33188771 | UNIFORM P-AXIS |
| 32 | 0.900 | 1000 | 30 | 180 | 2.0 | 1.0 | 2.00 | 2.00 | 91 | 16560 | 0.30000V | -0.07247309 | UNIFORM P-AXIS |
| 33 | 1.000 | 1000 | 30 | 180 | 2.0 | 1.0 | 2.00 | 2.00 | 59 | 10800 | 0.60000V | -0.08461177 | UNIFORM P-AXIS |
| | | | | | | | | Total Function Evaluations: | | 405780 | | | |
| 6 | 0.500 | 1000 | 30 | 60 | 2.0 | 1.0 | 2.00 | 2.00 | 45 | 2760 | 0.85000V | 0.00000000 | UNIFORM P-AXIS |

# F2

F2
Best Fitness = 0 returned by
Probe # 59 at Time Step 22

P59 coordinates:
1  0
2  0
3  0
4  0
5  0
6  0
7  0
8  0
9  0
10  0
11  0
12  0
13  0
14  0
15  0
16  0
17  0
18  0
19  0
20  0
21  0
22  0
23  0
24  0
25  0
26  0
27  0
28  0
29  0
30  0

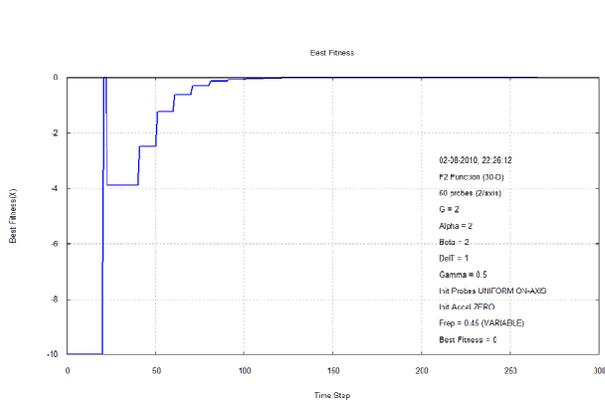

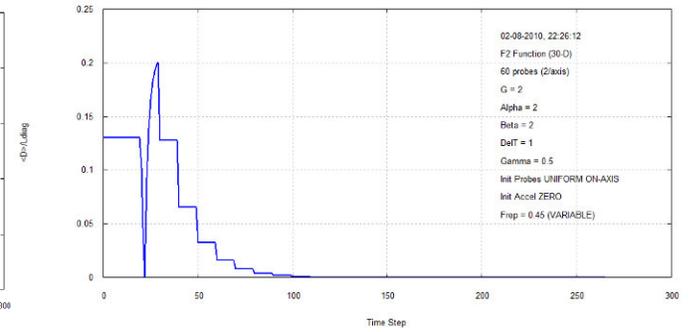

Run ID: 02-08-2010, 22:26:12 FUNCTION: F2

| Run # | Gamma | Nt | Nd | Np | G | DelT | Alpha | Beta | #Steps | Neval | Freq | Fitness | Initial Probes | |
|---|---|---|---|---|---|---|---|---|---|---|---|---|---|---|
| 0 | 0.000 | 1000 | 30 | 60 | 2.0 | 1.0 | 2.00 | 2.00 | 0 | 0 | 0.50000V | -9999.00000000 | UNIFORM | P-AXIS |
| 1 | 0.000 | 1000 | 30 | 60 | 2.0 | 1.0 | 2.00 | 2.00 | 45 | 2760 | 0.85000V | -280.00000000 | UNIFORM | P-AXIS |
| 2 | 0.100 | 1000 | 30 | 60 | 2.0 | 1.0 | 2.00 | 2.00 | 84 | 5100 | 0.90000V | -1.01746022 | UNIFORM | P-AXIS |
| 3 | 0.200 | 1000 | 30 | 60 | 2.0 | 1.0 | 2.00 | 2.00 | 88 | 5340 | 0.15000V | -0.58057928 | UNIFORM | P-AXIS |
| 4 | 0.300 | 1000 | 30 | 60 | 2.0 | 1.0 | 2.00 | 2.00 | 185 | 11160 | 0.25000V | -1.19099232 | UNIFORM | P-AXIS |
| 5 | 0.400 | 1000 | 30 | 60 | 2.0 | 1.0 | 2.00 | 2.00 | 186 | 11220 | 0.30000V | -2.45971267 | UNIFORM | P-AXIS |
| 6 | 0.500 | 1000 | 30 | 60 | 2.0 | 1.0 | 2.00 | 2.00 | 265 | 15960 | 0.45000V | 0.00000000 | UNIFORM | P-AXIS |
| 7 | 0.600 | 1000 | 30 | 60 | 2.0 | 1.0 | 2.00 | 2.00 | 132 | 7980 | 0.45000V | -2.68469072 | UNIFORM | P-AXIS |
| 8 | 0.700 | 1000 | 30 | 60 | 2.0 | 1.0 | 2.00 | 2.00 | 91 | 5520 | 0.30000V | -14.57506483 | UNIFORM | P-AXIS |
| 9 | 0.800 | 1000 | 30 | 60 | 2.0 | 1.0 | 2.00 | 2.00 | 45 | 2760 | 0.85000V | -174.00000000 | UNIFORM | P-AXIS |
| 10 | 0.900 | 1000 | 30 | 60 | 2.0 | 1.0 | 2.00 | 2.00 | 57 | 3480 | 0.50000V | -4.30049209 | UNIFORM | P-AXIS |
| 11 | 1.000 | 1000 | 30 | 60 | 2.0 | 1.0 | 2.00 | 2.00 | 65 | 3960 | 0.90000V | -101.25000000 | UNIFORM | P-AXIS |
| 12 | 0.000 | 1000 | 30 | 120 | 2.0 | 1.0 | 2.00 | 2.00 | 57 | 6960 | 0.50000V | -0.60665707 | UNIFORM | P-AXIS |
| 13 | 0.100 | 1000 | 30 | 120 | 2.0 | 1.0 | 2.00 | 2.00 | 93 | 11280 | 0.40000V | -0.04965977 | UNIFORM | P-AXIS |
| 14 | 0.200 | 1000 | 30 | 120 | 2.0 | 1.0 | 2.00 | 2.00 | 221 | 26640 | 0.15000V | -0.13613108 | UNIFORM | P-AXIS |
| 15 | 0.300 | 1000 | 30 | 120 | 2.0 | 1.0 | 2.00 | 2.00 | 54 | 6600 | 0.35000V | -0.31587227 | UNIFORM | P-AXIS |
| 16 | 0.400 | 1000 | 30 | 120 | 2.0 | 1.0 | 2.00 | 2.00 | 96 | 11640 | 0.55000V | -0.17798453 | UNIFORM | P-AXIS |
| 17 | 0.500 | 1000 | 30 | 120 | 2.0 | 1.0 | 2.00 | 2.00 | 35 | 4320 | 0.35000V | -0.06929315 | UNIFORM | P-AXIS |
| 18 | 0.600 | 1000 | 30 | 120 | 2.0 | 1.0 | 2.00 | 2.00 | 96 | 11640 | 0.55000V | -0.17798614 | UNIFORM | P-AXIS |
| 19 | 0.700 | 1000 | 30 | 120 | 2.0 | 1.0 | 2.00 | 2.00 | 53 | 6480 | 0.30000V | -0.66502585 | UNIFORM | P-AXIS |
| 20 | 0.800 | 1000 | 30 | 120 | 2.0 | 1.0 | 2.00 | 2.00 | 129 | 15600 | 0.30000V | -0.07154138 | UNIFORM | P-AXIS |
| 21 | 0.900 | 1000 | 30 | 120 | 2.0 | 1.0 | 2.00 | 2.00 | 86 | 10440 | 0.35000V | -0.11070850 | UNIFORM | P-AXIS |
| 22 | 1.000 | 1000 | 30 | 120 | 2.0 | 1.0 | 2.00 | 2.00 | 77 | 9360 | 0.55000V | -0.07847279 | UNIFORM | P-AXIS |
| 23 | 0.000 | 1000 | 30 | 180 | 2.0 | 1.0 | 2.00 | 2.00 | 35 | 6480 | 0.35000V | -0.48245739 | UNIFORM | P-AXIS |
| 24 | 0.100 | 1000 | 30 | 180 | 2.0 | 1.0 | 2.00 | 2.00 | 89 | 16200 | 0.20000V | -0.16951650 | UNIFORM | P-AXIS |
| 25 | 0.200 | 1000 | 30 | 180 | 2.0 | 1.0 | 2.00 | 2.00 | 118 | 21420 | 0.70000V | -0.00452866 | UNIFORM | P-AXIS |
| 26 | 0.300 | 1000 | 30 | 180 | 2.0 | 1.0 | 2.00 | 2.00 | 56 | 10260 | 0.45000V | -0.08109305 | UNIFORM | P-AXIS |
| 27 | 0.400 | 1000 | 30 | 180 | 2.0 | 1.0 | 2.00 | 2.00 | 53 | 9720 | 0.30000V | -0.16869854 | UNIFORM | P-AXIS |
| 28 | 0.500 | 1000 | 30 | 180 | 2.0 | 1.0 | 2.00 | 2.00 | 35 | 6480 | 0.35000V | -0.03716355 | UNIFORM | P-AXIS |
| 29 | 0.600 | 1000 | 30 | 180 | 2.0 | 1.0 | 2.00 | 2.00 | 53 | 9720 | 0.30000V | -0.16869854 | UNIFORM | P-AXIS |
| 30 | 0.700 | 1000 | 30 | 180 | 2.0 | 1.0 | 2.00 | 2.00 | 56 | 10260 | 0.45000V | -0.08109305 | UNIFORM | P-AXIS |
| 31 | 0.800 | 1000 | 30 | 180 | 2.0 | 1.0 | 2.00 | 2.00 | 118 | 21420 | 0.70000V | -0.00452866 | UNIFORM | P-AXIS |
| 32 | 0.900 | 1000 | 30 | 180 | 2.0 | 1.0 | 2.00 | 2.00 | 86 | 15660 | 0.05000V | -0.17709046 | UNIFORM | P-AXIS |
| 33 | 1.000 | 1000 | 30 | 180 | 2.0 | 1.0 | 2.00 | 2.00 | 35 | 6480 | 0.35000V | -0.46778102 | UNIFORM | P-AXIS |

Total Function Evaluations: 330300

| 6 | 0.500 | 1000 | 30 | 60 | 2.0 | 1.0 | 2.00 | 2.00 | 265 | 15960 | 0.45000V | 0.00000000 | UNIFORM | P-AXIS |

# F3

F3
Best Fitness = -.00003857 returned by
Probe # 42 at Time Step 72

P42 coordinates:
1  0
2  0
3  0
4  0
5  0
6  0
7  0
8  0
9  0
10  0
11  0
12  0
13  0
14  0
15  0
16  0
17  0
18  0
19  0
20  0
21  -.00196387
22  0
23  0
24  0
25  0
26  0
27  0
28  0
29  0
30  0

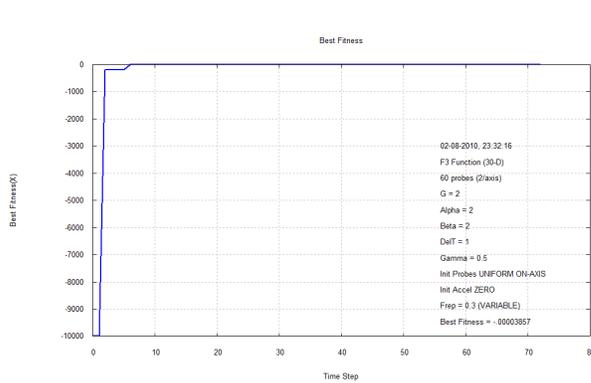

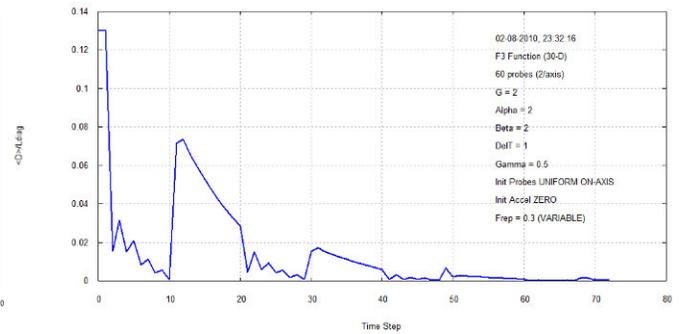

Run ID: 02-08-2010, 23:32:16 FUNCTION: F3

| Run # | Gamma | Nt | Nd | Np | G | DelT | Alpha | Beta | #Steps | Neval | Frep | Fitness | Initial Probes | |
|-------|-------|------|-----|-----|-----|------|-------|------|--------|-------|---------|--------------|---------|--------|
| 0 | 0.000 | 1000 | 30 | 60 | 2.0 | 1.0 | 2.00 | 2.00 | 0 | 0 | 0.50000V | -9999.0000000 | UNIFORM | P-AXIS |
| 1 | 0.000 | 1000 | 30 | 60 | 2.0 | 1.0 | 2.00 | 2.00 | 244 | 14700 | 0.35000V | -1293.66964377 | UNIFORM | P-AXIS |
| 2 | 0.100 | 1000 | 30 | 60 | 2.0 | 1.0 | 2.00 | 2.00 | 213 | 12840 | 0.70000V | -105.27390652 | UNIFORM | P-AXIS |
| 3 | 0.200 | 1000 | 30 | 60 | 2.0 | 1.0 | 2.00 | 2.00 | 246 | 14820 | 0.45000V | -63.28766254 | UNIFORM | P-AXIS |
| 4 | 0.300 | 1000 | 30 | 60 | 2.0 | 1.0 | 2.00 | 2.00 | 244 | 14700 | 0.35000V | -104.75646876 | UNIFORM | P-AXIS |
| 5 | 0.400 | 1000 | 30 | 60 | 2.0 | 1.0 | 2.00 | 2.00 | 224 | 13500 | 0.30000V | -92.05669991 | UNIFORM | P-AXIS |
| 6 | 0.500 | 1000 | 30 | 60 | 2.0 | 1.0 | 2.00 | 2.00 | 72 | 4380 | 0.30000V | -0.00003857 | UNIFORM | P-AXIS |
| 7 | 0.600 | 1000 | 30 | 60 | 2.0 | 1.0 | 2.00 | 2.00 | 260 | 15660 | 0.20000V | -93.54609885 | UNIFORM | P-AXIS |
| 8 | 0.700 | 1000 | 30 | 60 | 2.0 | 1.0 | 2.00 | 2.00 | 217 | 13080 | 0.90000V | -103.79458118 | UNIFORM | P-AXIS |
| 9 | 0.800 | 1000 | 30 | 60 | 2.0 | 1.0 | 2.00 | 2.00 | 225 | 13560 | 0.35000V | -63.30082093 | UNIFORM | P-AXIS |
| 10 | 0.900 | 1000 | 30 | 60 | 2.0 | 1.0 | 2.00 | 2.00 | 188 | 11340 | 0.40000V | -105.26164280 | UNIFORM | P-AXIS |
| 11 | 1.000 | 1000 | 30 | 60 | 2.0 | 1.0 | 2.00 | 2.00 | 249 | 15000 | 0.60000V | -1293.49476269 | UNIFORM | P-AXIS |
| 12 | 0.000 | 1000 | 30 | 120 | 2.0 | 1.0 | 2.00 | 2.00 | 224 | 27000 | 0.35000V | -109.23539687 | UNIFORM | P-AXIS |
| 13 | 0.100 | 1000 | 30 | 120 | 2.0 | 1.0 | 2.00 | 2.00 | 244 | 29400 | 0.35000V | -8.96993170 | UNIFORM | P-AXIS |
| 14 | 0.200 | 1000 | 30 | 120 | 2.0 | 1.0 | 2.00 | 2.00 | 206 | 24840 | 0.35000V | -72.07925119 | UNIFORM | P-AXIS |
| 15 | 0.300 | 1000 | 30 | 120 | 2.0 | 1.0 | 2.00 | 2.00 | 243 | 29280 | 0.35000V | -160.62007411 | UNIFORM | P-AXIS |
| 16 | 0.400 | 1000 | 30 | 120 | 2.0 | 1.0 | 2.00 | 2.00 | 215 | 25920 | 0.80000V | -82.62278254 | UNIFORM | P-AXIS |
| 17 | 0.500 | 1000 | 30 | 120 | 2.0 | 1.0 | 2.00 | 2.00 | 204 | 24600 | 0.25000V | -1.54061661 | UNIFORM | P-AXIS |
| 18 | 0.600 | 1000 | 30 | 120 | 2.0 | 1.0 | 2.00 | 2.00 | 219 | 26400 | 0.15000V | -82.65929106 | UNIFORM | P-AXIS |
| 19 | 0.700 | 1000 | 30 | 120 | 2.0 | 1.0 | 2.00 | 2.00 | 222 | 26760 | 0.20000V | -160.23211184 | UNIFORM | P-AXIS |
| 20 | 0.800 | 1000 | 30 | 120 | 2.0 | 1.0 | 2.00 | 2.00 | 202 | 24360 | 0.15000V | -72.03379381 | UNIFORM | P-AXIS |
| 21 | 0.900 | 1000 | 30 | 120 | 2.0 | 1.0 | 2.00 | 2.00 | 240 | 28920 | 0.15000V | -9.09313070 | UNIFORM | P-AXIS |
| 22 | 1.000 | 1000 | 30 | 120 | 2.0 | 1.0 | 2.00 | 2.00 | 231 | 27840 | 0.65000V | -111.44127606 | UNIFORM | P-AXIS |
| 23 | 0.000 | 1000 | 30 | 180 | 2.0 | 1.0 | 2.00 | 2.00 | 225 | 40680 | 0.35000V | -29.47040396 | UNIFORM | P-AXIS |
| 24 | 0.100 | 1000 | 30 | 180 | 2.0 | 1.0 | 2.00 | 2.00 | 231 | 41760 | 0.65000V | -10.88996770 | UNIFORM | P-AXIS |
| 25 | 0.200 | 1000 | 30 | 180 | 2.0 | 1.0 | 2.00 | 2.00 | 187 | 33840 | 0.35000V | -8.82879286 | UNIFORM | P-AXIS |
| 26 | 0.300 | 1000 | 30 | 180 | 2.0 | 1.0 | 2.00 | 2.00 | 225 | 40680 | 0.35000V | -152.23710641 | UNIFORM | P-AXIS |
| 27 | 0.400 | 1000 | 30 | 180 | 2.0 | 1.0 | 2.00 | 2.00 | 223 | 40320 | 0.25000V | -64.42608877 | UNIFORM | P-AXIS |
| 28 | 0.500 | 1000 | 30 | 180 | 2.0 | 1.0 | 2.00 | 2.00 | 196 | 35460 | 0.80000V | -0.27388078 | UNIFORM | P-AXIS |
| 29 | 0.600 | 1000 | 30 | 180 | 2.0 | 1.0 | 2.00 | 2.00 | 216 | 39060 | 0.85000V | -66.94232809 | UNIFORM | P-AXIS |
| 30 | 0.700 | 1000 | 30 | 180 | 2.0 | 1.0 | 2.00 | 2.00 | 207 | 37440 | 0.40000V | -150.72135845 | UNIFORM | P-AXIS |
| 31 | 0.800 | 1000 | 30 | 180 | 2.0 | 1.0 | 2.00 | 2.00 | 191 | 34560 | 0.55000V | -8.57923340 | UNIFORM | P-AXIS |
| 32 | 0.900 | 1000 | 30 | 180 | 2.0 | 1.0 | 2.00 | 2.00 | 206 | 37260 | 0.35000V | -9.08362761 | UNIFORM | P-AXIS |
| 33 | 1.000 | 1000 | 30 | 180 | 2.0 | 1.0 | 2.00 | 2.00 | 220 | 39780 | 0.10000V | -80.62274330 | UNIFORM | P-AXIS |

Total Function Evaluations:  859740

| 6 | 0.500 | 1000 | 30 | 60 | 2.0 | 1.0 | 2.00 | 2.00 | 72 | 4380 | 0.30000V | -0.00003857 | UNIFORM | P-AXIS |

## F4

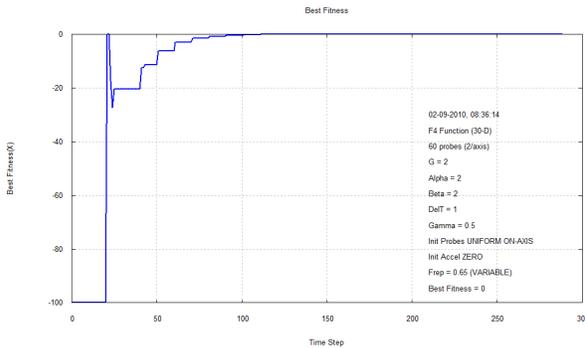

F4
Best Fitness = 0 returned by
Probe # 59 at Time Step 22

P59 coordinates:
1 0
2 0
3 0
4 0
5 0
6 0
7 0
8 0
9 0
10 0
11 0
12 0
13 0
14 0
15 0
16 0
17 0
18 0
19 0
20 0
21 0
22 0
23 0
24 0
25 0
26 0
27 0
28 0
29 0
30 0

02-09-2010, 08:36:14
F4 Function (30-D)
60 probes (2/axis)
G = 2
Alpha = 2
Beta = 2
DelT = 1
Init Probes UNIFORM ON-AXIS
Init Accel ZERO
Freq = 0.65 (VARIABLE)
Best Fitness = 0

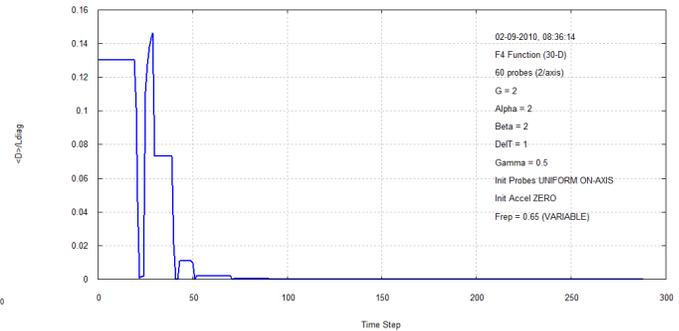

Average Distance of 59 Probes to Best Probe
Normalized to Size of Decision Space
[60 probes, 288 time steps]

02-09, 2010, 08:36:14
F4 Function (30-D)
60 probes (2/axis)
G = 2
Alpha = 2
Beta = 2
DelT = 1
Gamma = 0.5
Init Probes UNIFORM ON-AXIS
Init Accel ZERO
Freq = 0.65 (VARIABLE)

Run ID: 02-09-2010, 08:36:14 FUNCTION: F4

| Run # | Gamma | Nt | Nd | Np | G | DelT | Alpha | Beta | #Steps | Neval | Frep | Fitness | Initial Probes | |
|-------|-------|------|-----|-----|-----|------|-------|------|--------|-------|---------|--------------|---------|--------|
| 0 | 0.000 | 1000 | 30 | 60 | 2.0 | 1.0 | 2.00 | 2.00 | 35 | 2160 | 0.50000V | -9999.0000000 | UNIFORM | P-AXIS |
| 1 | 0.000 | 1000 | 30 | 60 | 2.0 | 1.0 | 2.00 | 2.00 | 35 | 2160 | 0.35000V | -100.00000000 | UNIFORM | P-AXIS |
| 2 | 0.100 | 1000 | 30 | 60 | 2.0 | 1.0 | 2.00 | 2.00 | 46 | 2820 | 0.90000V | -78.11005962 | UNIFORM | P-AXIS |
| 3 | 0.200 | 1000 | 30 | 60 | 2.0 | 1.0 | 2.00 | 2.00 | 47 | 2880 | 0.95000V | -53.13273282 | UNIFORM | P-AXIS |
| 4 | 0.300 | 1000 | 30 | 60 | 2.0 | 1.0 | 2.00 | 2.00 | 65 | 3960 | 0.90000V | -25.08520680 | UNIFORM | P-AXIS |
| 5 | 0.400 | 1000 | 30 | 60 | 2.0 | 1.0 | 2.00 | 2.00 | 49 | 3000 | 0.10000V | -17.71850321 | UNIFORM | P-AXIS |
| 6 | 0.500 | 1000 | 30 | 60 | 2.0 | 1.0 | 2.00 | 2.00 | 288 | 17340 | 0.65000V | 0.00000000 | UNIFORM | P-AXIS |
| 7 | 0.600 | 1000 | 30 | 60 | 2.0 | 1.0 | 2.00 | 2.00 | 45 | 2760 | 0.85000V | -20.00000000 | UNIFORM | P-AXIS |
| 8 | 0.700 | 1000 | 30 | 60 | 2.0 | 1.0 | 2.00 | 2.00 | 48 | 2940 | 0.95000V | -23.15228068 | UNIFORM | P-AXIS |
| 9 | 0.800 | 1000 | 30 | 60 | 2.0 | 1.0 | 2.00 | 2.00 | 46 | 2820 | 0.90000V | -55.29411765 | UNIFORM | P-AXIS |
| 10 | 0.900 | 1000 | 30 | 60 | 2.0 | 1.0 | 2.00 | 2.00 | 46 | 2820 | 0.90000V | -77.76470588 | UNIFORM | P-AXIS |
| 11 | 1.000 | 1000 | 30 | 60 | 2.0 | 1.0 | 2.00 | 2.00 | 50 | 3060 | 0.35000V | -10.00000000 | UNIFORM | P-AXIS |
| 12 | 0.000 | 1000 | 30 | 120 | 2.0 | 1.0 | 2.00 | 2.00 | 35 | 4320 | 0.35000V | -100.00000000 | UNIFORM | P-AXIS |
| 13 | 0.100 | 1000 | 30 | 120 | 2.0 | 1.0 | 2.00 | 2.00 | 35 | 4320 | 0.35000V | -69.33580190 | UNIFORM | P-AXIS |
| 14 | 0.200 | 1000 | 30 | 120 | 2.0 | 1.0 | 2.00 | 2.00 | 35 | 4320 | 0.35000V | -27.05570292 | UNIFORM | P-AXIS |
| 15 | 0.300 | 1000 | 30 | 120 | 2.0 | 1.0 | 2.00 | 2.00 | 50 | 6120 | 0.15000V | -5.57949024 | UNIFORM | P-AXIS |
| 16 | 0.400 | 1000 | 30 | 120 | 2.0 | 1.0 | 2.00 | 2.00 | 35 | 4320 | 0.35000V | -7.97580490 | UNIFORM | P-AXIS |
| 17 | 0.500 | 1000 | 30 | 120 | 2.0 | 1.0 | 2.00 | 2.00 | 35 | 4320 | 0.35000V | -0.69293152 | UNIFORM | P-AXIS |
| 18 | 0.600 | 1000 | 30 | 120 | 2.0 | 1.0 | 2.00 | 2.00 | 35 | 4320 | 0.35000V | -7.97580490 | UNIFORM | P-AXIS |
| 19 | 0.700 | 1000 | 30 | 120 | 2.0 | 1.0 | 2.00 | 2.00 | 50 | 6120 | 0.15000V | -5.57949024 | UNIFORM | P-AXIS |
| 20 | 0.800 | 1000 | 30 | 120 | 2.0 | 1.0 | 2.00 | 2.00 | 35 | 4320 | 0.35000V | -27.05570292 | UNIFORM | P-AXIS |
| 21 | 0.900 | 1000 | 30 | 120 | 2.0 | 1.0 | 2.00 | 2.00 | 35 | 4320 | 0.35000V | -69.33580190 | UNIFORM | P-AXIS |
| 22 | 1.000 | 1000 | 30 | 120 | 2.0 | 1.0 | 2.00 | 2.00 | 35 | 4320 | 0.35000V | -4.11445402 | UNIFORM | P-AXIS |
| 23 | 0.000 | 1000 | 30 | 180 | 2.0 | 1.0 | 2.00 | 2.00 | 35 | 6480 | 0.35000V | -100.00000000 | UNIFORM | P-AXIS |
| 24 | 0.100 | 1000 | 30 | 180 | 2.0 | 1.0 | 2.00 | 2.00 | 35 | 6480 | 0.35000V | -53.09337766 | UNIFORM | P-AXIS |
| 25 | 0.200 | 1000 | 30 | 180 | 2.0 | 1.0 | 2.00 | 2.00 | 50 | 9180 | 0.15000V | -2.16245242 | UNIFORM | P-AXIS |
| 26 | 0.300 | 1000 | 30 | 180 | 2.0 | 1.0 | 2.00 | 2.00 | 49 | 9000 | 0.10000V | -2.71235926 | UNIFORM | P-AXIS |
| 27 | 0.400 | 1000 | 30 | 180 | 2.0 | 1.0 | 2.00 | 2.00 | 35 | 6480 | 0.35000V | -0.83097051 | UNIFORM | P-AXIS |
| 28 | 0.500 | 1000 | 30 | 180 | 2.0 | 1.0 | 2.00 | 2.00 | 35 | 6480 | 0.35000V | -0.37144800 | UNIFORM | P-AXIS |
| 29 | 0.600 | 1000 | 30 | 180 | 2.0 | 1.0 | 2.00 | 2.00 | 35 | 6480 | 0.35000V | -0.73078479 | UNIFORM | P-AXIS |
| 30 | 0.700 | 1000 | 30 | 180 | 2.0 | 1.0 | 2.00 | 2.00 | 49 | 9000 | 0.10000V | -2.71235926 | UNIFORM | P-AXIS |
| 31 | 0.800 | 1000 | 30 | 180 | 2.0 | 1.0 | 2.00 | 2.00 | 48 | 8820 | 0.05000V | -0.67259144 | UNIFORM | P-AXIS |
| 32 | 0.900 | 1000 | 30 | 180 | 2.0 | 1.0 | 2.00 | 2.00 | 35 | 6480 | 0.35000V | -53.09339766 | UNIFORM | P-AXIS |
| 33 | 1.000 | 1000 | 30 | 180 | 2.0 | 1.0 | 2.00 | 2.00 | 35 | 6480 | 0.35000V | -1.87088294 | UNIFORM | P-AXIS |

Total Function Evaluations: 178140

| | | | | | | | | | | | | | | |
|---|---|---|---|---|---|---|---|---|---|---|---|---|---|---|
| 6 | 0.500 | 1000 | 30 | 60 | 2.0 | 1.0 | 2.00 | 2.00 | 288 | 17340 | 0.65000V | 0.00000000 | UNIFORM | P-AXIS |

# F5

F5
Best Fitness = -.00205081 returned by
Probe # 3 at Time Step 84

P3 coordinates:
1  1.00368807
2  1.00528707
3  1.00528319
4  1.00528319
5  1.00528319
6  1.00528319
7  1.00528319
8  1.00528319
9  1.00528319
10  1.00528319
11  1.00528319
12  1.00528319
13  1.00528319
14  1.00528319
15  1.00528319
16  1.00528319
17  1.00528319
18  1.00528319
19  1.00528319
20  1.00528319
21  1.00528319
22  1.00528319
23  1.00528319
24  1.00528319
25  1.00528319
26  1.00528319
27  1.00528319
28  1.00528319
29  1.00533606
30  1.02610996

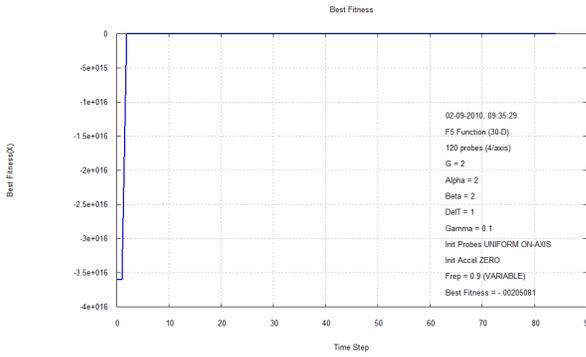

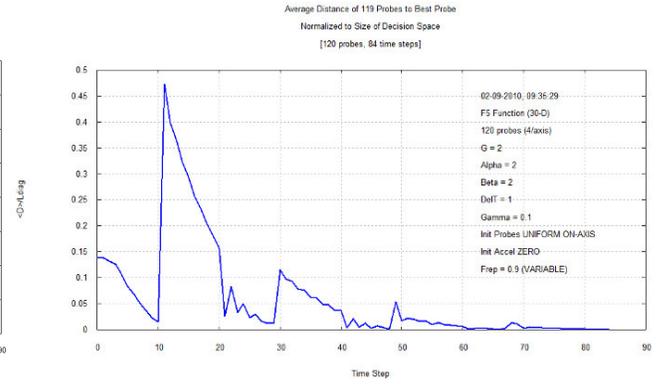

Run ID: 02-09-2010, 09:35:29 FUNCTION: F5

| Run # | Gamma | Nt | Nd | Np | G | DelT | Alpha | Beta | #Steps | Neval | Freq | Fitness | Initial Probes | |
|---|---|---|---|---|---|---|---|---|---|---|---|---|---|---|
| 0 | 0.000 | 1000 | 30 | 60 | | | | | 0 | 0 | 0.50000V | -9999.00000000 | | P-AXIS |
| 1 | 0.000 | 1000 | 30 | 60 | 2.0 | 1.0 | 2.00 | 2.00 | 176 | 10620 | 0.75000V | -0.01661844 | UNIFORM | P-AXIS |
| 2 | 0.100 | 1000 | 30 | 60 | 2.0 | 1.0 | 2.00 | 2.00 | 215 | 12960 | 0.80000V | -16.11043077 | UNIFORM | P-AXIS |
| 3 | 0.200 | 1000 | 30 | 60 | 2.0 | 1.0 | 2.00 | 2.00 | 35 | 2160 | 0.45000V | -7.35194119 | UNIFORM | P-AXIS |
| 4 | 0.300 | 1000 | 30 | 60 | 2.0 | 1.0 | 2.00 | 2.00 | 136 | 8220 | 0.65000V | -0.00589828 | UNIFORM | P-AXIS |
| 5 | 0.400 | 1000 | 30 | 60 | 2.0 | 1.0 | 2.00 | 2.00 | 100 | 6060 | 0.75000V | -76.16421846 | UNIFORM | P-AXIS |
| 6 | 0.500 | 1000 | 30 | 60 | 2.0 | 1.0 | 2.00 | 2.00 | 53 | 3240 | 0.30000V | -1940.35684876 | UNIFORM | P-AXIS |
| 7 | 0.600 | 1000 | 30 | 60 | 2.0 | 1.0 | 2.00 | 2.00 | 245 | 14760 | 0.40000V | -9477.17422201 | UNIFORM | P-AXIS |
| 8 | 0.700 | 1000 | 30 | 60 | 2.0 | 1.0 | 2.00 | 2.00 | 238 | 14340 | 0.05000V | -0.97607953 | UNIFORM | P-AXIS |
| 9 | 0.800 | 1000 | 30 | 60 | 2.0 | 1.0 | 2.00 | 2.00 | 264 | 15900 | 0.40000V | -8.29704934 | UNIFORM | P-AXIS |
| 10 | 0.900 | 1000 | 30 | 60 | 2.0 | 1.0 | 2.00 | 2.00 | 72 | 4380 | 0.30000V | -6.32865620 | UNIFORM | P-AXIS |
| 11 | 1.000 | 1000 | 30 | 60 | 2.0 | 1.0 | 2.00 | 2.00 | 35 | 2160 | 0.35000V | -44159.15850962 | UNIFORM | P-AXIS |
| 12 | 0.000 | 1000 | 30 | 120 | 2.0 | 1.0 | 2.00 | 2.00 | 243 | 29280 | 0.30000V | -0.06075909 | UNIFORM | P-AXIS |
| 13 | 0.100 | 1000 | 30 | 120 | 2.0 | 1.0 | 2.00 | 2.00 | 84 | 10200 | 0.90000V | -0.00205081 | UNIFORM | P-AXIS |
| 14 | 0.200 | 1000 | 30 | 120 | 2.0 | 1.0 | 2.00 | 2.00 | 156 | 18840 | 0.70000V | -0.00340017 | UNIFORM | P-AXIS |
| 15 | 0.300 | 1000 | 30 | 120 | 2.0 | 1.0 | 2.00 | 2.00 | 164 | 19800 | 0.15000V | -0.00330399 | UNIFORM | P-AXIS |
| 16 | 0.400 | 1000 | 30 | 120 | 2.0 | 1.0 | 2.00 | 2.00 | 208 | 25080 | 0.45000V | -0.72254709 | UNIFORM | P-AXIS |
| 17 | 0.500 | 1000 | 30 | 120 | 2.0 | 1.0 | 2.00 | 2.00 | 70 | 8520 | 0.20000V | -2.58493324 | UNIFORM | P-AXIS |
| 18 | 0.600 | 1000 | 30 | 120 | 2.0 | 1.0 | 2.00 | 2.00 | 244 | 29400 | 0.35000V | -0.70122336 | UNIFORM | P-AXIS |
| 19 | 0.700 | 1000 | 30 | 120 | 2.0 | 1.0 | 2.00 | 2.00 | 101 | 12240 | 0.80000V | -0.50169816 | UNIFORM | P-AXIS |
| 20 | 0.800 | 1000 | 30 | 120 | 2.0 | 1.0 | 2.00 | 2.00 | 78 | 9480 | 0.60000V | -0.22741523 | UNIFORM | P-AXIS |
| 21 | 0.900 | 1000 | 30 | 120 | 2.0 | 1.0 | 2.00 | 2.00 | 54 | 6600 | 0.35000V | -13.82601069 | UNIFORM | P-AXIS |
| 22 | 1.000 | 1000 | 30 | 120 | 2.0 | 1.0 | 2.00 | 2.00 | 35 | 4320 | 0.35000V | -45608.43461552 | UNIFORM | P-AXIS |
| 23 | 0.000 | 1000 | 30 | 180 | 2.0 | 1.0 | 2.00 | 2.00 | 72 | 13140 | 0.30000V | -1.81935832 | UNIFORM | P-AXIS |
| 24 | 0.100 | 1000 | 30 | 180 | 2.0 | 1.0 | 2.00 | 2.00 | 53 | 9720 | 0.30000V | -0.86881815 | UNIFORM | P-AXIS |
| 25 | 0.200 | 1000 | 30 | 180 | 2.0 | 1.0 | 2.00 | 2.00 | 53 | 9720 | 0.30000V | -18.44465545 | UNIFORM | P-AXIS |
| 26 | 0.300 | 1000 | 30 | 180 | 2.0 | 1.0 | 2.00 | 2.00 | 54 | 9900 | 0.35000V | -1.47748671 | UNIFORM | P-AXIS |
| 27 | 0.400 | 1000 | 30 | 180 | 2.0 | 1.0 | 2.00 | 2.00 | 92 | 16740 | 0.35000V | -1.54750724 | UNIFORM | P-AXIS |
| 28 | 0.500 | 1000 | 30 | 180 | 2.0 | 1.0 | 2.00 | 2.00 | 75 | 13680 | 0.45000V | -0.39827198 | UNIFORM | P-AXIS |
| 29 | 0.600 | 1000 | 30 | 180 | 2.0 | 1.0 | 2.00 | 2.00 | 241 | 43560 | 0.20000V | -9.56781290 | UNIFORM | P-AXIS |
| 30 | 0.700 | 1000 | 30 | 180 | 2.0 | 1.0 | 2.00 | 2.00 | 198 | 35820 | 0.90000V | -0.32453008 | UNIFORM | P-AXIS |
| 31 | 0.800 | 1000 | 30 | 180 | 2.0 | 1.0 | 2.00 | 2.00 | 52 | 9540 | 0.25000V | -2.50318155 | UNIFORM | P-AXIS |
| 32 | 0.900 | 1000 | 30 | 180 | 2.0 | 1.0 | 2.00 | 2.00 | 91 | 16560 | 0.30000V | -0.00229565 | UNIFORM | P-AXIS |
| 33 | 1.000 | 1000 | 30 | 180 | 2.0 | 1.0 | 2.00 | 2.00 | 169 | 30600 | 0.40000V | -26.42668352 | UNIFORM | P-AXIS |

Total Function Evaluations: 477540

| | | | | | | | | | | | | | | |
|---|---|---|---|---|---|---|---|---|---|---|---|---|---|---|
| 13 | 0.100 | 1000 | 30 | 120 | 2.0 | 1.0 | 2.00 | 2.00 | 84 | 10200 | 0.90000V | -0.00205081 | UNIFORM | P-AXIS |

# F6

F6
Best Fitness = 0 returned by
Probe #375 at Time Step 35

P375 coordinates:
1 -.44
2 -.44
3 -.44
4 -.44
5 -.44
6 -.44
7 -.44
8 -.44
9 -.44
10 -.44
11 -.44
12 -.44
13 -.44
14 -.44
15 -.44
16 -.44
17 -.44
18 -.44
19 -.44
20 -.44
21 -.44
22 -.44
23 -.44
24 -.44
25 -.44
26 -.44
27 -.44
28 -.44
29 -.44
30 -.44

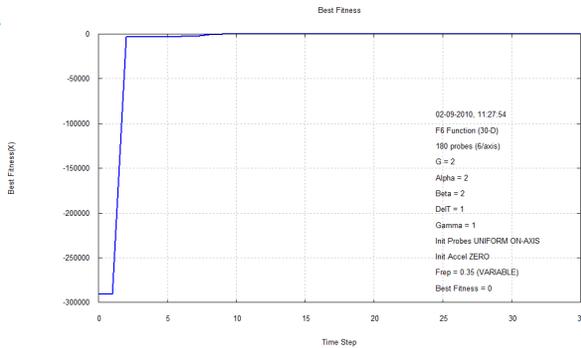

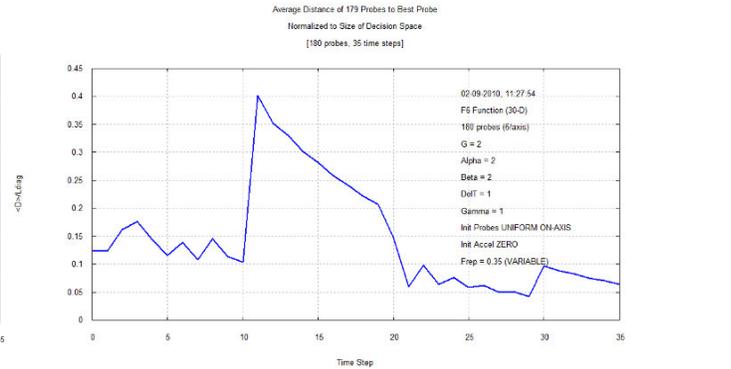

Run ID: 02-09-2010, 11:27:54 FUNCTION: F6

| Run # | Gamma | Nt | Nd | Np | G | DelT | Alpha | Beta | #Steps | Neval | Freq | Fitness | Initial | Probes |
|-------|-------|------|----|-----|-----|-----|------|------|--------|-------|----------|-------------------|---------|--------|
| 0 | 0.000 | 1000 | 30 | 60 | 2.0 | 1.0 | 2.00 | 2.00 | 0 | 0 | 0.50000V | -9999.00000000 | UNIFORM | P-AXIS |
| 1 | 0.000 | 1000 | 30 | 60 | 2.0 | 1.0 | 2.00 | 2.00 | 57 | 3480 | 0.50000V | -30.00000000 | UNIFORM | P-AXIS |
| 2 | 0.100 | 1000 | 30 | 60 | 2.0 | 1.0 | 2.00 | 2.00 | 58 | 3540 | 0.55000V | -291.00000000 | UNIFORM | P-AXIS |
| 3 | 0.200 | 1000 | 30 | 60 | 2.0 | 1.0 | 2.00 | 2.00 | 78 | 4740 | 0.60000V | -196.00000000 | UNIFORM | P-AXIS |
| 4 | 0.300 | 1000 | 30 | 60 | 2.0 | 1.0 | 2.00 | 2.00 | 48 | 2940 | 0.65000V | -29.00000000 | UNIFORM | P-AXIS |
| 5 | 0.400 | 1000 | 30 | 60 | 2.0 | 1.0 | 2.00 | 2.00 | 85 | 5160 | 0.95000V | -442.00000000 | UNIFORM | P-AXIS |
| 6 | 0.500 | 1000 | 30 | 60 | 2.0 | 1.0 | 2.00 | 2.00 | 45 | 2760 | 0.85000V | 0.00000000 | UNIFORM | P-AXIS |
| 7 | 0.600 | 1000 | 30 | 60 | 2.0 | 1.0 | 2.00 | 2.00 | 71 | 4320 | 0.25000V | -484.00000000 | UNIFORM | P-AXIS |
| 8 | 0.700 | 1000 | 30 | 60 | 2.0 | 1.0 | 2.00 | 2.00 | 50 | 3060 | 0.15000V | -569.00000000 | UNIFORM | P-AXIS |
| 9 | 0.800 | 1000 | 30 | 60 | 2.0 | 1.0 | 2.00 | 2.00 | 73 | 4440 | 0.35000V | -1486.00000000 | UNIFORM | P-AXIS |
| 10 | 0.900 | 1000 | 30 | 60 | 2.0 | 1.0 | 2.00 | 2.00 | 68 | 4140 | 0.10000V | -100.00000000 | UNIFORM | P-AXIS |
| 11 | 1.000 | 1000 | 30 | 60 | 2.0 | 1.0 | 2.00 | 2.00 | 65 | 3960 | 0.90000V | -480.00000000 | UNIFORM | P-AXIS |
| 12 | 0.000 | 1000 | 30 | 120 | 2.0 | 1.0 | 2.00 | 2.00 | 71 | 8640 | 0.25000V | -2.00000000 | UNIFORM | P-AXIS |
| 13 | 0.100 | 1000 | 30 | 120 | 2.0 | 1.0 | 2.00 | 2.00 | 65 | 7920 | 0.90000V | 0.00000000 | UNIFORM | P-AXIS |
| 14 | 0.200 | 1000 | 30 | 120 | 2.0 | 1.0 | 2.00 | 2.00 | 71 | 8640 | 0.25000V | 0.00000000 | UNIFORM | P-AXIS |
| 15 | 0.300 | 1000 | 30 | 120 | 2.0 | 1.0 | 2.00 | 2.00 | 71 | 8640 | 0.25000V | 0.00000000 | UNIFORM | P-AXIS |
| 16 | 0.400 | 1000 | 30 | 120 | 2.0 | 1.0 | 2.00 | 2.00 | 54 | 6600 | 0.35000V | 0.00000000 | UNIFORM | P-AXIS |
| 17 | 0.500 | 1000 | 30 | 120 | 2.0 | 1.0 | 2.00 | 2.00 | 53 | 6480 | 0.30000V | 0.00000000 | UNIFORM | P-AXIS |
| 18 | 0.600 | 1000 | 30 | 120 | 2.0 | 1.0 | 2.00 | 2.00 | 54 | 6600 | 0.35000V | 0.00000000 | UNIFORM | P-AXIS |
| 19 | 0.700 | 1000 | 30 | 120 | 2.0 | 1.0 | 2.00 | 2.00 | 53 | 6480 | 0.30000V | 0.00000000 | UNIFORM | P-AXIS |
| 20 | 0.800 | 1000 | 30 | 120 | 2.0 | 1.0 | 2.00 | 2.00 | 53 | 6480 | 0.30000V | -1.00000000 | UNIFORM | P-AXIS |
| 21 | 0.900 | 1000 | 30 | 120 | 2.0 | 1.0 | 2.00 | 2.00 | 54 | 6600 | 0.35000V | -1.00000000 | UNIFORM | P-AXIS |
| 22 | 1.000 | 1000 | 30 | 120 | 2.0 | 1.0 | 2.00 | 2.00 | 71 | 8640 | 0.25000V | -2.00000000 | UNIFORM | P-AXIS |
| 23 | 0.000 | 1000 | 30 | 180 | 2.0 | 1.0 | 2.00 | 2.00 | 35 | 6480 | 0.35000V | 0.00000000 | UNIFORM | P-AXIS |
| 24 | 0.100 | 1000 | 30 | 180 | 2.0 | 1.0 | 2.00 | 2.00 | 53 | 9720 | 0.30000V | -4.00000000 | UNIFORM | P-AXIS |
| 25 | 0.200 | 1000 | 30 | 180 | 2.0 | 1.0 | 2.00 | 2.00 | 53 | 9720 | 0.30000V | 0.00000000 | UNIFORM | P-AXIS |
| 26 | 0.300 | 1000 | 30 | 180 | 2.0 | 1.0 | 2.00 | 2.00 | 71 | 12960 | 0.25000V | -1.00000000 | UNIFORM | P-AXIS |
| 27 | 0.400 | 1000 | 30 | 180 | 2.0 | 1.0 | 2.00 | 2.00 | 53 | 9720 | 0.30000V | 0.00000000 | UNIFORM | P-AXIS |
| 28 | 0.500 | 1000 | 30 | 180 | 2.0 | 1.0 | 2.00 | 2.00 | 53 | 9720 | 0.30000V | -1.00000000 | UNIFORM | P-AXIS |
| 29 | 0.600 | 1000 | 30 | 180 | 2.0 | 1.0 | 2.00 | 2.00 | 35 | 6480 | 0.35000V | 0.00000000 | UNIFORM | P-AXIS |
| 30 | 0.700 | 1000 | 30 | 180 | 2.0 | 1.0 | 2.00 | 2.00 | 71 | 12960 | 0.25000V | -1.00000000 | UNIFORM | P-AXIS |
| 31 | 0.800 | 1000 | 30 | 180 | 2.0 | 1.0 | 2.00 | 2.00 | 53 | 9720 | 0.30000V | 0.00000000 | UNIFORM | P-AXIS |
| 32 | 0.900 | 1000 | 30 | 180 | 2.0 | 1.0 | 2.00 | 2.00 | 53 | 9720 | 0.30000V | -4.00000000 | UNIFORM | P-AXIS |
| 33 | 1.000 | 1000 | 30 | 180 | 2.0 | 1.0 | 2.00 | 2.00 | 35 | 6480 | 0.35000V | 0.00000000 | UNIFORM | P-AXIS |

Total Function Evaluations: 227940

| 6 | 0.500 | 1000 | 30 | 60 | 2.0 | 1.0 | 2.00 | 2.00 | 45 | 2760 | 0.85000V | 0.00000000 | UNIFORM | P-AXIS |

## F7

F7
Best Fitness = -.00222278 returned
Probe #137 at Time Step 95

P137 coordinates:
1  -.08051977
2  .00552789
3  -.0815247
4  -.06650312
5  -.01037271
6  -.0522252
7  -.25409365
8  .04311234
9  .00496216
10  -.0252135
11  .0551.9664
12  -.02505972
13  .01427634
14  -.08202646
15  -.05066876
16  -.00068353
17  .01009897
18  .01430978
19  -.00710511
20  -.00410024
21  -.00696877
22  -.07231775
23  .01375418
24  .05701515
25  -.01368012
26  -.00575963
27  .0413441
28  -.00122705
29  .05961814
30  .04209653

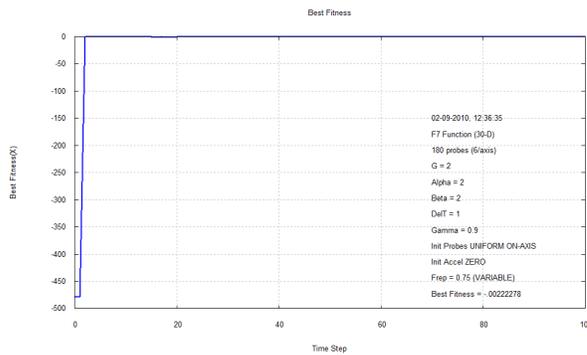

Best Fitness

02-09-2010, 12:36:36
F7 Function (30-D)
100 probes (6 axis)
G = 2
Alpha = 2
Beta = 2
DaT = 1
Gamma = 0.9
Init Probes UNIFORM ON-AXIS
Init Accel ZERO
Freq = 0.75 (VARIABLE)
Best Fitness = -.00222278

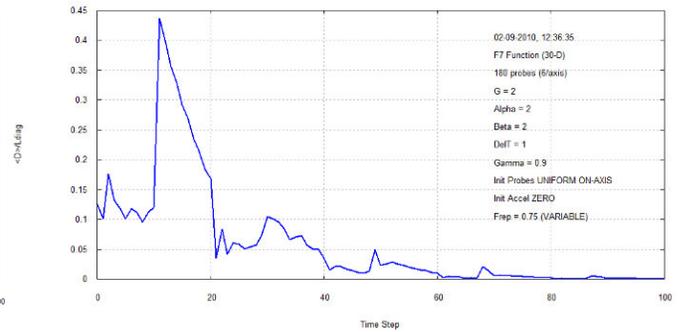

Average Distance of 179 Probes to Best Probe

Normalized to Size of Decision Space

[100 probes, 100 time steps]

02-09-2010, 12:36:36
F7 Function (30-D)
100 probes (6 axis)
G = 2
Alpha = 2
Beta = 2
DaT = 1
Gamma = 0.9
Init Probes UNIFORM ON-AXIS
Init Accel ZERO
Freq = 0.75 (VARIABLE)

Run ID: 02-09-2010, 12:36:35 FUNCTION: F7

| Run # | Gamma | Nt | Nd | Np | G | DelT | Alpha | Beta | #Steps | Neval | Freq | Fitness | Initial | Probes |
|-------|-------|-----|----|-----|-----|-----|------|------|--------|-------|----------|---------------|---------|--------|
| 0 | 0.000 | 100 | 30 | 60 | 2.0 | 1.0 | 2.00 | 2.00 | 0 | 0 | 0.50000V | -9999.00000000 | UNIFORM | P-AXIS |
| 1 | 0.000 | 100 | 30 | 60 | 2.0 | 1.0 | 2.00 | 2.00 | 100 | 6060 | 0.75000V | -0.00499514 | UNIFORM | P-AXIS |
| 2 | 0.100 | 100 | 30 | 60 | 2.0 | 1.0 | 2.00 | 2.00 | 100 | 6060 | 0.75000V | -0.00348562 | UNIFORM | P-AXIS |
| 3 | 0.200 | 100 | 30 | 60 | 2.0 | 1.0 | 2.00 | 2.00 | 100 | 6060 | 0.75000V | -0.00219622 | UNIFORM | P-AXIS |
| 4 | 0.300 | 100 | 30 | 60 | 2.0 | 1.0 | 2.00 | 2.00 | 100 | 6060 | 0.75000V | -0.00378328 | UNIFORM | P-AXIS |
| 5 | 0.400 | 100 | 30 | 60 | 2.0 | 1.0 | 2.00 | 2.00 | 100 | 6060 | 0.75000V | -0.00401040 | UNIFORM | P-AXIS |
| 6 | 0.500 | 100 | 30 | 60 | 2.0 | 1.0 | 2.00 | 2.00 | 100 | 6060 | 0.75000V | -0.00586654 | UNIFORM | P-AXIS |
| 7 | 0.600 | 100 | 30 | 60 | 2.0 | 1.0 | 2.00 | 2.00 | 100 | 6060 | 0.75000V | -0.00694166 | UNIFORM | P-AXIS |
| 8 | 0.700 | 100 | 30 | 60 | 2.0 | 1.0 | 2.00 | 2.00 | 100 | 6060 | 0.75000V | -0.00472027 | UNIFORM | P-AXIS |
| 9 | 0.800 | 100 | 30 | 60 | 2.0 | 1.0 | 2.00 | 2.00 | 100 | 6060 | 0.75000V | -0.00213190 | UNIFORM | P-AXIS |
| 10 | 0.900 | 100 | 30 | 60 | 2.0 | 1.0 | 2.00 | 2.00 | 100 | 6060 | 0.75000V | -0.00260706 | UNIFORM | P-AXIS |
| 11 | 1.000 | 100 | 30 | 60 | 2.0 | 1.0 | 2.00 | 2.00 | 100 | 6060 | 0.75000V | -0.01388331 | UNIFORM | P-AXIS |
| 12 | 0.000 | 100 | 30 | 120 | 2.0 | 1.0 | 2.00 | 2.00 | 100 | 12120 | 0.75000V | -0.00170746 | UNIFORM | P-AXIS |
| 13 | 0.100 | 100 | 30 | 120 | 2.0 | 1.0 | 2.00 | 2.00 | 100 | 12120 | 0.75000V | -0.00192496 | UNIFORM | P-AXIS |
| 14 | 0.200 | 100 | 30 | 120 | 2.0 | 1.0 | 2.00 | 2.00 | 100 | 12120 | 0.75000V | -0.00317438 | UNIFORM | P-AXIS |
| 15 | 0.300 | 100 | 30 | 120 | 2.0 | 1.0 | 2.00 | 2.00 | 100 | 12120 | 0.75000V | -0.00217893 | UNIFORM | P-AXIS |
| 16 | 0.400 | 100 | 30 | 120 | 2.0 | 1.0 | 2.00 | 2.00 | 100 | 12120 | 0.75000V | -0.00122809 | UNIFORM | P-AXIS |
| 17 | 0.500 | 100 | 30 | 120 | 2.0 | 1.0 | 2.00 | 2.00 | 100 | 12120 | 0.75000V | -0.00264547 | UNIFORM | P-AXIS |
| 18 | 0.600 | 100 | 30 | 120 | 2.0 | 1.0 | 2.00 | 2.00 | 100 | 12120 | 0.75000V | -0.00270660 | UNIFORM | P-AXIS |
| 19 | 0.700 | 100 | 30 | 120 | 2.0 | 1.0 | 2.00 | 2.00 | 100 | 12120 | 0.75000V | -0.00194405 | UNIFORM | P-AXIS |
| 20 | 0.800 | 100 | 30 | 120 | 2.0 | 1.0 | 2.00 | 2.00 | 100 | 12120 | 0.75000V | -0.00154875 | UNIFORM | P-AXIS |
| 21 | 0.900 | 100 | 30 | 120 | 2.0 | 1.0 | 2.00 | 2.00 | 100 | 12120 | 0.75000V | -0.00264599 | UNIFORM | P-AXIS |
| 22 | 1.000 | 100 | 30 | 120 | 2.0 | 1.0 | 2.00 | 2.00 | 100 | 12120 | 0.75000V | -0.00095676 | UNIFORM | P-AXIS |
| 23 | 0.000 | 100 | 30 | 180 | 2.0 | 1.0 | 2.00 | 2.00 | 100 | 18180 | 0.75000V | -0.00184171 | UNIFORM | P-AXIS |
| 24 | 0.100 | 100 | 30 | 180 | 2.0 | 1.0 | 2.00 | 2.00 | 100 | 18180 | 0.75000V | -0.00092270 | UNIFORM | P-AXIS |
| 25 | 0.200 | 100 | 30 | 180 | 2.0 | 1.0 | 2.00 | 2.00 | 100 | 18180 | 0.75000V | -0.00134199 | UNIFORM | P-AXIS |
| 26 | 0.300 | 100 | 30 | 180 | 2.0 | 1.0 | 2.00 | 2.00 | 100 | 18180 | 0.75000V | -0.00373414 | UNIFORM | P-AXIS |
| 27 | 0.400 | 100 | 30 | 180 | 2.0 | 1.0 | 2.00 | 2.00 | 100 | 18180 | 0.75000V | -0.00455917 | UNIFORM | P-AXIS |
| 28 | 0.500 | 100 | 30 | 180 | 2.0 | 1.0 | 2.00 | 2.00 | 100 | 18180 | 0.75000V | -0.00290262 | UNIFORM | P-AXIS |
| 29 | 0.600 | 100 | 30 | 180 | 2.0 | 1.0 | 2.00 | 2.00 | 100 | 18180 | 0.75000V | -0.00192635 | UNIFORM | P-AXIS |
| 30 | 0.700 | 100 | 30 | 180 | 2.0 | 1.0 | 2.00 | 2.00 | 100 | 18180 | 0.75000V | -0.00167796 | UNIFORM | P-AXIS |
| 31 | 0.800 | 100 | 30 | 180 | 2.0 | 1.0 | 2.00 | 2.00 | 100 | 18180 | 0.75000V | -0.00264478 | UNIFORM | P-AXIS |

Total Function Evaluations: 399960

---

| | 0.900 | 100 | 30 | 180 | 2.0 | 1.0 | 2.00 | 2.00 | 100 | 18180 | 0.75000V | −0.00023835 | UNIFORM | P-AXIS |

---

NOTE: THIS FILE HAS BEEN EDITED TO SHOW THE CORRECT BEST RUN ON THE BOTTOM LINE BECAUSE FUNCTION F7 IS RANDOM.

# F8

F8
Best Fitness = 12569.486616U
Probe # 78 at Time Step 61

F8 coordinates:
1  420.96802222
2  420.96802222
3  420.96802222
4  420.96802222
5  420.96802222
6  420.96802222
7  420.96802222
8  420.96802222
9  420.96802222
10 420.96802222
11 420.96802222
12 420.96802222
13 420.96802222
14 420.96802222
15 420.96802222
16 420.96802222
17 420.96802222
18 420.96802222
19 420.96802222
20 420.96819593
21 420.96802222
22 420.96802222
23 420.96802222
24 420.96802222
25 420.96802222
26 420.96802222
27 420.96802222
28 420.96802222
29 420.96802222
30 420.96802222

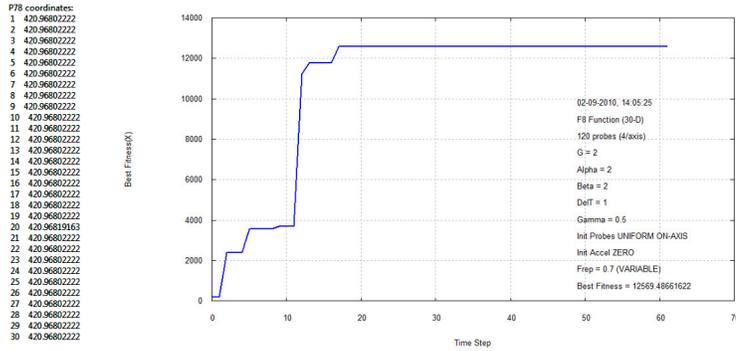

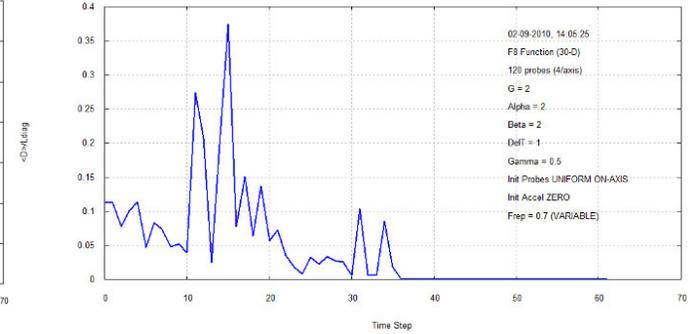

Run ID: 02-09-2010, 14:05:25 FUNCTION: F8

| Run # | Gamma | Nt | Nd | Np | G | DelT | Alpha | Beta | #Steps | Neval | Freq | Fitness | Initial Probes | |
|-------|-------|------|----|-----|-----|------|-------|------|--------|-------|----------|---------------|---------|-------|
| 0 | 0.000 | 1000 | 30 | 60 | 2.0 | 1.0 | 2.00 | 2.00 | 0 | 0 | 0.50000V | −9999.0000000 | UNIFORM | P-AXIS |
| 1 | 0.000 | 1000 | 30 | 60 | 2.0 | 1.0 | 2.00 | 2.00 | 97 | 5880 | 0.60000V | 5537.12371689 | UNIFORM | P-AXIS |
| 2 | 0.100 | 1000 | 30 | 60 | 2.0 | 1.0 | 2.00 | 2.00 | 99 | 6000 | 0.70000V | 12569.29322789 | UNIFORM | P-AXIS |
| 3 | 0.200 | 1000 | 30 | 60 | 2.0 | 1.0 | 2.00 | 2.00 | 63 | 3840 | 0.80000V | 12569.07685412 | UNIFORM | P-AXIS |
| 4 | 0.300 | 1000 | 30 | 60 | 2.0 | 1.0 | 2.00 | 2.00 | 62 | 3780 | 0.75000V | 12569.38585772 | UNIFORM | P-AXIS |
| 5 | 0.400 | 1000 | 30 | 60 | 2.0 | 1.0 | 2.00 | 2.00 | 62 | 3780 | 0.75000V | 12569.48453770 | UNIFORM | P-AXIS |
| 6 | 0.500 | 1000 | 30 | 60 | 2.0 | 1.0 | 2.00 | 2.00 | 125 | 7560 | 0.10000V | 12527.75480873 | UNIFORM | P-AXIS |
| 7 | 0.600 | 1000 | 30 | 60 | 2.0 | 1.0 | 2.00 | 2.00 | 114 | 6900 | 0.50000V | 12569.02032482 | UNIFORM | P-AXIS |
| 8 | 0.700 | 1000 | 30 | 60 | 2.0 | 1.0 | 2.00 | 2.00 | 165 | 9960 | 0.20000V | 9016.30715992 | UNIFORM | P-AXIS |
| 9 | 0.800 | 1000 | 30 | 60 | 2.0 | 1.0 | 2.00 | 2.00 | 88 | 5340 | 0.15000V | 12569.29470932 | UNIFORM | P-AXIS |
| 10 | 0.900 | 1000 | 30 | 60 | 2.0 | 1.0 | 2.00 | 2.00 | 41 | 2520 | 0.65000V | 12568.78874263 | UNIFORM | P-AXIS |
| 11 | 1.000 | 1000 | 30 | 60 | 2.0 | 1.0 | 2.00 | 2.00 | 65 | 3960 | 0.90000V | 8652.91905674 | UNIFORM | P-AXIS |
| 12 | 0.000 | 1000 | 30 | 120 | 2.0 | 1.0 | 2.00 | 2.00 | 223 | 26880 | 0.25000V | 8751.21949420 | UNIFORM | P-AXIS |
| 13 | 0.100 | 1000 | 30 | 120 | 2.0 | 1.0 | 2.00 | 2.00 | 80 | 9720 | 0.70000V | 12569.48495757 | UNIFORM | P-AXIS |
| 14 | 0.200 | 1000 | 30 | 120 | 2.0 | 1.0 | 2.00 | 2.00 | 69 | 8400 | 0.11000V | 12569.43708994 | UNIFORM | P-AXIS |
| 15 | 0.300 | 1000 | 30 | 120 | 2.0 | 1.0 | 2.00 | 2.00 | 62 | 7560 | 0.75000V | 12569.47312035 | UNIFORM | P-AXIS |
| 16 | 0.400 | 1000 | 30 | 120 | 2.0 | 1.0 | 2.00 | 2.00 | 60 | 7320 | 0.65000V | 12569.47738634 | UNIFORM | P-AXIS |
| 17 | 0.500 | 1000 | 30 | 120 | 2.0 | 1.0 | 2.00 | 2.00 | 61 | 7440 | 0.70000V | 12569.48661622 | UNIFORM | P-AXIS |
| 18 | 0.600 | 1000 | 30 | 120 | 2.0 | 1.0 | 2.00 | 2.00 | 64 | 7800 | 0.85000V | 12569.47542872 | UNIFORM | P-AXIS |
| 19 | 0.700 | 1000 | 30 | 120 | 2.0 | 1.0 | 2.00 | 2.00 | 80 | 9720 | 0.70000V | 12569.48061037 | UNIFORM | P-AXIS |
| 20 | 0.800 | 1000 | 30 | 120 | 2.0 | 1.0 | 2.00 | 2.00 | 48 | 5880 | 0.65000V | 8995.25108485 | UNIFORM | P-AXIS |
| 21 | 0.900 | 1000 | 30 | 120 | 2.0 | 1.0 | 2.00 | 2.00 | 61 | 7440 | 0.70000V | 12569.48656989 | UNIFORM | P-AXIS |
| 22 | 1.000 | 1000 | 30 | 120 | 2.0 | 1.0 | 2.00 | 2.00 | 99 | 12000 | 0.70000V | 12569.32153414 | UNIFORM | P-AXIS |
| 23 | 0.000 | 1000 | 30 | 180 | 2.0 | 1.0 | 2.00 | 2.00 | 98 | 17820 | 0.65000V | 12569.43502880 | UNIFORM | P-AXIS |
| 24 | 0.100 | 1000 | 30 | 180 | 2.0 | 1.0 | 2.00 | 2.00 | 102 | 18540 | 0.85000V | 12569.40212342 | UNIFORM | P-AXIS |
| 25 | 0.200 | 1000 | 30 | 180 | 2.0 | 1.0 | 2.00 | 2.00 | 63 | 11340 | 0.75000V | 12351.08839167 | UNIFORM | P-AXIS |
| 26 | 0.300 | 1000 | 30 | 180 | 2.0 | 1.0 | 2.00 | 2.00 | 80 | 14580 | 0.70000V | 12569.48182136 | UNIFORM | P-AXIS |
| 27 | 0.400 | 1000 | 30 | 180 | 2.0 | 1.0 | 2.00 | 2.00 | 48 | 8820 | 0.05000V | 12270.77098790 | UNIFORM | P-AXIS |
| 28 | 0.500 | 1000 | 30 | 180 | 2.0 | 1.0 | 2.00 | 2.00 | 64 | 11700 | 0.05000V | 12569.45025623 | UNIFORM | P-AXIS |
| 29 | 0.600 | 1000 | 30 | 180 | 2.0 | 1.0 | 2.00 | 2.00 | 65 | 11880 | 0.90000V | 12569.44913508 | UNIFORM | P-AXIS |
| 30 | 0.700 | 1000 | 30 | 180 | 2.0 | 1.0 | 2.00 | 2.00 | 114 | 20700 | 0.50000V | 12569.47931267 | UNIFORM | P-AXIS |
| 31 | 0.800 | 1000 | 30 | 180 | 2.0 | 1.0 | 2.00 | 2.00 | 70 | 12780 | 0.20000V | 12569.26333560 | UNIFORM | P-AXIS |
| 32 | 0.900 | 1000 | 30 | 180 | 2.0 | 1.0 | 2.00 | 2.00 | 65 | 11760 | 0.70000V | 12569.46219939 | UNIFORM | P-AXIS |
| 33 | 1.000 | 1000 | 30 | 180 | 2.0 | 1.0 | 2.00 | 2.00 | 98 | 17820 | 0.65000V | 12569.47962838 | UNIFORM | P-AXIS |

Total Function Evaluations: 326820

---

| 17 | 0.500 | 1000 | 30 | 120 | 2.0 | 1.0 | 2.00 | 2.00 | 61 | 7440 | 0.70000V | 12569.48661622 | UNIFORM | P-AXIS |

---

# F9

F9
Best Fitness = 0 returned by
Probe # 59 at Time Step 45

F9 coordinates:
1  0
2  0
3  0
4  0
5  0
6  0
7  0
8  0
9  0
10 0
11 0
12 0
13 0
14 0
15 0
16 0
17 0
18 0
19 0
20 0
21 0
22 0
23 0
24 0
25 0
26 0
27 0
28 0
29 0
30 0

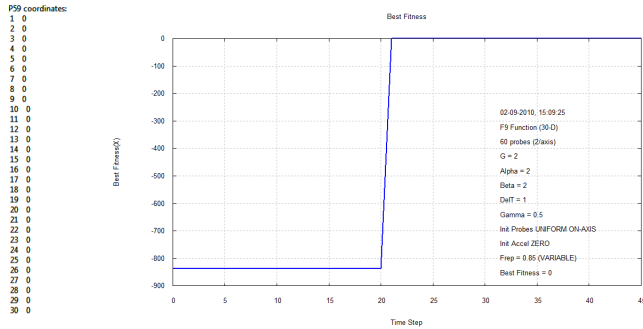

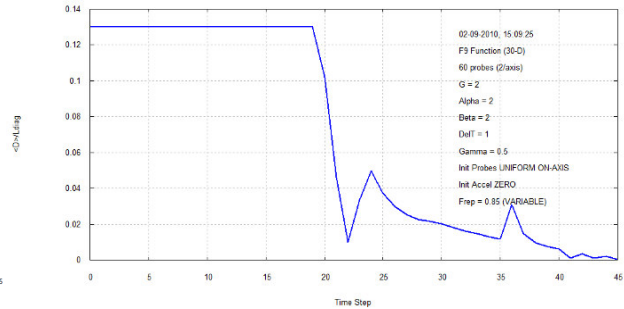

Run ID: 02-09-2010, 15:09:25 FUNCTION: F9

| Run # | Gamma | Nt | Nd | Np | G | DelT | Alpha | Beta | #Steps | Neval | Freq | Fitness | Initial Probes |
|-------|-------|------|----|-----|-----|------|-------|------|--------|-------|----------|-----------------|--------------|
| 0 | 0.000 | 1000 | 30 | 60 | 2.0 | 1.0 | 2.00 | 2.00 | 0 | 0 | 0.50000V | -9999.00000000 | UNIFORM P-AXIS |
| 1 | 0.000 | 1000 | 30 | 60 | 2.0 | 1.0 | 2.00 | 2.00 | 57 | 3480 | 0.50000V | -817.98078912 | UNIFORM P-AXIS |
| 2 | 0.100 | 1000 | 30 | 60 | 2.0 | 1.0 | 2.00 | 2.00 | 86 | 5220 | 0.55000V | -168.90806944 | UNIFORM P-AXIS |
| 3 | 0.200 | 1000 | 30 | 60 | 2.0 | 1.0 | 2.00 | 2.00 | 72 | 4380 | 0.30000V | -444.79080879 | UNIFORM P-AXIS |
| 4 | 0.300 | 1000 | 30 | 60 | 2.0 | 1.0 | 2.00 | 2.00 | 61 | 3720 | 0.90000V | -31.95278768 | UNIFORM P-AXIS |
| 5 | 0.400 | 1000 | 30 | 60 | 2.0 | 1.0 | 2.00 | 2.00 | 168 | 10140 | 0.35000V | -16.87213507 | UNIFORM P-AXIS |
| 6 | 0.500 | 1000 | 30 | 60 | 2.0 | 1.0 | 2.00 | 2.00 | 45 | 2760 | 0.85000V | 0.00000000 | UNIFORM P-AXIS |
| 7 | 0.600 | 1000 | 30 | 60 | 2.0 | 1.0 | 2.00 | 2.00 | 96 | 5820 | 0.55000V | -0.03076400 | UNIFORM P-AXIS |
| 8 | 0.700 | 1000 | 30 | 60 | 2.0 | 1.0 | 2.00 | 2.00 | 150 | 9060 | 0.40000V | -29.69863646 | UNIFORM P-AXIS |
| 9 | 0.800 | 1000 | 30 | 60 | 2.0 | 1.0 | 2.00 | 2.00 | 188 | 11340 | 0.40000V | -44.55147268 | UNIFORM P-AXIS |
| 10 | 0.900 | 1000 | 30 | 60 | 2.0 | 1.0 | 2.00 | 2.00 | 84 | 5100 | 0.90000V | -47.02033699 | UNIFORM P-AXIS |
| 11 | 1.000 | 1000 | 30 | 60 | 2.0 | 1.0 | 2.00 | 2.00 | 108 | 6540 | 0.20000V | -29.89546044 | UNIFORM P-AXIS |
| 12 | 0.000 | 1000 | 30 | 120 | 2.0 | 1.0 | 2.00 | 2.00 | 35 | 4320 | 0.35000V | -40.51140239 | UNIFORM P-AXIS |
| 13 | 0.100 | 1000 | 30 | 120 | 2.0 | 1.0 | 2.00 | 2.00 | 88 | 10680 | 0.15000V | -0.00016823 | UNIFORM P-AXIS |
| 14 | 0.200 | 1000 | 30 | 120 | 2.0 | 1.0 | 2.00 | 2.00 | 70 | 8520 | 0.20000V | -29.76575213 | UNIFORM P-AXIS |
| 15 | 0.300 | 1000 | 30 | 120 | 2.0 | 1.0 | 2.00 | 2.00 | 47 | 5760 | 0.95000V | -0.87005977 | UNIFORM P-AXIS |
| 16 | 0.400 | 1000 | 30 | 120 | 2.0 | 1.0 | 2.00 | 2.00 | 35 | 4320 | 0.35000V | -0.00120735 | UNIFORM P-AXIS |
| 17 | 0.500 | 1000 | 30 | 120 | 2.0 | 1.0 | 2.00 | 2.00 | 107 | 12960 | 0.15000V | -0.99727308 | UNIFORM P-AXIS |
| 18 | 0.600 | 1000 | 30 | 120 | 2.0 | 1.0 | 2.00 | 2.00 | 35 | 4320 | 0.35000V | -0.00120735 | UNIFORM P-AXIS |
| 19 | 0.700 | 1000 | 30 | 120 | 2.0 | 1.0 | 2.00 | 2.00 | 47 | 5760 | 0.95000V | -0.87005977 | UNIFORM P-AXIS |
| 20 | 0.800 | 1000 | 30 | 120 | 2.0 | 1.0 | 2.00 | 2.00 | 70 | 8520 | 0.20000V | -29.76575213 | UNIFORM P-AXIS |
| 21 | 0.900 | 1000 | 30 | 120 | 2.0 | 1.0 | 2.00 | 2.00 | 92 | 11160 | 0.35000V | -0.99091630 | UNIFORM P-AXIS |
| 22 | 1.000 | 1000 | 30 | 120 | 2.0 | 1.0 | 2.00 | 2.00 | 91 | 11040 | 0.30000V | -29.76828435 | UNIFORM P-AXIS |
| 23 | 0.000 | 1000 | 30 | 180 | 2.0 | 1.0 | 2.00 | 2.00 | 85 | 15480 | 0.95000V | -0.00006649 | UNIFORM P-AXIS |
| 24 | 0.100 | 1000 | 30 | 180 | 2.0 | 1.0 | 2.00 | 2.00 | 75 | 13680 | 0.45000V | -0.00001622 | UNIFORM P-AXIS |
| 25 | 0.200 | 1000 | 30 | 180 | 2.0 | 1.0 | 2.00 | 2.00 | 91 | 16560 | 0.30000V | -44.56106521 | UNIFORM P-AXIS |
| 26 | 0.300 | 1000 | 30 | 180 | 2.0 | 1.0 | 2.00 | 2.00 | 210 | 37980 | 0.55000V | -28.88260924 | UNIFORM P-AXIS |
| 27 | 0.400 | 1000 | 30 | 180 | 2.0 | 1.0 | 2.00 | 2.00 | 35 | 6480 | 0.35000V | -40.51140239 | UNIFORM P-AXIS |
| 28 | 0.500 | 1000 | 30 | 180 | 2.0 | 1.0 | 2.00 | 2.00 | 72 | 13140 | 0.30000V | -0.00021752 | UNIFORM P-AXIS |
| 29 | 0.600 | 1000 | 30 | 180 | 2.0 | 1.0 | 2.00 | 2.00 | 77 | 14040 | 0.55000V | -0.00000295 | UNIFORM P-AXIS |
| 30 | 0.700 | 1000 | 30 | 180 | 2.0 | 1.0 | 2.00 | 2.00 | 132 | 23940 | 0.45000V | -28.73588054 | UNIFORM P-AXIS |
| 31 | 0.800 | 1000 | 30 | 180 | 2.0 | 1.0 | 2.00 | 2.00 | 186 | 33660 | 0.30000V | -44.55183590 | UNIFORM P-AXIS |
| 32 | 0.900 | 1000 | 30 | 180 | 2.0 | 1.0 | 2.00 | 2.00 | 75 | 13680 | 0.45000V | -0.00001622 | UNIFORM P-AXIS |
| 33 | 1.000 | 1000 | 30 | 180 | 2.0 | 1.0 | 2.00 | 2.00 | 85 | 15480 | 0.95000V | -0.00006649 | UNIFORM P-AXIS |

Total Function Evaluations: 359040

| 6 | 0.500 | 1000 | 30 | 60 | 2.0 | 1.0 | 2.00 | 2.00 | 45 | 2760 | 0.85000V | 0.00000000 | UNIFORM P-AXIS |

## F10

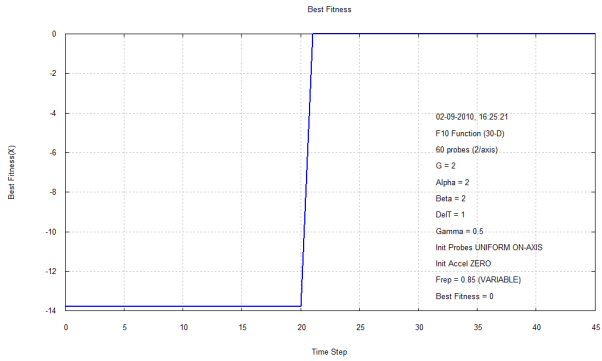

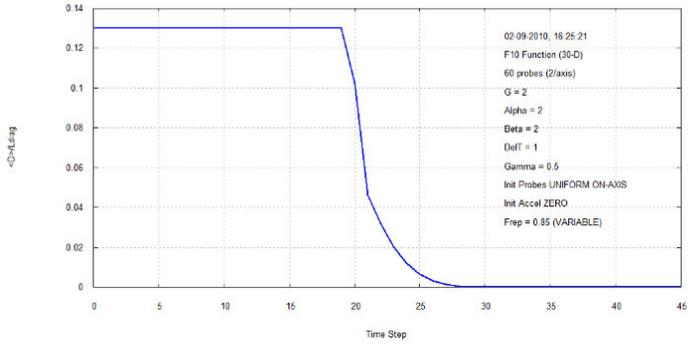

Run ID: 02-09-2010, 16:25:21 FUNCTION: F10

| Run # | Gamma | Nt | Nd | Np | G | DelT | Alpha | Beta | #Steps | Neval | Freq | Fitness | Initial Probes |
|-------|-------|------|----|-----|-----|------|-------|------|--------|-------|----------|-----------------|--------------|
| 0 | 0.000 | 1000 | 30 | 60 | 2.0 | 1.0 | 2.00 | 2.00 | 0 | 0 | 0.50000V | -9999.00000000 | UNIFORM P-AXIS |
| 1 | 0.000 | 1000 | 30 | 60 | 2.0 | 1.0 | 2.00 | 2.00 | 93 | 5640 | 0.40000V | -19.95554498 | UNIFORM P-AXIS |
| 2 | 0.100 | 1000 | 30 | 60 | 2.0 | 1.0 | 2.00 | 2.00 | 105 | 6360 | 0.05000V | -21.97959083 | UNIFORM P-AXIS |
| 3 | 0.200 | 1000 | 30 | 60 | 2.0 | 1.0 | 2.00 | 2.00 | 175 | 10560 | 0.70000V | -19.74824472 | UNIFORM P-AXIS |
| 4 | 0.300 | 1000 | 30 | 60 | 2.0 | 1.0 | 2.00 | 2.00 | 87 | 5280 | 0.10000V | -16.93358137 | UNIFORM P-AXIS |
| 5 | 0.400 | 1000 | 30 | 60 | 2.0 | 1.0 | 2.00 | 2.00 | 65 | 3960 | 0.90000V | -16.18811949 | UNIFORM P-AXIS |
| 6 | 0.500 | 1000 | 30 | 60 | 2.0 | 1.0 | 2.00 | 2.00 | 45 | 2760 | 0.85000V | 0.00000000 | UNIFORM P-AXIS |
| 7 | 0.600 | 1000 | 30 | 60 | 2.0 | 1.0 | 2.00 | 2.00 | 65 | 3960 | 0.90000V | -16.22923685 | UNIFORM P-AXIS |
| 8 | 0.700 | 1000 | 30 | 60 | 2.0 | 1.0 | 2.00 | 2.00 | 81 | 4920 | 0.75000V | -17.76008071 | UNIFORM P-AXIS |
| 9 | 0.800 | 1000 | 30 | 60 | 2.0 | 1.0 | 2.00 | 2.00 | 187 | 11280 | 0.35000V | -19.52444864 | UNIFORM P-AXIS |
| 10 | 0.900 | 1000 | 30 | 60 | 2.0 | 1.0 | 2.00 | 2.00 | 115 | 6960 | 0.55000V | -22.00875670 | UNIFORM P-AXIS |
| 11 | 1.000 | 1000 | 30 | 60 | 2.0 | 1.0 | 2.00 | 2.00 | 131 | 7920 | 0.40000V | -0.00067597 | UNIFORM P-AXIS |
| 12 | 0.000 | 1000 | 30 | 120 | 2.0 | 1.0 | 2.00 | 2.00 | 93 | 11280 | 0.40000V | -19.95554498 | UNIFORM P-AXIS |
| 13 | 0.100 | 1000 | 30 | 120 | 2.0 | 1.0 | 2.00 | 2.00 | 130 | 15720 | 0.35000V | -21.95522239 | UNIFORM P-AXIS |
| 14 | 0.200 | 1000 | 30 | 120 | 2.0 | 1.0 | 2.00 | 2.00 | 112 | 13560 | 0.40000V | -19.56322754 | UNIFORM P-AXIS |
| 15 | 0.300 | 1000 | 30 | 120 | 2.0 | 1.0 | 2.00 | 2.00 | 86 | 10320 | 0.95000V | -6.6133671O | UNIFORM P-AXIS |
| 16 | 0.400 | 1000 | 30 | 120 | 2.0 | 1.0 | 2.00 | 2.00 | 107 | 12960 | 0.15000V | -12.55536480 | UNIFORM P-AXIS |
| 17 | 0.500 | 1000 | 30 | 120 | 2.0 | 1.0 | 2.00 | 2.00 | 206 | 24840 | 0.35000V | -0.02837028 | UNIFORM P-AXIS |
| 18 | 0.600 | 1000 | 30 | 120 | 2.0 | 1.0 | 2.00 | 2.00 | 165 | 19920 | 0.20000V | -12.46390247 | UNIFORM P-AXIS |
| 19 | 0.700 | 1000 | 30 | 120 | 2.0 | 1.0 | 2.00 | 2.00 | 107 | 12960 | 0.15000V | -17.74769110 | UNIFORM P-AXIS |
| 20 | 0.800 | 1000 | 30 | 120 | 2.0 | 1.0 | 2.00 | 2.00 | 78 | 9480 | 0.60000V | -19.30852660 | UNIFORM P-AXIS |
| 21 | 0.900 | 1000 | 30 | 120 | 2.0 | 1.0 | 2.00 | 2.00 | 134 | 16200 | 0.55000V | -21.99722257 | UNIFORM P-AXIS |
| 22 | 1.000 | 1000 | 30 | 120 | 2.0 | 1.0 | 2.00 | 2.00 | 89 | 10800 | 0.20000V | -0.00463449 | UNIFORM P-AXIS |
| 23 | 0.000 | 1000 | 30 | 180 | 2.0 | 1.0 | 2.00 | 2.00 | 93 | 16920 | 0.40000V | -19.95554498 | UNIFORM P-AXIS |
| 24 | 0.100 | 1000 | 30 | 180 | 2.0 | 1.0 | 2.00 | 2.00 | 108 | 19620 | 0.20000V | -21.70151508 | UNIFORM P-AXIS |
| 25 | 0.200 | 1000 | 30 | 180 | 2.0 | 1.0 | 2.00 | 2.00 | 194 | 35100 | 0.70000V | -19.63371971 | UNIFORM P-AXIS |
| 26 | 0.300 | 1000 | 30 | 180 | 2.0 | 1.0 | 2.00 | 2.00 | 169 | 30420 | 0.35000V | -16.05132032 | UNIFORM P-AXIS |
| 27 | 0.400 | 1000 | 30 | 180 | 2.0 | 1.0 | 2.00 | 2.00 | 155 | 28080 | 0.65000V | -12.58872373 | UNIFORM P-AXIS |
| 28 | 0.500 | 1000 | 30 | 180 | 2.0 | 1.0 | 2.00 | 2.00 | 53 | 9720 | 0.30000V | -0.39610444 | UNIFORM P-AXIS |
| 29 | 0.600 | 1000 | 30 | 180 | 2.0 | 1.0 | 2.00 | 2.00 | 101 | 18360 | 0.80000V | -3.31261067 | UNIFORM P-AXIS |
| 30 | 0.700 | 1000 | 30 | 180 | 2.0 | 1.0 | 2.00 | 2.00 | 120 | 21780 | 0.80000V | -15.95693658 | UNIFORM P-AXIS |
| 31 | 0.800 | 1000 | 30 | 180 | 2.0 | 1.0 | 2.00 | 2.00 | 157 | 28440 | 0.75000V | -0.30534274 | UNIFORM P-AXIS |

| 32 | 0.900 | 1000 | 30 | 180 | 2.0 | 1.0 | 2.00 | 2.00 | 129 | 23400 | 0.30000V | -21.88600789 | UNIFORM | P-AXIS |
| 33 | 1.000 | 1000 | 30 | 180 | 2.0 | 1.0 | 2.00 | 2.00 | 105 | 19080 | 0.05000V | -0.00092027 | UNIFORM | P-AXIS |

Total Function Evaluations: 478560

| 6 | 0.500 | 1000 | 30 | 60 | 2.0 | 1.0 | 2.00 | 2.00 | 45 | 2760 | 0.85000V | 0.00000000 | UNIFORM | P-AXIS |

# F11

F11
Best Fitness = -.04419764 returned by
Probe # 99 at Time Step 195

P99 coordinates:
1   100.13753327
2   99.8139754
3   99.82927589
4   99.93043425
5   99.92512394
6   99.82747474
7   99.82286596
8   99.8225315
9   99.82210602
10  99.82225317
11  99.92215967
12  99.92150067
13  99.82227013
14  99.82236379
15  99.82247188
16  99.21298528
17  99.23273128
18  99.22328479
19  99.82298047
20  99.82311341
21  99.82326743
22  100.68078421
23  99.8225035
24  99.92362812
25  99.92377508
26  99.82388761
27  99.82398166
28  99.82409028
29  99.82419948

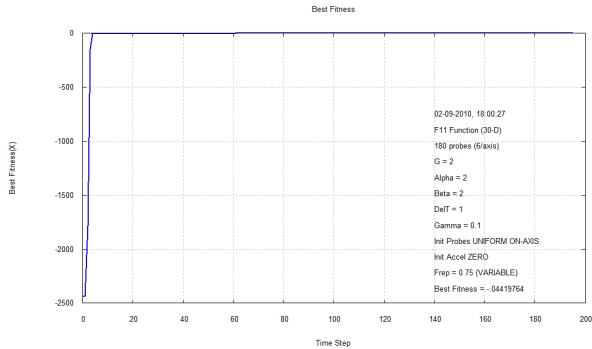

Run ID: 02-09-2010, 18:00:27 FUNCTION: F11

| Run # | Gamma | Nt | Nd | Np | G | DelT | Alpha | Beta | #Steps | Neval | Frep | Fitness | Initial Probes | |
|-------|-------|------|----|-----|-----|------|-------|------|--------|-------|----------|----------------|---------|--------|
| 0 | 0.000 | 1000 | 30 | 60 | 2.0 | 1.0 | 2.00 | 2.00 | 0 | 0 | 0.50000V | -9999.00000000 | UNIFORM | P-AXIS |
| 1 | 0.000 | 1000 | 30 | 60 | 2.0 | 1.0 | 2.00 | 2.00 | 46 | 2820 | 0.90000V | -3301.63525000 | UNIFORM | P-AXIS |
| 2 | 0.100 | 1000 | 30 | 60 | 2.0 | 1.0 | 2.00 | 2.00 | 46 | 2820 | 0.90000V | -2297.08769018 | UNIFORM | P-AXIS |
| 3 | 0.200 | 1000 | 30 | 60 | 2.0 | 1.0 | 2.00 | 2.00 | 54 | 3300 | 0.35000V | -1.23635432 | UNIFORM | P-AXIS |
| 4 | 0.300 | 1000 | 30 | 60 | 2.0 | 1.0 | 2.00 | 2.00 | 92 | 5580 | 0.35000V | -623.86779790 | UNIFORM | P-AXIS |
| 5 | 0.400 | 1000 | 30 | 60 | 2.0 | 1.0 | 2.00 | 2.00 | 45 | 2760 | 0.85000V | -336.38104615 | UNIFORM | P-AXIS |
| 6 | 0.500 | 1000 | 30 | 60 | 2.0 | 1.0 | 2.00 | 2.00 | 45 | 2760 | 0.85000V | -68.73220895 | UNIFORM | P-AXIS |
| 7 | 0.600 | 1000 | 30 | 60 | 2.0 | 1.0 | 2.00 | 2.00 | 246 | 14820 | 0.45000V | -4.00273011 | UNIFORM | P-AXIS |
| 8 | 0.700 | 1000 | 30 | 60 | 2.0 | 1.0 | 2.00 | 2.00 | 65 | 3960 | 0.90000V | -143.37903298 | UNIFORM | P-AXIS |
| 9 | 0.800 | 1000 | 30 | 60 | 2.0 | 1.0 | 2.00 | 2.00 | 65 | 3960 | 0.90000V | -480.33011229 | UNIFORM | P-AXIS |
| 10 | 0.900 | 1000 | 30 | 60 | 2.0 | 1.0 | 2.00 | 2.00 | 65 | 3960 | 0.90000V | -1010.47811863 | UNIFORM | P-AXIS |
| 11 | 1.000 | 1000 | 30 | 60 | 2.0 | 1.0 | 2.00 | 2.00 | 125 | 7560 | 0.10000V | -2.00126777 | UNIFORM | P-AXIS |
| 12 | 0.000 | 1000 | 30 | 120 | 2.0 | 1.0 | 2.00 | 2.00 | 54 | 6600 | 0.35000V | -109.92964955 | UNIFORM | P-AXIS |
| 13 | 0.100 | 1000 | 30 | 120 | 2.0 | 1.0 | 2.00 | 2.00 | 35 | 4320 | 0.35000V | -3.03975464 | UNIFORM | P-AXIS |
| 14 | 0.200 | 1000 | 30 | 120 | 2.0 | 1.0 | 2.00 | 2.00 | 35 | 4320 | 0.35000V | -2.57331594 | UNIFORM | P-AXIS |
| 15 | 0.300 | 1000 | 30 | 120 | 2.0 | 1.0 | 2.00 | 2.00 | 35 | 4320 | 0.35000V | -1.39086441 | UNIFORM | P-AXIS |
| 16 | 0.400 | 1000 | 30 | 120 | 2.0 | 1.0 | 2.00 | 2.00 | 35 | 4320 | 0.35000V | -5.67090092 | UNIFORM | P-AXIS |
| 17 | 0.500 | 1000 | 30 | 120 | 2.0 | 1.0 | 2.00 | 2.00 | 267 | 32160 | 0.55000V | -63.15584071 | UNIFORM | P-AXIS |
| 18 | 0.600 | 1000 | 30 | 120 | 2.0 | 1.0 | 2.00 | 2.00 | 35 | 4320 | 0.55000V | -6.40000000 | UNIFORM | P-AXIS |
| 19 | 0.700 | 1000 | 30 | 120 | 2.0 | 1.0 | 2.00 | 2.00 | 75 | 9120 | 0.45000V | -136.71068178 | UNIFORM | P-AXIS |
| 20 | 0.800 | 1000 | 30 | 120 | 2.0 | 1.0 | 2.00 | 2.00 | 35 | 4320 | 0.35000V | -445.43143407 | UNIFORM | P-AXIS |
| 21 | 0.900 | 1000 | 30 | 120 | 2.0 | 1.0 | 2.00 | 2.00 | 35 | 4320 | 0.35000V | -40.34010306 | UNIFORM | P-AXIS |
| 22 | 1.000 | 1000 | 30 | 120 | 2.0 | 1.0 | 2.00 | 2.00 | 35 | 4320 | 0.35000V | -4.05519280 | UNIFORM | P-AXIS |
| 23 | 0.000 | 1000 | 30 | 180 | 2.0 | 1.0 | 2.00 | 2.00 | 61 | 11160 | 0.70000V | -104.70613925 | UNIFORM | P-AXIS |
| 24 | 0.100 | 1000 | 30 | 180 | 2.0 | 1.0 | 2.00 | 2.00 | 195 | 35280 | 0.75000V | -0.04419764 | UNIFORM | P-AXIS |
| 25 | 0.200 | 1000 | 30 | 180 | 2.0 | 1.0 | 2.00 | 2.00 | 135 | 24480 | 0.60000V | -1.91298072 | UNIFORM | P-AXIS |
| 26 | 0.300 | 1000 | 30 | 180 | 2.0 | 1.0 | 2.00 | 2.00 | 35 | 6480 | 0.35000V | -4.42021222 | UNIFORM | P-AXIS |
| 27 | 0.400 | 1000 | 30 | 180 | 2.0 | 1.0 | 2.00 | 2.00 | 35 | 6480 | 0.35000V | -8.01901276 | UNIFORM | P-AXIS |
| 28 | 0.500 | 1000 | 30 | 180 | 2.0 | 1.0 | 2.00 | 2.00 | 35 | 6480 | 0.35000V | -68.05499564 | UNIFORM | P-AXIS |
| 29 | 0.600 | 1000 | 30 | 180 | 2.0 | 1.0 | 2.00 | 2.00 | 35 | 6480 | 0.35000V | -3.84611240 | UNIFORM | P-AXIS |
| 30 | 0.700 | 1000 | 30 | 180 | 2.0 | 1.0 | 2.00 | 2.00 | 75 | 13680 | 0.45000V | -122.29488564 | UNIFORM | P-AXIS |
| 31 | 0.800 | 1000 | 30 | 180 | 2.0 | 1.0 | 2.00 | 2.00 | 35 | 6480 | 0.35000V | -12.45943847 | UNIFORM | P-AXIS |
| 32 | 0.900 | 1000 | 30 | 180 | 2.0 | 1.0 | 2.00 | 2.00 | 35 | 6480 | 0.35000V | -3.88969420 | UNIFORM | P-AXIS |
| 33 | 1.000 | 1000 | 30 | 180 | 2.0 | 1.0 | 2.00 | 2.00 | 35 | 6480 | 0.35000V | -3.56479100 | UNIFORM | P-AXIS |

Total Function Evaluations: 266700

| 24 | 0.100 | 1000 | 30 | 180 | 2.0 | 1.0 | 2.00 | 2.00 | 195 | 35280 | 0.75000V | -0.04419764 | UNIFORM | P-AXIS |

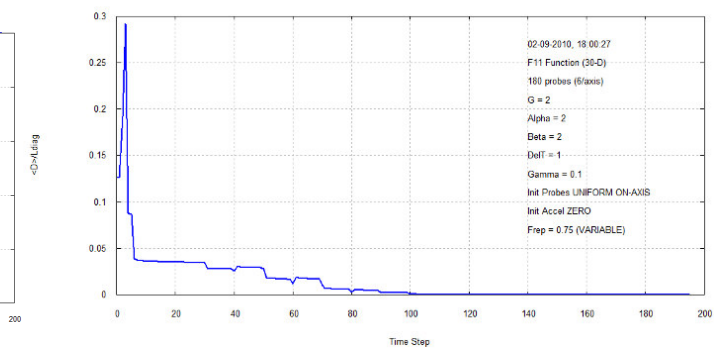

# F12

F12
Best Fitness = -.00002067 returned
Probe # 60 at Time Step 35

P60 coordinates:
1   -1.0046925
2   -1.0046925
3   -1.0046925
4   -1.0046925
5   -1.0046925
6   -1.0046925
7   -1.0046925
8   -1.0046925
9   -1.0046925
10  -1.0046925
11  -1.0046925
12  -1.0046925
13  -1.0046925
14  -1.0046925
15  -1.0046925
16  -1.0046925
17  -1.0046925
18  -1.0046925
19  -1.0046925
20  -1.0046925
21  -1.0046925
22  -1.0046925
23  -1.0046925
24  -1.0046925
25  -1.0046925
26  -1.0046925
27  -1.0046925
28  -1.0046925
29  -1.0046925
30  -.9813473

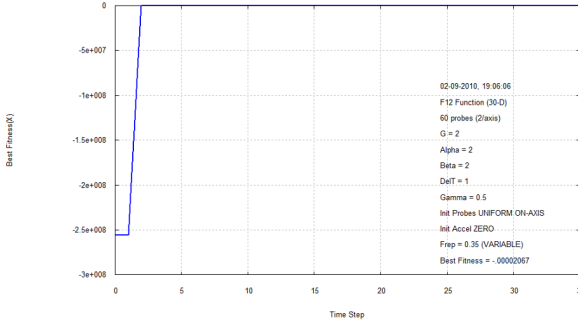

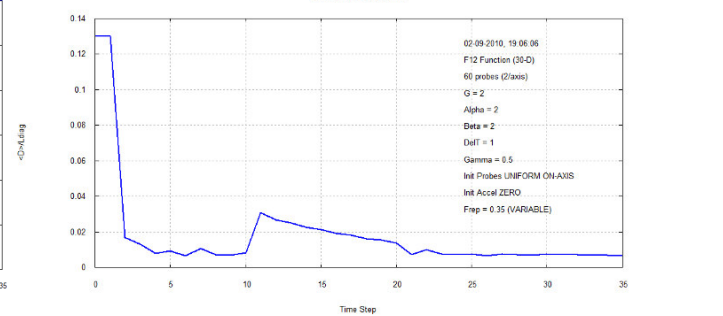

Run ID: 02-09-2010, 19:06:06 FUNCTION: F12

| Run # | Gamma | Nt | Nd | Np | G | DelT | Alpha | Beta | #Steps | Neval | Freq | Fitness | Initial Probes |
|---|---|---|---|---|---|---|---|---|---|---|---|---|---|
| 0 | 0.000 | 1000 | 30 | 60 | 2.00 | 1.0 | 2.00 | 2.00 | 0 | 0 | 0.50000V | -9999.00000000 | UNIFORM P-AXIS |
| 1 | 0.000 | 1000 | 30 | 60 | 2.00 | 1.0 | 2.00 | 2.00 | 50 | 3060 | 0.15000V | -0.39750628 | UNIFORM P-AXIS |
| 2 | 0.100 | 1000 | 30 | 60 | 2.00 | 1.0 | 2.00 | 2.00 | 180 | 10860 | 0.95000V | -0.43833828 | UNIFORM P-AXIS |
| 3 | 0.200 | 1000 | 30 | 60 | 2.00 | 1.0 | 2.00 | 2.00 | 49 | 3000 | 0.10000V | -0.76766783 | UNIFORM P-AXIS |
| 4 | 0.300 | 1000 | 30 | 60 | 2.00 | 1.0 | 2.00 | 2.00 | 63 | 3840 | 0.80000V | -0.66608776 | UNIFORM P-AXIS |
| 5 | 0.400 | 1000 | 30 | 60 | 2.00 | 1.0 | 2.00 | 2.00 | 64 | 3900 | 0.85000V | -0.05333970 | UNIFORM P-AXIS |
| 6 | 0.500 | 1000 | 30 | 60 | 2.00 | 1.0 | 2.00 | 2.00 | 35 | 2160 | 0.35000V | -0.00002067 | UNIFORM P-AXIS |
| 7 | 0.600 | 1000 | 30 | 60 | 2.00 | 1.0 | 2.00 | 2.00 | 35 | 2160 | 0.35000V | -0.24925810 | UNIFORM P-AXIS |
| 8 | 0.700 | 1000 | 30 | 60 | 2.00 | 1.0 | 2.00 | 2.00 | 69 | 4200 | 0.15000V | -0.52517658 | UNIFORM P-AXIS |
| 9 | 0.800 | 1000 | 30 | 60 | 2.00 | 1.0 | 2.00 | 2.00 | 50 | 3060 | 0.15000V | -0.44564592 | UNIFORM P-AXIS |
| 10 | 0.900 | 1000 | 30 | 60 | 2.00 | 1.0 | 2.00 | 2.00 | 65 | 3960 | 0.90000V | -0.04164606 | UNIFORM P-AXIS |
| 11 | 1.000 | 1000 | 30 | 60 | 2.00 | 1.0 | 2.00 | 2.00 | 35 | 2160 | 0.35000V | -1.74026604 | UNIFORM P-AXIS |
| 12 | 0.000 | 1000 | 30 | 120 | 2.00 | 1.0 | 2.00 | 2.00 | 65 | 7920 | 0.90000V | -0.13370701 | UNIFORM P-AXIS |
| 13 | 0.100 | 1000 | 30 | 120 | 2.00 | 1.0 | 2.00 | 2.00 | 35 | 4320 | 0.30000V | -0.00368253 | UNIFORM P-AXIS |
| 14 | 0.200 | 1000 | 30 | 120 | 2.00 | 1.0 | 2.00 | 2.00 | 37 | 4560 | 0.45000V | -0.25600092 | UNIFORM P-AXIS |
| 15 | 0.300 | 1000 | 30 | 120 | 2.00 | 1.0 | 2.00 | 2.00 | 67 | 8160 | 0.05000V | -0.09159413 | UNIFORM P-AXIS |
| 16 | 0.400 | 1000 | 30 | 120 | 2.00 | 1.0 | 2.00 | 2.00 | 35 | 4320 | 0.35000V | -0.29262880 | UNIFORM P-AXIS |
| 17 | 0.500 | 1000 | 30 | 120 | 2.00 | 1.0 | 2.00 | 2.00 | 175 | 21120 | 0.70000V | -0.00331292 | UNIFORM P-AXIS |
| 18 | 0.600 | 1000 | 30 | 120 | 2.00 | 1.0 | 2.00 | 2.00 | 85 | 10320 | 0.95000V | -0.25045678 | UNIFORM P-AXIS |
| 19 | 0.700 | 1000 | 30 | 120 | 2.00 | 1.0 | 2.00 | 2.00 | 35 | 4320 | 0.35000V | -0.45368588 | UNIFORM P-AXIS |
| 20 | 0.800 | 1000 | 30 | 120 | 2.00 | 1.0 | 2.00 | 2.00 | 35 | 4320 | 0.35000V | -0.02887759 | UNIFORM P-AXIS |
| 21 | 0.900 | 1000 | 30 | 120 | 2.00 | 1.0 | 2.00 | 2.00 | 49 | 6000 | 0.10000V | -0.01299175 | UNIFORM P-AXIS |
| 22 | 1.000 | 1000 | 30 | 120 | 2.00 | 1.0 | 2.00 | 2.00 | 35 | 4320 | 0.35000V | -0.07623230 | UNIFORM P-AXIS |
| 23 | 0.000 | 1000 | 30 | 180 | 2.00 | 1.0 | 2.00 | 2.00 | 54 | 9900 | 0.35000V | -0.01640636 | UNIFORM P-AXIS |
| 24 | 0.100 | 1000 | 30 | 180 | 2.00 | 1.0 | 2.00 | 2.00 | 35 | 6480 | 0.35000V | -0.00696541 | UNIFORM P-AXIS |
| 25 | 0.200 | 1000 | 30 | 180 | 2.00 | 1.0 | 2.00 | 2.00 | 35 | 6480 | 0.35000V | -0.07738490 | UNIFORM P-AXIS |
| 26 | 0.300 | 1000 | 30 | 180 | 2.00 | 1.0 | 2.00 | 2.00 | 73 | 13320 | 0.35000V | -0.22682589 | UNIFORM P-AXIS |
| 27 | 0.400 | 1000 | 30 | 180 | 2.00 | 1.0 | 2.00 | 2.00 | 35 | 6480 | 0.35000V | -0.03082085 | UNIFORM P-AXIS |
| 28 | 0.500 | 1000 | 30 | 180 | 2.00 | 1.0 | 2.00 | 2.00 | 155 | 28080 | 0.65000V | -0.00226701 | UNIFORM P-AXIS |
| 29 | 0.600 | 1000 | 30 | 180 | 2.00 | 1.0 | 2.00 | 2.00 | 70 | 12780 | 0.20000V | -0.33620253 | UNIFORM P-AXIS |
| 30 | 0.700 | 1000 | 30 | 180 | 2.00 | 1.0 | 2.00 | 2.00 | 35 | 6480 | 0.35000V | -0.00606971 | UNIFORM P-AXIS |
| 31 | 0.800 | 1000 | 30 | 180 | 2.00 | 1.0 | 2.00 | 2.00 | 35 | 6480 | 0.35000V | -0.01177648 | UNIFORM P-AXIS |
| 32 | 0.900 | 1000 | 30 | 180 | 2.00 | 1.0 | 2.00 | 2.00 | 35 | 6480 | 0.35000V | -0.01952312 | UNIFORM P-AXIS |
| 33 | 1.000 | 1000 | 30 | 180 | 2.00 | 1.0 | 2.00 | 2.00 | 45 | 8280 | 0.85000V | -0.26244867 | UNIFORM P-AXIS |

Total Function Evaluations: 233280

| 6 | 0.500 | 1000 | 30 | 60 | 2.0 | 1.0 | 2.00 | 2.00 | 35 | 2160 | 0.35000V | -0.00002067 | UNIFORM P-AXIS |

## F13

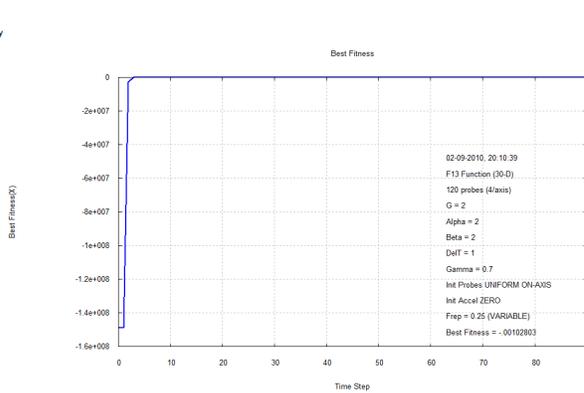

F13
Best Fitness = -.00102803 returned by
Probe # 2 at Time Step 90

P2 coordinates:
1  1.00714426
2  1.00080501
3  1.00080832
4  1.00080832
5  1.00080832
6  1.00080832
7  1.00080832
8  1.00080832
9  1.00080832
10 1.00080832
11 1.00080832
12 1.00080832
13 1.00080832
14 1.00080832
15 1.00080832
16 1.00080832
17 1.00080832
18 1.00080832
19 1.00080832
20 1.00080832
21 1.00080832
22 1.00080832
23 1.00080832
24 1.00080832
25 1.00080832
26 1.00080832
27 1.00082626
28 .99930179
29 .98441356
30 1.03234427

Best Fitness

02-09-2010, 20:10:39
F13 Function (30-D)
120 probes (4/axis)
G = 2
Alpha = 2
Beta = 2
DelT = 1
Gamma = 0.7
Init Probes UNIFORM ON-AXIS
Init Accel ZERO
Freq = 0.25 (VARIABLE)
Best Fitness = -.00102803

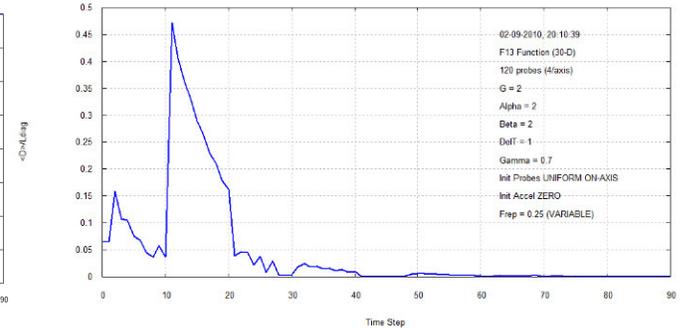

Average Distance of 119 Probes to Best Probe
Normalized to Size of Decision Space
[120 probes, 90 time steps]

02-09-2010, 20:10:39
F13 Function (30-D)
120 probes (4/axis)
G = 2
Alpha = 2
Beta = 2
DelT = 1
Gamma = 0.7
Init Probes UNIFORM ON-AXIS
Init Accel ZERO
Freq = 0.25 (VARIABLE)

Run ID: 02-09-2010, 20:10:39 FUNCTION: F13

| Run # | Gamma | Nt | Nd | Np | G | DelT | Alpha | Beta | #Steps | Neval | Freq | Fitness | Initial Probes |
|---|---|---|---|---|---|---|---|---|---|---|---|---|---|
| 0 | 0.000 | 1000 | 30 | 60 | 2.00 | 1.0 | 2.00 | 2.00 | 0 | 0 | 0.50000V | -9999.00000000 | UNIFORM P-AXIS |
| 1 | 0.000 | 1000 | 30 | 60 | 2.00 | 1.0 | 2.00 | 2.00 | 148 | 8940 | 0.30000V | -8.26885122 | UNIFORM P-AXIS |
| 2 | 0.100 | 1000 | 30 | 60 | 2.00 | 1.0 | 2.00 | 2.00 | 130 | 7860 | 0.35000V | -0.17447951 | UNIFORM P-AXIS |
| 3 | 0.200 | 1000 | 30 | 60 | 2.00 | 1.0 | 2.00 | 2.00 | 212 | 12780 | 0.65000V | -0.94453275 | UNIFORM P-AXIS |
| 4 | 0.300 | 1000 | 30 | 60 | 2.00 | 1.0 | 2.00 | 2.00 | 55 | 3360 | 0.65000V | -0.25838992 | UNIFORM P-AXIS |
| 5 | 0.400 | 1000 | 30 | 60 | 2.00 | 1.0 | 2.00 | 2.00 | 155 | 9360 | 0.65000V | -0.19569476 | UNIFORM P-AXIS |
| 6 | 0.500 | 1000 | 30 | 60 | 2.00 | 1.0 | 2.00 | 2.00 | 185 | 11160 | 0.25000V | -0.01793140 | UNIFORM P-AXIS |
| 7 | 0.600 | 1000 | 30 | 60 | 2.00 | 1.0 | 2.00 | 2.00 | 122 | 7380 | 0.90000V | -0.31891980 | UNIFORM P-AXIS |
| 8 | 0.700 | 1000 | 30 | 60 | 2.00 | 1.0 | 2.00 | 2.00 | 126 | 7620 | 0.15000V | -0.80812159 | UNIFORM P-AXIS |
| 9 | 0.800 | 1000 | 30 | 60 | 2.00 | 1.0 | 2.00 | 2.00 | 35 | 2160 | 0.35000V | -0.31643666 | UNIFORM P-AXIS |
| 10 | 0.900 | 1000 | 30 | 60 | 2.00 | 1.0 | 2.00 | 2.00 | 35 | 2160 | 0.35000V | -0.77660895 | UNIFORM P-AXIS |
| 11 | 1.000 | 1000 | 30 | 60 | 2.00 | 1.0 | 2.00 | 2.00 | 104 | 6300 | 0.95000V | -4.37182559 | UNIFORM P-AXIS |
| 12 | 0.000 | 1000 | 30 | 120 | 2.00 | 1.0 | 2.00 | 2.00 | 36 | 4440 | 0.40000V | -0.15002174 | UNIFORM P-AXIS |
| 13 | 0.100 | 1000 | 30 | 120 | 2.00 | 1.0 | 2.00 | 2.00 | 140 | 16920 | 0.85000V | -0.07366657 | UNIFORM P-AXIS |
| 14 | 0.200 | 1000 | 30 | 120 | 2.00 | 1.0 | 2.00 | 2.00 | 35 | 4320 | 0.35000V | -0.23509468 | UNIFORM P-AXIS |
| 15 | 0.300 | 1000 | 30 | 120 | 2.00 | 1.0 | 2.00 | 2.00 | 11 | 1040 | 0.30000V | -0.00395229 | UNIFORM P-AXIS |
| 16 | 0.400 | 1000 | 30 | 120 | 2.00 | 1.0 | 2.00 | 2.00 | 155 | 18720 | 0.65000V | -0.09332906 | UNIFORM P-AXIS |
| 17 | 0.500 | 1000 | 30 | 120 | 2.00 | 1.0 | 2.00 | 2.00 | 145 | 17520 | 0.15000V | -0.14889261 | UNIFORM P-AXIS |
| 18 | 0.600 | 1000 | 30 | 120 | 2.00 | 1.0 | 2.00 | 2.00 | 66 | 8040 | 0.95000V | -0.04850906 | UNIFORM P-AXIS |
| 19 | 0.700 | 1000 | 30 | 120 | 2.00 | 1.0 | 2.00 | 2.00 | 90 | 10920 | 0.25000V | -0.00102803 | UNIFORM P-AXIS |
| 20 | 0.800 | 1000 | 30 | 120 | 2.00 | 1.0 | 2.00 | 2.00 | 36 | 4440 | 0.40000V | -0.41136061 | UNIFORM P-AXIS |
| 21 | 0.900 | 1000 | 30 | 120 | 2.00 | 1.0 | 2.00 | 2.00 | 35 | 4320 | 0.35000V | -0.50216139 | UNIFORM P-AXIS |
| 22 | 1.000 | 1000 | 30 | 120 | 2.00 | 1.0 | 2.00 | 2.00 | 36 | 4440 | 0.40000V | -0.25179744 | UNIFORM P-AXIS |
| 23 | 0.000 | 1000 | 30 | 180 | 2.00 | 1.0 | 2.00 | 2.00 | 35 | 6480 | 0.35000V | -0.01299025 | UNIFORM P-AXIS |
| 24 | 0.100 | 1000 | 30 | 180 | 2.00 | 1.0 | 2.00 | 2.00 | 35 | 6480 | 0.35000V | -0.13062351 | UNIFORM P-AXIS |
| 25 | 0.200 | 1000 | 30 | 180 | 2.00 | 1.0 | 2.00 | 2.00 | 71 | 12960 | 0.25000V | -0.11071910 | UNIFORM P-AXIS |
| 26 | 0.300 | 1000 | 30 | 180 | 2.00 | 1.0 | 2.00 | 2.00 | 35 | 6480 | 0.35000V | -0.12164685 | UNIFORM P-AXIS |
| 27 | 0.400 | 1000 | 30 | 180 | 2.00 | 1.0 | 2.00 | 2.00 | 35 | 6480 | 0.35000V | -0.42130507 | UNIFORM P-AXIS |
| 28 | 0.500 | 1000 | 30 | 180 | 2.00 | 1.0 | 2.00 | 2.00 | 175 | 31680 | 0.70000V | -0.00226882 | UNIFORM P-AXIS |
| 29 | 0.600 | 1000 | 30 | 180 | 2.00 | 1.0 | 2.00 | 2.00 | 155 | 28080 | 0.65000V | -0.12170854 | UNIFORM P-AXIS |
| 30 | 0.700 | 1000 | 30 | 180 | 2.00 | 1.0 | 2.00 | 2.00 | 35 | 6480 | 0.35000V | -0.11071039 | UNIFORM P-AXIS |

Total Function Evaluations:    317280

-------------------------------------------------------------------------------

| 19 | 0.700 | 1000 | 30 | 120 | 2.0 | 1.0 | 2.00 | 2.00 | 90 | 10920 | 0.25000V | -0.00102803 | UNIFORM | P-AXIS |

# F14

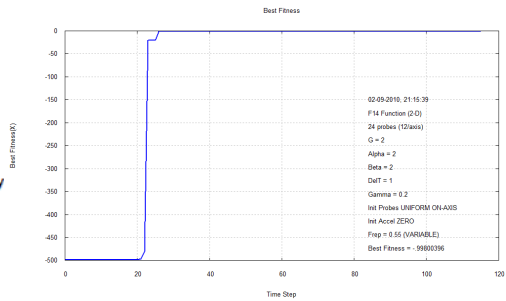

**F14**
**Best Fitness = -.99800396 returned by**
**Probe # 21 at Time Step 112**

**P21 coordinates:**
1   -32.06987068
2   -32.01456105

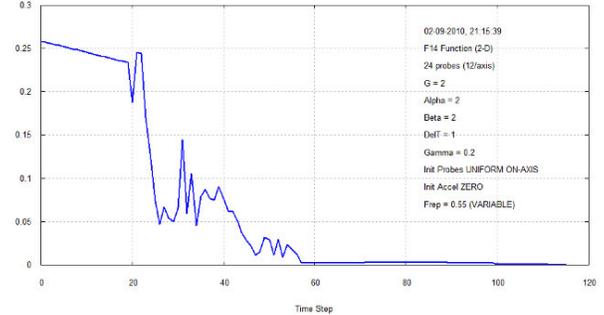

Run ID: 02-09-2010, 21:15:39 FUNCTION: F14

| Run # | Gamma | Nt | Nd | Np | G | DelT | Alpha | Beta | #Steps | Neval | Frep | Fitness | Initial Probes |
|-------|-------|----|----|----|---|------|-------|------|--------|-------|------|---------|----------------|
| 0 | 0.000 | 1000 | 2 | 4 | 2.0 | 1.0 | 2.00 | 2.00 | 0 | 0 | 0.50000V | -9999.00000000 | UNIFORM | P-AXIS |
| 1 | 0.000 | 1000 | 2 | 4 | 2.0 | 1.0 | 2.00 | 2.00 | 55 | 224 | 0.40000V | -1.40642307 | UNIFORM | P-AXIS |
| 2 | 0.100 | 1000 | 2 | 4 | 2.0 | 1.0 | 2.00 | 2.00 | 86 | 348 | 0.05000V | -1.08004550 | UNIFORM | P-AXIS |
| 3 | 0.200 | 1000 | 2 | 4 | 2.0 | 1.0 | 2.00 | 2.00 | 105 | 424 | 0.05000V | -416.57350238 | UNIFORM | P-AXIS |
| 4 | 0.300 | 1000 | 2 | 4 | 2.0 | 1.0 | 2.00 | 2.00 | 84 | 340 | 0.90000V | -2.05928607 | UNIFORM | P-AXIS |
| 5 | 0.400 | 1000 | 2 | 4 | 2.0 | 1.0 | 2.00 | 2.00 | 95 | 384 | 0.50000V | -7.87339681 | UNIFORM | P-AXIS |
| 6 | 0.500 | 1000 | 2 | 4 | 2.0 | 1.0 | 2.00 | 2.00 | 70 | 284 | 0.20000V | -10.76318244 | UNIFORM | P-AXIS |
| 7 | 0.600 | 1000 | 2 | 4 | 2.0 | 1.0 | 2.00 | 2.00 | 46 | 188 | 0.90000V | -12.67050581 | UNIFORM | P-AXIS |
| 8 | 0.700 | 1000 | 2 | 4 | 2.0 | 1.0 | 2.00 | 2.00 | 75 | 304 | 0.45000V | -11.71979621 | UNIFORM | P-AXIS |
| 9 | 0.800 | 1000 | 2 | 4 | 2.0 | 1.0 | 2.00 | 2.00 | 71 | 288 | 0.25000V | -8.84196997 | UNIFORM | P-AXIS |
| 10 | 0.900 | 1000 | 2 | 4 | 2.0 | 1.0 | 2.00 | 2.00 | 81 | 328 | 0.75000V | -13.61872972 | UNIFORM | P-AXIS |
| 11 | 1.000 | 1000 | 2 | 4 | 2.0 | 1.0 | 2.00 | 2.00 | 105 | 424 | 0.05000V | -490.41174276 | UNIFORM | P-AXIS |
| 12 | 0.000 | 1000 | 2 | 8 | 2.0 | 1.0 | 2.00 | 2.00 | 78 | 632 | 0.60000V | -1.02471512 | UNIFORM | P-AXIS |
| 13 | 0.100 | 1000 | 2 | 8 | 2.0 | 1.0 | 2.00 | 2.00 | 136 | 1096 | 0.65000V | -202.44979417 | UNIFORM | P-AXIS |
| 14 | 0.200 | 1000 | 2 | 8 | 2.0 | 1.0 | 2.00 | 2.00 | 70 | 568 | 0.20000V | -2.98221228 | UNIFORM | P-AXIS |
| 15 | 0.300 | 1000 | 2 | 8 | 2.0 | 1.0 | 2.00 | 2.00 | 48 | 392 | 0.05000V | -6.90333595 | UNIFORM | P-AXIS |
| 16 | 0.400 | 1000 | 2 | 8 | 2.0 | 1.0 | 2.00 | 2.00 | 53 | 432 | 0.30000V | -6.90336530 | UNIFORM | P-AXIS |
| 17 | 0.500 | 1000 | 2 | 8 | 2.0 | 1.0 | 2.00 | 2.00 | 69 | 560 | 0.15000V | -1.99362076 | UNIFORM | P-AXIS |
| 18 | 0.600 | 1000 | 2 | 8 | 2.0 | 1.0 | 2.00 | 2.00 | 45 | 368 | 0.85000V | -18.31709957 | UNIFORM | P-AXIS |
| 19 | 0.700 | 1000 | 2 | 8 | 2.0 | 1.0 | 2.00 | 2.00 | 58 | 472 | 0.55000V | -18.30771797 | UNIFORM | P-AXIS |
| 20 | 0.800 | 1000 | 2 | 8 | 2.0 | 1.0 | 2.00 | 2.00 | 68 | 552 | 0.10000V | -9.80389913 | UNIFORM | P-AXIS |
| 21 | 0.900 | 1000 | 2 | 8 | 2.0 | 1.0 | 2.00 | 2.00 | 104 | 840 | 0.95000V | -213.07705759 | UNIFORM | P-AXIS |
| 22 | 1.000 | 1000 | 2 | 8 | 2.0 | 1.0 | 2.00 | 2.00 | 58 | 472 | 0.55000V | -12.67050685 | UNIFORM | P-AXIS |
| 23 | 0.000 | 1000 | 2 | 12 | 2.0 | 1.0 | 2.00 | 2.00 | 85 | 1032 | 0.95000V | -1.99316040 | UNIFORM | P-AXIS |
| 24 | 0.100 | 1000 | 2 | 12 | 2.0 | 1.0 | 2.00 | 2.00 | 70 | 852 | 0.20000V | -1.00244850 | UNIFORM | P-AXIS |
| 25 | 0.200 | 1000 | 2 | 12 | 2.0 | 1.0 | 2.00 | 2.00 | 70 | 852 | 0.20000V | -2.98335908 | UNIFORM | P-AXIS |
| 26 | 0.300 | 1000 | 2 | 12 | 2.0 | 1.0 | 2.00 | 2.00 | 66 | 804 | 0.95000V | -6.90333670 | UNIFORM | P-AXIS |
| 27 | 0.400 | 1000 | 2 | 12 | 2.0 | 1.0 | 2.00 | 2.00 | 48 | 588 | 0.95000V | -5.92949534 | UNIFORM | P-AXIS |
| 28 | 0.500 | 1000 | 2 | 12 | 2.0 | 1.0 | 2.00 | 2.00 | 66 | 804 | 0.95000V | -2.98211825 | UNIFORM | P-AXIS |
| 29 | 0.600 | 1000 | 2 | 12 | 2.0 | 1.0 | 2.00 | 2.00 | 58 | 708 | 0.55000V | -6.95096960 | UNIFORM | P-AXIS |
| 30 | 0.700 | 1000 | 2 | 12 | 2.0 | 1.0 | 2.00 | 2.00 | 54 | 660 | 0.35000V | -8.84083729 | UNIFORM | P-AXIS |
| 31 | 0.800 | 1000 | 2 | 12 | 2.0 | 1.0 | 2.00 | 2.00 | 60 | 732 | 0.65000V | -0.99811146 | UNIFORM | P-AXIS |
| 32 | 0.900 | 1000 | 2 | 12 | 2.0 | 1.0 | 2.00 | 2.00 | 90 | 1092 | 0.25000V | -18.33845734 | UNIFORM | P-AXIS |
| 33 | 1.000 | 1000 | 2 | 12 | 2.0 | 1.0 | 2.00 | 2.00 | 63 | 768 | 0.80000V | -12.67172208 | UNIFORM | P-AXIS |
| 34 | 0.000 | 1000 | 2 | 16 | 2.0 | 1.0 | 2.00 | 2.00 | 71 | 1152 | 0.25000V | -1.99360589 | UNIFORM | P-AXIS |
| 35 | 0.100 | 1000 | 2 | 16 | 2.0 | 1.0 | 2.00 | 2.00 | 67 | 1088 | 0.05000V | -1.99203464 | UNIFORM | P-AXIS |
| 36 | 0.200 | 1000 | 2 | 16 | 2.0 | 1.0 | 2.00 | 2.00 | 53 | 864 | 0.30000V | -1.99263283 | UNIFORM | P-AXIS |
| 37 | 0.300 | 1000 | 2 | 16 | 2.0 | 1.0 | 2.00 | 2.00 | 54 | 880 | 0.35000V | -3.96825029 | UNIFORM | P-AXIS |
| 38 | 0.400 | 1000 | 2 | 16 | 2.0 | 1.0 | 2.00 | 2.00 | 35 | 576 | 0.35000V | -13.61860920 | UNIFORM | P-AXIS |
| 39 | 0.500 | 1000 | 2 | 16 | 2.0 | 1.0 | 2.00 | 2.00 | 50 | 816 | 0.15000V | -6.90333605 | UNIFORM | P-AXIS |
| 40 | 0.600 | 1000 | 2 | 16 | 2.0 | 1.0 | 2.00 | 2.00 | 35 | 576 | 0.35000V | -8.85576762 | UNIFORM | P-AXIS |
| 41 | 0.700 | 1000 | 2 | 16 | 2.0 | 1.0 | 2.00 | 2.00 | 61 | 992 | 0.70000V | -4.95092057 | UNIFORM | P-AXIS |
| 42 | 0.800 | 1000 | 2 | 16 | 2.0 | 1.0 | 2.00 | 2.00 | 54 | 880 | 0.35000V | -8.84083596 | UNIFORM | P-AXIS |
| 43 | 0.900 | 1000 | 2 | 16 | 2.0 | 1.0 | 2.00 | 2.00 | 70 | 1136 | 0.20000V | -13.62150695 | UNIFORM | P-AXIS |
| 44 | 1.000 | 1000 | 2 | 16 | 2.0 | 1.0 | 2.00 | 2.00 | 45 | 736 | 0.85000V | -100.92549919 | UNIFORM | P-AXIS |
| 45 | 0.000 | 1000 | 2 | 20 | 2.0 | 1.0 | 2.00 | 2.00 | 75 | 1520 | 0.45000V | -5.92930017 | UNIFORM | P-AXIS |
| 46 | 0.100 | 1000 | 2 | 20 | 2.0 | 1.0 | 2.00 | 2.00 | 53 | 1080 | 0.30000V | -1.99220800 | UNIFORM | P-AXIS |
| 47 | 0.200 | 1000 | 2 | 20 | 2.0 | 1.0 | 2.00 | 2.00 | 76 | 1540 | 0.50000V | -3.28650962 | UNIFORM | P-AXIS |
| 48 | 0.300 | 1000 | 2 | 20 | 2.0 | 1.0 | 2.00 | 2.00 | 35 | 720 | 0.35000V | -2.00373774 | UNIFORM | P-AXIS |
| 49 | 0.400 | 1000 | 2 | 20 | 2.0 | 1.0 | 2.00 | 2.00 | 35 | 720 | 0.35000V | -12.67050581 | UNIFORM | P-AXIS |
| 50 | 0.500 | 1000 | 2 | 20 | 2.0 | 1.0 | 2.00 | 2.00 | 52 | 1060 | 0.25000V | -1.99362004 | UNIFORM | P-AXIS |
| 51 | 0.600 | 1000 | 2 | 20 | 2.0 | 1.0 | 2.00 | 2.00 | 53 | 1080 | 0.30000V | -8.84115801 | UNIFORM | P-AXIS |
| 52 | 0.700 | 1000 | 2 | 20 | 2.0 | 1.0 | 2.00 | 2.00 | 68 | 1380 | 0.10000V | -5.93231258 | UNIFORM | P-AXIS |
| 53 | 0.800 | 1000 | 2 | 20 | 2.0 | 1.0 | 2.00 | 2.00 | 68 | 1380 | 0.10000V | -13.61862500 | UNIFORM | P-AXIS |
| 54 | 0.900 | 1000 | 2 | 20 | 2.0 | 1.0 | 2.00 | 2.00 | 67 | 1360 | 0.05000V | -19.25918532 | UNIFORM | P-AXIS |
| 55 | 1.000 | 1000 | 2 | 20 | 2.0 | 1.0 | 2.00 | 2.00 | 70 | 1420 | 0.20000V | -7.87435050 | UNIFORM | P-AXIS |
| 56 | 0.000 | 1000 | 2 | 24 | 2.0 | 1.0 | 2.00 | 2.00 | 66 | 1608 | 0.95000V | -0.99800405 | UNIFORM | P-AXIS |
| 57 | 0.100 | 1000 | 2 | 24 | 2.0 | 1.0 | 2.00 | 2.00 | 66 | 1608 | 0.95000V | -5.92886921 | UNIFORM | P-AXIS |
| 58 | 0.200 | 1000 | 2 | 24 | 2.0 | 1.0 | 2.00 | 2.00 | 115 | 2784 | 0.55000V | -0.99800396 | UNIFORM | P-AXIS |
| 59 | 0.300 | 1000 | 2 | 24 | 2.0 | 1.0 | 2.00 | 2.00 | 35 | 864 | 0.35000V | -12.67050583 | UNIFORM | P-AXIS |
| 60 | 0.400 | 1000 | 2 | 24 | 2.0 | 1.0 | 2.00 | 2.00 | 42 | 1032 | 0.70000V | -20.00049330 | UNIFORM | P-AXIS |
| 61 | 0.500 | 1000 | 2 | 24 | 2.0 | 1.0 | 2.00 | 2.00 | 56 | 1368 | 0.45000V | -2.98210858 | UNIFORM | P-AXIS |
| 62 | 0.600 | 1000 | 2 | 24 | 2.0 | 1.0 | 2.00 | 2.00 | 70 | 1704 | 0.20000V | -3.97401795 | UNIFORM | P-AXIS |
| 63 | 0.700 | 1000 | 2 | 24 | 2.0 | 1.0 | 2.00 | 2.00 | 35 | 864 | 0.35000V | -12.67050581 | UNIFORM | P-AXIS |
| 64 | 0.800 | 1000 | 2 | 24 | 2.0 | 1.0 | 2.00 | 2.00 | 70 | 1704 | 0.20000V | -13.61923447 | UNIFORM | P-AXIS |
| 65 | 0.900 | 1000 | 2 | 24 | 2.0 | 1.0 | 2.00 | 2.00 | 115 | 2784 | 0.55000V | -18.30430953 | UNIFORM | P-AXIS |
| 66 | 1.000 | 1000 | 2 | 24 | 2.0 | 1.0 | 2.00 | 2.00 | 71 | 1728 | 0.25000V | -1.99208977 | UNIFORM | P-AXIS |
| 67 | 0.000 | 1000 | 2 | 28 | 2.0 | 1.0 | 2.00 | 2.00 | 91 | 2576 | 0.30000V | -5.92957967 | UNIFORM | P-AXIS |
| 68 | 0.100 | 1000 | 2 | 28 | 2.0 | 1.0 | 2.00 | 2.00 | 68 | 1932 | 0.10000V | -5.92982307 | UNIFORM | P-AXIS |

```
69    0.200   1000    2    28    2.0   1.0    2.00    2.00    52     1484    0.25000V      -0.99800500    UNIFORM  P-AXIS
70    0.300   1000    2    28    2.0   1.0    2.00    2.00    53     1512    0.30000V     -12.68239387    UNIFORM  P-AXIS
71    0.400   1000    2    28    2.0   1.0    2.00    2.00    53     1484    0.25000V      -5.92884534    UNIFORM  P-AXIS
72    0.500   1000    2    28    2.0   1.0    2.00    2.00    35     1008    0.35000V      -7.87399305    UNIFORM  P-AXIS
73    0.600   1000    2    28    2.0   1.0    2.00    2.00    54     1540    0.35000V      -7.87399303    UNIFORM  P-AXIS
74    0.700   1000    2    28    2.0   1.0    2.00    2.00    53     1512    0.30000V     -12.67050650    UNIFORM  P-AXIS
75    0.800   1000    2    28    2.0   1.0    2.00    2.00    53     1512    0.30000V      -6.90348942    UNIFORM  P-AXIS
76    0.900   1000    2    28    2.0   1.0    2.00    2.00    83     2352    0.85000V     -13.61898868    UNIFORM  P-AXIS
77    1.000   1000    2    28    2.0   1.0    2.00    2.00    85     2408    0.95000V     -11.78451401    UNIFORM  P-AXIS

                                         Total Function Evaluations:    79136
```

------------------------------------------------------------------------------

```
58    0.200   1000    2    24    2.0   1.0    2.00    2.00   115     2784    0.55000V      -0.99800396    UNIFORM  P-AXIS
```

## __F15__

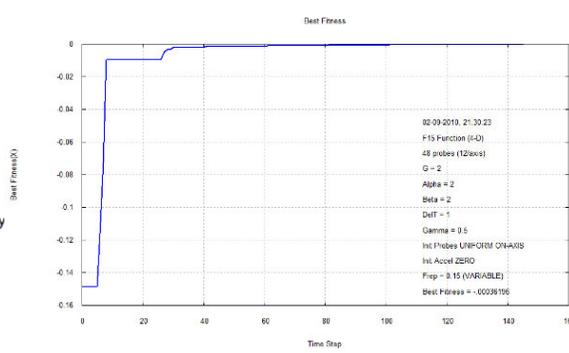

**F15**
**Best Fitness = -.00036196 returned by**
**Probe # 23 at Time Step 145**

**P23 coordinates:**
1   .19599896
2   .26677096
3   .20238446
4   .17093415

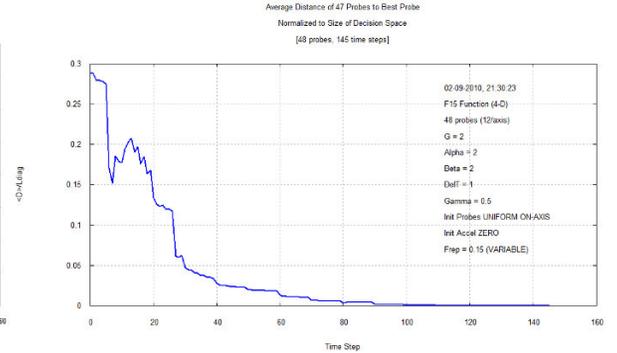

```
Run ID: 02-09-2010, 21:30:23 FUNCTION: F15

Run #   Gamma    Nt     Nd   Np    G    DelT   Alpha   Beta   #Steps   Neval     Frep        Fitness       Initial Probes
-----   ------   ----   --   --   ---   ----   -----   ----   ------   -----   ------      -------------   --------------
0       0.000    1000    4   16   2.0   1.0    2.00     0      0       288    0.50000V   -9999.00000000   UNIFORM  P-AXIS
-------------------------------------------------------------------------------------------------------------------------
1       0.000    1000    4    8   2.0   1.0    2.00    2.00    35      288    0.35000V      -0.05626396    UNIFORM  P-AXIS
2       0.100    1000    4    8   2.0   1.0    2.00    2.00    35      288    0.35000V      -0.05629426    UNIFORM  P-AXIS
3       0.200    1000    4    8   2.0   1.0    2.00    2.00    35      288    0.35000V      -0.03272933    UNIFORM  P-AXIS
4       0.300    1000    4    8   2.0   1.0    2.00    2.00    35      288    0.35000V      -0.13010491    UNIFORM  P-AXIS
5       0.400    1000    4    8   2.0   1.0    2.00    2.00    35      288    0.35000V      -0.13669559    UNIFORM  P-AXIS
6       0.500    1000    4    8   2.0   1.0    2.00    2.00    35      288    0.35000V      -0.08532319    UNIFORM  P-AXIS
7       0.600    1000    4    8   2.0   1.0    2.00    2.00    35      288    0.35000V      -0.13758964    UNIFORM  P-AXIS
8       0.700    1000    4    8   2.0   1.0    2.00    2.00    35      288    0.35000V      -0.09075532    UNIFORM  P-AXIS
9       0.800    1000    4    8   2.0   1.0    2.00    2.00    35      288    0.35000V      -0.08183848    UNIFORM  P-AXIS
10      0.900    1000    4    8   2.0   1.0    2.00    2.00   115      928    0.55000V      -0.01919626    UNIFORM  P-AXIS
11      1.000    1000    4    8   2.0   1.0    2.00    2.00    35      288    0.35000V      -0.05799290    UNIFORM  P-AXIS
12      0.000    1000    4   16   2.0   1.0    2.00    2.00    35      576    0.35000V      -0.00969087    UNIFORM  P-AXIS
13      0.100    1000    4   16   2.0   1.0    2.00    2.00    35      576    0.35000V      -0.04147696    UNIFORM  P-AXIS
14      0.200    1000    4   16   2.0   1.0    2.00    2.00    55      896    0.40000V      -0.01229134    UNIFORM  P-AXIS
15      0.300    1000    4   16   2.0   1.0    2.00    2.00   158     2544    0.80000V      -0.01353189    UNIFORM  P-AXIS
16      0.400    1000    4   16   2.0   1.0    2.00    2.00    35      576    0.35000V      -0.04063055    UNIFORM  P-AXIS
17      0.500    1000    4   16   2.0   1.0    2.00    2.00    56      912    0.45000V      -0.00575910    UNIFORM  P-AXIS
18      0.600    1000    4   16   2.0   1.0    2.00    2.00    35      576    0.35000V      -0.12870474    UNIFORM  P-AXIS
19      0.700    1000    4   16   2.0   1.0    2.00    2.00    35      576    0.35000V      -0.11830207    UNIFORM  P-AXIS
20      0.800    1000    4   16   2.0   1.0    2.00    2.00    35      576    0.35000V      -0.06996965    UNIFORM  P-AXIS
21      0.900    1000    4   16   2.0   1.0    2.00    2.00    35      576    0.35000V      -0.05833726    UNIFORM  P-AXIS
22      1.000    1000    4   16   2.0   1.0    2.00    2.00    95     1536    0.50000V      -0.01224564    UNIFORM  P-AXIS
23      0.000    1000    4   24   2.0   1.0    2.00    2.00    42     1032    0.70000V      -0.01602030    UNIFORM  P-AXIS
24      0.100    1000    4   24   2.0   1.0    2.00    2.00    35      864    0.35000V      -0.00973842    UNIFORM  P-AXIS
25      0.200    1000    4   24   2.0   1.0    2.00    2.00    35      864    0.35000V      -0.01306855    UNIFORM  P-AXIS
26      0.300    1000    4   24   2.0   1.0    2.00    2.00    35      864    0.35000V      -0.02651240    UNIFORM  P-AXIS
27      0.400    1000    4   24   2.0   1.0    2.00    2.00    35      864    0.35000V      -0.04798806    UNIFORM  P-AXIS
28      0.500    1000    4   24   2.0   1.0    2.00    2.00    35      864    0.35000V      -0.03475317    UNIFORM  P-AXIS
29      0.600    1000    4   24   2.0   1.0    2.00    2.00    35      864    0.35000V      -0.09216877    UNIFORM  P-AXIS
30      0.700    1000    4   24   2.0   1.0    2.00    2.00    95     2304    0.50000V      -0.00099094    UNIFORM  P-AXIS
31      0.800    1000    4   24   2.0   1.0    2.00    2.00   143     3456    0.75000V      -0.04055102    UNIFORM  P-AXIS
32      0.900    1000    4   24   2.0   1.0    2.00    2.00    66     1608    0.95000V      -0.00730714    UNIFORM  P-AXIS
33      1.000    1000    4   24   2.0   1.0    2.00    2.00   105     2544    0.05000V      -0.01337984    UNIFORM  P-AXIS
34      0.000    1000    4   32   2.0   1.0    2.00    2.00    56     1824    0.45000V      -0.02228527    UNIFORM  P-AXIS
35      0.100    1000    4   32   2.0   1.0    2.00    2.00    56     1824    0.45000V      -0.00361589    UNIFORM  P-AXIS
36      0.200    1000    4   32   2.0   1.0    2.00    2.00    65     2112    0.90000V      -0.01734236    UNIFORM  P-AXIS
37      0.300    1000    4   32   2.0   1.0    2.00    2.00    35     1152    0.35000V      -0.05553851    UNIFORM  P-AXIS
38      0.400    1000    4   32   2.0   1.0    2.00    2.00    35     1152    0.35000V      -0.02345614    UNIFORM  P-AXIS
39      0.500    1000    4   32   2.0   1.0    2.00    2.00    56     1824    0.45000V      -0.00543764    UNIFORM  P-AXIS
40      0.600    1000    4   32   2.0   1.0    2.00    2.00    55     1792    0.40000V      -0.01873877    UNIFORM  P-AXIS
41      0.700    1000    4   32   2.0   1.0    2.00    2.00    47     1536    0.95000V      -0.02297385    UNIFORM  P-AXIS
42      0.800    1000    4   32   2.0   1.0    2.00    2.00   144     4640    0.15000V      -0.03414598    UNIFORM  P-AXIS
43      0.900    1000    4   32   2.0   1.0    2.00    2.00    35     1152    0.35000V      -0.09252196    UNIFORM  P-AXIS
44      1.000    1000    4   32   2.0   1.0    2.00    2.00   145     4672    0.15000V      -0.01358199    UNIFORM  P-AXIS
45      0.000    1000    4   40   2.0   1.0    2.00    2.00    35     1440    0.35000V      -0.00760705    UNIFORM  P-AXIS
46      0.100    1000    4   40   2.0   1.0    2.00    2.00    35     1440    0.35000V      -0.04211797    UNIFORM  P-AXIS
47      0.200    1000    4   40   2.0   1.0    2.00    2.00    35     1440    0.35000V      -0.02822103    UNIFORM  P-AXIS
48      0.300    1000    4   40   2.0   1.0    2.00    2.00    35     1440    0.35000V      -0.07442684    UNIFORM  P-AXIS
49      0.400    1000    4   40   2.0   1.0    2.00    2.00    47     1920    0.95000V      -0.05345329    UNIFORM  P-AXIS
50      0.500    1000    4   40   2.0   1.0    2.00    2.00    55     2240    0.40000V      -0.01253059    UNIFORM  P-AXIS
51      0.600    1000    4   40   2.0   1.0    2.00    2.00    35     1440    0.35000V      -0.04223306    UNIFORM  P-AXIS
52      0.700    1000    4   40   2.0   1.0    2.00    2.00    35     1440    0.35000V      -0.03074483    UNIFORM  P-AXIS
53      0.800    1000    4   40   2.0   1.0    2.00    2.00   154     6200    0.60000V      -0.02014413    UNIFORM  P-AXIS
54      0.900    1000    4   40   2.0   1.0    2.00    2.00    53     2160    0.30000V      -0.00839958    UNIFORM  P-AXIS
55      1.000    1000    4   40   2.0   1.0    2.00    2.00   105     4240    0.05000V      -0.00177246    UNIFORM  P-AXIS
56      0.000    1000    4   48   2.0   1.0    2.00    2.00    69     3360    0.15000V      -0.00276244    UNIFORM  P-AXIS
57      0.100    1000    4   48   2.0   1.0    2.00    2.00    63     3072    0.80000V      -0.00339446    UNIFORM  P-AXIS
58      0.200    1000    4   48   2.0   1.0    2.00    2.00    96     4656    0.55000V      -0.00103590    UNIFORM  P-AXIS
```

```
59    0.300   1000    4   48    2.0   1.0   2.00   2.00    154    7440   0.60000V    -0.00562573   UNIFORM  P-AXIS
60    0.400   1000    4   48    2.0   1.0   2.00   2.00     35    1728   0.35000V    -0.05265765   UNIFORM  P-AXIS
61    0.500   1000    4   48    2.0   1.0   2.00   2.00    145    7008   0.15000V    -0.00036196   UNIFORM  P-AXIS
62    0.600   1000    4   48    2.0   1.0   2.00   2.00     35    1728   0.35000V    -0.03430604   UNIFORM  P-AXIS
63    0.700   1000    4   48    2.0   1.0   2.00   2.00     55    2688   0.40000V    -0.01100842   UNIFORM  P-AXIS
64    0.800   1000    4   48    2.0   1.0   2.00   2.00     65    3168   0.90000V    -0.02418517   UNIFORM  P-AXIS
65    0.900   1000    4   48    2.0   1.0   2.00   2.00     65    3168   0.90000V    -0.00063838   UNIFORM  P-AXIS
66    1.000   1000    4   48    2.0   1.0   2.00   2.00    125    6048   0.10000V    -0.01333495   UNIFORM  P-AXIS
67    0.000   1000    4   56    2.0   1.0   2.00   2.00     98    5544   0.65000V    -0.00248142   UNIFORM  P-AXIS
68    0.100   1000    4   56    2.0   1.0   2.00   2.00     68    3864   0.10000V    -0.00294772   UNIFORM  P-AXIS
69    0.200   1000    4   56    2.0   1.0   2.00   2.00     35    2016   0.35000V    -0.01356890   UNIFORM  P-AXIS
70    0.300   1000    4   56    2.0   1.0   2.00   2.00     35    2016   0.35000V    -0.03136321   UNIFORM  P-AXIS
71    0.400   1000    4   56    2.0   1.0   2.00   2.00     35    2016   0.35000V    -0.05285727   UNIFORM  P-AXIS
72    0.500   1000    4   56    2.0   1.0   2.00   2.00    154    8680   0.60000V    -0.00743062   UNIFORM  P-AXIS
73    0.600   1000    4   56    2.0   1.0   2.00   2.00    155    8736   0.65000V    -0.00442334   UNIFORM  P-AXIS
74    0.700   1000    4   56    2.0   1.0   2.00   2.00     85    4816   0.95000V    -0.01751023   UNIFORM  P-AXIS
75    0.800   1000    4   56    2.0   1.0   2.00   2.00     55    3136   0.40000V    -0.01071065   UNIFORM  P-AXIS
76    0.900   1000    4   56    2.0   1.0   2.00   2.00     65    3696   0.90000V    -0.01040173   UNIFORM  P-AXIS
77    1.000   1000    4   56    2.0   1.0   2.00   2.00     65    3696   0.90000V    -0.00961650   UNIFORM  P-AXIS

                                          Total Function Evaluations:    171216

------------------------------------------------------------------------------------------------------------------
61    0.500   1000    4   48    2.0   1.0   2.00   2.00    145    7008   0.15000V    -0.00036196   UNIFORM  P-AXIS
```

# **F16**

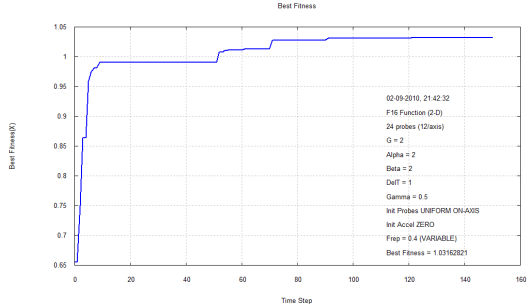

F16
Best Fitness = 1.03162821 returned by
Probe # 21 at Time Step 132

P21 coordinates:
1    .09007732
2    -.71273396

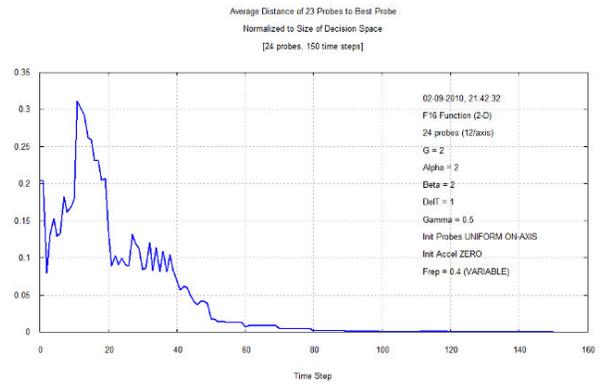

```
Run ID: 02-09-2010, 21:42:32 FUNCTION: F16

Run #   Gamma    Nt    Nd   Np    G    DelT   Alpha   Beta   #Steps   Neval    Frep        Fitness       Initial Probes
-----   -----   ----   --   ---  ----  ----   -----   ----   ------   -----   -------   --------------   --------------
  0     0.000   1000    2    8   2.0   1.0   2.00   2.00        0        0   0.50000V   -9999.00000000   UNIFORM  P-AXIS
-----------------------------------------------------------------------------------------------------------------------
  1     0.000   1000    2    4   2.0   1.0   2.00   2.00       50      204   0.15000V      0.00123418     UNIFORM  P-AXIS
  2     0.100   1000    2    4   2.0   1.0   2.00   2.00       50      204   0.15000V      0.95897275     UNIFORM  P-AXIS
  3     0.200   1000    2    4   2.0   1.0   2.00   2.00       95      384   0.50000V      0.73126193     UNIFORM  P-AXIS
  4     0.300   1000    2    4   2.0   1.0   2.00   2.00       45      184   0.85000V      0.02364336     UNIFORM  P-AXIS
  5     0.400   1000    2    4   2.0   1.0   2.00   2.00       57      232   0.50000V      0.86962651     UNIFORM  P-AXIS
  6     0.500   1000    2    4   2.0   1.0   2.00   2.00       65      264   0.90000V      0.99552303     UNIFORM  P-AXIS
  7     0.600   1000    2    4   2.0   1.0   2.00   2.00       35      144   0.35000V     -2.24278937     UNIFORM  P-AXIS
  8     0.700   1000    2    4   2.0   1.0   2.00   2.00       45      184   0.85000V      0.02364336     UNIFORM  P-AXIS
  9     0.800   1000    2    4   2.0   1.0   2.00   2.00       95      384   0.50000V      0.73126193     UNIFORM  P-AXIS
 10     0.900   1000    2    4   2.0   1.0   2.00   2.00       50      204   0.15000V      0.95897275     UNIFORM  P-AXIS
 11     1.000   1000    2    4   2.0   1.0   2.00   2.00      464   55000   0.55000V      0.95343805     UNIFORM  P-AXIS
 12     0.000   1000    2    8   2.0   1.0   2.00   2.00       53      432   0.30000V      1.02587776     UNIFORM  P-AXIS
 13     0.100   1000    2    8   2.0   1.0   2.00   2.00       67      544   0.05000V      1.03158983     UNIFORM  P-AXIS
 14     0.200   1000    2    8   2.0   1.0   2.00   2.00       35      288   0.35000V      0.99305889     UNIFORM  P-AXIS
 15     0.300   1000    2    8   2.0   1.0   2.00   2.00       85      688   0.95000V      1.03153738     UNIFORM  P-AXIS
 16     0.400   1000    2    8   2.0   1.0   2.00   2.00       49      400   0.10000V      1.03090469     UNIFORM  P-AXIS
 17     0.500   1000    2    8   2.0   1.0   2.00   2.00       55      448   0.40000V      1.00998353     UNIFORM  P-AXIS
 18     0.600   1000    2    8   2.0   1.0   2.00   2.00       49      400   0.10000V      1.03090469     UNIFORM  P-AXIS
 19     0.700   1000    2    8   2.0   1.0   2.00   2.00       85      688   0.95000V      1.03153738     UNIFORM  P-AXIS
 20     0.800   1000    2    8   2.0   1.0   2.00   2.00       35      288   0.35000V      0.99305889     UNIFORM  P-AXIS
 21     0.900   1000    2    8   2.0   1.0   2.00   2.00       67      544   0.05000V      1.03158983     UNIFORM  P-AXIS
 22     1.000   1000    2    8   2.0   1.0   2.00   2.00       53      432   0.30000V      1.02587776     UNIFORM  P-AXIS
 23     0.000   1000    2   12   2.0   1.0   2.00   2.00       68      828   0.10000V      1.03102797     UNIFORM  P-AXIS
 24     0.100   1000    2   12   2.0   1.0   2.00   2.00       68      828   0.10000V      1.02588029     UNIFORM  P-AXIS
 25     0.200   1000    2   12   2.0   1.0   2.00   2.00       52      636   0.25000V      1.03161663     UNIFORM  P-AXIS
 26     0.300   1000    2   12   2.0   1.0   2.00   2.00      135     1632   0.60000V      1.03158533     UNIFORM  P-AXIS
 27     0.400   1000    2   12   2.0   1.0   2.00   2.00       35      432   0.35000V      1.01947425     UNIFORM  P-AXIS
 28     0.500   1000    2   12   2.0   1.0   2.00   2.00       35      432   0.35000V      0.99998652     UNIFORM  P-AXIS
 29     0.600   1000    2   12   2.0   1.0   2.00   2.00       35      432   0.35000V      1.01947425     UNIFORM  P-AXIS
 30     0.700   1000    2   12   2.0   1.0   2.00   2.00      163     1968   0.10000V      1.03085997     UNIFORM  P-AXIS
 31     0.800   1000    2   12   2.0   1.0   2.00   2.00       35      432   0.35000V      1.03031188     UNIFORM  P-AXIS
 32     0.900   1000    2   12   2.0   1.0   2.00   2.00       68      828   0.10000V      1.02588029     UNIFORM  P-AXIS
 33     1.000   1000    2   12   2.0   1.0   2.00   2.00       68      828   0.10000V      1.03102797     UNIFORM  P-AXIS
 34     0.000   1000    2   16   2.0   1.0   2.00   2.00       51      832   0.20000V      1.01646464     UNIFORM  P-AXIS
 35     0.100   1000    2   16   2.0   1.0   2.00   2.00       69     1120   0.15000V      1.02661434     UNIFORM  P-AXIS
 36     0.200   1000    2   16   2.0   1.0   2.00   2.00      145     2336   0.15000V      1.02794010     UNIFORM  P-AXIS
 37     0.300   1000    2   16   2.0   1.0   2.00   2.00       55      896   0.40000V      1.02675307     UNIFORM  P-AXIS
 38     0.400   1000    2   16   2.0   1.0   2.00   2.00       35      576   0.35000V      1.00131504     UNIFORM  P-AXIS
 39     0.500   1000    2   16   2.0   1.0   2.00   2.00       84     1360   0.90000V      1.03004500     UNIFORM  P-AXIS
 40     0.600   1000    2   16   2.0   1.0   2.00   2.00       35      576   0.35000V      1.00131504     UNIFORM  P-AXIS
 41     0.700   1000    2   16   2.0   1.0   2.00   2.00       55      896   0.40000V      1.02675307     UNIFORM  P-AXIS
 42     0.800   1000    2   16   2.0   1.0   2.00   2.00      125     2016   0.10000V      1.02785242     UNIFORM  P-AXIS
 43     0.900   1000    2   16   2.0   1.0   2.00   2.00       69     1120   0.15000V      1.02661434     UNIFORM  P-AXIS
```

| Run # | Gamma | Nt | Nd | Np | G | DelT | Alpha | Beta | #Steps | Neval | Freq | Fitness | Initial Probes | |
|---|---|---|---|---|---|---|---|---|---|---|---|---|---|---|
| 44 | 1.000 | 1000 | 2 | 16 | 2.0 | 1.0 | 2.00 | 2.00 | 35 | 576 | 0.35000V | 1.01938921 | UNIFORM | P-AXIS |
| 45 | 0.000 | 1000 | 2 | 20 | 2.0 | 1.0 | 2.00 | 2.00 | 35 | 720 | 0.35000V | 0.98563739 | UNIFORM | P-AXIS |
| 46 | 0.100 | 1000 | 2 | 20 | 2.0 | 1.0 | 2.00 | 2.00 | 35 | 720 | 0.35000V | 1.02425714 | UNIFORM | P-AXIS |
| 47 | 0.200 | 1000 | 2 | 20 | 2.0 | 1.0 | 2.00 | 2.00 | 85 | 1720 | 0.95000V | 1.03118668 | UNIFORM | P-AXIS |
| 48 | 0.300 | 1000 | 2 | 20 | 2.0 | 1.0 | 2.00 | 2.00 | 70 | 1420 | 0.20000V | 1.03097639 | UNIFORM | P-AXIS |
| 49 | 0.400 | 1000 | 2 | 20 | 2.0 | 1.0 | 2.00 | 2.00 | 53 | 1080 | 0.30000V | 1.03120923 | UNIFORM | P-AXIS |
| 50 | 0.500 | 1000 | 2 | 20 | 2.0 | 1.0 | 2.00 | 2.00 | 77 | 1560 | 0.55000V | 1.01863350 | UNIFORM | P-AXIS |
| 51 | 0.600 | 1000 | 2 | 20 | 2.0 | 1.0 | 2.00 | 2.00 | 53 | 1080 | 0.30000V | 1.03120923 | UNIFORM | P-AXIS |
| 52 | 0.700 | 1000 | 2 | 20 | 2.0 | 1.0 | 2.00 | 2.00 | 70 | 1420 | 0.20000V | 1.03097639 | UNIFORM | P-AXIS |
| 53 | 0.800 | 1000 | 2 | 20 | 2.0 | 1.0 | 2.00 | 2.00 | 85 | 1720 | 0.95000V | 1.03118668 | UNIFORM | P-AXIS |
| 54 | 0.900 | 1000 | 2 | 20 | 2.0 | 1.0 | 2.00 | 2.00 | 35 | 720 | 0.35000V | 1.02425714 | UNIFORM | P-AXIS |
| 55 | 1.000 | 1000 | 2 | 20 | 2.0 | 1.0 | 2.00 | 2.00 | 35 | 720 | 0.35000V | 0.02062108 | UNIFORM | P-AXIS |
| 56 | 0.000 | 1000 | 2 | 24 | 2.0 | 1.0 | 2.00 | 2.00 | 35 | 864 | 0.35000V | 1.02305918 | UNIFORM | P-AXIS |
| 57 | 0.100 | 1000 | 2 | 24 | 2.0 | 1.0 | 2.00 | 2.00 | 67 | 1632 | 0.05000V | 1.02991226 | UNIFORM | P-AXIS |
| 58 | 0.200 | 1000 | 2 | 24 | 2.0 | 1.0 | 2.00 | 2.00 | 65 | 1584 | 0.35000V | 1.01778685 | UNIFORM | P-AXIS |
| 59 | 0.300 | 1000 | 2 | 24 | 2.0 | 1.0 | 2.00 | 2.00 | 35 | 864 | 0.35000V | 1.02692623 | UNIFORM | P-AXIS |
| 60 | 0.400 | 1000 | 2 | 24 | 2.0 | 1.0 | 2.00 | 2.00 | 49 | 1200 | 0.10000V | 0.95570586 | UNIFORM | P-AXIS |
| 61 | 0.500 | 1000 | 2 | 24 | 2.0 | 1.0 | 2.00 | 2.00 | 150 | 3624 | 0.40000V | 1.03162821 | UNIFORM | P-AXIS |
| 62 | 0.600 | 1000 | 2 | 24 | 2.0 | 1.0 | 2.00 | 2.00 | 70 | 1704 | 0.20000V | 1.03159618 | UNIFORM | P-AXIS |
| 63 | 0.700 | 1000 | 2 | 24 | 2.0 | 1.0 | 2.00 | 2.00 | 35 | 864 | 0.35000V | 0.99808281 | UNIFORM | P-AXIS |
| 64 | 0.800 | 1000 | 2 | 24 | 2.0 | 1.0 | 2.00 | 2.00 | 35 | 864 | 0.35000V | 1.00883536 | UNIFORM | P-AXIS |
| 65 | 0.900 | 1000 | 2 | 24 | 2.0 | 1.0 | 2.00 | 2.00 | 73 | 1776 | 0.30000V | 1.02454227 | UNIFORM | P-AXIS |
| 66 | 1.000 | 1000 | 2 | 24 | 2.0 | 1.0 | 2.00 | 2.00 | 35 | 864 | 0.35000V | 1.01912590 | UNIFORM | P-AXIS |
| 67 | 0.000 | 1000 | 2 | 28 | 2.0 | 1.0 | 2.00 | 2.00 | 35 | 1008 | 0.35000V | 1.02614496 | UNIFORM | P-AXIS |
| 68 | 0.100 | 1000 | 2 | 28 | 2.0 | 1.0 | 2.00 | 2.00 | 49 | 1400 | 0.10000V | 1.03161154 | UNIFORM | P-AXIS |
| 69 | 0.200 | 1000 | 2 | 28 | 2.0 | 1.0 | 2.00 | 2.00 | 70 | 1988 | 0.20000V | 1.03147156 | UNIFORM | P-AXIS |
| 70 | 0.300 | 1000 | 2 | 28 | 2.0 | 1.0 | 2.00 | 2.00 | 79 | 2240 | 0.65000V | 1.03080624 | UNIFORM | P-AXIS |
| 71 | 0.400 | 1000 | 2 | 28 | 2.0 | 1.0 | 2.00 | 2.00 | 35 | 1008 | 0.35000V | 1.02949203 | UNIFORM | P-AXIS |
| 72 | 0.500 | 1000 | 2 | 28 | 2.0 | 1.0 | 2.00 | 2.00 | 35 | 1008 | 0.35000V | 0.99911920 | UNIFORM | P-AXIS |
| 73 | 0.600 | 1000 | 2 | 28 | 2.0 | 1.0 | 2.00 | 2.00 | 35 | 1008 | 0.35000V | 1.02949203 | UNIFORM | P-AXIS |
| 74 | 0.700 | 1000 | 2 | 28 | 2.0 | 1.0 | 2.00 | 2.00 | 79 | 2240 | 0.65000V | 1.03080624 | UNIFORM | P-AXIS |
| 75 | 0.800 | 1000 | 2 | 28 | 2.0 | 1.0 | 2.00 | 2.00 | 70 | 1988 | 0.20000V | 1.03147156 | UNIFORM | P-AXIS |
| 76 | 0.900 | 1000 | 2 | 28 | 2.0 | 1.0 | 2.00 | 2.00 | 49 | 1400 | 0.10000V | 1.03161154 | UNIFORM | P-AXIS |
| 77 | 1.000 | 1000 | 2 | 28 | 2.0 | 1.0 | 2.00 | 2.00 | 69 | 1960 | 0.15000V | 1.03108369 | UNIFORM | P-AXIS |

```
                                         Total Function Evaluations:    74832
```

| 61 | 0.500 | 1000 | 2 | 24 | 2.0 | 1.0 | 2.00 | 2.00 | 150 | 3624 | 0.40000V | 1.03162821 | UNIFORM | P-AXIS |

# F17

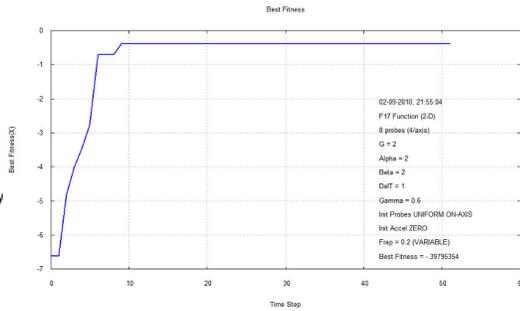

**F17**
Best Fitness = -.39795354 returned by Probe # 5 at Time Step 51

P5 coordinates:
1  3.14008701
2  2.28361107

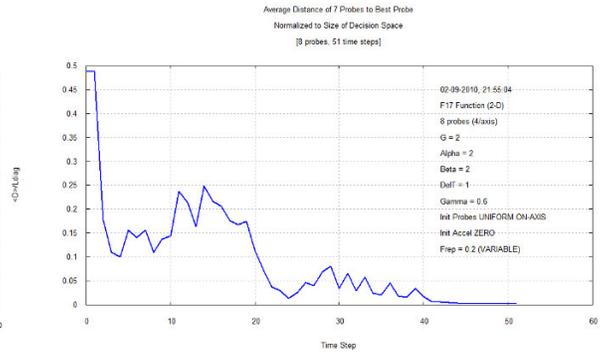

Run ID: 02-09-2010, 21:55:04 FUNCTION: F17

| Run # | Gamma | Nt | Nd | Np | G | DelT | Alpha | Beta | #Steps | Neval | Freq | Fitness | Initial Probes | |
|---|---|---|---|---|---|---|---|---|---|---|---|---|---|---|
| 0 | 0.000 | 1000 | 2 | 8 | 2.0 | 1.0 | 2.00 | 2.00 | 0 | 0 | 0.50000V | -9999.00000000 | UNIFORM | P-AXIS |
| 1 | 0.000 | 1000 | 2 | 4 | 2.0 | 1.0 | 2.00 | 2.00 | 51 | 208 | 0.20000V | -0.92814909 | UNIFORM | P-AXIS |
| 2 | 0.100 | 1000 | 2 | 4 | 2.0 | 1.0 | 2.00 | 2.00 | 96 | 388 | 0.55000V | -0.44924574 | UNIFORM | P-AXIS |
| 3 | 0.200 | 1000 | 2 | 4 | 2.0 | 1.0 | 2.00 | 2.00 | 101 | 408 | 0.80000V | -0.47281396 | UNIFORM | P-AXIS |
| 4 | 0.300 | 1000 | 2 | 4 | 2.0 | 1.0 | 2.00 | 2.00 | 61 | 248 | 0.70000V | -0.71270873 | UNIFORM | P-AXIS |
| 5 | 0.400 | 1000 | 2 | 4 | 2.0 | 1.0 | 2.00 | 2.00 | 65 | 264 | 0.90000V | -0.44349466 | UNIFORM | P-AXIS |
| 6 | 0.500 | 1000 | 2 | 4 | 2.0 | 1.0 | 2.00 | 2.00 | 35 | 144 | 0.35000V | -0.42898989 | UNIFORM | P-AXIS |
| 7 | 0.600 | 1000 | 2 | 4 | 2.0 | 1.0 | 2.00 | 2.00 | 66 | 268 | 0.95000V | -0.53816322 | UNIFORM | P-AXIS |
| 8 | 0.700 | 1000 | 2 | 4 | 2.0 | 1.0 | 2.00 | 2.00 | 56 | 228 | 0.45000V | -0.43544206 | UNIFORM | P-AXIS |
| 9 | 0.800 | 1000 | 2 | 4 | 2.0 | 1.0 | 2.00 | 2.00 | 47 | 192 | 0.95000V | -0.56156336 | UNIFORM | P-AXIS |
| 10 | 0.900 | 1000 | 2 | 4 | 2.0 | 1.0 | 2.00 | 2.00 | 57 | 232 | 0.50000V | -0.81096049 | UNIFORM | P-AXIS |
| 11 | 1.000 | 1000 | 2 | 4 | 2.0 | 1.0 | 2.00 | 2.00 | 51 | 208 | 0.20000V | -0.92814909 | UNIFORM | P-AXIS |
| 12 | 0.000 | 1000 | 2 | 8 | 2.0 | 1.0 | 2.00 | 2.00 | 65 | 528 | 0.90000V | -0.45920358 | UNIFORM | P-AXIS |
| 13 | 0.100 | 1000 | 2 | 8 | 2.0 | 1.0 | 2.00 | 2.00 | 56 | 456 | 0.45000V | -0.44911166 | UNIFORM | P-AXIS |
| 14 | 0.200 | 1000 | 2 | 8 | 2.0 | 1.0 | 2.00 | 2.00 | 73 | 592 | 0.35000V | -0.41800912 | UNIFORM | P-AXIS |
| 15 | 0.300 | 1000 | 2 | 8 | 2.0 | 1.0 | 2.00 | 2.00 | 75 | 608 | 0.45000V | -0.44492251 | UNIFORM | P-AXIS |
| 16 | 0.400 | 1000 | 2 | 8 | 2.0 | 1.0 | 2.00 | 2.00 | 85 | 688 | 0.45000V | -0.41240435 | UNIFORM | P-AXIS |
| 17 | 0.500 | 1000 | 2 | 8 | 2.0 | 1.0 | 2.00 | 2.00 | 56 | 456 | 0.45000V | -0.40104700 | UNIFORM | P-AXIS |
| 18 | 0.600 | 1000 | 2 | 8 | 2.0 | 1.0 | 2.00 | 2.00 | 51 | 416 | 0.20000V | -0.39795354 | UNIFORM | P-AXIS |
| 19 | 0.700 | 1000 | 2 | 8 | 2.0 | 1.0 | 2.00 | 2.00 | 35 | 288 | 0.35000V | -0.44406252 | UNIFORM | P-AXIS |
| 20 | 0.800 | 1000 | 2 | 8 | 2.0 | 1.0 | 2.00 | 2.00 | 52 | 424 | 0.25000V | -0.41692733 | UNIFORM | P-AXIS |
| 21 | 0.900 | 1000 | 2 | 8 | 2.0 | 1.0 | 2.00 | 2.00 | 70 | 568 | 0.20000V | -0.42993690 | UNIFORM | P-AXIS |
| 22 | 1.000 | 1000 | 2 | 8 | 2.0 | 1.0 | 2.00 | 2.00 | 50 | 404 | 0.25000V | -0.42345094 | UNIFORM | P-AXIS |
| 23 | 0.000 | 1000 | 2 | 12 | 2.0 | 1.0 | 2.00 | 2.00 | 75 | 612 | 0.15000V | -0.45149075 | UNIFORM | P-AXIS |
| 24 | 0.100 | 1000 | 2 | 12 | 2.0 | 1.0 | 2.00 | 2.00 | 68 | 828 | 0.10000V | -0.41823788 | UNIFORM | P-AXIS |
| 25 | 0.200 | 1000 | 2 | 12 | 2.0 | 1.0 | 2.00 | 2.00 | 35 | 432 | 0.35000V | -1.10161915 | UNIFORM | P-AXIS |
| 26 | 0.300 | 1000 | 2 | 12 | 2.0 | 1.0 | 2.00 | 2.00 | 35 | 432 | 0.35000V | -0.40024733 | UNIFORM | P-AXIS |
| 27 | 0.400 | 1000 | 2 | 12 | 2.0 | 1.0 | 2.00 | 2.00 | 35 | 432 | 0.35000V | -0.44377180 | UNIFORM | P-AXIS |
| 28 | 0.500 | 1000 | 2 | 12 | 2.0 | 1.0 | 2.00 | 2.00 | 35 | 432 | 0.35000V | -0.47539348 | UNIFORM | P-AXIS |
| 29 | 0.600 | 1000 | 2 | 12 | 2.0 | 1.0 | 2.00 | 2.00 | 35 | 432 | 0.35000V | -0.42645784 | UNIFORM | P-AXIS |
| 30 | 0.700 | 1000 | 2 | 12 | 2.0 | 1.0 | 2.00 | 2.00 | 47 | 576 | 0.90000V | -0.39934165 | UNIFORM | P-AXIS |
| 31 | 0.800 | 1000 | 2 | 12 | 2.0 | 1.0 | 2.00 | 2.00 | 35 | 432 | 0.35000V | -0.46538381 | UNIFORM | P-AXIS |
| 32 | 0.900 | 1000 | 2 | 12 | 2.0 | 1.0 | 2.00 | 2.00 | 65 | 792 | 0.90000V | -0.40679395 | UNIFORM | P-AXIS |

| 33 | 1.000 | 1000 | 2 | 12 | 2.0 | 1.0 | 2.00 | 2.00 | 57 | 696 | 0.50000V | -0.45435903 | UNIFORM P-AXIS |
| 34 | 0.000 | 1000 | 2 | 16 | 2.0 | 1.0 | 2.00 | 2.00 | 35 | 576 | 0.35000V | -0.41178661 | UNIFORM P-AXIS |
| 35 | 0.100 | 1000 | 2 | 16 | 2.0 | 1.0 | 2.00 | 2.00 | 57 | 928 | 0.50000V | -0.40753775 | UNIFORM P-AXIS |
| 36 | 0.200 | 1000 | 2 | 16 | 2.0 | 1.0 | 2.00 | 2.00 | 49 | 800 | 0.10000V | -0.40267535 | UNIFORM P-AXIS |
| 37 | 0.300 | 1000 | 2 | 16 | 2.0 | 1.0 | 2.00 | 2.00 | 53 | 864 | 0.30000V | -0.42989266 | UNIFORM P-AXIS |
| 38 | 0.400 | 1000 | 2 | 16 | 2.0 | 1.0 | 2.00 | 2.00 | 68 | 1104 | 0.10000V | -0.41342074 | UNIFORM P-AXIS |
| 39 | 0.500 | 1000 | 2 | 16 | 2.0 | 1.0 | 2.00 | 2.00 | 35 | 576 | 0.35000V | -0.42514730 | UNIFORM P-AXIS |
| 40 | 0.600 | 1000 | 2 | 16 | 2.0 | 1.0 | 2.00 | 2.00 | 66 | 1072 | 0.95000V | -0.41017339 | UNIFORM P-AXIS |
| 41 | 0.700 | 1000 | 2 | 16 | 2.0 | 1.0 | 2.00 | 2.00 | 95 | 1536 | 0.50000V | -0.40172536 | UNIFORM P-AXIS |
| 42 | 0.800 | 1000 | 2 | 16 | 2.0 | 1.0 | 2.00 | 2.00 | 50 | 816 | 0.15000V | -0.41725378 | UNIFORM P-AXIS |
| 43 | 0.900 | 1000 | 2 | 16 | 2.0 | 1.0 | 2.00 | 2.00 | 35 | 576 | 0.35000V | -0.40134318 | UNIFORM P-AXIS |
| 44 | 1.000 | 1000 | 2 | 16 | 2.0 | 1.0 | 2.00 | 2.00 | 50 | 816 | 0.15000V | -0.40508094 | UNIFORM P-AXIS |
| 45 | 0.000 | 1000 | 2 | 20 | 2.0 | 1.0 | 2.00 | 2.00 | 53 | 1080 | 0.40000V | -0.40090871 | UNIFORM P-AXIS |
| 46 | 0.100 | 1000 | 2 | 20 | 2.0 | 1.0 | 2.00 | 2.00 | 35 | 720 | 0.35000V | -0.40240145 | UNIFORM P-AXIS |
| 47 | 0.200 | 1000 | 2 | 20 | 2.0 | 1.0 | 2.00 | 2.00 | 66 | 1340 | 0.95000V | -0.39812262 | UNIFORM P-AXIS |
| 48 | 0.300 | 1000 | 2 | 20 | 2.0 | 1.0 | 2.00 | 2.00 | 77 | 1560 | 0.55000V | -0.40079759 | UNIFORM P-AXIS |
| 49 | 0.400 | 1000 | 2 | 20 | 2.0 | 1.0 | 2.00 | 2.00 | 75 | 1520 | 0.45000V | -0.40365150 | UNIFORM P-AXIS |
| 50 | 0.500 | 1000 | 2 | 20 | 2.0 | 1.0 | 2.00 | 2.00 | 51 | 1040 | 0.20000V | -0.40212945 | UNIFORM P-AXIS |
| 51 | 0.600 | 1000 | 2 | 20 | 2.0 | 1.0 | 2.00 | 2.00 | 35 | 720 | 0.35000V | -0.39908474 | UNIFORM P-AXIS |
| 52 | 0.700 | 1000 | 2 | 20 | 2.0 | 1.0 | 2.00 | 2.00 | 85 | 1720 | 0.95000V | -0.39869163 | UNIFORM P-AXIS |
| 53 | 0.800 | 1000 | 2 | 20 | 2.0 | 1.0 | 2.00 | 2.00 | 57 | 1160 | 0.50000V | -0.39842162 | UNIFORM P-AXIS |
| 54 | 0.900 | 1000 | 2 | 20 | 2.0 | 1.0 | 2.00 | 2.00 | 70 | 1420 | 0.20000V | -0.39835055 | UNIFORM P-AXIS |
| 55 | 1.000 | 1000 | 2 | 20 | 2.0 | 1.0 | 2.00 | 2.00 | 49 | 1000 | 0.10000V | -0.40068500 | UNIFORM P-AXIS |
| 56 | 0.000 | 1000 | 2 | 24 | 2.0 | 1.0 | 2.00 | 2.00 | 35 | 864 | 0.35000V | -0.40124974 | UNIFORM P-AXIS |
| 57 | 0.100 | 1000 | 2 | 24 | 2.0 | 1.0 | 2.00 | 2.00 | 49 | 1200 | 0.10000V | -0.39913945 | UNIFORM P-AXIS |
| 58 | 0.200 | 1000 | 2 | 24 | 2.0 | 1.0 | 2.00 | 2.00 | 69 | 1680 | 0.15000V | -0.39937003 | UNIFORM P-AXIS |
| 59 | 0.300 | 1000 | 2 | 24 | 2.0 | 1.0 | 2.00 | 2.00 | 70 | 1704 | 0.20000V | -0.41058613 | UNIFORM P-AXIS |
| 60 | 0.400 | 1000 | 2 | 24 | 2.0 | 1.0 | 2.00 | 2.00 | 47 | 1152 | 0.95000V | -0.40220074 | UNIFORM P-AXIS |
| 61 | 0.500 | 1000 | 2 | 24 | 2.0 | 1.0 | 2.00 | 2.00 | 54 | 1320 | 0.35000V | -0.40045620 | UNIFORM P-AXIS |
| 62 | 0.600 | 1000 | 2 | 24 | 2.0 | 1.0 | 2.00 | 2.00 | 35 | 864 | 0.35000V | -0.55824776 | UNIFORM P-AXIS |
| 63 | 0.700 | 1000 | 2 | 24 | 2.0 | 1.0 | 2.00 | 2.00 | 49 | 1200 | 0.10000V | -0.41082990 | UNIFORM P-AXIS |
| 64 | 0.800 | 1000 | 2 | 24 | 2.0 | 1.0 | 2.00 | 2.00 | 69 | 1680 | 0.15000V | -0.40176137 | UNIFORM P-AXIS |
| 65 | 0.900 | 1000 | 2 | 24 | 2.0 | 1.0 | 2.00 | 2.00 | 35 | 864 | 0.35000V | -0.42194353 | UNIFORM P-AXIS |
| 66 | 1.000 | 1000 | 2 | 24 | 2.0 | 1.0 | 2.00 | 2.00 | 39 | 960 | 0.35000V | -0.40267556 | UNIFORM P-AXIS |
| 67 | 0.000 | 1000 | 2 | 28 | 2.0 | 1.0 | 2.00 | 2.00 | 92 | 2604 | 0.35000V | -0.46232203 | UNIFORM P-AXIS |
| 68 | 0.100 | 1000 | 2 | 28 | 2.0 | 1.0 | 2.00 | 2.00 | 105 | 2968 | 0.05000V | -0.39839103 | UNIFORM P-AXIS |
| 69 | 0.200 | 1000 | 2 | 28 | 2.0 | 1.0 | 2.00 | 2.00 | 59 | 1680 | 0.60000V | -0.39811300 | UNIFORM P-AXIS |
| 70 | 0.300 | 1000 | 2 | 28 | 2.0 | 1.0 | 2.00 | 2.00 | 64 | 1820 | 0.85000V | -0.39868736 | UNIFORM P-AXIS |
| 71 | 0.400 | 1000 | 2 | 28 | 2.0 | 1.0 | 2.00 | 2.00 | 35 | 1008 | 0.35000V | -0.40399916 | UNIFORM P-AXIS |
| 72 | 0.500 | 1000 | 2 | 28 | 2.0 | 1.0 | 2.00 | 2.00 | 170 | 4788 | 0.45000V | -0.39809228 | UNIFORM P-AXIS |
| 73 | 0.600 | 1000 | 2 | 28 | 2.0 | 1.0 | 2.00 | 2.00 | 35 | 1008 | 0.35000V | -0.40740570 | UNIFORM P-AXIS |
| 74 | 0.700 | 1000 | 2 | 28 | 2.0 | 1.0 | 2.00 | 2.00 | 75 | 2128 | 0.45000V | -0.40093417 | UNIFORM P-AXIS |
| 75 | 0.800 | 1000 | 2 | 28 | 2.0 | 1.0 | 2.00 | 2.00 | 63 | 1792 | 0.80000V | -0.39929152 | UNIFORM P-AXIS |
| 76 | 0.900 | 1000 | 2 | 28 | 2.0 | 1.0 | 2.00 | 2.00 | 68 | 1932 | 0.10000V | -0.39854677 | UNIFORM P-AXIS |
| 77 | 1.000 | 1000 | 2 | 28 | 2.0 | 1.0 | 2.00 | 2.00 | 35 | 1008 | 0.35000V | -0.40202732 | UNIFORM P-AXIS |

```
                                Total Function Evaluations:   73444
```

| 18 | 0.600 | 1000 | 2 | 8 | 2.0 | 1.0 | 2.00 | 2.00 | 51 | 416 | 0.20000V | -0.39795354 | UNIFORM P-AXIS |

## **F18**

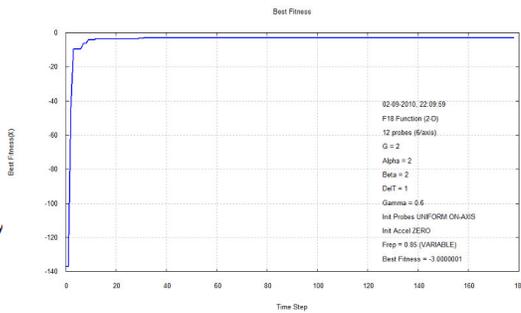

F18
Best Fitness = -3.0000001 returned by
Probe # 3 at Time Step 178

P3 coordinates:
1  .00001703
2  -.99998668

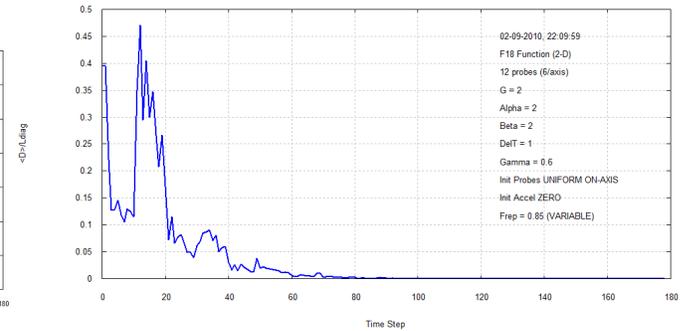

Run ID: 02-09-2010, 22:09:59 FUNCTION: F18

| Run # | Gamma | Nt | Nd | Np | G | DelT | Alpha | Beta | #Steps | Neval | Freq | Fitness | Initial Probes |
|-------|-------|------|----|----|-----|------|-------|------|--------|-------|----------|--------------|----------------|
| 0 | 0.000 | 1000 | 2 | 4 | 2.0 | 1.0 | 2.00 | 2.00 | 0 | 0 | 0.50000V | -9999.00000000 | UNIFORM P-AXIS |
| 1 | 0.000 | 1000 | 2 | 4 | 2.0 | 1.0 | 2.00 | 2.00 | 35 | 144 | 0.35000V | -31.41571104 | UNIFORM P-AXIS |
| 2 | 0.100 | 1000 | 2 | 4 | 2.0 | 1.0 | 2.00 | 2.00 | 74 | 300 | 0.40000V | -30.00714716 | UNIFORM P-AXIS |
| 3 | 0.200 | 1000 | 2 | 4 | 2.0 | 1.0 | 2.00 | 2.00 | 35 | 144 | 0.35000V | -3.00076414 | UNIFORM P-AXIS |
| 4 | 0.300 | 1000 | 2 | 4 | 2.0 | 1.0 | 2.00 | 2.00 | 104 | 420 | 0.95000V | -3.00220928 | UNIFORM P-AXIS |
| 5 | 0.400 | 1000 | 2 | 4 | 2.0 | 1.0 | 2.00 | 2.00 | 86 | 348 | 0.05000V | -3.00030997 | UNIFORM P-AXIS |
| 6 | 0.500 | 1000 | 2 | 4 | 2.0 | 1.0 | 2.00 | 2.00 | 115 | 464 | 0.55000V | -3.00038565 | UNIFORM P-AXIS |
| 7 | 0.600 | 1000 | 2 | 4 | 2.0 | 1.0 | 2.00 | 2.00 | 74 | 300 | 0.40000V | -3.00099008 | UNIFORM P-AXIS |
| 8 | 0.700 | 1000 | 2 | 4 | 2.0 | 1.0 | 2.00 | 2.00 | 35 | 144 | 0.35000V | -5.83465371 | UNIFORM P-AXIS |
| 9 | 0.800 | 1000 | 2 | 4 | 2.0 | 1.0 | 2.00 | 2.00 | 94 | 380 | 0.45000V | -4.60021841 | UNIFORM P-AXIS |
| 10 | 0.900 | 1000 | 2 | 4 | 2.0 | 1.0 | 2.00 | 2.00 | 54 | 220 | 0.35000V | -3.00488045 | UNIFORM P-AXIS |
| 11 | 1.000 | 1000 | 2 | 4 | 2.0 | 1.0 | 2.00 | 2.00 | 135 | 544 | 0.60000V | -3.00151384 | UNIFORM P-AXIS |
| 12 | 0.000 | 1000 | 2 | 8 | 2.0 | 1.0 | 2.00 | 2.00 | 35 | 288 | 0.35000V | -5.70277793 | UNIFORM P-AXIS |
| 13 | 0.100 | 1000 | 2 | 8 | 2.0 | 1.0 | 2.00 | 2.00 | 87 | 704 | 0.10000V | -3.00011972 | UNIFORM P-AXIS |
| 14 | 0.200 | 1000 | 2 | 8 | 2.0 | 1.0 | 2.00 | 2.00 | 89 | 720 | 0.20000V | -3.00019821 | UNIFORM P-AXIS |
| 15 | 0.300 | 1000 | 2 | 8 | 2.0 | 1.0 | 2.00 | 2.00 | 35 | 288 | 0.35000V | -4.42204710 | UNIFORM P-AXIS |
| 16 | 0.400 | 1000 | 2 | 8 | 2.0 | 1.0 | 2.00 | 2.00 | 106 | 856 | 0.10000V | -3.00100293 | UNIFORM P-AXIS |
| 17 | 0.500 | 1000 | 2 | 8 | 2.0 | 1.0 | 2.00 | 2.00 | 73 | 592 | 0.35000V | -3.00167285 | UNIFORM P-AXIS |
| 18 | 0.600 | 1000 | 2 | 8 | 2.0 | 1.0 | 2.00 | 2.00 | 71 | 576 | 0.25000V | -3.00128899 | UNIFORM P-AXIS |
| 19 | 0.700 | 1000 | 2 | 8 | 2.0 | 1.0 | 2.00 | 2.00 | 125 | 1008 | 0.10000V | -3.00105224 | UNIFORM P-AXIS |
| 20 | 0.800 | 1000 | 2 | 8 | 2.0 | 1.0 | 2.00 | 2.00 | 120 | 968 | 0.80000V | -3.00018709 | UNIFORM P-AXIS |

| | Gamma | Nt | Nd | Np | G | DelT | Alpha | Beta | #Steps | Neval | Frep | Fitness | Initial | Probes |
|---|---|---|---|---|---|---|---|---|---|---|---|---|---|---|
| 21 | 0.900 | 1000 | 2 | 8 | 2.0 | 1.0 | 2.00 | 2.00 | 35 | 288 | 0.35000V | -85.59879086 | UNIFORM | P-AXIS |
| 22 | 1.000 | 1000 | 2 | 8 | 2.0 | 1.0 | 2.00 | 2.00 | 87 | 704 | 0.10000V | -3.00022386 | UNIFORM | P-AXIS |
| 23 | 0.000 | 1000 | 2 | 12 | 2.0 | 1.0 | 2.00 | 2.00 | 72 | 876 | 0.30000V | -3.00601488 | UNIFORM | P-AXIS |
| 24 | 0.100 | 1000 | 2 | 12 | 2.0 | 1.0 | 2.00 | 2.00 | 125 | 1512 | 0.10000V | -3.00013640 | UNIFORM | P-AXIS |
| 25 | 0.200 | 1000 | 2 | 12 | 2.0 | 1.0 | 2.00 | 2.00 | 35 | 432 | 0.35000V | -3.01970178 | UNIFORM | P-AXIS |
| 26 | 0.300 | 1000 | 2 | 12 | 2.0 | 1.0 | 2.00 | 2.00 | 35 | 432 | 0.35000V | -3.03496345 | UNIFORM | P-AXIS |
| 27 | 0.400 | 1000 | 2 | 12 | 2.0 | 1.0 | 2.00 | 2.00 | 107 | 1296 | 0.15000V | -3.00034853 | UNIFORM | P-AXIS |
| 28 | 0.500 | 1000 | 2 | 12 | 2.0 | 1.0 | 2.00 | 2.00 | 120 | 1452 | 0.80000V | -3.00087309 | UNIFORM | P-AXIS |
| 29 | 0.600 | 1000 | 2 | 12 | 2.0 | 1.0 | 2.00 | 2.00 | 178 | 2148 | 0.85000V | -3.00000010 | UNIFORM | P-AXIS |
| 30 | 0.700 | 1000 | 2 | 12 | 2.0 | 1.0 | 2.00 | 2.00 | 107 | 1296 | 0.15000V | -3.00015260 | UNIFORM | P-AXIS |
| 31 | 0.800 | 1000 | 2 | 12 | 2.0 | 1.0 | 2.00 | 2.00 | 107 | 1296 | 0.15000V | -3.00048375 | UNIFORM | P-AXIS |
| 32 | 0.900 | 1000 | 2 | 12 | 2.0 | 1.0 | 2.00 | 2.00 | 120 | 1452 | 0.30000V | -3.00001380 | UNIFORM | P-AXIS |
| 33 | 1.000 | 1000 | 2 | 12 | 2.0 | 1.0 | 2.00 | 2.00 | 137 | 1656 | 0.35.32348051 | -35.32348051 | UNIFORM | P-AXIS |
| 34 | 0.000 | 1000 | 2 | 16 | 2.0 | 1.0 | 2.00 | 2.00 | 108 | 1744 | 0.20000V | -3.00006674 | UNIFORM | P-AXIS |
| 35 | 0.100 | 1000 | 2 | 16 | 2.0 | 1.0 | 2.00 | 2.00 | 70 | 1136 | 0.30000V | -3.00008450 | UNIFORM | P-AXIS |
| 36 | 0.200 | 1000 | 2 | 16 | 2.0 | 1.0 | 2.00 | 2.00 | 71 | 1152 | 0.25000V | -3.00022399 | UNIFORM | P-AXIS |
| 37 | 0.300 | 1000 | 2 | 16 | 2.0 | 1.0 | 2.00 | 2.00 | 73 | 1184 | 0.35000V | -3.00092605 | UNIFORM | P-AXIS |
| 38 | 0.400 | 1000 | 2 | 16 | 2.0 | 1.0 | 2.00 | 2.00 | 107 | 1728 | 0.15000V | -3.00027300 | UNIFORM | P-AXIS |
| 39 | 0.500 | 1000 | 2 | 16 | 2.0 | 1.0 | 2.00 | 2.00 | 79 | 1280 | 0.65000V | -3.00013010 | UNIFORM | P-AXIS |
| 40 | 0.600 | 1000 | 2 | 16 | 2.0 | 1.0 | 2.00 | 2.00 | 72 | 1168 | 0.30000V | -3.00004427 | UNIFORM | P-AXIS |
| 41 | 0.700 | 1000 | 2 | 16 | 2.0 | 1.0 | 2.00 | 2.00 | 122 | 1968 | 0.90000V | -3.00006030 | UNIFORM | P-AXIS |
| 42 | 0.800 | 1000 | 2 | 16 | 2.0 | 1.0 | 2.00 | 2.00 | 91 | 1472 | 0.30000V | -3.00000334 | UNIFORM | P-AXIS |
| 43 | 0.900 | 1000 | 2 | 16 | 2.0 | 1.0 | 2.00 | 2.00 | 117 | 1888 | 0.65000V | -3.00001143 | UNIFORM | P-AXIS |
| 44 | 1.000 | 1000 | 2 | 16 | 2.0 | 1.0 | 2.00 | 2.00 | 91 | 1472 | 0.30000V | -3.00020979 | UNIFORM | P-AXIS |
| 45 | 0.000 | 1000 | 2 | 20 | 2.0 | 1.0 | 2.00 | 2.00 | 108 | 2180 | 0.20000V | -3.00016941 | UNIFORM | P-AXIS |
| 46 | 0.100 | 1000 | 2 | 20 | 2.0 | 1.0 | 2.00 | 2.00 | 35 | 720 | 0.35000V | -3.03274596 | UNIFORM | P-AXIS |
| 47 | 0.200 | 1000 | 2 | 20 | 2.0 | 1.0 | 2.00 | 2.00 | 35 | 720 | 0.35000V | -3.08225372 | UNIFORM | P-AXIS |
| 48 | 0.300 | 1000 | 2 | 20 | 2.0 | 1.0 | 2.00 | 2.00 | 76 | 1540 | 0.50000V | -3.00035313 | UNIFORM | P-AXIS |
| 49 | 0.400 | 1000 | 2 | 20 | 2.0 | 1.0 | 2.00 | 2.00 | 71 | 1440 | 0.25000V | -3.00004988 | UNIFORM | P-AXIS |
| 50 | 0.500 | 1000 | 2 | 20 | 2.0 | 1.0 | 2.00 | 2.00 | 121 | 2440 | 0.85000V | -3.00000025 | UNIFORM | P-AXIS |
| 51 | 0.600 | 1000 | 2 | 20 | 2.0 | 1.0 | 2.00 | 2.00 | 35 | 720 | 0.35000V | -4.40118278 | UNIFORM | P-AXIS |
| 52 | 0.700 | 1000 | 2 | 20 | 2.0 | 1.0 | 2.00 | 2.00 | 106 | 2140 | 0.10000V | -3.00007038 | UNIFORM | P-AXIS |
| 53 | 0.800 | 1000 | 2 | 20 | 2.0 | 1.0 | 2.00 | 2.00 | 110 | 2220 | 0.30000V | -3.00003484 | UNIFORM | P-AXIS |
| 54 | 0.900 | 1000 | 2 | 20 | 2.0 | 1.0 | 2.00 | 2.00 | 67 | 1360 | 0.05000V | -3.01137298 | UNIFORM | P-AXIS |
| 55 | 1.000 | 1000 | 2 | 20 | 2.0 | 1.0 | 2.00 | 2.00 | 57 | 1160 | 0.50000V | -3.00023093 | UNIFORM | P-AXIS |
| 56 | 0.000 | 1000 | 2 | 24 | 2.0 | 1.0 | 2.00 | 2.00 | 35 | 864 | 0.35000V | -3.76690180 | UNIFORM | P-AXIS |
| 57 | 0.100 | 1000 | 2 | 24 | 2.0 | 1.0 | 2.00 | 2.00 | 35 | 864 | 0.35000V | -3.00574094 | UNIFORM | P-AXIS |
| 58 | 0.200 | 1000 | 2 | 24 | 2.0 | 1.0 | 2.00 | 2.00 | 62 | 1512 | 0.75000V | -3.00123133 | UNIFORM | P-AXIS |
| 59 | 0.300 | 1000 | 2 | 24 | 2.0 | 1.0 | 2.00 | 2.00 | 62 | 1512 | 0.35000V | -3.00061768 | UNIFORM | P-AXIS |
| 60 | 0.400 | 1000 | 2 | 24 | 2.0 | 1.0 | 2.00 | 2.00 | 35 | 864 | 0.35000V | -3.01365482 | UNIFORM | P-AXIS |
| 61 | 0.500 | 1000 | 2 | 24 | 2.0 | 1.0 | 2.00 | 2.00 | 35 | 864 | 0.35000V | -3.39524141 | UNIFORM | P-AXIS |
| 62 | 0.600 | 1000 | 2 | 24 | 2.0 | 1.0 | 2.00 | 2.00 | 121 | 2928 | 0.85000V | -3.00005383 | UNIFORM | P-AXIS |
| 63 | 0.700 | 1000 | 2 | 24 | 2.0 | 1.0 | 2.00 | 2.00 | 88 | 2136 | 0.15000V | -3.00114899 | UNIFORM | P-AXIS |
| 64 | 0.800 | 1000 | 2 | 24 | 2.0 | 1.0 | 2.00 | 2.00 | 112 | 2712 | 0.40000V | -3.00006149 | UNIFORM | P-AXIS |
| 65 | 0.900 | 1000 | 2 | 24 | 2.0 | 1.0 | 2.00 | 2.00 | 178 | 4296 | 0.85000V | -3.00000034 | UNIFORM | P-AXIS |
| 66 | 1.000 | 1000 | 2 | 24 | 2.0 | 1.0 | 2.00 | 2.00 | 91 | 2208 | 0.30000V | -3.00000308 | UNIFORM | P-AXIS |
| 67 | 0.000 | 1000 | 2 | 28 | 2.0 | 1.0 | 2.00 | 2.00 | 35 | 1008 | 0.35000V | -3.83584197 | UNIFORM | P-AXIS |
| 68 | 0.100 | 1000 | 2 | 28 | 2.0 | 1.0 | 2.00 | 2.00 | 35 | 1008 | 0.35000V | -3.00904692 | UNIFORM | P-AXIS |
| 69 | 0.200 | 1000 | 2 | 28 | 2.0 | 1.0 | 2.00 | 2.00 | 77 | 2184 | 0.55000V | -3.00005967 | UNIFORM | P-AXIS |
| 70 | 0.300 | 1000 | 2 | 28 | 2.0 | 1.0 | 2.00 | 2.00 | 35 | 1008 | 0.35000V | -3.05557167 | UNIFORM | P-AXIS |
| 71 | 0.400 | 1000 | 2 | 28 | 2.0 | 1.0 | 2.00 | 2.00 | 35 | 1008 | 0.35000V | -3.06202214 | UNIFORM | P-AXIS |
| 72 | 0.500 | 1000 | 2 | 28 | 2.0 | 1.0 | 2.00 | 2.00 | 35 | 1008 | 0.35000V | -3.38354074 | UNIFORM | P-AXIS |
| 73 | 0.600 | 1000 | 2 | 28 | 2.0 | 1.0 | 2.00 | 2.00 | 63 | 1792 | 0.80000V | -3.00028262 | UNIFORM | P-AXIS |
| 74 | 0.700 | 1000 | 2 | 28 | 2.0 | 1.0 | 2.00 | 2.00 | 67 | 1904 | 0.05000V | -3.00083602 | UNIFORM | P-AXIS |
| 75 | 0.800 | 1000 | 2 | 28 | 2.0 | 1.0 | 2.00 | 2.00 | 53 | 1512 | 0.30000V | -3.01209386 | UNIFORM | P-AXIS |
| 76 | 0.900 | 1000 | 2 | 28 | 2.0 | 1.0 | 2.00 | 2.00 | 92 | 2604 | 0.35000V | -3.00003890 | UNIFORM | P-AXIS |
| 77 | 1.000 | 1000 | 2 | 28 | 2.0 | 1.0 | 2.00 | 2.00 | 63 | 1792 | 0.80000V | -3.00050814 | UNIFORM | P-AXIS |

Total Function Evaluations: 94668

| 29 | 0.600 | 1000 | 2 | 12 | 2.0 | 1.0 | 2.00 | 2.00 | 178 | 2148 | 0.85000V | -3.00000010 | UNIFORM | P-AXIS |

# F19

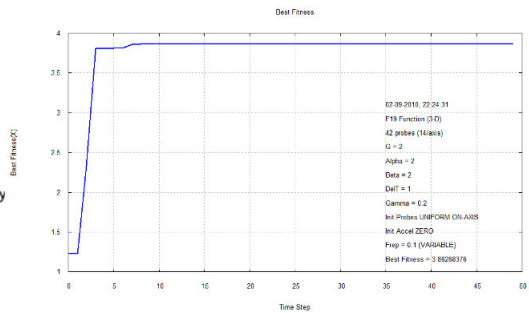

F19
Best Fitness = 3.86268376 returned by
Probe # 25 at Time Step 49

P25 coordinates:
1   .12462115
2   .55547346
3   .85190248

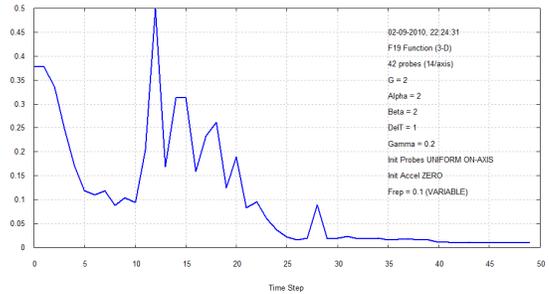

Run ID: 02-09-2010, 22:24:31 UNCTION: F19

| Run # | Gamma | Nt | Nd | Np | G | DelT | Alpha | Beta | #Steps | Neval | Frep | Fitness | Initial | Probes |
|---|---|---|---|---|---|---|---|---|---|---|---|---|---|---|
| 0 | 0.000 | 1000 | 3 | 12 | 2.0 | 1.0 | 2.00 | 2.00 | 0 | | 0.50000V | -9999.00000000 | UNIFORM | P-AXIS |
| 1 | 0.000 | 1000 | 3 | 6 | 2.0 | 1.0 | 2.00 | 2.00 | 50 | 306 | 0.15000V | 3.63800485 | UNIFORM | P-AXIS |
| 2 | 0.100 | 1000 | 3 | 6 | 2.0 | 1.0 | 2.00 | 2.00 | 58 | 354 | 0.55000V | 0.92269376 | UNIFORM | P-AXIS |
| 3 | 0.200 | 1000 | 3 | 6 | 2.0 | 1.0 | 2.00 | 2.00 | 55 | 336 | 0.40000V | 0.99931705 | UNIFORM | P-AXIS |
| 4 | 0.300 | 1000 | 3 | 6 | 2.0 | 1.0 | 2.00 | 2.00 | 35 | 216 | 0.35000V | 3.84976952 | UNIFORM | P-AXIS |
| 5 | 0.400 | 1000 | 3 | 6 | 2.0 | 1.0 | 2.00 | 2.00 | 65 | 396 | 0.90000V | 3.82910303 | UNIFORM | P-AXIS |
| 6 | 0.500 | 1000 | 3 | 6 | 2.0 | 1.0 | 2.00 | 2.00 | 35 | 216 | 0.35000V | 3.85244716 | UNIFORM | P-AXIS |
| 7 | 0.600 | 1000 | 3 | 6 | 2.0 | 1.0 | 2.00 | 2.00 | 35 | 216 | 0.35000V | 3.82995472 | UNIFORM | P-AXIS |
| 8 | 0.700 | 1000 | 3 | 6 | 2.0 | 1.0 | 2.00 | 2.00 | 35 | 216 | 0.35000V | 3.78901265 | UNIFORM | P-AXIS |
| 9 | 0.800 | 1000 | 3 | 6 | 2.0 | 1.0 | 2.00 | 2.00 | 50 | 306 | 0.15000V | 3.70101987 | UNIFORM | P-AXIS |

```
10    0.900   1000   3    6    2.0   1.0   2.00   2.00    48     294    0.05000V   3.81328117   UNIFORM   P-AXIS
11    1.000   1000   3    6    2.0   1.0   2.00   2.00    60     366    0.65000V   3.81473371   UNIFORM   P-AXIS
12    0.000   1000   3   12    2.0   1.0   2.00   2.00   105    1272    0.05000V   3.77509213   UNIFORM   P-AXIS
13    0.100   1000   3   12    2.0   1.0   2.00   2.00   144    1740    0.10000V   3.80792241   UNIFORM   P-AXIS
14    0.200   1000   3   12    2.0   1.0   2.00   2.00    48     588    0.05000V   3.85201064   UNIFORM   P-AXIS
15    0.300   1000   3   12    2.0   1.0   2.00   2.00   145    1752    0.15000V   3.83268353   UNIFORM   P-AXIS
16    0.400   1000   3   12    2.0   1.0   2.00   2.00   152    1836    0.50000V   3.86258768   UNIFORM   P-AXIS
17    0.500   1000   3   12    2.0   1.0   2.00   2.00    77     936    0.55000V   3.85073825   UNIFORM   P-AXIS
18    0.600   1000   3   12    2.0   1.0   2.00   2.00    49     600    0.15000V   3.83938507   UNIFORM   P-AXIS
19    0.700   1000   3   12    2.0   1.0   2.00   2.00    35     432    0.35000V   3.80501343   UNIFORM   P-AXIS
20    0.800   1000   3   12    2.0   1.0   2.00   2.00    35     432    0.35000V   3.84711642   UNIFORM   P-AXIS
21    0.900   1000   3   12    2.0   1.0   2.00   2.00    85     600    0.95000V   3.84645830   UNIFORM   P-AXIS
22    1.000   1000   3   12    2.0   1.0   2.00   2.00    49     600    0.10000V   3.70557750   UNIFORM   P-AXIS
23    0.000   1000   3   18    2.0   1.0   2.00   2.00    95    1728    0.50000V   3.80693408   UNIFORM   P-AXIS
24    0.100   1000   3   18    2.0   1.0   2.00   2.00    66    1206    0.95000V   3.84901551   UNIFORM   P-AXIS
25    0.200   1000   3   18    2.0   1.0   2.00   2.00   135    2448    0.60000V   3.84617586   UNIFORM   P-AXIS
26    0.300   1000   3   18    2.0   1.0   2.00   2.00    50     918    0.15000V   3.85932604   UNIFORM   P-AXIS
27    0.400   1000   3   18    2.0   1.0   2.00   2.00    69    1260    0.15000V   3.83458308   UNIFORM   P-AXIS
28    0.500   1000   3   18    2.0   1.0   2.00   2.00    52     954    0.25000V   3.85075229   UNIFORM   P-AXIS
29    0.600   1000   3   18    2.0   1.0   2.00   2.00    50     918    0.15000V   3.84323855   UNIFORM   P-AXIS
30    0.700   1000   3   18    2.0   1.0   2.00   2.00    35     648    0.35000V   3.84577666   UNIFORM   P-AXIS
31    0.800   1000   3   18    2.0   1.0   2.00   2.00    71    1296    0.25000V   3.80442837   UNIFORM   P-AXIS
32    0.900   1000   3   18    2.0   1.0   2.00   2.00   147    2664    0.25000V   3.84347708   UNIFORM   P-AXIS
33    1.000   1000   3   18    2.0   1.0   2.00   2.00    64    1170    0.85000V   3.81361784   UNIFORM   P-AXIS
34    0.000   1000   3   24    2.0   1.0   2.00   2.00    75    1824    0.45000V   3.77931411   UNIFORM   P-AXIS
35    0.100   1000   3   24    2.0   1.0   2.00   2.00    35     864    0.35000V   3.85575532   UNIFORM   P-AXIS
36    0.200   1000   3   24    2.0   1.0   2.00   2.00    35     864    0.35000V   3.85827634   UNIFORM   P-AXIS
37    0.300   1000   3   24    2.0   1.0   2.00   2.00    69    1680    0.15000V   3.86018061   UNIFORM   P-AXIS
38    0.400   1000   3   24    2.0   1.0   2.00   2.00    36     888    0.40000V   3.80982656   UNIFORM   P-AXIS
39    0.500   1000   3   24    2.0   1.0   2.00   2.00    35     864    0.35000V   3.85363011   UNIFORM   P-AXIS
40    0.600   1000   3   24    2.0   1.0   2.00   2.00    67    1632    0.05000V   3.81143976   UNIFORM   P-AXIS
41    0.700   1000   3   24    2.0   1.0   2.00   2.00    50    1224    0.15000V   3.84748287   UNIFORM   P-AXIS
42    0.800   1000   3   24    2.0   1.0   2.00   2.00    48    1176    0.05000V   3.85396008   UNIFORM   P-AXIS
43    0.900   1000   3   24    2.0   1.0   2.00   2.00    66    1608    0.95000V   3.85520640   UNIFORM   P-AXIS
44    1.000   1000   3   24    2.0   1.0   2.00   2.00    68    1656    0.10000V   3.85818968   UNIFORM   P-AXIS
45    0.000   1000   3   30    2.0   1.0   2.00   2.00    95    2880    0.50000V   3.84325797   UNIFORM   P-AXIS
46    0.100   1000   3   30    2.0   1.0   2.00   2.00   147    4440    0.25000V   3.85589860   UNIFORM   P-AXIS
47    0.200   1000   3   30    2.0   1.0   2.00   2.00    56    1710    0.45000V   3.82667121   UNIFORM   P-AXIS
48    0.300   1000   3   30    2.0   1.0   2.00   2.00    53    1590    0.25000V   3.85474673   UNIFORM   P-AXIS
49    0.400   1000   3   30    2.0   1.0   2.00   2.00    35    1080    0.35000V   3.86095502   UNIFORM   P-AXIS
50    0.500   1000   3   30    2.0   1.0   2.00   2.00    35    1080    0.35000V   3.83565648   UNIFORM   P-AXIS
51    0.600   1000   3   30    2.0   1.0   2.00   2.00   125    3780    0.10000V   3.81607827   UNIFORM   P-AXIS
52    0.700   1000   3   30    2.0   1.0   2.00   2.00    83    2070    0.10000V   3.87335680   UNIFORM   P-AXIS
53    0.800   1000   3   30    2.0   1.0   2.00   2.00   140    4230    0.85000V   3.83573357   UNIFORM   P-AXIS
54    0.900   1000   3   30    2.0   1.0   2.00   2.00    35    1080    0.35000V   3.85079537   UNIFORM   P-AXIS
55    1.000   1000   3   30    2.0   1.0   2.00   2.00   123    3720    0.95000V   3.85372578   UNIFORM   P-AXIS
56    0.000   1000   3   36    2.0   1.0   2.00   2.00    35    1296    0.35000V   3.84685849   UNIFORM   P-AXIS
57    0.100   1000   3   36    2.0   1.0   2.00   2.00    35    1296    0.35000V   3.85060158   UNIFORM   P-AXIS
58    0.200   1000   3   36    2.0   1.0   2.00   2.00    35    1296    0.35000V   3.84095115   UNIFORM   P-AXIS
59    0.300   1000   3   36    2.0   1.0   2.00   2.00    73    2664    0.35000V   3.84583259   UNIFORM   P-AXIS
60    0.400   1000   3   36    2.0   1.0   2.00   2.00    69    2520    0.15000V   3.84601948   UNIFORM   P-AXIS
61    0.500   1000   3   36    2.0   1.0   2.00   2.00    76    2772    0.50000V   3.84331442   UNIFORM   P-AXIS
62    0.600   1000   3   36    2.0   1.0   2.00   2.00    80    2916    0.70000V   3.85148128   UNIFORM   P-AXIS
63    0.700   1000   3   36    2.0   1.0   2.00   2.00    67    2448    0.35000V   3.85311094   UNIFORM   P-AXIS
64    0.800   1000   3   36    2.0   1.0   2.00   2.00    70    2556    0.20000V   3.85031329   UNIFORM   P-AXIS
65    0.900   1000   3   36    2.0   1.0   2.00   2.00    35    1296    0.35000V   3.84613483   UNIFORM   P-AXIS
66    1.000   1000   3   36    2.0   1.0   2.00   2.00   116    4176    0.55000V   3.79798510   UNIFORM   P-AXIS
67    0.100   1000   3   36    2.0   1.0   2.00   2.00    53    2268    0.30000V   3.84343914   UNIFORM   P-AXIS
68    0.100   1000   3   42    2.0   1.0   2.00   2.00    71    3024    0.25000V   3.83050298   UNIFORM   P-AXIS
69    0.200   1000   3   42    2.0   1.0   2.00   2.00    49    2100    0.10000V   3.86268376   UNIFORM   P-AXIS
70    0.300   1000   3   42    2.0   1.0   2.00   2.00    75    3234    0.50000V   3.85634646   UNIFORM   P-AXIS
71    0.400   1000   3   42    2.0   1.0   2.00   2.00   130    5502    0.35000V   3.83140283   UNIFORM   P-AXIS
72    0.500   1000   3   42    2.0   1.0   2.00   2.00    35    1512    0.35000V   3.85432960   UNIFORM   P-AXIS
73    0.600   1000   3   42    2.0   1.0   2.00   2.00    35    1512    0.35000V   3.82781037   UNIFORM   P-AXIS
74    0.700   1000   3   42    2.0   1.0   2.00   2.00    35    1512    0.35000V   3.85431079   UNIFORM   P-AXIS
75    0.800   1000   3   42    2.0   1.0   2.00   2.00    68    2898    0.10000V   3.85455250   UNIFORM   P-AXIS
76    0.900   1000   3   42    2.0   1.0   2.00   2.00    51    2184    0.20000V   3.85186330   UNIFORM   P-AXIS
77    1.000   1000   3   42    2.0   1.0   2.00   2.00   135    5712    0.60000V   3.84463440   UNIFORM   P-AXIS

                              Total Function Evaluations:   128286
---------------------------------------------------------------------------------------------
69    0.200   1000   3   42    2.0   1.0   2.00   2.00    49    2100    0.10000V   3.86268376   UNIFORM   P-AXIS
```

## F20

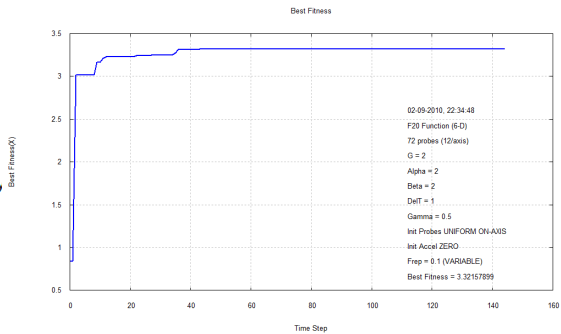

F20
Best Fitness = 3.32157899 returned by
Probe # 28 at Time Step 144

P28 coordinates:
1   .20030625
2   .15122217
3   .47422158
4   .27665257
5   .31191696
6   .65605313

Best Fitness

02-09-2010, 22:34:48
F20 Function (6-D)
72 probes (12/axis)
G = 2
Alpha = 2
Beta = 2
DelT = 1
Gamma = 0.5
Init Probes UNIFORM ON-AXIS
Init Accel ZERO
Frep = 0.1 (VARIABLE)
Best Fitness = 3.32157899

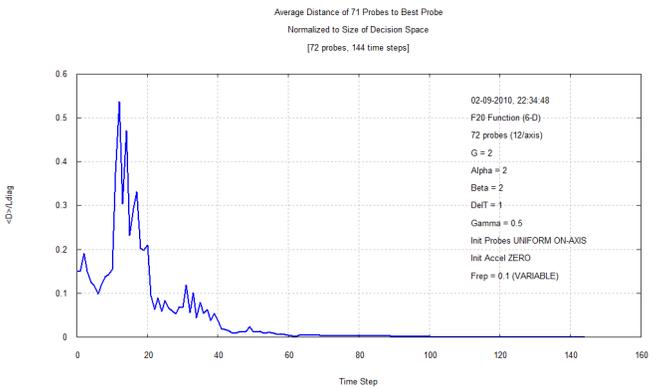

Average Distance of 71 Probes to Best Probe
Normalized to Size of Decision Space
[72 probes, 144 time steps]

02-09-2010, 22:34:48
F20 Function (6-D)
72 probes (12/axis)
G = 2
Alpha = 2
Beta = 2
DelT = 1
Gamma = 0.5
Init Probes UNIFORM ON-AXIS
Init Accel ZERO
Freq = 0.1 (VARIABLE)

Run ID: 02-09-2010, 22:34:4 UNCTION: F20

| Run # | Gamma | Nt | Nd | Np | G | DelT | Alpha | Beta | #Steps | Neval | Frep | Fitness | Initial Probes |
|-------|-------|----|----|----|---|------|-------|------|--------|-------|------|---------|----------------|
| 0 | 0.000 | 1000 | 6 | 24 | 2.0 | 1.0 | 2.00 | 2.00 | 0 | 0 | 0.50000V | -9999.0000000 | UNIFORM  P-AXIS |
| 1 | 0.000 | 1000 | 6 | 12 | 2.0 | 1.0 | 2.00 | 2.00 | 176 | 2124 | 0.75000V | 3.00110692 | UNIFORM  P-AXIS |
| 2 | 0.100 | 1000 | 6 | 12 | 2.0 | 1.0 | 2.00 | 2.00 | 113 | 1368 | 0.45000V | 3.26644185 | UNIFORM  P-AXIS |
| 3 | 0.200 | 1000 | 6 | 12 | 2.0 | 1.0 | 2.00 | 2.00 | 96 | 1164 | 0.55000V | 3.29273975 | UNIFORM P-AXIS |
| 4 | 0.300 | 1000 | 6 | 12 | 2.0 | 1.0 | 2.00 | 2.00 | 50 | 612 | 0.15000V | 3.30673597 | UNIFORM P-AXIS |
| 5 | 0.400 | 1000 | 6 | 12 | 2.0 | 1.0 | 2.00 | 2.00 | 124 | 1500 | 0.05000V | 3.21084307 | UNIFORM P-AXIS |
| 6 | 0.500 | 1000 | 6 | 12 | 2.0 | 1.0 | 2.00 | 2.00 | 122 | 1476 | 0.90000V | 3.22812173 | UNIFORM P-AXIS |
| 7 | 0.600 | 1000 | 6 | 12 | 2.0 | 1.0 | 2.00 | 2.00 | 125 | 1512 | 0.10000V | 3.10029116 | UNIFORM P-AXIS |
| 8 | 0.700 | 1000 | 6 | 12 | 2.0 | 1.0 | 2.00 | 2.00 | 108 | 1308 | 0.20000V | 3.12325233 | UNIFORM P-AXIS |
| 9 | 0.800 | 1000 | 6 | 12 | 2.0 | 1.0 | 2.00 | 2.00 | 86 | 1044 | 0.05000V | 0.28390268 | UNIFORM P-AXIS |
| 10 | 0.900 | 1000 | 6 | 12 | 2.0 | 1.0 | 2.00 | 2.00 | 150 | 1812 | 0.40000V | 0.96018454 | UNIFORM P-AXIS |
| 11 | 1.000 | 1000 | 6 | 12 | 2.0 | 1.0 | 2.00 | 2.00 | 125 | 1512 | 0.10000V | 3.03486357 | UNIFORM P-AXIS |
| 12 | 0.000 | 1000 | 6 | 24 | 2.0 | 1.0 | 2.00 | 2.00 | 80 | 1944 | 0.70000V | 3.25492906 | UNIFORM P-AXIS |
| 13 | 0.100 | 1000 | 6 | 24 | 2.0 | 1.0 | 2.00 | 2.00 | 92 | 2232 | 0.35000V | 3.30964348 | UNIFORM P-AXIS |
| 14 | 0.200 | 1000 | 6 | 24 | 2.0 | 1.0 | 2.00 | 2.00 | 105 | 2544 | 0.05000V | 3.31490965 | UNIFORM P-AXIS |
| 15 | 0.300 | 1000 | 6 | 24 | 2.0 | 1.0 | 2.00 | 2.00 | 96 | 2328 | 0.55000V | 3.29714583 | UNIFORM P-AXIS |
| 16 | 0.400 | 1000 | 6 | 24 | 2.0 | 1.0 | 2.00 | 2.00 | 104 | 2520 | 0.95000V | 3.31350060 | UNIFORM P-AXIS |
| 17 | 0.500 | 1000 | 6 | 24 | 2.0 | 1.0 | 2.00 | 2.00 | 72 | 1752 | 0.30000V | 3.31924736 | UNIFORM P-AXIS |
| 18 | 0.600 | 1000 | 6 | 24 | 2.0 | 1.0 | 2.00 | 2.00 | 105 | 2544 | 0.05000V | 3.28697060 | UNIFORM P-AXIS |
| 19 | 0.700 | 1000 | 6 | 24 | 2.0 | 1.0 | 2.00 | 2.00 | 77 | 1872 | 0.55000V | 3.13461051 | UNIFORM P-AXIS |
| 20 | 0.800 | 1000 | 6 | 24 | 2.0 | 1.0 | 2.00 | 2.00 | 114 | 2760 | 0.50000V | 3.24306717 | UNIFORM P-AXIS |
| 21 | 0.900 | 1000 | 6 | 24 | 2.0 | 1.0 | 2.00 | 2.00 | 129 | 3120 | 0.30000V | 1.22812569 | UNIFORM P-AXIS |
| 22 | 1.000 | 1000 | 6 | 24 | 2.0 | 1.0 | 2.00 | 2.00 | 82 | 1992 | 0.80000V | 3.29883050 | UNIFORM P-AXIS |
| 23 | 0.000 | 1000 | 6 | 36 | 2.0 | 1.0 | 2.00 | 2.00 | 114 | 4140 | 0.50000V | 3.31647943 | UNIFORM P-AXIS |
| 24 | 0.100 | 1000 | 6 | 36 | 2.0 | 1.0 | 2.00 | 2.00 | 121 | 4392 | 0.85000V | 3.31670159 | UNIFORM P-AXIS |
| 25 | 0.200 | 1000 | 6 | 36 | 2.0 | 1.0 | 2.00 | 2.00 | 155 | 5616 | 0.65000V | 3.31081731 | UNIFORM P-AXIS |
| 26 | 0.300 | 1000 | 6 | 36 | 2.0 | 1.0 | 2.00 | 2.00 | 35 | 1296 | 0.35000V | 3.27857040 | UNIFORM P-AXIS |
| 27 | 0.400 | 1000 | 6 | 36 | 2.0 | 1.0 | 2.00 | 2.00 | 87 | 3168 | 0.10000V | 3.29204492 | UNIFORM P-AXIS |
| 28 | 0.500 | 1000 | 6 | 36 | 2.0 | 1.0 | 2.00 | 2.00 | 95 | 3456 | 0.50000V | 3.31925222 | UNIFORM P-AXIS |
| 29 | 0.600 | 1000 | 6 | 36 | 2.0 | 1.0 | 2.00 | 2.00 | 96 | 3492 | 0.55000V | 3.08535746 | UNIFORM P-AXIS |
| 30 | 0.700 | 1000 | 6 | 36 | 2.0 | 1.0 | 2.00 | 2.00 | 135 | 4896 | 0.60000V | 3.12242034 | UNIFORM P-AXIS |
| 31 | 0.800 | 1000 | 6 | 36 | 2.0 | 1.0 | 2.00 | 2.00 | 95 | 3456 | 0.50000V | 3.31063202 | UNIFORM P-AXIS |
| 32 | 0.900 | 1000 | 6 | 36 | 2.0 | 1.0 | 2.00 | 2.00 | 75 | 2736 | 0.45000V | 0.75036445 | UNIFORM P-AXIS |
| 33 | 1.000 | 1000 | 6 | 36 | 2.0 | 1.0 | 2.00 | 2.00 | 150 | 5436 | 0.40000V | 3.28899932 | UNIFORM P-AXIS |
| 34 | 0.000 | 1000 | 6 | 48 | 2.0 | 1.0 | 2.00 | 2.00 | 135 | 6528 | 0.40000V | 3.31988479 | UNIFORM P-AXIS |
| 35 | 0.100 | 1000 | 6 | 48 | 2.0 | 1.0 | 2.00 | 2.00 | 103 | 4992 | 0.90000V | 3.31852743 | UNIFORM P-AXIS |
| 36 | 0.200 | 1000 | 6 | 48 | 2.0 | 1.0 | 2.00 | 2.00 | 85 | 4128 | 0.95000V | 3.31735441 | UNIFORM P-AXIS |
| 37 | 0.300 | 1000 | 6 | 48 | 2.0 | 1.0 | 2.00 | 2.00 | 35 | 1728 | 0.35000V | 3.25677057 | UNIFORM P-AXIS |
| 38 | 0.400 | 1000 | 6 | 48 | 2.0 | 1.0 | 2.00 | 2.00 | 56 | 2736 | 0.45000V | 3.31505600 | UNIFORM P-AXIS |
| 39 | 0.500 | 1000 | 6 | 48 | 2.0 | 1.0 | 2.00 | 2.00 | 150 | 7248 | 0.40000V | 3.31941449 | UNIFORM P-AXIS |
| 40 | 0.600 | 1000 | 6 | 48 | 2.0 | 1.0 | 2.00 | 2.00 | 129 | 6240 | 0.30000V | 3.28492664 | UNIFORM P-AXIS |
| 41 | 0.700 | 1000 | 6 | 48 | 2.0 | 1.0 | 2.00 | 2.00 | 87 | 4224 | 0.10000V | 3.17021423 | UNIFORM P-AXIS |
| 42 | 0.800 | 1000 | 6 | 48 | 2.0 | 1.0 | 2.00 | 2.00 | 71 | 3456 | 0.25000V | 3.31924942 | UNIFORM P-AXIS |
| 43 | 0.900 | 1000 | 6 | 48 | 2.0 | 1.0 | 2.00 | 2.00 | 113 | 5472 | 0.45000V | 0.35100320 | UNIFORM P-AXIS |
| 44 | 1.000 | 1000 | 6 | 48 | 2.0 | 1.0 | 2.00 | 2.00 | 99 | 4800 | 0.70000V | 3.16154131 | UNIFORM P-AXIS |
| 45 | 0.000 | 1000 | 6 | 60 | 2.0 | 1.0 | 2.00 | 2.00 | 167 | 10080 | 0.35000V | 3.31221066 | UNIFORM P-AXIS |
| 46 | 0.100 | 1000 | 6 | 60 | 2.0 | 1.0 | 2.00 | 2.00 | 104 | 6300 | 0.95000V | 3.31834692 | UNIFORM P-AXIS |
| 47 | 0.200 | 1000 | 6 | 60 | 2.0 | 1.0 | 2.00 | 2.00 | 127 | 7680 | 0.20000V | 3.31302815 | UNIFORM P-AXIS |
| 48 | 0.300 | 1000 | 6 | 60 | 2.0 | 1.0 | 2.00 | 2.00 | 135 | 8160 | 0.60000V | 3.31978776 | UNIFORM P-AXIS |
| 49 | 0.400 | 1000 | 6 | 60 | 2.0 | 1.0 | 2.00 | 2.00 | 35 | 2160 | 0.35000V | 3.28507694 | UNIFORM P-AXIS |
| 50 | 0.500 | 1000 | 6 | 60 | 2.0 | 1.0 | 2.00 | 2.00 | 111 | 6720 | 0.35000V | 3.30685529 | UNIFORM P-AXIS |
| 51 | 0.600 | 1000 | 6 | 60 | 2.0 | 1.0 | 2.00 | 2.00 | 128 | 7740 | 0.25000V | 3.31089479 | UNIFORM P-AXIS |
| 52 | 0.700 | 1000 | 6 | 60 | 2.0 | 1.0 | 2.00 | 2.00 | 126 | 7620 | 0.15000V | 3.16403629 | UNIFORM P-AXIS |
| 53 | 0.800 | 1000 | 6 | 60 | 2.0 | 1.0 | 2.00 | 2.00 | 105 | 6360 | 0.05000V | 3.31829960 | UNIFORM P-AXIS |
| 54 | 0.900 | 1000 | 6 | 60 | 2.0 | 1.0 | 2.00 | 2.00 | 73 | 4440 | 0.35000V | 0.61425506 | UNIFORM P-AXIS |
| 55 | 1.000 | 1000 | 6 | 60 | 2.0 | 1.0 | 2.00 | 2.00 | 149 | 9000 | 0.35000V | 3.28106954 | UNIFORM P-AXIS |
| 56 | 0.000 | 1000 | 6 | 72 | 2.0 | 1.0 | 2.00 | 2.00 | 96 | 6984 | 0.55000V | 3.30940732 | UNIFORM P-AXIS |
| 57 | 0.100 | 1000 | 6 | 72 | 2.0 | 1.0 | 2.00 | 2.00 | 134 | 9720 | 0.70000V | 3.31046899 | UNIFORM P-AXIS |
| 58 | 0.200 | 1000 | 6 | 72 | 2.0 | 1.0 | 2.00 | 2.00 | 125 | 9072 | 0.10000V | 3.31326150 | UNIFORM P-AXIS |
| 59 | 0.300 | 1000 | 6 | 72 | 2.0 | 1.0 | 2.00 | 2.00 | 124 | 9000 | 0.05000V | 3.32047139 | UNIFORM P-AXIS |
| 60 | 0.400 | 1000 | 6 | 72 | 2.0 | 1.0 | 2.00 | 2.00 | 35 | 2592 | 0.35000V | 3.23213556 | UNIFORM P-AXIS |
| <span style="color:red">61</span> | <span style="color:red">0.500</span> | <span style="color:red">1000</span> | <span style="color:red">6</span> | <span style="color:red">72</span> | <span style="color:red">2.0</span> | <span style="color:red">1.0</span> | <span style="color:red">2.00</span> | <span style="color:red">2.00</span> | <span style="color:red">144</span> | <span style="color:red">10440</span> | <span style="color:red">0.10000V</span> | <span style="color:red">3.32157899</span> | <span style="color:red">UNIFORM P-AXIS</span> |
| 62 | 0.600 | 1000 | 6 | 72 | 2.0 | 1.0 | 2.00 | 2.00 | 105 | 7632 | 0.05000V | 3.31735712 | UNIFORM P-AXIS |
| 63 | 0.700 | 1000 | 6 | 72 | 2.0 | 1.0 | 2.00 | 2.00 | 174 | 12672 | 0.70000V | 3.17826324 | UNIFORM P-AXIS |
| 64 | 0.800 | 1000 | 6 | 72 | 2.0 | 1.0 | 2.00 | 2.00 | 124 | 9000 | 0.05000V | 3.32017848 | UNIFORM P-AXIS |
| 65 | 0.900 | 1000 | 6 | 72 | 2.0 | 1.0 | 2.00 | 2.00 | 121 | 8784 | 0.85000V | 3.15497318 | UNIFORM P-AXIS |
| 66 | 1.000 | 1000 | 6 | 72 | 2.0 | 1.0 | 2.00 | 2.00 | 115 | 8352 | 0.55000V | 3.30828208 | UNIFORM P-AXIS |
| 67 | 0.000 | 1000 | 6 | 84 | 2.0 | 1.0 | 2.00 | 2.00 | 121 | 10248 | 0.85000V | 3.31957793 | UNIFORM P-AXIS |
| 68 | 0.100 | 1000 | 6 | 84 | 2.0 | 1.0 | 2.00 | 2.00 | 98 | 8316 | 0.65000V | 3.31688022 | UNIFORM P-AXIS |
| 69 | 0.200 | 1000 | 6 | 84 | 2.0 | 1.0 | 2.00 | 2.00 | 125 | 10584 | 0.10000V | 3.30946639 | UNIFORM P-AXIS |
| 70 | 0.300 | 1000 | 6 | 84 | 2.0 | 1.0 | 2.00 | 2.00 | 68 | 5796 | 0.10000V | 3.32143938 | UNIFORM P-AXIS |
| 71 | 0.400 | 1000 | 6 | 84 | 2.0 | 1.0 | 2.00 | 2.00 | 124 | 10500 | 0.05000V | 3.31975870 | UNIFORM P-AXIS |
| 72 | 0.500 | 1000 | 6 | 84 | 2.0 | 1.0 | 2.00 | 2.00 | 105 | 8904 | 0.05000V | 3.30609285 | UNIFORM P-AXIS |
| 73 | 0.600 | 1000 | 6 | 84 | 2.0 | 1.0 | 2.00 | 2.00 | 98 | 8316 | 0.65000V | 3.30492736 | UNIFORM P-AXIS |
| 74 | 0.700 | 1000 | 6 | 84 | 2.0 | 1.0 | 2.00 | 2.00 | 111 | 9408 | 0.35000V | 3.16684849 | UNIFORM P-AXIS |
| 75 | 0.800 | 1000 | 6 | 84 | 2.0 | 1.0 | 2.00 | 2.00 | 130 | 11004 | 0.35000V | 3.30406021 | UNIFORM P-AXIS |
| 76 | 0.900 | 1000 | 6 | 84 | 2.0 | 1.0 | 2.00 | 2.00 | 135 | 11424 | 0.60000V | 3.30049716 | UNIFORM P-AXIS |
| 77 | 1.000 | 1000 | 6 | 84 | 2.0 | 1.0 | 2.00 | 2.00 | 99 | 8400 | 0.70000V | 3.31805211 | UNIFORM P-AXIS |

Total Function Evaluations:  408084

<span style="color:red">61  0.500  1000  6  72  2.0  1.0  2.00  2.00  144  10440  0.10000V  3.32157899  UNIFORM P-AXIS</span>

## F21

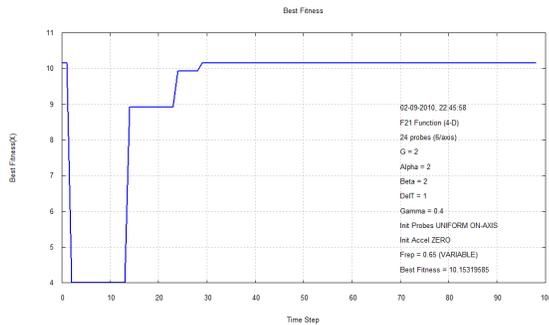

F21
Best Fitness = 10.15319585 returned by
Probe # 21 at Time Step 1

P21 coordinates:
1  4
2  4
3  4
4  4

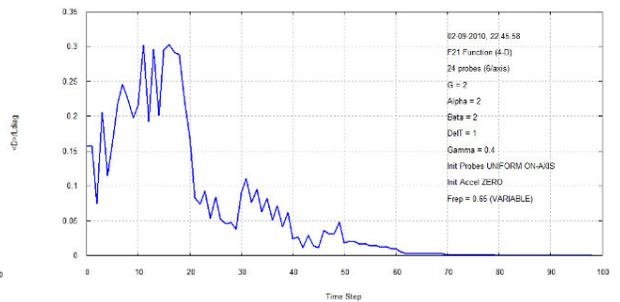

| Run # | Gamma | Nt | Nd | Np | G | DelT | Alpha | Beta | #Steps | Neval | Frep | Fitness | Initial Probes | |
|---|---|---|---|---|---|---|---|---|---|---|---|---|---|---|
| 0 | 0.000 | 1000 | 4 | 16 | 2.0 | 1.0 | 2.00 | 2.00 | 0 | 0 | 0.50000V | -9999.00000000 | UNIFORM | P-AXIS |
| 1 | 0.000 | 1000 | 4 | 8 | 2.0 | 1.0 | 2.00 | 2.00 | 35 | 288 | 0.35000V | 0.27311534 | UNIFORM | P-AXIS |
| 2 | 0.100 | 1000 | 4 | 8 | 2.0 | 1.0 | 2.00 | 2.00 | 84 | 680 | 0.90000V | 0.04992953 | UNIFORM | P-AXIS |
| 3 | 0.200 | 1000 | 4 | 8 | 2.0 | 1.0 | 2.00 | 2.00 | 75 | 608 | 0.45000V | 0.43005482 | UNIFORM | P-AXIS |
| 4 | 0.300 | 1000 | 4 | 8 | 2.0 | 1.0 | 2.00 | 2.00 | 95 | 768 | 0.50000V | 0.61032898 | UNIFORM | P-AXIS |
| 5 | 0.400 | 1000 | 4 | 8 | 2.0 | 1.0 | 2.00 | 2.00 | 105 | 848 | 0.05000V | 10.15239952 | UNIFORM | P-AXIS |
| 6 | 0.500 | 1000 | 4 | 8 | 2.0 | 1.0 | 2.00 | 2.00 | 85 | 688 | 0.95000V | 0.59797768 | UNIFORM | P-AXIS |
| 7 | 0.600 | 1000 | 4 | 8 | 2.0 | 1.0 | 2.00 | 2.00 | 82 | 664 | 0.80000V | 2.66392078 | UNIFORM | P-AXIS |
| 8 | 0.700 | 1000 | 4 | 8 | 2.0 | 1.0 | 2.00 | 2.00 | 105 | 848 | 0.05000V | 0.55777866 | UNIFORM | P-AXIS |
| 9 | 0.800 | 1000 | 4 | 8 | 2.0 | 1.0 | 2.00 | 2.00 | 105 | 848 | 0.05000V | 5.09983295 | UNIFORM | P-AXIS |
| 10 | 0.900 | 1000 | 4 | 8 | 2.0 | 1.0 | 2.00 | 2.00 | 85 | 688 | 0.95000V | 0.6758059 | UNIFORM | P-AXIS |
| 11 | 1.000 | 1000 | 4 | 8 | 2.0 | 1.0 | 2.00 | 2.00 | 125 | 1008 | 0.10000V | 5.01275110 | UNIFORM | P-AXIS |
| 12 | 0.000 | 1000 | 4 | 16 | 2.0 | 1.0 | 2.00 | 2.00 | 35 | 576 | 0.35000V | 0.27311534 | UNIFORM | P-AXIS |
| 13 | 0.100 | 1000 | 4 | 16 | 2.0 | 1.0 | 2.00 | 2.00 | 96 | 1552 | 0.55000V | 5.05477755 | UNIFORM | P-AXIS |
| 14 | 0.200 | 1000 | 4 | 16 | 2.0 | 1.0 | 2.00 | 2.00 | 75 | 1216 | 0.45000V | 0.43912346 | UNIFORM | P-AXIS |
| 15 | 0.300 | 1000 | 4 | 16 | 2.0 | 1.0 | 2.00 | 2.00 | 76 | 1232 | 0.50000V | 0.66324069 | UNIFORM | P-AXIS |
| 16 | 0.400 | 1000 | 4 | 16 | 2.0 | 1.0 | 2.00 | 2.00 | 110 | 1776 | 0.30000V | 10.15154874 | UNIFORM | P-AXIS |
| 17 | 0.500 | 1000 | 4 | 16 | 2.0 | 1.0 | 2.00 | 2.00 | 85 | 1376 | 0.95000V | 0.61701220 | UNIFORM | P-AXIS |
| 18 | 0.600 | 1000 | 4 | 16 | 2.0 | 1.0 | 2.00 | 2.00 | 118 | 1904 | 0.70000V | 0.15221591 | UNIFORM | P-AXIS |
| 19 | 0.700 | 1000 | 4 | 16 | 2.0 | 1.0 | 2.00 | 2.00 | 35 | 576 | 0.35000V | 0.53087283 | UNIFORM | P-AXIS |
| 20 | 0.800 | 1000 | 4 | 16 | 2.0 | 1.0 | 2.00 | 2.00 | 103 | 1664 | 0.90000V | 5.08959588 | UNIFORM | P-AXIS |
| 21 | 0.900 | 1000 | 4 | 16 | 2.0 | 1.0 | 2.00 | 2.00 | 90 | 1456 | 0.25000V | 0.74990177 | UNIFORM | P-AXIS |
| 22 | 1.000 | 1000 | 4 | 16 | 2.0 | 1.0 | 2.00 | 2.00 | 107 | 1728 | 0.15000V | 10.14746927 | UNIFORM | P-AXIS |
| 23 | 0.000 | 1000 | 4 | 24 | 2.0 | 1.0 | 2.00 | 2.00 | 103 | 2496 | 0.90000V | 4.97572338 | UNIFORM | P-AXIS |
| 24 | 0.100 | 1000 | 4 | 24 | 2.0 | 1.0 | 2.00 | 2.00 | 115 | 2784 | 0.55000V | 5.05471853 | UNIFORM | P-AXIS |
| 25 | 0.200 | 1000 | 4 | 24 | 2.0 | 1.0 | 2.00 | 2.00 | 93 | 2256 | 0.40000V | 10.15252419 | UNIFORM | P-AXIS |
| 26 | 0.300 | 1000 | 4 | 24 | 2.0 | 1.0 | 2.00 | 2.00 | 75 | 1824 | 0.45000V | 2.62584999 | UNIFORM | P-AXIS |
| 27 | 0.400 | 1000 | 4 | 24 | 2.0 | 1.0 | 2.00 | 2.00 | 98 | 2376 | 0.65000V | 10.15319585 | UNIFORM | P-AXIS |
| 28 | 0.500 | 1000 | 4 | 24 | 2.0 | 1.0 | 2.00 | 2.00 | 35 | 864 | 0.35000V | 0.57297331 | UNIFORM | P-AXIS |
| 29 | 0.600 | 1000 | 4 | 24 | 2.0 | 1.0 | 2.00 | 2.00 | 86 | 2088 | 0.05000V | 0.54568877 | UNIFORM | P-AXIS |
| 30 | 0.700 | 1000 | 4 | 24 | 2.0 | 1.0 | 2.00 | 2.00 | 95 | 2304 | 0.50000V | 0.53155269 | UNIFORM | P-AXIS |
| 31 | 0.800 | 1000 | 4 | 24 | 2.0 | 1.0 | 2.00 | 2.00 | 106 | 2568 | 0.10000V | 0.15289253 | UNIFORM | P-AXIS |
| 32 | 0.900 | 1000 | 4 | 24 | 2.0 | 1.0 | 2.00 | 2.00 | 106 | 2568 | 0.10000V | 2.92615580 | UNIFORM | P-AXIS |
| 33 | 1.000 | 1000 | 4 | 24 | 2.0 | 1.0 | 2.00 | 2.00 | 109 | 2640 | 0.25000V | 10.15290252 | UNIFORM | P-AXIS |
| 34 | 0.000 | 1000 | 4 | 32 | 2.0 | 1.0 | 2.00 | 2.00 | 116 | 3744 | 0.60000V | 5.05409964 | UNIFORM | P-AXIS |
| 35 | 0.100 | 1000 | 4 | 32 | 2.0 | 1.0 | 2.00 | 2.00 | 46 | 1504 | 0.90000V | 5.05426392 | UNIFORM | P-AXIS |
| 36 | 0.200 | 1000 | 4 | 32 | 2.0 | 1.0 | 2.00 | 2.00 | 85 | 1472 | 0.80000V | 0.40447200 | UNIFORM | P-AXIS |
| 37 | 0.300 | 1000 | 4 | 32 | 2.0 | 1.0 | 2.00 | 2.00 | 118 | 3808 | 0.70000V | 0.87498816 | UNIFORM | P-AXIS |
| 38 | 0.400 | 1000 | 4 | 32 | 2.0 | 1.0 | 2.00 | 2.00 | 35 | 1152 | 0.30000V | 0.11951712 | UNIFORM | P-AXIS |
| 39 | 0.500 | 1000 | 4 | 32 | 2.0 | 1.0 | 2.00 | 2.00 | 35 | 1152 | 0.35000V | 0.58456509 | UNIFORM | P-AXIS |
| 40 | 0.600 | 1000 | 4 | 32 | 2.0 | 1.0 | 2.00 | 2.00 | 74 | 2400 | 0.40000V | 4.56208143 | UNIFORM | P-AXIS |
| 41 | 0.700 | 1000 | 4 | 32 | 2.0 | 1.0 | 2.00 | 2.00 | 35 | 1152 | 0.35000V | 0.53137912 | UNIFORM | P-AXIS |
| 42 | 0.800 | 1000 | 4 | 32 | 2.0 | 1.0 | 2.00 | 2.00 | 50 | 1632 | 0.15000V | 0.02571078 | UNIFORM | P-AXIS |
| 43 | 0.900 | 1000 | 4 | 32 | 2.0 | 1.0 | 2.00 | 2.00 | 76 | 2464 | 0.50000V | 0.57328180 | UNIFORM | P-AXIS |
| 44 | 1.000 | 1000 | 4 | 32 | 2.0 | 1.0 | 2.00 | 2.00 | 71 | 2304 | 0.25000V | 0.08743193 | UNIFORM | P-AXIS |
| 45 | 0.000 | 1000 | 4 | 40 | 2.0 | 1.0 | 2.00 | 2.00 | 125 | 5040 | 0.10000V | 0.05091124 | UNIFORM | P-AXIS |
| 46 | 0.100 | 1000 | 4 | 40 | 2.0 | 1.0 | 2.00 | 2.00 | 65 | 2640 | 0.90000V | 5.05453431 | UNIFORM | P-AXIS |
| 47 | 0.200 | 1000 | 4 | 40 | 2.0 | 1.0 | 2.00 | 2.00 | 45 | 1840 | 0.85000V | 0.40841261 | UNIFORM | P-AXIS |
| 48 | 0.300 | 1000 | 4 | 40 | 2.0 | 1.0 | 2.00 | 2.00 | 137 | 5520 | 0.70000V | 8.88590310 | UNIFORM | P-AXIS |
| 49 | 0.400 | 1000 | 4 | 40 | 2.0 | 1.0 | 2.00 | 2.00 | 69 | 2800 | 0.15000V | 0.14419268 | UNIFORM | P-AXIS |
| 50 | 0.500 | 1000 | 4 | 40 | 2.0 | 1.0 | 2.00 | 2.00 | 35 | 1440 | 0.35000V | 0.58586398 | UNIFORM | P-AXIS |
| 51 | 0.600 | 1000 | 4 | 40 | 2.0 | 1.0 | 2.00 | 2.00 | 107 | 4320 | 0.15000V | 0.15278578 | UNIFORM | P-AXIS |
| 52 | 0.700 | 1000 | 4 | 40 | 2.0 | 1.0 | 2.00 | 2.00 | 35 | 1440 | 0.35000V | 0.53087283 | UNIFORM | P-AXIS |
| 53 | 0.800 | 1000 | 4 | 40 | 2.0 | 1.0 | 2.00 | 2.00 | 45 | 1840 | 0.85000V | 0.11758227 | UNIFORM | P-AXIS |
| 54 | 0.900 | 1000 | 4 | 40 | 2.0 | 1.0 | 2.00 | 2.00 | 182 | 7320 | 0.10000V | 5.09826513 | UNIFORM | P-AXIS |
| 55 | 1.000 | 1000 | 4 | 40 | 2.0 | 1.0 | 2.00 | 2.00 | 124 | 5000 | 0.05000V | 10.15297599 | UNIFORM | P-AXIS |
| 56 | 0.000 | 1000 | 4 | 48 | 2.0 | 1.0 | 2.00 | 2.00 | 120 | 5808 | 0.80000V | 0.05490710 | UNIFORM | P-AXIS |
| 57 | 0.100 | 1000 | 4 | 48 | 2.0 | 1.0 | 2.00 | 2.00 | 86 | 4176 | 0.05000V | 0.15013272 | UNIFORM | P-AXIS |
| 58 | 0.200 | 1000 | 4 | 48 | 2.0 | 1.0 | 2.00 | 2.00 | 84 | 4080 | 0.90000V | 0.41909769 | UNIFORM | P-AXIS |
| 59 | 0.300 | 1000 | 4 | 48 | 2.0 | 1.0 | 2.00 | 2.00 | 149 | 7200 | 0.35000V | 7.33104820 | UNIFORM | P-AXIS |
| 60 | 0.400 | 1000 | 4 | 48 | 2.0 | 1.0 | 2.00 | 2.00 | 126 | 6096 | 0.15000V | 10.15294403 | UNIFORM | P-AXIS |
| 61 | 0.500 | 1000 | 4 | 48 | 2.0 | 1.0 | 2.00 | 2.00 | 35 | 1728 | 0.35000V | 0.58526200 | UNIFORM | P-AXIS |
| 62 | 0.600 | 1000 | 4 | 48 | 2.0 | 1.0 | 2.00 | 2.00 | 136 | 6576 | 0.65000V | 10.15307253 | UNIFORM | P-AXIS |
| 63 | 0.700 | 1000 | 4 | 48 | 2.0 | 1.0 | 2.00 | 2.00 | 35 | 1728 | 0.35000V | 0.53140202 | UNIFORM | P-AXIS |
| 64 | 0.800 | 1000 | 4 | 48 | 2.0 | 1.0 | 2.00 | 2.00 | 76 | 3696 | 0.50000V | 0.15318435 | UNIFORM | P-AXIS |
| 65 | 0.900 | 1000 | 4 | 48 | 2.0 | 1.0 | 2.00 | 2.00 | 112 | 5424 | 0.05000V | 2.73855600 | UNIFORM | P-AXIS |
| 66 | 1.000 | 1000 | 4 | 48 | 2.0 | 1.0 | 2.00 | 2.00 | 48 | 2352 | 0.05000V | 0.15035294 | UNIFORM | P-AXIS |
| 67 | 0.000 | 1000 | 4 | 56 | 2.0 | 1.0 | 2.00 | 2.00 | 89 | 5040 | 0.20000V | 5.05059878 | UNIFORM | P-AXIS |
| 68 | 0.100 | 1000 | 4 | 56 | 2.0 | 1.0 | 2.00 | 2.00 | 53 | 3024 | 0.30000V | 5.05389514 | UNIFORM | P-AXIS |
| 69 | 0.200 | 1000 | 4 | 56 | 2.0 | 1.0 | 2.00 | 2.00 | 94 | 5320 | 0.45000V | 0.48545567 | UNIFORM | P-AXIS |
| 70 | 0.300 | 1000 | 4 | 56 | 2.0 | 1.0 | 2.00 | 2.00 | 129 | 7280 | 0.30000V | 8.79527986 | UNIFORM | P-AXIS |
| 71 | 0.400 | 1000 | 4 | 56 | 2.0 | 1.0 | 2.00 | 2.00 | 35 | 2016 | 0.35000V | 0.11776011 | UNIFORM | P-AXIS |
| 72 | 0.500 | 1000 | 4 | 56 | 2.0 | 1.0 | 2.00 | 2.00 | 35 | 2016 | 0.35000V | 0.58434919 | UNIFORM | P-AXIS |
| 73 | 0.600 | 1000 | 4 | 56 | 2.0 | 1.0 | 2.00 | 2.00 | 125 | 7056 | 0.15000V | 10.15277405 | UNIFORM | P-AXIS |
| 74 | 0.700 | 1000 | 4 | 56 | 2.0 | 1.0 | 2.00 | 2.00 | 35 | 2016 | 0.35000V | 0.53103979 | UNIFORM | P-AXIS |
| 75 | 0.800 | 1000 | 4 | 56 | 2.0 | 1.0 | 2.00 | 2.00 | 76 | 4312 | 0.50000V | 10.14325960 | UNIFORM | P-AXIS |
| 76 | 0.900 | 1000 | 4 | 56 | 2.0 | 1.0 | 2.00 | 2.00 | 171 | 9632 | 0.05000V | 5.10052005 | UNIFORM | P-AXIS |
| 77 | 1.000 | 1000 | 4 | 56 | 2.0 | 1.0 | 2.00 | 2.00 | 70 | 3976 | 0.20000V | 10.15294129 | UNIFORM | P-AXIS |
| | | | | | | | | | Total Function Evaluations: | | 210296 | | | |
| 27 | 0.400 | 1000 | 4 | 24 | 2.0 | 1.0 | 2.00 | 2.00 | 98 | 2376 | 0.65000V | 10.15319585 | UNIFORM | P-AXIS |

## F22

Best Fitness

Average Distance of 39 Probes to Best Probe
Normalized to Size of Decision Space
[40 probes; 190 time steps]

F22
Best Fitness = 10.4029108 returned by
Probe # 36 at Time Step 181

P36 coordinates:
1    4.00051312
2    4.00060817
3    3.99912463
4    3.99920506

| Run # | Gamma | Nt | Nd | Np | G | DelT | Alpha | Beta | #Steps | Neval | Frep | Fitness | Initial Probes |
|-------|-------|------|----|----|-----|------|-------|------|--------|-------|----------|----------------|-------------|
| 0 | 0.000 | 1000 | 4 | 16 | 2.0 | 1.0 | 2.00 | 2.00 | 0 | 0 | 0.50000V | -9999.00000000 | UNIFORM P-AXIS |
| 1 | 0.000 | 1000 | 4 | 8 | 2.0 | 1.0 | 2.00 | 2.00 | 35 | 288 | 0.35000V | 0.29361829 | UNIFORM P-AXIS |
| 2 | 0.100 | 1000 | 4 | 8 | 2.0 | 1.0 | 2.00 | 2.00 | 145 | 1168 | 0.15000V | 5.08679975 | UNIFORM P-AXIS |
| 3 | 0.200 | 1000 | 4 | 8 | 2.0 | 1.0 | 2.00 | 2.00 | 75 | 608 | 0.45000V | 0.48267098 | UNIFORM P-AXIS |
| 4 | 0.300 | 1000 | 4 | 8 | 2.0 | 1.0 | 2.00 | 2.00 | 95 | 768 | 0.15000V | 0.85439165 | UNIFORM P-AXIS |
| 5 | 0.400 | 1000 | 4 | 8 | 2.0 | 1.0 | 2.00 | 2.00 | 168 | 1352 | 0.35000V | 10.39908269 | UNIFORM P-AXIS |
| 6 | 0.500 | 1000 | 4 | 8 | 2.0 | 1.0 | 2.00 | 2.00 | 75 | 608 | 0.45000V | 0.91841106 | UNIFORM P-AXIS |
| 7 | 0.600 | 1000 | 4 | 8 | 2.0 | 1.0 | 2.00 | 2.00 | 79 | 640 | 0.65000V | 2.70371306 | UNIFORM P-AXIS |
| 8 | 0.700 | 1000 | 4 | 8 | 2.0 | 1.0 | 2.00 | 2.00 | 124 | 1000 | 0.05000V | 0.59760416 | UNIFORM P-AXIS |
| 9 | 0.800 | 1000 | 4 | 8 | 2.0 | 1.0 | 2.00 | 2.00 | 106 | 856 | 0.10000V | 5.11885171 | UNIFORM P-AXIS |
| 10 | 0.900 | 1000 | 4 | 8 | 2.0 | 1.0 | 2.00 | 2.00 | 85 | 688 | 0.95000V | 6.69633673 | UNIFORM P-AXIS |
| 11 | 1.000 | 1000 | 4 | 8 | 2.0 | 1.0 | 2.00 | 2.00 | 75 | 608 | 0.45000V | 5.08234051 | UNIFORM P-AXIS |
| 12 | 0.000 | 1000 | 4 | 16 | 2.0 | 1.0 | 2.00 | 2.00 | 75 | 608 | 0.35000V | 0.29361829 | UNIFORM P-AXIS |
| 13 | 0.100 | 1000 | 4 | 16 | 2.0 | 1.0 | 2.00 | 2.00 | 88 | 1424 | 0.15000V | 5.07610809 | UNIFORM P-AXIS |
| 14 | 0.200 | 1000 | 4 | 16 | 2.0 | 1.0 | 2.00 | 2.00 | 75 | 1216 | 0.45000V | 0.48192089 | UNIFORM P-AXIS |
| 15 | 0.300 | 1000 | 4 | 16 | 2.0 | 1.0 | 2.00 | 2.00 | 109 | 1760 | 0.25000V | 1.27391214 | UNIFORM P-AXIS |
| 16 | 0.400 | 1000 | 4 | 16 | 2.0 | 1.0 | 2.00 | 2.00 | 109 | 1760 | 0.25000V | 10.40160104 | UNIFORM P-AXIS |
| 17 | 0.500 | 1000 | 4 | 16 | 2.0 | 1.0 | 2.00 | 2.00 | 105 | 1696 | 0.05000V | 0.96194302 | UNIFORM P-AXIS |
| 18 | 0.600 | 1000 | 4 | 16 | 2.0 | 1.0 | 2.00 | 2.00 | 106 | 1712 | 0.10000V | 5.12831412 | UNIFORM P-AXIS |
| 19 | 0.700 | 1000 | 4 | 16 | 2.0 | 1.0 | 2.00 | 2.00 | 35 | 576 | 0.35000V | 0.57542458 | UNIFORM P-AXIS |
| 20 | 0.800 | 1000 | 4 | 16 | 2.0 | 1.0 | 2.00 | 2.00 | 53 | 864 | 0.30000V | 5.05132629 | UNIFORM P-AXIS |
| 21 | 0.900 | 1000 | 4 | 16 | 2.0 | 1.0 | 2.00 | 2.00 | 92 | 1488 | 0.35000V | 0.79299630 | UNIFORM P-AXIS |
| 22 | 1.000 | 1000 | 4 | 16 | 2.0 | 1.0 | 2.00 | 2.00 | 126 | 2032 | 0.15000V | 10.40201326 | UNIFORM P-AXIS |
| 23 | 0.000 | 1000 | 4 | 24 | 2.0 | 1.0 | 2.00 | 2.00 | 103 | 2496 | 0.90000V | 5.08532477 | UNIFORM P-AXIS |
| 24 | 0.100 | 1000 | 4 | 24 | 2.0 | 1.0 | 2.00 | 2.00 | 51 | 1248 | 0.20000V | 5.08608697 | UNIFORM P-AXIS |
| 25 | 0.200 | 1000 | 4 | 24 | 2.0 | 1.0 | 2.00 | 2.00 | 90 | 2184 | 0.25000V | 10.40102633 | UNIFORM P-AXIS |
| 26 | 0.300 | 1000 | 4 | 24 | 2.0 | 1.0 | 2.00 | 2.00 | 95 | 2304 | 0.50000V | 2.06725642 | UNIFORM P-AXIS |
| 27 | 0.400 | 1000 | 4 | 24 | 2.0 | 1.0 | 2.00 | 2.00 | 185 | 4464 | 0.25000V | 10.40289926 | UNIFORM P-AXIS |
| 28 | 0.500 | 1000 | 4 | 24 | 2.0 | 1.0 | 2.00 | 2.00 | 100 | 2424 | 0.75000V | 0.90063630 | UNIFORM P-AXIS |
| 29 | 0.600 | 1000 | 4 | 24 | 2.0 | 1.0 | 2.00 | 2.00 | 77 | 1872 | 0.55000V | 10.37643998 | UNIFORM P-AXIS |
| 30 | 0.700 | 1000 | 4 | 24 | 2.0 | 1.0 | 2.00 | 2.00 | 85 | 2064 | 0.95000V | 0.59302699 | UNIFORM P-AXIS |
| 31 | 0.800 | 1000 | 4 | 24 | 2.0 | 1.0 | 2.00 | 2.00 | 66 | 1608 | 0.95000V | 0.40286053 | UNIFORM P-AXIS |
| 32 | 0.900 | 1000 | 4 | 24 | 2.0 | 1.0 | 2.00 | 2.00 | 106 | 2568 | 0.10000V | 2.94874733 | UNIFORM P-AXIS |
| 33 | 1.000 | 1000 | 4 | 24 | 2.0 | 1.0 | 2.00 | 2.00 | 119 | 2880 | 0.75000V | 10.40216100 | UNIFORM P-AXIS |
| 34 | 0.000 | 1000 | 4 | 32 | 2.0 | 1.0 | 2.00 | 2.00 | 105 | 3392 | 0.05000V | 5.08017080 | UNIFORM P-AXIS |
| 35 | 0.100 | 1000 | 4 | 32 | 2.0 | 1.0 | 2.00 | 2.00 | 74 | 2400 | 0.40000V | 10.39334404 | UNIFORM P-AXIS |
| 36 | 0.200 | 1000 | 4 | 32 | 2.0 | 1.0 | 2.00 | 2.00 | 45 | 1472 | 0.85000V | 4.45553914 | UNIFORM P-AXIS |
| 37 | 0.300 | 1000 | 4 | 32 | 2.0 | 1.0 | 2.00 | 2.00 | 78 | 2528 | 0.60000V | 0.86258604 | UNIFORM P-AXIS |
| 38 | 0.400 | 1000 | 4 | 32 | 2.0 | 1.0 | 2.00 | 2.00 | 35 | 1152 | 0.35000V | 10.35599982 | UNIFORM P-AXIS |
| 39 | 0.500 | 1000 | 4 | 32 | 2.0 | 1.0 | 2.00 | 2.00 | 105 | 3392 | 0.05000V | 0.85988903 | UNIFORM P-AXIS |
| 40 | 0.600 | 1000 | 4 | 32 | 2.0 | 1.0 | 2.00 | 2.00 | 49 | 1600 | 0.10000V | 5.12582418 | UNIFORM P-AXIS |
| 41 | 0.700 | 1000 | 4 | 32 | 2.0 | 1.0 | 2.00 | 2.00 | 35 | 1152 | 0.35000V | 0.57306071 | UNIFORM P-AXIS |
| 42 | 0.800 | 1000 | 4 | 32 | 2.0 | 1.0 | 2.00 | 2.00 | 201 | 6464 | 0.10000V | 10.40251424 | UNIFORM P-AXIS |
| 43 | 0.900 | 1000 | 4 | 32 | 2.0 | 1.0 | 2.00 | 2.00 | 76 | 2464 | 0.50000V | 0.57161391 | UNIFORM P-AXIS |
| 44 | 1.000 | 1000 | 4 | 32 | 2.0 | 1.0 | 2.00 | 2.00 | 95 | 3072 | 0.50000V | 10.40256809 | UNIFORM P-AXIS |
| 45 | 0.000 | 1000 | 4 | 40 | 2.0 | 1.0 | 2.00 | 2.00 | 175 | 7040 | 0.70000V | 5.08765859 | UNIFORM P-AXIS |
| 46 | 0.100 | 1000 | 4 | 40 | 2.0 | 1.0 | 2.00 | 2.00 | 68 | 2760 | 0.10000V | 10.40057107 | UNIFORM P-AXIS |
| 47 | 0.200 | 1000 | 4 | 40 | 2.0 | 1.0 | 2.00 | 2.00 | 75 | 3040 | 0.45000V | 0.46440424 | UNIFORM P-AXIS |
| 48 | 0.300 | 1000 | 4 | 40 | 2.0 | 1.0 | 2.00 | 2.00 | 105 | 4240 | 0.05000V | 1.62906434 | UNIFORM P-AXIS |
| 49 | 0.400 | 1000 | 4 | 40 | 2.0 | 1.0 | 2.00 | 2.00 | 190 | 7640 | 0.50000V | 10.40291080 | UNIFORM P-AXIS |
| 50 | 0.500 | 1000 | 4 | 40 | 2.0 | 1.0 | 2.00 | 2.00 | 122 | 4920 | 0.90000V | 0.84479109 | UNIFORM P-AXIS |
| 51 | 0.600 | 1000 | 4 | 40 | 2.0 | 1.0 | 2.00 | 2.00 | 117 | 4720 | 0.65000V | 10.40276936 | UNIFORM P-AXIS |
| 52 | 0.700 | 1000 | 4 | 40 | 2.0 | 1.0 | 2.00 | 2.00 | 35 | 1440 | 0.35000V | 0.57542458 | UNIFORM P-AXIS |
| 53 | 0.800 | 1000 | 4 | 40 | 2.0 | 1.0 | 2.00 | 2.00 | 50 | 2040 | 0.15000V | 10.40200744 | UNIFORM P-AXIS |
| 54 | 0.900 | 1000 | 4 | 40 | 2.0 | 1.0 | 2.00 | 2.00 | 174 | 7000 | 0.65000V | 5.11523803 | UNIFORM P-AXIS |
| 55 | 1.000 | 1000 | 4 | 40 | 2.0 | 1.0 | 2.00 | 2.00 | 185 | 7440 | 0.25000V | 10.40256326 | UNIFORM P-AXIS |
| 56 | 0.000 | 1000 | 4 | 48 | 2.0 | 1.0 | 2.00 | 2.00 | 119 | 5760 | 0.75000V | 5.08630865 | UNIFORM P-AXIS |
| 57 | 0.100 | 1000 | 4 | 48 | 2.0 | 1.0 | 2.00 | 2.00 | 50 | 2448 | 0.15000V | 5.07238621 | UNIFORM P-AXIS |
| 58 | 0.200 | 1000 | 4 | 48 | 2.0 | 1.0 | 2.00 | 2.00 | 75 | 3648 | 0.45000V | 0.46815144 | UNIFORM P-AXIS |
| 59 | 0.300 | 1000 | 4 | 48 | 2.0 | 1.0 | 2.00 | 2.00 | 111 | 5376 | 0.35000V | 3.61420456 | UNIFORM P-AXIS |
| 60 | 0.400 | 1000 | 4 | 48 | 2.0 | 1.0 | 2.00 | 2.00 | 80 | 3888 | 0.70000V | 10.39348222 | UNIFORM P-AXIS |
| 61 | 0.500 | 1000 | 4 | 48 | 2.0 | 1.0 | 2.00 | 2.00 | 105 | 5088 | 0.05000V | 1.38401242 | UNIFORM P-AXIS |
| 62 | 0.600 | 1000 | 4 | 48 | 2.0 | 1.0 | 2.00 | 2.00 | 188 | 9072 | 0.40000V | 10.40196476 | UNIFORM P-AXIS |
| 63 | 0.700 | 1000 | 4 | 48 | 2.0 | 1.0 | 2.00 | 2.00 | 35 | 1728 | 0.35000V | 0.57429457 | UNIFORM P-AXIS |
| 64 | 0.800 | 1000 | 4 | 48 | 2.0 | 1.0 | 2.00 | 2.00 | 74 | 3600 | 0.40000V | 10.37337226 | UNIFORM P-AXIS |
| 65 | 0.900 | 1000 | 4 | 48 | 2.0 | 1.0 | 2.00 | 2.00 | 155 | 7488 | 0.65000V | 2.71776260 | UNIFORM P-AXIS |
| 66 | 1.000 | 1000 | 4 | 48 | 2.0 | 1.0 | 2.00 | 2.00 | 194 | 9360 | 0.70000V | 10.40290339 | UNIFORM P-AXIS |
| 67 | 0.000 | 1000 | 4 | 56 | 2.0 | 1.0 | 2.00 | 2.00 | 165 | 9296 | 0.20000V | 5.08483250 | UNIFORM P-AXIS |
| 68 | 0.100 | 1000 | 4 | 56 | 2.0 | 1.0 | 2.00 | 2.00 | 127 | 7168 | 0.20000V | 10.40290343 | UNIFORM P-AXIS |
| 69 | 0.200 | 1000 | 4 | 56 | 2.0 | 1.0 | 2.00 | 2.00 | 83 | 4704 | 0.85000V | 0.47792814 | UNIFORM P-AXIS |
| 70 | 0.300 | 1000 | 4 | 56 | 2.0 | 1.0 | 2.00 | 2.00 | 130 | 7336 | 0.35000V | 6.68089941 | UNIFORM P-AXIS |
| 71 | 0.400 | 1000 | 4 | 56 | 2.0 | 1.0 | 2.00 | 2.00 | 110 | 6216 | 0.30000V | 10.40264284 | UNIFORM P-AXIS |
| 72 | 0.500 | 1000 | 4 | 56 | 2.0 | 1.0 | 2.00 | 2.00 | 101 | 5712 | 0.80000V | 1.30969368 | UNIFORM P-AXIS |
| 73 | 0.600 | 1000 | 4 | 56 | 2.0 | 1.0 | 2.00 | 2.00 | 188 | 10584 | 0.40000V | 10.40234703 | UNIFORM P-AXIS |
| 74 | 0.700 | 1000 | 4 | 56 | 2.0 | 1.0 | 2.00 | 2.00 | 35 | 2016 | 0.35000V | 0.57352478 | UNIFORM P-AXIS |
| 75 | 0.800 | 1000 | 4 | 56 | 2.0 | 1.0 | 2.00 | 2.00 | 57 | 3248 | 0.50000V | 10.39187412 | UNIFORM P-AXIS |
| 76 | 0.900 | 1000 | 4 | 56 | 2.0 | 1.0 | 2.00 | 2.00 | 175 | 9856 | 0.70000V | 5.10174780 | UNIFORM P-AXIS |
| 77 | 1.000 | 1000 | 4 | 56 | 2.0 | 1.0 | 2.00 | 2.00 | 35 | 2016 | 0.35000V | 10.38954738 | UNIFORM P-AXIS |

Total Function Evaluations: 256776

| 49 | 0.400 | 1000 | 4 | 40 | 2.0 | 1.0 | 2.00 | 2.00 | 190 | 7640 | 0.50000V | 10.40291080 | UNIFORM P-AXIS |

## F23

F23
Best Fitness = 10.53633734 returned by Probe # 32 at Time Step 192

P32 coordinates:
1   4.00047771
2   4.00127084
3   3.99922443
4   3.99952773

Best Fitness

Average Distance of 47 Probes to Best Probe
Normalized to Size of Decision Space
[48 probes, 192 time steps]

| Run # | Gamma | Nt | Nd | Np | G | DelT | Alpha | Beta | #Steps | Neval | Frep | Fitness | Initial Probes | |
|-------|-------|------|----|----|-----|------|-------|------|--------|-------|----------|----------------|---------|--------|
| 0 | 0.000 | 1000 | 4 | 16 | 2.0 | 1.0 | 2.00 | 2.00 | 0 | 0 | 0.50000V | -9999.00000000 | UNIFORM | P-AXIS |
| 1 | 0.000 | 1000 | 4 | 8 | 2.0 | 1.0 | 2.00 | 2.00 | 35 | 288 | 0.35000V | 0.32172905 | UNIFORM | P-AXIS |
| 2 | 0.100 | 1000 | 4 | 8 | 2.0 | 1.0 | 2.00 | 2.00 | 135 | 1088 | 0.60000V | 5.12378086 | UNIFORM | P-AXIS |
| 3 | 0.200 | 1000 | 4 | 8 | 2.0 | 1.0 | 2.00 | 2.00 | 75 | 608 | 0.45000V | 0.53509181 | UNIFORM | P-AXIS |
| 4 | 0.300 | 1000 | 4 | 8 | 2.0 | 1.0 | 2.00 | 2.00 | 95 | 768 | 0.50000V | 0.97247524 | UNIFORM | P-AXIS |
| 5 | 0.400 | 1000 | 4 | 8 | 2.0 | 1.0 | 2.00 | 2.00 | 155 | 1248 | 0.65000V | 10.53413608 | UNIFORM | P-AXIS |
| 6 | 0.500 | 1000 | 4 | 8 | 2.0 | 1.0 | 2.00 | 2.00 | 75 | 608 | 0.45000V | 1.06454224 | UNIFORM | P-AXIS |
| 7 | 0.600 | 1000 | 4 | 8 | 2.0 | 1.0 | 2.00 | 2.00 | 45 | 368 | 0.85000V | 1.13128949 | UNIFORM | P-AXIS |
| 8 | 0.700 | 1000 | 4 | 8 | 2.0 | 1.0 | 2.00 | 2.00 | 95 | 768 | 0.50000V | 0.65397567 | UNIFORM | P-AXIS |
| 9 | 0.800 | 1000 | 4 | 8 | 2.0 | 1.0 | 2.00 | 2.00 | 96 | 776 | 0.55000V | 5.10657412 | UNIFORM | P-AXIS |
| 10 | 0.900 | 1000 | 4 | 8 | 2.0 | 1.0 | 2.00 | 2.00 | 85 | 688 | 0.95000V | 0.73466659 | UNIFORM | P-AXIS |
| 11 | 1.000 | 1000 | 4 | 8 | 2.0 | 1.0 | 2.00 | 2.00 | 69 | 560 | 0.15000V | 5.12396635 | UNIFORM | P-AXIS |
| 12 | 0.000 | 1000 | 4 | 16 | 2.0 | 1.0 | 2.00 | 2.00 | 35 | 576 | 0.35000V | 0.32172905 | UNIFORM | P-AXIS |
| 13 | 0.100 | 1000 | 4 | 16 | 2.0 | 1.0 | 2.00 | 2.00 | 127 | 2048 | 0.20000V | 10.53615386 | UNIFORM | P-AXIS |
| 14 | 0.200 | 1000 | 4 | 16 | 2.0 | 1.0 | 2.00 | 2.00 | 75 | 1216 | 0.45000V | 0.53147397 | UNIFORM | P-AXIS |
| 15 | 0.300 | 1000 | 4 | 16 | 2.0 | 1.0 | 2.00 | 2.00 | 102 | 1648 | 0.85000V | 1.40214975 | UNIFORM | P-AXIS |
| 16 | 0.400 | 1000 | 4 | 16 | 2.0 | 1.0 | 2.00 | 2.00 | 110 | 1776 | 0.30000V | 10.53568027 | UNIFORM | P-AXIS |
| 17 | 0.500 | 1000 | 4 | 16 | 2.0 | 1.0 | 2.00 | 2.00 | 105 | 1696 | 0.05000V | 1.11059146 | UNIFORM | P-AXIS |
| 18 | 0.600 | 1000 | 4 | 16 | 2.0 | 1.0 | 2.00 | 2.00 | 38 | 624 | 0.50000V | 2.86623321 | UNIFORM | P-AXIS |
| 19 | 0.700 | 1000 | 4 | 16 | 2.0 | 1.0 | 2.00 | 2.00 | 35 | 576 | 0.35000V | 0.65227664 | UNIFORM | P-AXIS |
| 20 | 0.800 | 1000 | 4 | 16 | 2.0 | 1.0 | 2.00 | 2.00 | 125 | 2016 | 0.10000V | 5.17422682 | UNIFORM | P-AXIS |
| 21 | 0.900 | 1000 | 4 | 16 | 2.0 | 1.0 | 2.00 | 2.00 | 112 | 1808 | 0.40000V | 1.06501432 | UNIFORM | P-AXIS |
| 22 | 1.000 | 1000 | 4 | 16 | 2.0 | 1.0 | 2.00 | 2.00 | 78 | 1264 | 0.60000V | 5.12696956 | UNIFORM | P-AXIS |
| 23 | 0.000 | 1000 | 4 | 24 | 2.0 | 1.0 | 2.00 | 2.00 | 145 | 3504 | 0.15000V | 5.08712015 | UNIFORM | P-AXIS |
| 24 | 0.100 | 1000 | 4 | 24 | 2.0 | 1.0 | 2.00 | 2.00 | 115 | 2784 | 0.55000V | 5.12517258 | UNIFORM | P-AXIS |
| 25 | 0.200 | 1000 | 4 | 24 | 2.0 | 1.0 | 2.00 | 2.00 | 44 | 1080 | 0.80000V | 5.10013673 | UNIFORM | P-AXIS |
| 26 | 0.300 | 1000 | 4 | 24 | 2.0 | 1.0 | 2.00 | 2.00 | 104 | 2520 | 0.95000V | 2.54576820 | UNIFORM | P-AXIS |
| 27 | 0.400 | 1000 | 4 | 24 | 2.0 | 1.0 | 2.00 | 2.00 | 88 | 2136 | 0.15000V | 10.53628373 | UNIFORM | P-AXIS |
| 28 | 0.500 | 1000 | 4 | 24 | 2.0 | 1.0 | 2.00 | 2.00 | 95 | 2304 | 0.50000V | 1.04472452 | UNIFORM | P-AXIS |
| 29 | 0.600 | 1000 | 4 | 24 | 2.0 | 1.0 | 2.00 | 2.00 | 110 | 2664 | 0.30000V | 10.53539823 | UNIFORM | P-AXIS |
| 30 | 0.700 | 1000 | 4 | 24 | 2.0 | 1.0 | 2.00 | 2.00 | 105 | 2544 | 0.05000V | 0.71390918 | UNIFORM | P-AXIS |
| 31 | 0.800 | 1000 | 4 | 24 | 2.0 | 1.0 | 2.00 | 2.00 | 73 | 1776 | 0.35000V | 0.53633517 | UNIFORM | P-AXIS |
| 32 | 0.900 | 1000 | 4 | 24 | 2.0 | 1.0 | 2.00 | 2.00 | 169 | 4080 | 0.40000V | 2.95815330 | UNIFORM | P-AXIS |
| 33 | 1.000 | 1000 | 4 | 24 | 2.0 | 1.0 | 2.00 | 2.00 | 129 | 3120 | 0.30000V | 10.53600674 | UNIFORM | P-AXIS |
| 34 | 0.000 | 1000 | 4 | 32 | 2.0 | 1.0 | 2.00 | 2.00 | 114 | 3680 | 0.50000V | 5.12758927 | UNIFORM | P-AXIS |
| 35 | 0.100 | 1000 | 4 | 32 | 2.0 | 1.0 | 2.00 | 2.00 | 87 | 2816 | 0.10000V | 10.53571214 | UNIFORM | P-AXIS |
| 36 | 0.200 | 1000 | 4 | 32 | 2.0 | 1.0 | 2.00 | 2.00 | 55 | 1792 | 0.40000V | 0.51380012 | UNIFORM | P-AXIS |
| 37 | 0.300 | 1000 | 4 | 32 | 2.0 | 1.0 | 2.00 | 2.00 | 131 | 4224 | 0.40000V | 1.05203434 | UNIFORM | P-AXIS |
| 38 | 0.400 | 1000 | 4 | 32 | 2.0 | 1.0 | 2.00 | 2.00 | 35 | 1152 | 0.35000V | 10.47302654 | UNIFORM | P-AXIS |
| 39 | 0.500 | 1000 | 4 | 32 | 2.0 | 1.0 | 2.00 | 2.00 | 105 | 3392 | 0.05000V | 1.00716472 | UNIFORM | P-AXIS |
| 40 | 0.600 | 1000 | 4 | 32 | 2.0 | 1.0 | 2.00 | 2.00 | 50 | 1632 | 0.15000V | 1.23364082 | UNIFORM | P-AXIS |
| 41 | 0.700 | 1000 | 4 | 32 | 2.0 | 1.0 | 2.00 | 2.00 | 35 | 1152 | 0.35000V | 0.64610891 | UNIFORM | P-AXIS |
| 42 | 0.800 | 1000 | 4 | 32 | 2.0 | 1.0 | 2.00 | 2.00 | 56 | 1824 | 0.45000V | 10.49097850 | UNIFORM | P-AXIS |
| 43 | 0.900 | 1000 | 4 | 32 | 2.0 | 1.0 | 2.00 | 2.00 | 76 | 2464 | 0.50000V | 0.63249072 | UNIFORM | P-AXIS |
| 44 | 1.000 | 1000 | 4 | 32 | 2.0 | 1.0 | 2.00 | 2.00 | 123 | 3968 | 0.95000V | 10.53588503 | UNIFORM | P-AXIS |
| 45 | 0.000 | 1000 | 4 | 40 | 2.0 | 1.0 | 2.00 | 2.00 | 175 | 7040 | 0.70000V | 5.12819747 | UNIFORM | P-AXIS |
| 46 | 0.100 | 1000 | 4 | 40 | 2.0 | 1.0 | 2.00 | 2.00 | 93 | 3760 | 0.40000V | 10.53491728 | UNIFORM | P-AXIS |
| 47 | 0.200 | 1000 | 4 | 40 | 2.0 | 1.0 | 2.00 | 2.00 | 55 | 2240 | 0.40000V | 0.51944788 | UNIFORM | P-AXIS |
| 48 | 0.300 | 1000 | 4 | 40 | 2.0 | 1.0 | 2.00 | 2.00 | 111 | 4480 | 0.35000V | 2.13581640 | UNIFORM | P-AXIS |
| 49 | 0.400 | 1000 | 4 | 40 | 2.0 | 1.0 | 2.00 | 2.00 | 136 | 5480 | 0.65000V | 10.53593107 | UNIFORM | P-AXIS |
| 50 | 0.500 | 1000 | 4 | 40 | 2.0 | 1.0 | 2.00 | 2.00 | 106 | 4280 | 0.10000V | 0.99205603 | UNIFORM | P-AXIS |
| 51 | 0.600 | 1000 | 4 | 40 | 2.0 | 1.0 | 2.00 | 2.00 | 64 | 2600 | 0.85000V | 5.17536737 | UNIFORM | P-AXIS |
| 52 | 0.700 | 1000 | 4 | 40 | 2.0 | 1.0 | 2.00 | 2.00 | 35 | 1440 | 0.35000V | 0.65227664 | UNIFORM | P-AXIS |
| 53 | 0.800 | 1000 | 4 | 40 | 2.0 | 1.0 | 2.00 | 2.00 | 89 | 3600 | 0.20000V | 10.53484926 | UNIFORM | P-AXIS |
| 54 | 0.900 | 1000 | 4 | 40 | 2.0 | 1.0 | 2.00 | 2.00 | 88 | 3560 | 0.15000V | 5.17269921 | UNIFORM | P-AXIS |
| 55 | 1.000 | 1000 | 4 | 40 | 2.0 | 1.0 | 2.00 | 2.00 | 48 | 1960 | 0.05000V | 10.37269288 | UNIFORM | P-AXIS |
| 56 | 0.000 | 1000 | 4 | 48 | 2.0 | 1.0 | 2.00 | 2.00 | 120 | 5808 | 0.90000V | 5.12830155 | UNIFORM | P-AXIS |
| 57 | 0.100 | 1000 | 4 | 48 | 2.0 | 1.0 | 2.00 | 2.00 | 101 | 4896 | 0.80000V | 10.53546484 | UNIFORM | P-AXIS |
| 58 | 0.200 | 1000 | 4 | 48 | 2.0 | 1.0 | 2.00 | 2.00 | 75 | 3648 | 0.45000V | 0.52315570 | UNIFORM | P-AXIS |
| 59 | 0.300 | 1000 | 4 | 48 | 2.0 | 1.0 | 2.00 | 2.00 | 141 | 6816 | 0.90000V | 0.45882950 | UNIFORM | P-AXIS |
| 60 | 0.400 | 1000 | 4 | 48 | 2.0 | 1.0 | 2.00 | 2.00 | 125 | 6048 | 0.10000V | 10.53627116 | UNIFORM | P-AXIS |
| 61 | 0.500 | 1000 | 4 | 48 | 2.0 | 1.0 | 2.00 | 2.00 | 98 | 4752 | 0.65000V | 1.52823207 | UNIFORM | P-AXIS |
| 62 | 0.600 | 1000 | 4 | 48 | 2.0 | 1.0 | 2.00 | 2.00 | 81 | 3936 | 0.75000V | 6.06351786 | UNIFORM | P-AXIS |
| 63 | 0.700 | 1000 | 4 | 48 | 2.0 | 1.0 | 2.00 | 2.00 | 35 | 1728 | 0.35000V | 0.65302072 | UNIFORM | P-AXIS |
| 64 | 0.800 | 1000 | 4 | 48 | 2.0 | 1.0 | 2.00 | 2.00 | 192 | 9264 | 0.60000V | 10.53615531 | UNIFORM | P-AXIS |
| 65 | 0.900 | 1000 | 4 | 48 | 2.0 | 1.0 | 2.00 | 2.00 | 120 | 5808 | 0.80000V | 2.96848532 | UNIFORM | P-AXIS |
| <span style="color:red">66</span> | <span style="color:red">1.000</span> | <span style="color:red">1000</span> | <span style="color:red">4</span> | <span style="color:red">48</span> | <span style="color:red">2.0</span> | <span style="color:red">1.0</span> | <span style="color:red">2.00</span> | <span style="color:red">2.00</span> | <span style="color:red">192</span> | <span style="color:red">9264</span> | <span style="color:red">0.60000V</span> | <span style="color:red">10.53633734</span> | <span style="color:red">UNIFORM</span> | <span style="color:red">P-AXIS</span> |
| 67 | 0.000 | 1000 | 4 | 56 | 2.0 | 1.0 | 2.00 | 2.00 | 179 | 10080 | 0.90000V | 5.12842344 | UNIFORM | P-AXIS |
| 68 | 0.100 | 1000 | 4 | 56 | 2.0 | 1.0 | 2.00 | 2.00 | 192 | 10808 | 0.60000V | 10.53625567 | UNIFORM | P-AXIS |
| 69 | 0.200 | 1000 | 4 | 56 | 2.0 | 1.0 | 2.00 | 2.00 | 55 | 3136 | 0.40000V | 0.51424723 | UNIFORM | P-AXIS |
| 70 | 0.300 | 1000 | 4 | 56 | 2.0 | 1.0 | 2.00 | 2.00 | 136 | 7672 | 0.65000V | 6.36433520 | UNIFORM | P-AXIS |
| 71 | 0.400 | 1000 | 4 | 56 | 2.0 | 1.0 | 2.00 | 2.00 | 74 | 4200 | 0.40000V | 10.53540430 | UNIFORM | P-AXIS |
| 72 | 0.500 | 1000 | 4 | 56 | 2.0 | 1.0 | 2.00 | 2.00 | 103 | 5824 | 0.90000V | 1.25538481 | UNIFORM | P-AXIS |
| 73 | 0.600 | 1000 | 4 | 56 | 2.0 | 1.0 | 2.00 | 2.00 | 54 | 3080 | 0.35000V | 10.51784265 | UNIFORM | P-AXIS |
| 74 | 0.700 | 1000 | 4 | 56 | 2.0 | 1.0 | 2.00 | 2.00 | 35 | 2016 | 0.35000V | 0.64916466 | UNIFORM | P-AXIS |
| 75 | 0.800 | 1000 | 4 | 56 | 2.0 | 1.0 | 2.00 | 2.00 | 53 | 3024 | 0.30000V | 10.52739086 | UNIFORM | P-AXIS |
| 76 | 0.900 | 1000 | 4 | 56 | 2.0 | 1.0 | 2.00 | 2.00 | 145 | 8176 | 0.15000V | 5.12031227 | UNIFORM | P-AXIS |
| 77 | 1.000 | 1000 | 4 | 56 | 2.0 | 1.0 | 2.00 | 2.00 | 120 | 2800 | 0.10000V | 10.47520626 | UNIFORM | P-AXIS |

Total Function Evaluations: 242848

| <span style="color:red">66</span> | <span style="color:red">1.000</span> | <span style="color:red">1000</span> | <span style="color:red">4</span> | <span style="color:red">48</span> | <span style="color:red">2.0</span> | <span style="color:red">1.0</span> | <span style="color:red">2.00</span> | <span style="color:red">2.00</span> | <span style="color:red">192</span> | <span style="color:red">9264</span> | <span style="color:red">0.60000V</span> | <span style="color:red">10.53633734</span> | <span style="color:red">UNIFORM</span> | <span style="color:red">P-AXIS</span> |

# Appendix 3.  CFO Source Code

**All results reported in this note were generated using the following program.**

```
'Program 'CFO_02-07-2010.BAS' compiled with
'Power Basic/Windows Compiler 9.03 (www.PowerBasic.com).

'THIS IS THE PROGRAM USED TO GENERATE DATA FOR arXiv PAPER
'"Comparative Results: Group Search Optimizer and Central Force Optimization,"
'February 2010.

'LAST MOD 02-09-2010 ~2113 HRS EST

'CHANGES MADE TO "CFO_11-26-09.BAS" TO CREATE THIS VERSION
'WHICH IS USED FOR arXiv CFO PSEUDORANDOMNESS PAPER VER #2:
'   (1).  OUTER LOOP Np/Nd = 2 to MaxProbesPerAxis by 2 ADDED
'   (2).  Np IGNORED & ARRAYS REDIMENSIONED AS REQUIRED BY (1).

'ADDITIONAL CHANGES MADE TO CFO_12-25-09.BAS TO CREATE THIS VERSION:
'   (1).  LOOPS ADDED TO VARY G and ALPHA
'   (2).  SHRINK DS EVERY ?? STEPS INSTEAD OF 20
'   (3).  TERMINATE EARLY USING ?? STEP AVERAGE INSTEAD OF 50?

'IMPORTANT CHANGE MADE 01-15-10: MOVED REDIM STATEMENT INTO SUB CFO().
'THIS IS NECESSARY TO PROPERLY RE-INITIALIZE ALL MATRICES...

'SEPARATE ROSENBROCK & SPHERE FUNCTIONS ADDED 02-02-10.

'PROBE RETRIEVAL SCHEME MODIFIED 02-07-2010 PER MAHAMED OMRAN's SUGGESTION.

'# STEPS FOR FITNESS SATURATION = 25, 02-07-2010.

'NOTE: ALL PBM FUNCTIONS HAVE WIRE RADIUS SET TO 0.00001LAMBDA
'=================================================================

'THIS PROGRAM IMPLEMENTS A SIMPLE VERSION OF "CENTRAL
'FORCE OPTIMIZATION."  IT IS DISTRIBUTED FREE OF CHARGE
'TO INCREASE AWARENESS OF CFO AND TO ENCOURAGE EXPERI-
'MENTATION WITH THE ALGORITHM.

'CFO IS A MULTIDIMENSIONAL SEARCH AND OPTIMIZATION
'ALGORITHM THAT LOCATES THE GLOBAL MAXIMA OF A FUNCTION.
'UNLIKE MOST OTHER ALGORITHMS, CFO IS COMPLETELY DETERMIN-
'ISTIC, SO THAT EVERY RUN WITH THE SAME SETUP PRODUCES
'THE SAME RESULTS.

'Please email questions, comments, and significant
'results to: CFO_questions@yahoo.com.  Thanks!

'(c) 2006-2009 Richard A. Formato

'ALL RIGHTS RESERVED WORLDWIDE

'THIS PROGRAM IS FREEWARE.  IT MAY BE COPIED AND
'DISTRIBUTED WITHOUT LIMITATION AS LONG AS THIS
'COPYRIGHT NOTICE AND THE GNUPLOT AND REFERENCE
'INFORMATION BELOW ARE INCLUDED WITHOUT MODIFICATION,
'AND AS LONG AS NO FEE OR COMPENSATION IS CHARGED,
'INCLUDING "TIE-IN" OR "BUNDLING" FEES CHARGED FOR
'OTHER PRODUCTS.

'=================================================================

'THIS PROGRAM REQUIRES wgnuplot.exe TO DISPLAY PLOTS.
'Gnuplot is a copyrighted freeware plotting program
'available at http://www.gnuplot.info/index.html.

'IT ALSO REQUIRES A VERSION OF THE Numerical Electromagnetics
'Code (NEC) in order to run the PBM benchmarks.  If this file
'is not present, a runtime error occurs.  Remove the code
'that checks for the NEC EXE is there is no interest in the
'PBM functions.

'=================================================================

'REFERENCES
'----------
'Suggestion:  Read references 1 and 2 first (both available online at no charge)...

'1. "Central Force Optimization: A New Metaheuristic with Applications in Applied Electromagnetics,"
'    Progress in Electromagnetics Research, PIER 77, 425-491, 2007 (online). DOI:10.2528/PIER07082403
'Abstract - Central Force Optimization (CFO) is a new deterministic multi-dimensional search
'metaheuristic based on the metaphor of gravitational kinematics. It models "probes" that "fly"
'through the decision space by analogy to masses moving under the influence of gravity.  Equations
'are developed for the probes' positions and accelerations using the analogy of particle motion
'in a gravitational field.  In the physical universe, objects traveling through three-dimensional
'space become trapped in close orbits around highly gravitating masses, which is analogous to
'locating the maximum value of an objective function.  In the CFO metaphor, "mass" is a user-defined
'function of the value of the objective function to be maximized.  CFO is readily implemented in a
'compact computer program, and sample pseudocode is presented.  As tests of CFO's effectiveness,
'an equalizer is designed for the well-known Fano load, and a 32-element linear array is synthesized.
'CFO results are compared to several other optimization methods.

'2. "Pseudorandomness in Central Force Optimization"
'    arXiv:1001.0317v1[cs.NE], online: www.arXiv.org
'Abstract - Central Force Optimization is a deterministic metaheuristic for an evolutionary algorithm
'that searches a decision space by flying probes whose trajectories are computed using a gravitational
'metaphor.  CFO benefits substantially from the inclusion of a pseudorandom component (a numerical
'sequence that is precisely known by specification or calculation but otherwise arbitrary).  The essential
'requirement is that the sequence is uncorrelated with the decision space topology, so that its
'effect is to pseudorandomly distribute probes throughout the landscape.  While this process may appear
'to be similar to the randomness in an inherently stochastic algorithm, it is in fact fundamentally
'different because CFO remains deterministic at every step.  Three pseudorandom methods are discussed
'(initial probe distribution, repositioning factor, and decision space adaptation).  A sample problem
'is presented in detail and summary data included for a 23-function benchmark suite.  CFO's performance
'is quite good compared to other highly developed, state-of-the-art algorithms.

'3. "Central Force Optimization: A New Nature Inspired Computational Framework for Multidimensional
'    Search and Optimization," Studies in Computational Intelligence (SCI), vol. 129, 221-238 (2008),
'    www.springerlink.com, Springer-Verlag Berlin Heidelberg 2008.
'Abstract - This paper presents Central Force Optimization, a novel, nature inspired, deterministic
'search metaheuristic for constrained multi-dimensional optimization.  CFO is based on the metaphor
```

'of gravitational kinematics.  Equations are presented for the positions and accelerations experienced
'by "probes" that "fly" through the decision space by analogy to masses moving under the influence of
'gravity.  In the physical universe, probe satellites become trapped in close orbits around highly
'gravitating masses.  In the CFO analogy, "mass" corresponds to a user-defined function of the value
'of an objective function to be maximized.  CFO is a simple algorithm that is easily implemented in
'a compact computer program.  A typical CFO implementation is applied to several test functions.
'CFO exhibits very good performance, suggesting that it merits further study.

'4.  Are Near Earth Objects the Key to Optimization Theory?
'    arXiv:0912.1394v1[astro-ph.EP], online: www.arXiv.org
'Abstract - This note suggests that near earth objects and Central Force Optimization have something in common,
'that NEO theory may hold the key to solving some vexing problems in deterministic optimization: local trapping
'and proof of convergence.  CFO analogizes Newton's laws to locate the global maxima of a function.  The NEO-CFO
'nexus is the striking similarity between CFO's  and an NEO's  curves.  Both exhibit oscillatory plateau-like
'regions connected by jumps, suggesting that CFO's metaphorical "gravity" indeed behaves like real gravity,
'thereby connecting NEOs and CFO and being the basis for speculating that NEO theory may address difficult
'issues in optimization.

'5. "Central Force Optimisation: A New Gradient-Like Metaheuristic for Multidimensional Search and
'    Optimisation," International Journal of Bio-Inspired Computation, vol. 1, no. 4, 217-238 (2009).
'    DOI: 10.1504/IJBIC.2009.024721
'Abstract: This paper introduces Central Force Optimization, a novel, Nature-inspired, deterministic
'search metaheuristic for constrained multidimensional optimization in highly multimodal, smooth, or
'discontinuous decision spaces.  CFO is based on the metaphor of gravitational kinematics.  The
'algorithm searches a decision space by "flying" its "probes" through the space by analogy to masses
'moving through physical space under the influence of gravity.  Equations are developed for the probes'
'positions and accelerations using the gravitational metaphor.  Small objects in our Universe can become
'trapped in close orbits around highly gravitating masses.  In "CFO space" probes are attracted to "masses"
'created by a user-defined function of the value of an objective function to be maximized.  CFO may be
'thought of in terms of a vector "force field" or, loosely, as a "generalized gradient" methodology
'because the force of gravity can be computed as the gradient of a scalar potential.  The CFO algorithm
'is simple and easily implemented in a compact computer program.  Its effectiveness is demonstrated by
'running CFO against several widely used benchmark functions.  The algorithm exhibits very good performance,
'suggesting that it merits further study.

'6. "Central Force Optimization: A New Gradient-Like Optimization Metaheuristic," OPSEARCH, Journal
'    of the Operational Research Society of India, vol. 46, no. 1, 25-51, 2009, Springer India.
'    DOI: 10.1007/s12597-009-0003-4
'Abstract:  This paper introduces Central Force Optimization as a new, Nature-inspired metaheuristic
'for multidimensional search and optimization based on the metaphor of gravitational kinematics.  CFO is a
'"gradient-like" deterministic algorithm that explores a decision space by "flying" a group of "probes"
'whose trajectories are governed by equations analogous to the equations of gravitational motion in the
'physical Universe.  This paper suggests the possibility of creating a new "hyperspace directional derivative"
'using the Unit Step function t7o create positive-definite "masses" in "CFO space."  A simple CFO implementation
'is tested against several recognized benchmark functions with excellent results, suggesting that CFO merits
'further investigation.

'7. "Synthesis of Antenna Arrays Using Central Force Optimization," Mosharaka International Conference
'    on Communications, Computers and Applications, MIC-CPE 2009. (Gubran Mohammad and Nihad Dib)
'Abstract: Central force optimization (CFO) technique is a new deterministic multi-dimensional
'search metaheuristic based on an analogy to classical particle kinematics in a gravitational
'field. CFO is a simple technique and is still in its infancy. To enhance its global search
'ability while keeping its simplicity, a new selection part is introduced in this paper. CFO
'is briefly presented and applied in the design of linear antenna array and the modified CFO
'is applied to the design of circular array. The results are compared with those obtained using
'other evolutionary optimization techniques.

'8. "Central Force Optimization and NEOs - First Cousins?" Journal of Multiple Valued Logic and Soft
'    Computing (in press)
'Abstract: Central Force Optimization is a new deterministic multidimensional search and optimization algorithm
'based on the metaphor of gravitational kinematics.  This paper describes CFO and suggests some possible
'directions for its future development.  Because CFO is deterministic, it is more computationally efficient
'than stochastic algorithms and may lend itself well to "parameter tuning" implementations.  But, like
'all deterministic algorithms, CFO is prone to local trapping.  Oscillation in CFO's Davg curve appears to
'be a reliable harbinger of trapping.  And there seems to be a reasonable basis for believing that
'trapping can be handled deterministically and using the theory of gravitationally trapped Near Earth Objects.
'Deterministic mitigation of local trapping would be a major step forward in optimization theory.  Finally,
'CFO may be thought of as a "gradient-like" algorithm utilizing the Unit Step function as a critical element,
'and it is suggested that a useful, new derivative-like mathematical construct might be defined based on the Unit
'Step.

'9. "Antenna Benchmark Performance and Array Synthesis Using Central Force Optimization," IET Microwaves,
'    Antennas & Propagation (UK) (in press).  DOI: 10.1049/iet-map.2009.0147 (with G. M. Qubati and N. I. Dib)
'Abstract - Central force optimization (CFO) is a new deterministic multi-dimensional search metaheuristic
'based on an analogy to classical particle kinematics in a gravitational field. CFO is a simple technique
'that is still in its infancy.  This paper evaluates CFO's performance and provides further examples of its
'effectiveness by applying it to a set of "real world" antenna benchmarks and to pattern synthesis for linear
'and circular array antennas. A new selection scheme is introduced that enhances CFO's global search ability
'while maintaining its simplicity. The improved CFO algorithm is applied to the design of a circular array
'with very good results.  CFO's performance on the antenna benchmarks and the synthesis problems is compared
'to that of other evolutionary optimization techniques.
'================================================================================================================

#COMPILE EXE

#DIM ALL

%USEMACROS = 1

#INCLUDE "Win32API.inc"

DEFEXT A-Z

'------ EQUATES -----

%IDC_FRAME1    = 101
%IDC_FRAME2    = 102

%IDC_Function_Number1  = 121
%IDC_Function_Number2  = 122
%IDC_Function_Number3  = 123
%IDC_Function_Number4  = 124
%IDC_Function_Number5  = 125
%IDC_Function_Number6  = 126
%IDC_Function_Number7  = 127
%IDC_Function_Number8  = 128
%IDC_Function_Number9  = 129
%IDC_Function_Number10 = 130
%IDC_Function_Number11 = 131
%IDC_Function_Number12 = 132
%IDC_Function_Number13 = 133
%IDC_Function_Number14 = 134
%IDC_Function_Number15 = 135
%IDC_Function_Number16 = 136

```
%IDC_Function_Number17 = 137
%IDC_Function_Number18 = 138
%IDC_Function_Number19 = 139
%IDC_Function_Number20 = 140
%IDC_Function_Number21 = 141
%IDC_Function_Number22 = 142
%IDC_Function_Number23 = 143
%IDC_Function_Number24 = 144
%IDC_Function_Number25 = 145
%IDC_Function_Number26 = 146
%IDC_Function_Number27 = 147
%IDC_Function_Number28 = 148
%IDC_Function_Number29 = 149
%IDC_Function_Number30 = 150
%IDC_Function_Number31 = 151
%IDC_Function_Number32 = 152
%IDC_Function_Number33 = 153
%IDC_Function_Number34 = 154
%IDC_Function_Number35 = 155
%IDC_Function_Number36 = 156
%IDC_Function_Number37 = 157
%IDC_Function_Number38 = 158
%IDC_Function_Number39 = 159
%IDC_Function_Number40 = 160
%IDC_Function_Number41 = 161
%IDC_Function_Number42 = 162
%IDC_Function_Number43 = 163
%IDC_Function_Number44 = 164
%IDC_Function_Number45 = 165
%IDC_Function_Number46 = 166
%IDC_Function_Number47 = 167
%IDC_Function_Number48 = 168
%IDC_Function_Number49 = 169
%IDC_Function_Number50 = 170

'------------------------------ GLOBAL CONSTANTS & SYMBOLS ---------------------------

GLOBAL Aij() AS EXT 'array for Shekel's Foxholes function

GLOBAL EulerConst, Pi, Pi2, Pi4, TwoPi, FourPi, e, Root2 AS EXT 'mathematical constants

GLOBAL Alphabet$, Digits$, RunID$  'upper/lower case alphabet, digits 0-9 & Run ID

GLOBAL Quote$, SpecialCharacters$  'quotation mark & special symbols

GLOBAL Mu0, Eps0, c, eta0 AS EXT   'E&M constants

GLOBAL Rad2Deg, Deg2Rad, Feet2Meters, Meters2Feet, Inches2Meters, Meters2Inches AS EXT 'conversion factors

GLOBAL Miles2Meters, Meters2Miles, NautMi2Meters, Meters2NautMi AS EXT              'conversion factors

GLOBAL ScreenWidth&, ScreenHeight& 'screen width & height

GLOBAL xOffset&, yOffset&          'offsets for probe plot windows

GLOBAL FunctionNumber%

GLOBAL AddNoiseToPBM2$

'------------------------------ TEST FUNCTION DECLARATIONS ---------------------------

DECLARE FUNCTION F1(R(),Nd%,p%,j&)        'F1 (n-D)

DECLARE FUNCTION F2(R(),Nd%,p%,j&)        'F2(n-D)

DECLARE FUNCTION F3(R(),Nd%,p%,j&)        'F3 (n-D)

DECLARE FUNCTION F4(R(),Nd%,p%,j&)        'F4 (n-D)

DECLARE FUNCTION F5(R(),Nd%,p%,j&)        'F5 (n-D)

DECLARE FUNCTION F6(R(),Nd%,p%,j&)        'F6 (n-D)

DECLARE FUNCTION F7(R(),Nd%,p%,j&)        'F7 (n-D)

DECLARE FUNCTION F8(R(),Nd%,p%,j&)        'F8 (n-D)

DECLARE FUNCTION F9(R(),Nd%,p%,j&)        'F9 (n-D)

DECLARE FUNCTION F10(R(),Nd%,p%,j&)       'F10 (n-D)

DECLARE FUNCTION F11(R(),Nd%,p%,j&)       'F11 (n-D)

DECLARE FUNCTION F12(R(),Nd%,p%,j&)       'F12 (n-D)

DECLARE FUNCTION u(Xi,a,k,m)              'Auxiliary function for F12 & F13

DECLARE FUNCTION F13(R(),Nd%,p%,j&)       'F13 (n-D)

DECLARE FUNCTION F14(R(),Nd%,p%,j&)       'F14 (n-D)

DECLARE FUNCTION F15(R(),Nd%,p%,j&)       'F15 (n-D)

DECLARE FUNCTION F16(R(),Nd%,p%,j&)       'F16 (n-D)

DECLARE FUNCTION F17(R(),Nd%,p%,j&)       'F17 (n-D)

DECLARE FUNCTION F18(R(),Nd%,p%,j&)       'F18 (n-D)

DECLARE FUNCTION F19(R(),Nd%,p%,j&)       'F19 (n-D)

DECLARE FUNCTION F20(R(),Nd%,p%,j&)       'F20 (n-D)

DECLARE FUNCTION F21(R(),Nd%,p%,j&)       'F21 (n-D)

DECLARE FUNCTION F22(R(),Nd%,p%,j&)       'F22 (n-D)

DECLARE FUNCTION F23(R(),Nd%,p%,j&)       'F23 (n-D)

DECLARE FUNCTION F24(R(),Nd%,p%,j&)       'F24 (n-D)

DECLARE FUNCTION F25(R(),Nd%,p%,j&)       'F25 (n-D)

DECLARE FUNCTION F26(R(),Nd%,p%,j&)       'F26 (n-D)
```

```
DECLARE FUNCTION F27(R(),Nd%,p%,j&)              'F27 (n-D)

DECLARE FUNCTION ParrottF4(R(),Nd%,p%,j&)        'Parrott F4 (1-D)

DECLARE FUNCTION SGO(R(),Nd%,p%,j&)              'SGO Function (2-D)

DECLARE FUNCTION GoldsteinPrice(R(),Nd%,p%,j&)   'Goldstein-Price Function (2-D)

DECLARE FUNCTION StepFunction(R(),Nd%,p%,j&)     'Step Function (n-D)

DECLARE FUNCTION Schwefel226(R(),Nd%,p%,j&)      'Schwefel Prob. 2.26 (n-D)

DECLARE FUNCTION Colville(R(),Nd%,p%,j&)         'Colville Function (4-D)

DECLARE FUNCTION Griewank(R(),Nd%,p%,j&)         'Griewank (n-D)

DECLARE FUNCTION Himmelblau(R(),Nd%,p%,j&)       'Himmelblau (2-D)

DECLARE FUNCTION Rosenbrock(R(),Nd%,p%,j&)       'Rosenbrock (n-D)

DECLARE FUNCTION Sphere(R(),Nd%,p%,j&)           'Sphere (n-D)

DECLARE FUNCTION PBM_1(R(),Nd%,p%,j&)            'PBM Benchmark #1

DECLARE FUNCTION PBM_2(R(),Nd%,p%,j&)            'PBM Benchmark #2

DECLARE FUNCTION PBM_3(R(),Nd%,p%,j&)            'PBM Benchmark #3

DECLARE FUNCTION PBM_4(R(),Nd%,p%,j&)            'PBM Benchmark #4

DECLARE FUNCTION PBM_5(R(),Nd%,p%,j&)            'PBM Benchmark #5

'---------------------------------- SUB DECLARATIONS ------------------------------------

DECLARE SUB CheckNECFiles(NECfileError$)

DECLARE SUB GetTestFunctionNumber(FunctionName$)

DECLARE SUB FillArrayAij

DECLARE SUB ResetDecisionSpaceBoundaries(Nd%,XiMin(),XiMax(),StartingXiMin(),StartingXiMax())

DECLARE SUB Plot3DbestProbeTrajectories(NumTrajectories%,M(),R(),XiMin(),XiMax(),Np%,Nd%,LastStep&,FunctionName$)

DECLARE SUB Plot2DbestProbeTrajectories(NumTrajectories%,M(),R(),XiMin(),XiMax(),Np%,Nd%,LastStep&,FunctionName$)

DECLARE SUB Plot2DindividualProbeTrajectories(NumTrajectories%,M(),R(),XiMin(),XiMax(),Np%,Nd%,LastStep&,FunctionName$)

DECLARE SUB SelectTestFunction(FunctionName$)

DECLARE SUB Show2Dprobes(R(),Np%,Nt&,j&,XiMin(),XiMax(),Frep,BestFitness,BestProbeNumber%,BestTimeStep&,FunctionName$,RepositionFactor$,Gamma)

DECLARE SUB StatusWindow(FunctionName$,StatusWindowHandle???)

DECLARE SUB
DisplayRunParameters(FunctionName$,Nd%,Np%,Nt&,G,DeltaT,Alpha,Beta,Frep,R(),A(),M(),PlaceInitialProbes$,InitialAcceleration$,RepositionFactor$,RunCFO$,ShrinkDS$,CheckForEarlyTermination$)

DECLARE SUB GetBestFitness(M(),Np%,StepNumber&,BestFitness,BestProbeNumber%,BestTimeStep&)

DECLARE SUB
Tabulate1DprobeCoordinates(Max1DprobesPlotted%,Nd%,Np%,LastStep&,G,DeltaT,Alpha,Beta,Frep,R(),M(),PlaceInitialProbes$,InitialAcceleration$,RepositionFactor$,FunctionName$,Gamma)

DECLARE SUB GetPlotAnnotation(PlotAnnotation$,Nd%,Np%,Nt&,G,DeltaT,Alpha,Beta,Frep,M(),PlaceInitialProbes$,InitialAcceleration$,RepositionFactor$,FunctionName$,Gamma)

DECLARE SUB ChangeRunParameters(NumProbesPerDimension%,Np%,Nd%,Nt&,G,Alpha,Beta,DeltaT,Frep,PlaceInitialProbes$,InitialAcceleration$,RepositionFactor$,FunctionName$)

DECLARE SUB CLEANUP

DECLARE SUB DisplayBestFitness(Np%,Nd%,LastStep&,M(),R(),BestFitnessProbeNumber%,BestFitnessTimeStep&,FunctionName$)

DECLARE SUB
Plot1DprobePositions(Max1DprobesPlotted%,Nd%,Np%,LastStep&,G,DeltaT,Alpha,Beta,Frep,R(),M(),PlaceInitialProbes$,InitialAcceleration$,RepositionFactor$,FunctionName$,Gamma)

DECLARE SUB DisplayMmatrix(Np%,Nt&,M())

DECLARE SUB DisplayMmatrixThisTimeStep(Np%,j&,M())

DECLARE SUB DisplayAmatrix(Np%,Nd%,Nt&,A())

DECLARE SUB DisplayAmatrixThisTimeStep(Np%,Nd%,j&,A())

DECLARE SUB DisplayRmatrix(Np%,Nd%,Nt&,R())

DECLARE SUB DisplayRmatrixThisTimeStep(Np%,Nd%,j&,R(),Gamma)

DECLARE SUB DisplayXiMinMax(Nd%,XiMin(),XiMax())

DECLARE SUB DisplayRunParameters2(FunctionName$,Nd%,Np%,Nt&,G,DeltaT,Alpha,Beta,Frep,PlaceInitialProbes$,InitialAcceleration$,RepositionFactor$)

DECLARE SUB PlotBestProbeVsTimeStep(Nd%,Np%,LastStep&,G,DeltaT,Alpha,Beta,Frep,M(),PlaceInitialProbes$,InitialAcceleration$,RepositionFactor$,FunctionName$,Gamma)

DECLARE SUB PlotBestFitnessEvolution(Nd%,Np%,LastStep&,G,DeltaT,Alpha,Beta,Frep,M(),PlaceInitialProbes$,InitialAcceleration$,RepositionFactor$,FunctionName$,Gamma)

DECLARE SUB PlotAverageDistance(Nd%,Np%,LastStep&,G,DeltaT,Alpha,Beta,Frep,M(),PlaceInitialProbes$,InitialAcceleration$,RepositionFactor$,FunctionName$,R(),DiagLength,Gamma)

DECLARE SUB Plot2Dfunction(FunctionName$,XiMin(),XiMax(),R())

DECLARE SUB Plot1Dfunction(FunctionName$,XiMin(),XiMax(),R())

DECLARE SUB GetFunctionRunParameters(FunctionName$,Nd%,Np%,Nt&,G,DeltaT,Alpha,Beta,Frep,R(),A(),M(),XiMin(),XiMax(),StartingXiMin(),StartingXiMax(),_
                                     DiagLength,PlaceInitialProbes$,InitialAcceleration$,RepositionFactor$)

DECLARE SUB InitialProbeDistribution(Np%,Nd%,Nt&,XiMin(),XiMax(),R(),PlaceInitialProbes$,Gamma)

DECLARE SUB InitialProbeAccelerations(Np%,Nd%,A(),InitialAcceleration$,MaxInitialRandomAcceleration,MaxInitialFixedAcceleration)

DECLARE SUB RetrieveErrantProbes(Np%,Nd%,j&,XiMin(),XiMax(),R(),M(),RepositionFactor$,Frep)

DECLARE SUB CFO(Nd%,Np%,Nt&,G,DeltaT,Alpha,Beta,Frep,R(),A(),M(),XiMin(),XiMax(),DiagLength,PlaceInitialProbes$,InitialAcceleration$,_

RepositionFactor$,FunctionName$,LastStep&,CheckForEarlyTermination$,BestFitnessThisRun,GammaRunNumber%,NumGammaRuns%,Gamma,BestOverallFitness,ShrinkDS$,MaxProbesPerAxis%,MinProbesPerAxis%,_
```

```
                AbsoluteRunNumber%,TotalRuns%)

DECLARE SUB ThreeDplot(PlotFileName$,PlotTitle$,Annotation$,xCoord$,yCoord$,zCoord$, _
                    Xaxislabel$,Yaxislabel$,Zaxislabel$,zMin$,zMax$,GnuPlotEXE$,A$)

DECLARE SUB ThreeDplot2(PlotFileName$,PlotTitle$,Annotation$,xCoord$,yCoord$,zCoord$,Xaxislabel$,_
                    Yaxislabel$,Zaxislabel$,zMin$,zMax$,GnuPlotEXE$,A$,xStart$,xStop$,yStart$,yStop$)

DECLARE SUB TwoDplot(PlotFileName$,PlotTitle$,xCoord$,yCoord$,Xaxislabel$,Yaxislabel$, _
                    LogXaxis$,LogYaxis$,xMin$,xMax$,yMin$,yMax$,xTics$,yTics$,GnuPlotEXE$,LineType$,Annotation$)

DECLARE SUB TwoDplot2Curves(PlotFileName1$,PlotTitle$,PlotFileName2$,PlotTitle2$,Annotation$,xCoord$,yCoord$,Xaxislabel$,Yaxislabel$, _
                    LogXaxis$,LogYaxis$,xMin$,xMax$,yMin$,yMax$,xTics$,yTics$,GnuPlotEXE$,LineSize)

DECLARE SUB TwoDplot3curves(NumCurves%,PlotFileName1$,PlotFileName2$,PlotTitle$,PlotFileName3$,PlotTitle$,Annotation$,xCoord$,yCoord$,Xaxislabel$,Yaxislabel$, _
                    LogXaxis$,LogYaxis$,xMin$,xMax$,yMin$,yMax$,xTics$,yTics$,GnuPlotEXE$)

DECLARE SUB CreateGNUplotINIfile(PlotWindowULC_X%,PlotWindowULC_Y%,PlotWindowWidth%,PlotWindowHeight%)

DECLARE SUB Delay(NumSecs)

DECLARE SUB MathematicalConstants

DECLARE SUB AlphabetAndDigits

DECLARE SUB SpecialSymbols

DECLARE SUB EMconstants

DECLARE SUB ConversionFactors

DECLARE SUB ShowConstants

'------ FUNCTION DECLARATIONS -------

DECLARE FUNCTION SlopeRatio(M(),Np%,StepNumber&)

DECLARE CALLBACK FUNCTION DlgProc

DECLARE FUNCTION HasFITNESSsaturated$(Nsteps&,j&,Np%,Nd%,M(),R(),DiagLength)

DECLARE FUNCTION HasDAVGsaturated$(Nsteps&,j&,Np%,Nd%,M(),R(),DiagLength)

DECLARE FUNCTION OscillationInDavg$(j&,Np%,Nd%,M(),R(),DiagLength)

DECLARE FUNCTION DavgThisStep(j&,Np%,Nd%,M(),R(),DiagLength)

DECLARE FUNCTION NoSpaces$(X,NumDigits%)

DECLARE FUNCTION FormatFP$(X,Ndigits%)

DECLARE FUNCTION FormatInteger$(M%)

DECLARE FUNCTION TerminateNowForSaturation$(j&,Nd%,Np%,Nt&,G,DeltaT,Alpha,Beta,R(),A(),M())

DECLARE FUNCTION MagVector(V(),N%)

DECLARE FUNCTION UniformDeviate(u&&)

DECLARE FUNCTION RandomNum(a,b)

DECLARE FUNCTION GaussianDeviate(Mu,Sigma)

DECLARE FUNCTION UnitStep(X)

DECLARE FUNCTION Fibonacci&&(N%)

DECLARE FUNCTION ObjectiveFunction(R(),Nd%,p%,j&,FunctionName$)

DECLARE FUNCTION UnitStep(X)

'================================================================================================

'----- MAIN PROGRAM ------

FUNCTION PBMAIN () AS LONG

'    ------ CFO Parameters -----

    LOCAL Nd%, Np%, Nt&

    LOCAL G, DeltaT, Alpha, Beta, Frep AS EXT

    LOCAL PlaceInitialProbes$, InitialAcceleration$, RepositionFactor$

    LOCAL R(), A(), M() AS EXT    'position, acceleration & fitness matrices

    LOCAL XiMin(), XiMax(), StartingXiMin(), StartingXiMax() AS EXT 'decision space boundaries

    LOCAL FunctionName$          'name of objective function

    LOCAL DiagLength AS EXT

'    ----------- Miscellaneous Setup Parameters --------------

    LOCAL StartingG, StartingDeltaT, StartingAlpha, StartingBeta, StartingFrep AS EXT

    LOCAL N%, i%, YN&

    LOCAL A$

    LOCAL NumGammaRuns%, GammaRunNumber%, BestRunNumber%, TotalFunctionEvaluations&&, AbsoluteRunNumber%, AbsoluteBestRunNumber%

    LOCAL Gamma, StartingGamma, StoppingGamma, BestFitnessThisRun, BestOverallFitness AS EXT

    LOCAL MinAlpha, MaxAlpha, MinG, MaxG, BestAlpha, BestG AS EXT

    LOCAL NumAlphaRuns%, NumGruns%, GrunNumber%, AlphaRunNumber%

    LOCAL NumTrajectories%

    LOCAL Max1DprobesPlotted%

    LOCAL BestFitnessProbeNumber%, BestFitnessTimeStep&, BestNumProbes%, MaxProbesPerAxis%, MinProbesPerAxis%, NumProbesPerAxis%, NumProbeRuns%, TotalRuns%
```

```
    LOCAL RunCFO$, CFOversion$

    LOCAL StatusWindowHandle???

    LOCAL LastStep&

    LOCAL CheckForEarlyTermination$ 'early termination checking? (YES/NO)

    LOCAL ShrinkDS$ 'adaptively shrink DS? (YES/NO)

    LOCAL NECfileError$
'    -------------------- Global Constants --------------------
    REDIM Aij(1 TO 2, 1 TO 25) '(GLOBAL array for Shekel's Foxholes function)

    CALL FillArrayAij

    CALL MathematicalConstants 'NOTE: Calling order is important!!

    CALL AlphabetAndDigits

    CALL SpecialSymbols

    CALL EMconstants

    CALL ConversionFactors          ': CALL ShowConstants 'to verify constants have been set

    xOffset& = 20 : yOffset& = 20 'offsets for successive probe position plots
'    --------------------------- General Setup ----------------------------
    CFOversion$ = "CFO Ver. 02-07-2010"

    RANDOMIZE TIMER 'seed random number generator with program start time

    DESKTOP GET SIZE TO ScreenWidth&, ScreenHeight& 'get screen size (global variables)

    IF DIR$("wgnuplot.exe") = "" THEN

        MSGBOX("WARNING! 'wgnuplot.exe' not found. Run terminated.") : EXIT FUNCTION

    END IF

    IF DIR$("ProbeCoordinates.DAT") <> "" THEN KILL "ProbeCoordinates.DAT" 'get rid of this file to avoid confusion

    IF DIR$("Frep.DAT") <> "" THEN KILL "Frep.DAT" 'ditto
'    ------------------------------------------------------------------------------------ CFO RUN PARAMETERS ------------------------------------------------------------------------------------
    Max1DprobesPlotted% = 15 : Max1DprobesPlotted% = MIN(15,Max1DprobesPlotted%) '15 MAX!

    CALL GetTestFunctionNumber(FunctionName$)' : exit function 'DEBUG

    CALL GetFunctionRunParameters(FunctionName$,Nd%,Np%,Nt&,G,DeltaT,Alpha,Beta,Frep,R(),A(),M(),XiMin(),XiMax(),StartingXiMin(),StartingXiMax(),_
                                  DiagLength,PlaceInitialProbes$,InitialAcceleration$,RepositionFactor$)

' !! NOTE !!  IN THIS VERSION Np% IS IGNORED BECAUSE THE PROGRAM LOOPS ON Np/Nd (= 2-8 by 2) AND COMPUTES Np FROM Np = (Np/Nd)*Nd.
'            BUT THE PROGRAM STILL RETURNS Np FROM THIS SUBROUTINE (TOO MUCH OF A BOTHER TO TAKE IT OUT ...).

    StartingG = G : StartingDeltaT = DeltaT : StartingAlpha = Alpha : StartingBeta = Beta : StartingFrep = Frep

'    IMPORTANT NOTE: Arrays XiMin() & XiMax() are dimensioned (1 TO Nd%) in SUB GetFunctionRunParameters

    REDIM R(1 TO Np%, 1 TO Nd%, 0 TO Nt&), A(1 TO Np%, 1 TO Nd%, 0 TO Nt&), M(1 TO Np%, 1 TO Nd%, 0 TO Nt&) 'position, acceleration & fitness matrices
'    -------- PLOT 1D and 2D FUNCTIONS ON-SCREEN FOR VISUALIZATION --------
    IF Nd% = 2 AND INSTR(FunctionName$,"PRM_") > 0 THEN

        CALL CheckNECFiles(NECfileError$)

        IF NECfileError$ = "YES" THEN
            EXIT FUNCTION
        ELSE
            MSGBOX("Begin computing plot of function "+FunctionName$+"?  May take a while - be patient...")
        END IF

    END IF

    SELECT CASE Nd%
        CASE 1 : CALL Plot1Dfunction(FunctionName$,XiMin(),XiMax(),R()) : REDIM R(1 TO Np%, 1 TO Nd%, 0 TO Nt&) 'erases coordinate data in R()used to plot function
        CASE 2 : CALL Plot2Dfunction(FunctionName$,XiMin(),XiMax(),R()) : REDIM R(1 TO Np%, 1 TO Nd%, 0 TO Nt&) 'ditto
    END SELECT
'    ------------------------------------------------------------------- DISPLAY PARAMETERS & RUN CFO -------------------------------------------------------------------
    CALL
DisplayRunParameters(FunctionName$,Nd%,Np%,Nt&,G,DeltaT,Alpha,Beta,Frep,R(),A(),M(),PlaceInitialProbes$,InitialAcceleration$,RepositionFactor$,RunCFO$,ShrinkDS$,CheckForEarlyTermi
nation$)

    IF RunCFO$ = "YES" THEN 'run CFO

        CALL StatusWindow(FunctionName$+", "+CFOversion$,StatusWindowHandle???)

        MinAlpha = 2## : MaxAlpha = 4## : NumAlphaRuns% = 1 'ALPHA LOOP

        MinG    = 2## : MaxG     = 3## : NumGruns%    = 1 'G LOOP

        MinProbesPerAxis% = 2 : MaxProbesPerAxis% = 14 '(by 2) 'Np/Nd LOOP (Initial Probe Dist'n)

        NumProbeRuns% = 1 + (MaxProbesPerAxis%-MinProbesPerAxis%)\2

        StartingGamma = 0## : StoppingGamma = 1## : NumGammaRuns% = 11 'GAMMA LOOP

        TotalRuns% = NumAlphaRuns%*NumGruns%*NumProbeRuns%*NumGammaRuns%

        MSGBOX("Tot Runs="+STR$(TotalRuns%)") : EXIT FUNCTION
'        --------------- Output Data File & Header ---------------
        N% = FREEFILE : OPEN FunctionName$+".DAT" FOR OUTPUT AS #N%
```

```
        PRINT #N%, "Run ID: "+RunID$+CHR$(13)+CHR$(13)+"FUNCTION: "+UCASE$(FunctionName$)+CHR$(13)+CHR$(13)

        PRINT #N%, _
        "Run #      Gamma     Nt    Nd    Np     G      DelT   Alpha    Beta   #Steps   Neval     Frep       Fitness     Initial Probes" + CHR$(13) +_
        "-----    -------   ----  ----  ----  ----  ----.-  --.--   --.--  ------  -------   -------  .####\\ ######.######   \          \"
        A$ = _
        "####    ###.###  ####  ####  ####  ###.#  ##.#  ##.##   ##.##  ######  #######  .#####\\ ######.######   \                        \"
'                                                                                          UNIFORM ON-AXIS
        PRINT #N%, USING$(A$,0,StartingGamma,Nt&,Nd%,Np%,StartingG,StartingDeltaT,StartingAlpha,StartingBeta,0,0,StartingFrep,LEFT$(RepositionFactor$,1),-9999,PlaceInitialProbes$)
'header

        PRINT #N%,_
        "-----------------------------------------------------------------------------------------------------------------------------------------"

        BestOverallFitness = -1E4200 'very large negative number...

        TotalFunctionEvaluations&& = 0

        AbsoluteRunNumber% = 0

'        ---------- ALPHA LOOP ---------
        FOR AlphaRunNumber% = 1 TO NumAlphaRuns%

            Alpha = MinAlpha + (AlphaRunNumber%-1)*(MaxAlpha-MinAlpha)/(NumAlphaRuns%-1)

            IF NumAlphaRuns% = 1 THEN Alpha = StartingAlpha

'        ------ Gravitational Constant, G LOOP ------
        FOR GrunNumber% = 1 TO NumGruns%

            G = MinG + (GrunNumber%-1)*(MaxG-MinG)/(NumGruns%-1)

            IF NumGruns% = 1 THEN G = StartingG

'        ------------------------- Np/Nd LOOP --------------------
        FOR NumProbesPerAxis% = MinProbesPerAxis% TO MaxProbesPerAxis% STEP 2 '2/4/6/.../MaxProbesPerAxis% probes per axis

        Np% = NumProbesPerAxis%*Nd% 'sets Np% regardless of what Sub GetFunctionRunParameters() returns.

'        ------------------------- GAMMA LOOP --------------------
        FOR GammaRunNumber% = 1 TO NumGammaRuns%

            INCR AbsoluteRunNumber%

            Gamma = StartingGamma + (GammaRunNumber%-1)*(StoppingGamma-StartingGamma)/(NumGammaRuns%-1)

            IF NumGammaRuns% = 1 THEN Gamma = StartingGamma

            'G = StartingG
            DeltaT = StartingDeltaT
            'Alpha = StartingAlpha
            Beta = StartingBeta
            Frep = StartingFrep

            CALL ResetDecisionSpaceBoundaries(Nd%,XiMin(),XiMax(),StartingXiMin(),StartingXiMax())

            CALL CFO(Nd%,Np%,Nt&,G,DeltaT,Alpha,Beta,Frep,R(),A(),M(),XiMin(),XiMax(),DiagLength,PlaceInitialProbes$,InitialAcceleration$,_
                        RepositionFactor$,FunctionName$,LastStep&,CheckForEarlyTermination$,BestFitnessThisRun,GammaRunNumber%,NumGammaRuns%,Gamma,BestOverallFitness,ShrinkDS$,MaxProbesPerAxis%,MinProbes
PerAxis%,_
                        AbsoluteRunNumber%,TotalRuns$)

            IF BestFitnessThisRun >= BestOverallFitness THEN

                AbsoluteBestRunNumber% = AbsoluteRunNumber% : BestOverallFitness = BestFitnessThisRun : BestRunNumber% = GammaRunNumber% : BestNumProbes% = Np% : BestG = G :
BestAlpha = Alpha

            END IF

            TotalFunctionEvaluations&& = TotalFunctionEvaluations&& + (LastStep&+1)*Np%

            PRINT #N%, USING$(A$,AbsoluteRunNumber%,Gamma,Nt&,Nd%,Np%,G,DeltaT,Alpha,Beta,LastStep&,(Laststep&+1)*Np%,Frep,_
                        LEFT$(RepositionFactor$,1),BestFitnessThisRun,PlaceInitialProbes$)

        NEXT GammaRunNumber%  'GAMMA LOOP

        NEXT NumProbesPerAxis% 'Np/Nd LOOP

        NEXT GrunNumber%   'G LOOP

        NEXT AlphaRunNumber%  'ALPHA LOOP

        PRINT #N%, CHR$(13)+USING$("                                                  Total Function Evaluations: ##########",TotalFunctionEvaluations&&)+CHR$(13)

'        ---------------------------------------------- Re-Run Best Run for Final Results ----------------------------------------------

        Gamma = StartingGamma + (BestRunNumber%-1)*(StoppingGamma-StartingGamma)/(NumGammaRuns%-1)

        MSGBOX("  Best Run#="+STR$(AbsoluteBestRunNumber%)+"   Gamma="+STR$(Gamma))

        'G = StartingG
        DeltaT = StartingDeltaT
        'Alpha = StartingAlpha
        Beta = StartingBeta
        Frep = StartingFrep
        Np% = BestNumProbes%
        G = BestG
        Alpha = BestAlpha

'       REDIM R(1 TO Np%, 1 TO Nd%, 0 TO Nt&), A(1 TO Np%, 1 TO Nd%, 0 TO Nt&), M(1 TO Np%, 0 TO Nt&) 'must redim these matrices because Np has changed

        CALL ResetDecisionSpaceBoundaries(Nd%,XiMin(),XiMax(),StartingXiMin(),StartingXiMax())

        MSGBOX("Best run input parameters:"+CHR$(13)+CHR$(13)+_
                "Nd="+STR$(Nd%)+"  Np="+STR$(Np%)+"  Nt="+STR$(Nt&)+"  G="+STR$(G)+"  DelT="+STR$(DeltaT)+"  Alpha="+STR$(Alpha)+"  Beta="+STR$(Beta)+"  Frep="+STR$(Frep)+"
Gamma="+STR$(Gamma)+CHR$(13))

        CALL CFO(Nd%,Np%,Nt&,G,DeltaT,Alpha,Beta,Frep,R(),A(),M(),XiMin(),XiMax(),DiagLength,PlaceInitialProbes$,InitialAcceleration$,_
```

```
RepositionFactor$,FunctionName$,LastStep&,CheckForEarlyTermination$,BestFitnessThisRun,BestRunNumber%,NumGammaRuns%,Gamma,BestOverallFitness,ShrinkDS$,MaxProbesPerAxis%,MinProbesP
erAxis%,_
                AbsoluteRunNumber%,TotalRuns%)

        MSGBOX("Best CFO run completed...")

        CALL ResetDecisionSpaceBoundaries(Nd%,XiMin(),XiMax(),StartingXiMin(),StartingXiMax())

        PRINT #N%,"-------------------------------------------------------------------------------------------------------------------------------------------------"

        PRINT #N%, USING$(A$,AbsoluteBestRunNumber%,Gamma,Nt&,Nd%,BestNumProbes%,G,DeltaT,Alpha,Beta,LastStep&,(Laststep&+1)*Np%,Frep,_
                    LEFT$(RepositionFactor$,1),BestFitnessThisRun,PlaceInitialProbes$)
        CLOSE #N%

    ELSE

        GOTO ExitPBMAIN

    END IF

'   -------------------------------- Display Best Fitness -----------------------------------------

    CALL DisplayBestFitness(Np%,Nd%,LastStep&,M(),R(),BestFitnessProbeNumber%,BestFitnessTimeStep&,FunctionName$)

'   ----------------------------------------------- PLOT EVOLUTION OF BEST FITNESS, AVG DISTANCE & BEST PROBE # --------------------------------------------------------

    CALL PlotBestFitnessEvolution(Nd%,Np%,LastStep&,G,DeltaT,Alpha,Beta,Frep,M(),PlaceInitialProbes%,InitialAcceleration$,RepositionFactor$,FunctionName$,Gamma)

    CALL PlotAverageDistance(Nd%,Np%,LastStep&,G,DeltaT,Alpha,Beta,Frep,M(),PlaceInitialProbes%,InitialAcceleration$,RepositionFactor$,DiagLength,Gamma)

    CALL BestProbeVsTimeStep(Nd%,Np%,LastStep&,G,DeltaT,Alpha,Beta,Frep,M(),PlaceInitialProbes%,InitialAcceleration$,RepositionFactor$,FunctionName$,Gamma)

'   ------------------------------------------------ PLOT TRAJECTORIES OF BEST PROBES FOR 2/3-D FUNCTIONS ----------------------------------------------------

    IF Nd% = 2 THEN

        NumTrajectories% = 10 : CALL Plot2DbestProbeTrajectories(NumTrajectories%,M(),R(),XiMin(),XiMax(),Np%,Nd%,LastStep&,FunctionName$)

        NumTrajectories% = 16 : CALL Plot2DindividualProbeTrajectories(NumTrajectories%,M(),R(),XiMin(),XiMax(),Np%,Nd%,LastStep&,FunctionName$)

    END IF

    IF Nd% = 3 THEN

        NumTrajectories% = 4 : CALL Plot3DbestProbeTrajectories(NumTrajectories%,M(),R(),XiMin(),XiMax(),Np%,Nd%,LastStep&,FunctionName$)

    END IF

'   ----------- For 1-D Objective Functions, Tabulate Probe Coordinates & if Np% =< Max1DprobesPlotted% Plot Evolution of Probe Positions ------------

    IF Nd% = 1 THEN

        CALL
Tabulate1DprobeCoordinates(Max1DprobesPlotted%,Nd%,Np%,LastStep&,G,DeltaT,Alpha,Beta,Frep,R(),M(),PlaceInitialProbes%,InitialAcceleration$,RepositionFactor$,FunctionName$,Gamma)

        IF Np% =< Max1DprobesPlotted% THEN CALL
Plot1DprobePositions(Max1DprobesPlotted%,Nd%,Np%,LastStep&,G,DeltaT,Alpha,Beta,Frep,R(),M(),PlaceInitialProbes%,InitialAcceleration$,RepositionFactor$,FunctionName$,Gamma)

        CALL CLEANUP 'delete probe coordinate plot files, if any

    END IF

ExitPBMAIN:

END FUNCTION 'PBMAIN()

'=================================================================================== CFO SUBROUTINE
====================================================================================

SUB CFO(Nd%,Np%,Nt&,G,DeltaT,Alpha,Beta,Frep,R(),A(),M(),XiMin(),XiMax(),DiagLength,PlaceInitialProbes%,InitialAcceleration$,_

RepositionFactor$,FunctionName$,LastStep&,CheckForEarlyTermination$,BestFitnessThisRun,GammaRunNumber%,NumGammaRuns%,Gamma,BestOverallFitness,ShrinkDS$,MaxProbesPerAxis%,MinProbes
PerAxis%,_
            AbsoluteRunNumber%,TotalRuns%)

LOCAL p%, i%, j& 'Standard Indices: Probe #, Coordinate #, Time Step #

LOCAL k%, L&      'Dummy summation indices

LOCAL SumSQ, Denom, Numerator, MaxInitialRandomAcceleration, MaxInitialFixedAcceleration AS EXT

LOCAL StepNumber&, BestProbeNumber%, BestTimeStep&

LOCAL BestFitness AS EXT

LOCAL DavgOscillation$, DavgSaturation$, FitnessSaturation$

LOCAL NstepsDavgSat&, NstepsFitnessSat&

LOCAL RatioOfSuccessiveSlopes AS EXT

REDIM R(1 TO Np%, 1 TO Nd%, 0 TO Nt&), A(1 TO Np%, 1 TO Nd%, 0 TO Nt&), M(1 TO Np%, 0 TO Nt&) 're-initializes Position Vetor/Acceleration/Fitness matrices to zero

'STEP (A1) ------------- Compute Initial Probe Distribution (Step 0)----------------

    CALL InitialProbeDistribution(Np%,Nd%,Nt&,XiMin(),XiMax(),R(),PlaceInitialProbes%,Gamma)

    IF Nd% = 2 THEN 'plot 2-D initial probe distribution at Step 0

        CALL CreateGNUplotINIfile(0.1##*ScreenWidth&,0.25##*ScreenHeight&,0.6##*ScreenWidth&,0.6##*ScreenHeight&)

        CALL Show2Dprobes(R(),Np%,Nt&,0,XiMin(),XiMax(),Frep,0##,1,0,FunctionName$,RepositionFactor$,Gamma) 'show initial probes for 2-D functions

    END IF

'STEP (A2) ------------- Compute Initial Fitness Matrix (Step 0) -------------

    FOR p% = 1 TO Np% : M(p%,0) = ObjectiveFunction(R(),Np%,Nd%,p%,0,Gamma) : NEXT p%

'STEP (A3) ------------- Compute Initial Probe Accelerations (Step 0)---------------

    MaxInitialRandomAcceleration =  2## 'maximum value for random initial acceleration: Random[0-MAX]
```

```
    MaxInitialFixedAcceleration  = 10## 'maximum value for fixed initial acceleration: [0.001-MAX]

    CALL InitialProbeAccelerations(Np%,Nd%,A(),InitialAcceleration$,MaxInitialRandomAcceleration,MaxInitialFixedAcceleration)

'  ======================================= LOOP ON TIME STEPS STARTING AT STEP #1 ===========================================

    LastStep& = Nt& 'unless run is terminated earlier

    BestFitnessThisRun = M(1,0)

    FOR j& = 1 TO Nt&

'STEP (B) ---------- Compute Probe Position Vectors for this Time Step --------

        FOR p% = 1 TO Np% : FOR i% = 1 TO Nd% : R(p%,i%,j&) = R(p%,i%,j&-1) + 0.5##*A(p%,i%,j&-1)*DeltaT^2 : NEXT i% : NEXT p%

'STEP (C) ---------- Retrieve Errant Probes --------------

        CALL RetrieveErrantProbes(Np%,Nd%,j&,XiMin(),XiMax(),R(),M(),RepositionFactor$,Frep)

'STEP (D) ---------- Compute Fitness Matrix for Current Probe Distribution ---------

        FOR p% = 1 TO Np% : M(p%,j&) = ObjectiveFunction(R(),Nd%,p%,j&,FunctionName$) : NEXT p%

'STEP (E) ---------- Compute Accelerations Based on Current Probe Distribution & Fitnesses ---------------

        FOR p% = 1 TO Np%

            FOR i% = 1 TO Nd%

                A(p%,i%,j&) = 0

                FOR k% = 1 TO Np%

                    IF k% <> p% THEN

                        SumSQ = 0##

                        FOR L% = 1 TO Nd%  : SumSQ = SumSQ + (R(k%,L%,j&)-R(p%,L%,j&))^2 : NEXT L% 'dummy index

                        Denom = SQR(SumSQ) : Numerator = UnitStep((M(k%,j&)-M(p%,j&)))*(M(k%,j&)-M(p%,j&))

                        A(p%,i%,j&) = A(p%,i%,j&) + G*(R(k%,i%,j&)-R(p%,i%,j&))*Numerator^Alpha/Denom^Beta

                    END IF

                NEXT k% 'dummy index

            NEXT i% 'coord (dimension) #

        NEXT p% 'probe #

'   --------- Get Best Fitness & Corresponding Probe # and Time Step ---------

    CALL GetBestFitness(M(),Np%,j&,BestFitness,BestProbeNumber%,BestTimeStep&)

    IF BestFitness >= BestFitnessThisRun THEN BestFitnessThisRun = BestFitness

'   ---------------------- Check for Davg OSC and Fitness/Davg SAT &  Display Run Status ------------------------

    DavgOscillation$  = OscillationInDavg$(j&,Np%,Nd%,M(),R(),DiagLength)              'check for oscillation in Davg

    NstepsDavgSat&    = 25 'CHANGED TO 25 ON 02-07-2010 '50 '25 'for 100D tests 'used 50 for new paper '15 '# steps for averaging Davg to test for saturation

    DavgSaturation$   = HasDAVGsaturated$(NstepsDavgSat&,j&,Np%,Nd%,M(),R(),DiagLength)    'check for saturation of Davg

    NstepsFitnessSat& = 25 'CHANGED ON 02-07-2010 '50 '# steps for averaging Fitness to test for saturation

    FitnessSaturation$ = HasFITNESSsaturated$(NstepsFitnessSat&,j&,Np%,Nd%,M(),R())  'check for saturation of best fitness

    GRAPHIC SET PIXEL (35,15) : GRAPHIC PRINT "Run #"  + REMOVE$(STR$(AbsoluteRunNumber%),ANY" ") + "/" + REMOVE$(STR$(TotalRuns%),ANY" ") + ", " +_
                                              "Step #" + REMOVE$(STR$(j&),ANY" ")        + "/" + REMOVE$(STR$(Nt&),ANY" ")         + ", Frep=" + REMOVE$(STR$(Frep,4),ANY" ") + STRING$(10," ")

    GRAPHIC SET PIXEL (35,35) : GRAPHIC PRINT "Gamma=" + REMOVE$(STR$(ROUND(Gamma,4)),ANY" ")     + ", Np/Nd=" + REMOVE$(STR$(Np%\Nd%),ANY" ") + STRING$(10," ")

    GRAPHIC SET PIXEL (35,55) : GRAPHIC PRINT "G="     + REMOVE$(STR$(ROUND(G,4)),ANY" ")         + ", Alpha=" + REMOVE$(STR$(ROUND(Alpha,4)),ANY" ") + STRING$(10," ")

    GRAPHIC SET PIXEL (35,75) : GRAPHIC PRINT "Best Fitness This Run = " + REMOVE$(STR$(ROUND(BestFitness,8)),ANY" ") +_
                                              " @ Probe #" + REMOVE$(STR$(BestProbeNumber%),ANY" ") + ", Step #" + REMOVE$(STR$(BestTimeStep&),ANY" ") + STRING$(15," ")

    IF AbsoluteRunNumber% = 1 THEN

        GRAPHIC SET PIXEL (35,95) : GRAPHIC PRINT "Best Fitness Overall = N/A @ Run #1" + STRING$(15," ")

    ELSE

        GRAPHIC SET PIXEL (35,95) : GRAPHIC PRINT "Best Fitness Overall = " + REMOVE$(STR$(ROUND(BestOverallFitness,8)),ANY" ") + STRING$(15," ")

    END IF

    GRAPHIC SET PIXEL (35,115) : GRAPHIC PRINT "Osc Davg? " + DavgOscillation$ + ",  Sat Davg? " + DavgSaturation$ +",  Sat Fitness? " + FitnessSaturation$ + STRING$(10," "):
GRAPHIC REDRAW

'   ------------------------------------------ If Variable Frep, Adjust Value-----------------------------------------

    IF RepositionFactor$ = "VARIABLE" THEN

        Frep = Frep + 0.05##

        IF Frep > 1## THEN Frep = 0.05## 'keep Frep in range [0.05,1]

    END IF

    IF RepositionFactor$ = "RANDOM" THEN Frep = RandomNum(0##,1##) 'set new Frep value randomly

'   ---------- Starting at Step #20 Shrink Decision Space Around Best Probe Every 10th Step -----------

    IF ShrinkDS$ = "YES" THEN

        IF j& MOD 10 = 0 AND j& >= 20 THEN 'SHRINK DS EVERY 10th STEP STARTING AT STEP #20 'ADDED 02-07-2010

'           --------------------------------------------------------- Shrink Decision Space Around Best Probe up to This Point in Run ---------------------------------------------
----------
```

```
        FOR i% = 1 TO Nd% : XiMin(i%) = XiMin(i%)+(R(BestProbeNumber%,i%,BestTimeStep&)-XiMin(i%))/2## : XiMax(i%) = XiMax(i%)-(XiMax(i%)-R(BestProbeNumber%,i%,BestTimeStep&))/2##
: NEXT i%

        CALL RetrieveErrantProbes(Np%,Nd%,j&,XiMin(),XiMax(),R(),M()),RepositionFactor$,Frep) 'TO RETRIEVE PROBES LYING OUTSIDE SHRUNKEN DS 'ADDED 02-07-2010

        END IF

    END IF

'STEP (F) -------------------- Check for Early Run Termination ---------------------

    IF CheckForEarlyTermination$ = "YES" THEN 'insert termination test

        IF FitnessSaturation = "YES" THEN 'terminate run

            LastStep& = j& : EXIT FOR  'time step loop

        END IF

    END IF

    NEXT j& 'END OF TIME STEP LOOP

END SUB 'CFO()

'====================================================================================================================================

SUB CheckNECFiles(NECfileError$)

LOCAL N%

    NECfileError$ = "NO"

'    ------------------- NEC Files Required for PBM Antenna Benchmarks -------------------

    IF DIR$("n41_2k1.exe") = "" THEN

        MSGBOX("WARNING!  'n41_2k1.exe' not found.  Run terminated.") : NECfileError$ = "YES" : EXIT SUB

    END IF

    N% = FREEFILE : OPEN "ENDERR.DAT" FOR OUTPUT AS #N%   : PRINT #N%, "NO"      : CLOSE #N%

    N% = FREEFILE : OPEN "FILE_MSG.DAT" FOR OUTPUT AS #N% : PRINT #N%, "NO"      : CLOSE #N%

    N% = FREEFILE : OPEN "NHSCALE.DAT"  FOR OUTPUT AS #N% : PRINT #N%, "0.00001" : CLOSE #N%

END SUB

'------

FUNCTION SlopeRatio(M(),Np%,StepNumber&)

LOCAL p% 'probe #

LOCAL BestFitnessAtStepNumber, BestFitnessAtStepNumberMinus1, BestFitnessAtStepNumberMinus2, Z AS EXT

    IF StepNumber& < 5 THEN 'at least five steps to perform this test

        Z = 1## : GOTO ExitSlopeRatio 'assumes no slope change

    END IF

    BestFitnessAtStepNumber       = M(1,StepNumber&)   : FOR p% = 1 TO Np% : IF M(p%,StepNumber&)   >= BestFitnessAtStepNumber       THEN BestFitnessAtStepNumber       =
M(p%,StepNumber&)   : NEXT p%

    BestFitnessAtStepNumberMinus1 = M(1,StepNumber&-1) : FOR p% = 1 TO Np% : IF M(p%,StepNumber&-1) >= BestFitnessAtStepNumberMinus1 THEN BestFitnessAtStepNumberMinus1 =
M(p%,StepNumber&-1) : NEXT p%

    BestFitnessAtStepNumberMinus2 = M(1,StepNumber&-2) : FOR p% = 1 TO Np% : IF M(p%,StepNumber&-2) >= BestFitnessAtStepNumberMinus2 THEN BestFitnessAtStepNumberMinus2 =
M(p%,StepNumber&-2) : NEXT p%

    Z = (BestFitnessAtStepNumber-BestFitnessAtStepNumberMinus1)/(BestFitnessAtStepNumberMinus1-BestFitnessAtStepNumberMinus2)

ExitSlopeRatio:

    SlopeRatio = Z

END FUNCTION

'------

SUB GetBestFitness(M(),Np%,StepNumber&,BestFitness,BestProbeNumber%,BestTimeStep&)

LOCAL p%, i&, A$

    BestFitness = M(1,0)

    FOR i& = 0 TO StepNumber&

        FOR p% = 1 TO Np%

            IF M(p%,i&) >= BestFitness THEN

                BestFitness = M(p%,i&) : BestProbeNumber% = p% : BestTimeStep& = i&

            END IF

        NEXT p%

    NEXT i&

END SUB

'=============================================================== FUNCTION DEFINITIONS ===============================================================

SUB SelectTestFunction(FunctionName$)

LOCAL A$

    A$ = INPUTBOX$("Which Function?"+CHR$(13)+"1 - Parrott F4 (1-D)"+CHR$(13)+"2 - SGO (2-D)"+CHR$(13)+"3 - GP (2-D)"+CHR$(13)+"4 - Step (n-D)"+CHR$(13)+ _
                "5 - Schwefel 2.26 (n-D)"+CHR$(13)+"6 - Colville (4-D)"+CHR$(13)+"7 - RESERVED"+CHR$(13)+"8 - more","SELECT OBJECTIVE FUNCTION","1")
```

```
        IF VAL(A$) < 1 OR VAL(A$) > 8 THEN A$ = "1"

        SELECT CASE VAL(A$)

            CASE 1 : FunctionName$ = "ParrottF4"
            CASE 2 : FunctionName$ = "SGO"
            CASE 3 : FunctionName$ = "GP"
            CASE 4 : FunctionName$ = "STEP"
            CASE 5 : FunctionName$ = "SCHWEFEL_226"
            CASE 6 : FunctionName$ = "COLVILLE"
            CASE 7 : FunctionName$ = "GRIEWANK"
            CASE 8 : FunctionName$ = "more"

        END SELECT

        IF FunctionName$ = "more" THEN

            A$ = INPUTBOX$("Which Function?"+CHR$(13)+"1 - F1"+CHR$(13)+"2 - F2"+CHR$(13)+"3 - F3"+CHR$(13)+"4 - F4"+CHR$(13)+ _
                "5 - F5"+CHR$(13)+"6 - F6"+CHR$(13)+"7 - F7"+CHR$(13)+"8 - more","SELECT OBJECTIVE FUNCTION","1")

            IF VAL(A$) < 1 OR VAL(A$) > 8 THEN A$ = "1"

            SELECT CASE VAL(A$)

                CASE 1 : FunctionName$ = "F1"
                CASE 2 : FunctionName$ = "F2"
                CASE 3 : FunctionName$ = "F3"
                CASE 4 : FunctionName$ = "F4"
                CASE 5 : FunctionName$ = "F5"
                CASE 6 : FunctionName$ = "F6"
                CASE 7 : FunctionName$ = "F7"
                CASE 8 : FunctionName$ = "more"

            END SELECT

        END IF

            IF FunctionName$ = "more" THEN

            A$ = INPUTBOX$("Which Function?"+CHR$(13)+"1 - F8"+CHR$(13)+"2 - F9"+CHR$(13)+"3 - F10"+CHR$(13)+"4 - F11"+CHR$(13)+ _
                    "5 - F12"+CHR$(13)+"6 - F13"+CHR$(13)+"7 - F14"+CHR$(13)+"8 - more","SELECT OBJECTIVE FUNCTION","1")

            IF VAL(A$) < 1 OR VAL(A$) > 8 THEN A$ = "1"

            SELECT CASE VAL(A$)

                CASE 1 : FunctionName$ = "F8"
                CASE 2 : FunctionName$ = "F9"
                CASE 3 : FunctionName$ = "F10"
                CASE 4 : FunctionName$ = "F11"
                CASE 5 : FunctionName$ = "F12"
                CASE 6 : FunctionName$ = "F13"
                CASE 7 : FunctionName$ = "F14"
                CASE 8 : FunctionName$ = "more"

            END SELECT

        END IF

        IF FunctionName$ = "more" THEN

            A$ = INPUTBOX$("Which Function?"+CHR$(13)+"1 - F15"+CHR$(13)+"2 - F16"+CHR$(13)+"3 - F17"+CHR$(13)+"4 - F18"+CHR$(13)+ _
                    "5 - F19"+CHR$(13)+"6 - F20"+CHR$(13)+"7 - F21"+CHR$(13)+"8 - more","SELECT OBJECTIVE FUNCTION","1")

            IF VAL(A$) < 1 OR VAL(A$) > 8 THEN A$ = "1"

            SELECT CASE VAL(A$)

                CASE 1 : FunctionName$ = "F15"
                CASE 2 : FunctionName$ = "F16"
                CASE 3 : FunctionName$ = "F17"
                CASE 4 : FunctionName$ = "F18"
                CASE 5 : FunctionName$ = "F19"
                CASE 6 : FunctionName$ = "F20"
                CASE 7 : FunctionName$ = "F21"
                CASE 8 : FunctionName$ = "more"

            END SELECT

        END IF

        IF FunctionName$ = "more" THEN

            A$ = INPUTBOX$("Which Function?"+CHR$(13)+"1 - F22"+CHR$(13)+"2 - F23"+CHR$(13)+"3 - F24"+CHR$(13)+"4 - F25"+CHR$(13)+ _
                    "5 - F26"+CHR$(13)+"6 - F27"+CHR$(13)+"7 - F28"+CHR$(13)+"8 - more","SELECT OBJECTIVE FUNCTION","1")

            IF VAL(A$) < 1 OR VAL(A$) > 2 THEN A$ = "1"

            SELECT CASE VAL(A$)

                CASE 1 : FunctionName$ = "F22"
                CASE 2 : FunctionName$ = "F23"
                CASE 3 : FunctionName$ = "F24"
                CASE 4 : FunctionName$ = "F25"
                CASE 5 : FunctionName$ = "F26"
                CASE 6 : FunctionName$ = "F27"
                CASE 7 : FunctionName$ = "F28"
                CASE 8 : FunctionName$ = "more"

            END SELECT

        END IF

END SUB 'SelectTestFunction()

'------

FUNCTION ObjectiveFunction(R(),Nd%,p%,j&,FunctionName$) 'Objective function to be MAXIMIZED is defined here

    SELECT CASE FunctionName$

        CASE "ParrottF4"    : ObjectiveFunction = ParrottF4(R(),Nd%,p%,j&)      'Parrott F4 (1-D)

        CASE "SGO"          : ObjectiveFunction = SGO(R(),Nd%,p%,j&)            'SGO Function (2-D)
```

```
        CASE "GP"            : ObjectiveFunction = GoldsteinPrice(R(),Nd%,p%,j&) 'Goldstein-Price Function (2-D)

        CASE "STEP"          : ObjectiveFunction = StepFunction(R(),Nd%,p%,j&)   'Step Function (n-D)

        CASE "SCHWEFEL_226"  : ObjectiveFunction = Schwefel226(R(),Nd%,p%,j&)    'Schwefel Prob. 2.26 (n-D)

        CASE "COLVILLE"      : ObjectiveFunction = Colville(R(),Nd%,p%,j&)       'Colville Function (4-D)

        CASE "GRIEWANK"      : ObjectiveFunction = Griewank(R(),Nd%,p%,j&)       'Griewank Function (n-D)

        CASE "HIMMELBLAU"    : ObjectiveFunction = Himmelblau(R(),Nd%,p%,j&)     'Himmelblau Function (2-D)

        CASE "ROSENBROCK"    : ObjectiveFunction = Rosenbrock(R(),Nd%,p%,j&)     'Rosenbrock Function (n-D)

        CASE "SPHERE"        : ObjectiveFunction = Sphere(R(),Nd%,p%,j&)         'Sphere Function (n-D)

'       ------------------------ GSO Paper Benchmark Functions ----------------------------
        CASE "F1"            : ObjectiveFunction = F1(R(),Nd%,p%,j&)             'F1  (n-D)
        CASE "F2"            : ObjectiveFunction = F2(R(),Nd%,p%,j&)             'F2  (n-D)
        CASE "F3"            : ObjectiveFunction = F3(R(),Nd%,p%,j&)             'F3  (n-D)
        CASE "F4"            : ObjectiveFunction = F4(R(),Nd%,p%,j&)             'F4  (n-D)
        CASE "F5"            : ObjectiveFunction = F5(R(),Nd%,p%,j&)             'F5  (n-D)
        CASE "F6"            : ObjectiveFunction = F6(R(),Nd%,p%,j&)             'F6  (n-D)
        CASE "F7"            : ObjectiveFunction = F7(R(),Nd%,p%,j&)             'F7  (n-D)
        CASE "F8"            : ObjectiveFunction = F8(R(),Nd%,p%,j&)             'F8  (n-D)
        CASE "F9"            : ObjectiveFunction = F9(R(),Nd%,p%,j&)             'F9  (n-D)
        CASE "F10"           : ObjectiveFunction = F10(R(),Nd%,p%,j&)           'F10 (n-D)
        CASE "F11"           : ObjectiveFunction = F11(R(),Nd%,p%,j&)           'F11 (n-D)
        CASE "F12"           : ObjectiveFunction = F12(R(),Nd%,p%,j&)           'F12 (n-D)
        CASE "F13"           : ObjectiveFunction = F13(R(),Nd%,p%,j&)           'F13 (n-D)
        CASE "F14"           : ObjectiveFunction = F14(R(),Nd%,p%,j&)           'F14 (2-D)
        CASE "F15"           : ObjectiveFunction = F15(R(),Nd%,p%,j&)           'F15 (4-D)
        CASE "F16"           : ObjectiveFunction = F16(R(),Nd%,p%,j&)           'F16 (2-D)
        CASE "F17"           : ObjectiveFunction = F17(R(),Nd%,p%,j&)           'F17 (2-D)
        CASE "F18"           : ObjectiveFunction = F18(R(),Nd%,p%,j&)           'F18 (2-D)
        CASE "F19"           : ObjectiveFunction = F19(R(),Nd%,p%,j&)           'F19 (3-D)
        CASE "F20"           : ObjectiveFunction = F20(R(),Nd%,p%,j&)           'F20 (6-D)
        CASE "F21"           : ObjectiveFunction = F21(R(),Nd%,p%,j&)           'F21 (4-D)
        CASE "F22"           : ObjectiveFunction = F22(R(),Nd%,p%,j&)           'F22 (4-D)
        CASE "F23"           : ObjectiveFunction = F23(R(),Nd%,p%,j&)           'F23 (4-D)

'       ------------------------- PBM Antenna Benchmarks ----------------------------
        CASE "PBM_1"         : ObjectiveFunction = PBM_1(R(),Nd%,p%,j&)         'PBM_1 (2-D)
        CASE "PBM_2"         : ObjectiveFunction = PBM_2(R(),Nd%,p%,j&)         'PBM_2 (2-D)
        CASE "PBM_3"         : ObjectiveFunction = PBM_3(R(),Nd%,p%,j&)         'PBM_3 (2-D)
        CASE "PBM_4"         : ObjectiveFunction = PBM_4(R(),Nd%,p%,j&)         'PBM_4 (2-D)
        CASE "PBM_5"         : ObjectiveFunction = PBM_5(R(),Nd%,p%,j&)         'PBM_5 (2-D)

    END SELECT

END FUNCTION 'ObjectiveFunction()

'-----------

SUB ResetDecisionSpaceBoundaries(Nd%,XiMin(),XiMax(),StartingXiMin(),StartingXiMax())

LOCAL i%

    FOR i% = 1 TO Nd% : XiMin(i%) = StartingXiMin(i%) : XiMax(i%) = StartingXiMax(i%) : NEXT i%

END SUB

'------

SUB GetFunctionRunParameters(FunctionName$,Nd%,Np%,Nt&,G,DeltaT,Alpha,Beta,Frep,R(),A(),M(),XiMin(),XiMax(),StartingXiMin(),StartingXiMax(),_
                             DiagLength,PlaceInitialProbes$,InitialAcceleration$,RepositionFactor$)

LOCAL i%, NumProbesPerDimension%, NN%, NumCollinearElements%

    SELECT CASE FunctionName$

        CASE "ParrottF4"

            Nd%                   = 1
            NumProbesPerDimension% = 3
            Np%                   = NumProbesPerDimension%*Nd%

            Nt&    = 500
            G      = 2##
            Alpha  = 2##
            Beta   = 2##
            DeltaT = 1##
            Frep   = 0.9##

            PlaceInitialProbes$   = "UNIFORM ON-AXIS"
            InitialAcceleration$  = "ZERO"
            RepositionFactor$     = "FIXED"

            CALL ChangeRunParameters(NumProbesPerDimension%,Np%,Nd%,Nt&,G,Alpha,Beta,DeltaT,Frep,PlaceInitialProbes$,InitialAcceleration$,RepositionFactor$,FunctionName$)

            Nd% = 1 'cannot change dimensionality of Parrott F4 function!

            NumProbesPerDimension% = MAX(NumProbesPerDimension%,3) 'at least three for 1-D functions

            Np% = NumProbesPerDimension%*Nd%

            REDIM XiMin(1 TO Nd%), XiMax(1 TO Nd%) : XiMin(1) = 0## : XiMax(1) = 1##

            REDIM StartingXiMin(1 TO Nd%), StartingXiMax(1 TO Nd%) : FOR i% = 1 TO Nd% : StartingXiMin(i%) = XiMin(i%) : StartingXiMax(i%) = XiMax(i%) : NEXT i%

        CASE "SGO"

            Nd%                   = 2
            NumProbesPerDimension% = 4 '10 ^4
            Np%                   = NumProbesPerDimension%*Nd%

            Nt&    = 500
            G      = 2##
            Alpha  = 2##
            Beta   = 2##
            DeltaT = 1##
            Frep   = 0.5##
```

```
        PlaceInitialProbes$  = "UNIFORM ON-AXIS" '"2D GRID"
        InitialAcceleration$ = "ZERO"
        RepositionFactor$    = "VARIABLE"

        CALL ChangeRunParameters(NumProbesPerDimension%,Np%,Nd%,Nt&,G,Alpha,Beta,DeltaT,Frep,PlaceInitialProbes$,InitialAcceleration$,RepositionFactor$,FunctionName$)

        Nd% = 2 'cannot change dimensionality of SGO function!

        Np% = NumProbesPerDimension%*Nd%

        REDIM XiMin(1 TO Nd%), XiMax(1 TO Nd%) : FOR i% = 1 TO Nd% : XiMin(i%) = -50## : XiMax(i%) = 50## : NEXT i%

        REDIM StartingXiMin(1 TO Nd%), StartingXiMax(1 TO Nd%) : FOR i% = 1 TO Nd% : StartingXiMin(i%) = XiMin(i%) : StartingXiMax(i%) = XiMax(i%) : NEXT i%

        IF PlaceInitialProbes$ = "2D GRID" THEN

            Np% = NumProbesPerDimension%^2 : REDIM R(1 TO Np%, 1 TO Nd%, 0 TO Nt&) 'to create (Np/Nd) x (Np/Nd) grid

        END IF

    CASE "GP"

        Nd%                  = 2
        NumProbesPerDimension% = 4 '10
        Np%                  = NumProbesPerDimension%*Nd%

        Nt&     = 500
        G       = 2##
        Alpha   = 2## '0.2##
        Beta    = 2##
        DeltaT  = 1##
        Frep    = 0.5## '0.8## '0.9##

        PlaceInitialProbes$  = "UNIFORM ON-AXIS" '"2D GRID"
        InitialAcceleration$ = "ZERO"
        RepositionFactor$    = "VARIABLE"

        CALL ChangeRunParameters(NumProbesPerDimension%,Np%,Nd%,Nt&,G,Alpha,Beta,DeltaT,Frep,PlaceInitialProbes$,InitialAcceleration$,RepositionFactor$,FunctionName$)

        Nd% = 2 'cannot change dimensionality of GP function!

        Np% = NumProbesPerDimension%*Nd%

        REDIM XiMin(1 TO Nd%), XiMax(1 TO Nd%) : FOR i% = 1 TO Nd% : XiMin(i%) = -100## : XiMax(i%) = 100## : NEXT i%

        REDIM StartingXiMin(1 TO Nd%), StartingXiMax(1 TO Nd%) : FOR i% = 1 TO Nd% : StartingXiMin(i%) = XiMin(i%) : StartingXiMax(i%) = XiMax(i%) : NEXT i%

        IF PlaceInitialProbes$ = "2D GRID" THEN

            Np% = NumProbesPerDimension%^2 : REDIM R(1 TO Np%, 1 TO Nd%, 0 TO Nt&) 'to create (Np/Nd) x (Np/Nd) grid

        END IF

    CASE "STEP"

        Nd%                  = 2
        NumProbesPerDimension% = 4 '20
        Np%                  = NumProbesPerDimension%*Nd%

        Nt&     = 500
        G       = 2##
        Alpha   = 2##
        Beta    = 2##
        DeltaT  = 1##
        Frep    = 0.5##

        PlaceInitialProbes$  = "UNIFORM ON-AXIS"
        InitialAcceleration$ = "ZERO"
        RepositionFactor$    = "VARIABLE" '"FIXED"

        CALL ChangeRunParameters(NumProbesPerDimension%,Np%,Nd%,Nt&,G,Alpha,Beta,DeltaT,Frep,PlaceInitialProbes$,InitialAcceleration$,RepositionFactor$,FunctionName$)

        Np% = NumProbesPerDimension%*Nd%

        REDIM XiMin(1 TO Nd%), XiMax(1 TO Nd%) : FOR i% = 1 TO Nd% : XiMin(i%) = -100## : XiMax(i%) = 100## : NEXT i%
'       REDIM XiMin(1 TO Nd%), XiMax(1 TO Nd%) : XiMin(1) = 72## : XiMax(1) = 78## : XiMin(2) = 27## : XiMax(2) = 33## 'use this to plot STEP detail

        REDIM StartingXiMin(1 TO Nd%), StartingXiMax(1 TO Nd%) : FOR i% = 1 TO Nd% : StartingXiMin(i%) = XiMin(i%) : StartingXiMax(i%) = XiMax(i%) : NEXT i%

        IF PlaceInitialProbes$ = "2D GRID" THEN

            Np% = NumProbesPerDimension%^2 : REDIM R(1 TO Np%, 1 TO Nd%, 0 TO Nt&) 'to create (Np/Nd) x (Np/Nd) grid

        END IF 'STEP

    CASE "SCHWEFEL_226"

        Nd%                  = 30
        NumProbesPerDimension% = 4
        Np%                  = NumProbesPerDimension%*Nd%

        Nt&     = 500
        G       = 2##
        Alpha   = 2##
        Beta    = 2##
        DeltaT  = 1##
        Frep    = 0.5##

        PlaceInitialProbes$  = "UNIFORM ON-AXIS"
        InitialAcceleration$ = "ZERO"
        RepositionFactor$    = "VARIABLE"

        CALL ChangeRunParameters(NumProbesPerDimension%,Np%,Nd%,Nt&,G,Alpha,Beta,DeltaT,Frep,PlaceInitialProbes$,InitialAcceleration$,RepositionFactor$,FunctionName$)

        Np% = NumProbesPerDimension%*Nd%

        REDIM XiMin(1 TO Nd%), XiMax(1 TO Nd%) : FOR i% = 1 TO Nd% : XiMin(i%) = -500## : XiMax(i%) = 500## : NEXT i%

        REDIM StartingXiMin(1 TO Nd%), StartingXiMax(1 TO Nd%) : FOR i% = 1 TO Nd% : StartingXiMin(i%) = XiMin(i%) : StartingXiMax(i%) = XiMax(i%) : NEXT i%

        IF PlaceInitialProbes$ = "2D GRID" THEN
```

```
            Np% = NumProbesPerDimension%^2 : REDIM R(1 TO Np%, 1 TO Nd%, 0 TO Nt&) 'to create (Np/Nd) x (Np/Nd) grid

        END IF

    CASE "COLVILLE"

        Nd%                  = 4
        NumProbesPerDimension% = 4 '14
        Np%                  = NumProbesPerDimension%*Nd%

        Nt&      = 500
        G        = 2##
        Alpha    = 2##
        Beta     = 2##
        DeltaT   = 1##
        Frep     = 0.5##

        PlaceInitialProbes$  = "UNIFORM ON-AXIS"
        InitialAcceleration$ = "ZERO"
        RepositionFactor$    = "VARIABLE"

        CALL ChangeRunParameters(NumProbesPerDimension%,Np%,Nd%,Nt&,G,Alpha,Beta,DeltaT,Frep,PlaceInitialProbes$,InitialAcceleration$,RepositionFactor$,FunctionName$)

        Nd% = 4 'cannot change dimensionality of Colville function!

        Np% = NumProbesPerDimension%*Nd%

        REDIM XiMin(1 TO Nd%), XiMax(1 TO Nd%) : FOR i% = 1 TO Nd% : XiMin(i%) = -10## : XiMax(i%) = 10## : NEXT i%

        REDIM StartingXiMin(1 TO Nd%), StartingXiMax(1 TO Nd%) : FOR i% = 1 TO Nd% : StartingXiMin(i%) = XiMin(i%) : StartingXiMax(i%) = XiMax(i%) : NEXT i%

        IF PlaceInitialProbes$ = "2D GRID" THEN

            Np% = NumProbesPerDimension%^2 : REDIM R(1 TO Np%, 1 TO Nd%, 0 TO Nt&) 'to create (Np/Nd) x (Np/Nd) grid

        END IF

    CASE "GRIEWANK"

        Nd%                  = 2
        NumProbesPerDimension% = 4 '14
        Np%                  = NumProbesPerDimension%*Nd%

        Nt&      = 500
        G        = 2##
        Alpha    = 2##
        Beta     = 2##
        DeltaT   = 1##
        Frep     = 0.5##

        PlaceInitialProbes$  = "UNIFORM ON-AXIS"
        InitialAcceleration$ = "ZERO"
        RepositionFactor$    = "VARIABLE"

        CALL ChangeRunParameters(NumProbesPerDimension%,Np%,Nd%,Nt&,G,Alpha,Beta,DeltaT,Frep,PlaceInitialProbes$,InitialAcceleration$,RepositionFactor$,FunctionName$)

        Np% = NumProbesPerDimension%*Nd%

        REDIM XiMin(1 TO Nd%), XiMax(1 TO Nd%) : FOR i% = 1 TO Nd% : XiMin(i%) = -600## : XiMax(i%) = 600## : NEXT i%

        REDIM StartingXiMin(1 TO Nd%), StartingXiMax(1 TO Nd%) : FOR i% = 1 TO Nd% : StartingXiMin(i%) = XiMin(i%) : StartingXiMax(i%) = XiMax(i%) : NEXT i%

        IF PlaceInitialProbes$ = "2D GRID" THEN

            Np% = NumProbesPerDimension%^2 : REDIM R(1 TO Np%, 1 TO Nd%, 0 TO Nt&) 'to create (Np/Nd) x (Np/Nd) grid

        END IF

    CASE "HIMMELBLAU"

        Nd%                  = 2
        NumProbesPerDimension% = 4 '14
        Np%                  = NumProbesPerDimension%*Nd%

        Nt&      = 500
        G        = 2##
        Alpha    = 2##
        Beta     = 2##
        DeltaT   = 1##
        Frep     = 0.5##

        PlaceInitialProbes$  = "UNIFORM ON-AXIS"
        InitialAcceleration$ = "ZERO"
        RepositionFactor$    = "VARIABLE"

        CALL ChangeRunParameters(NumProbesPerDimension%,Np%,Nd%,Nt&,G,Alpha,Beta,DeltaT,Frep,PlaceInitialProbes$,InitialAcceleration$,RepositionFactor$,FunctionName$)

        Nd% = 2 'cannot change dimensionality of Himmelblau function!

        Np% = NumProbesPerDimension%*Nd%

        REDIM XiMin(1 TO Nd%), XiMax(1 TO Nd%) : FOR i% = 1 TO Nd% : XiMin(i%) = -6## : XiMax(i%) = 6## : NEXT i%

        REDIM StartingXiMin(1 TO Nd%), StartingXiMax(1 TO Nd%) : FOR i% = 1 TO Nd% : StartingXiMin(i%) = XiMin(i%) : StartingXiMax(i%) = XiMax(i%) : NEXT i%

        IF PlaceInitialProbes$ = "2D GRID" THEN

            Np% = NumProbesPerDimension%^2 : REDIM R(1 TO Np%, 1 TO Nd%, 0 TO Nt&) 'to create (Np/Nd) x (Np/Nd) grid

        END IF 'Himmelblau

    CASE "ROSENBROCK" '(n-D)

        Nd%                  = 2'30
        NumProbesPerDimension% = 4
        Np%                  = NumProbesPerDimension%*Nd%

        Nt&      = 250
        G        = 2##
        Alpha    = 2##
        Beta     = 2##
        DeltaT   = 1##
        Frep     = 0.5##
```

```
            PlaceInitialProbes$  = "UNIFORM ON-AXIS"
            InitialAcceleration$ = "ZERO"
            RepositionFactor$    = "VARIABLE"

            CALL ChangeRunParameters(NumProbesPerDimension%,Np%,Nd%,Nt&,G,Alpha,Beta,DeltaT,Frep,PlaceInitialProbes$,InitialAcceleration$,RepositionFactor$,FunctionName$)

            Np% = NumProbesPerDimension%*Nd%

            REDIM XiMin(1 TO Nd%), XiMax(1 TO Nd%) : FOR i% = 1 TO Nd% : XiMin(i%) = -2## : XiMax(i%) = 2## : NEXT i% :'XiMin(i%) = -6## : XiMax(i%) = 6## : NEXT i%

            REDIM StartingXiMin(1 TO Nd%), StartingXiMax(1 TO Nd%) : FOR i% = 1 TO Nd% : StartingXiMin(i%) = XiMin(i%) : StartingXiMax(i%) = XiMax(i%) : NEXT i%

            IF PlaceInitialProbes$ = "2D GRID" THEN

                Np% = NumProbesPerDimension%^2 : REDIM R(1 TO Np%, 1 TO Nd%, 0 TO Np%) 'to create (Np/Nd) x (Np/Nd) grid

            END IF 'ROSENBROCK

    CASE "SPHERE" '(n-D)

            Nd%                   = 2*30
            NumProbesPerDimension% = 4
            Np%                   = NumProbesPerDimension%*Nd%

            Nt&      = 250
            G        = 2##
            Alpha    = 2##
            Beta     = 2##
            DeltaT   = 1##
            Frep     = 0.5##

            PlaceInitialProbes$  = "UNIFORM ON-AXIS"
            InitialAcceleration$ = "ZERO"
            RepositionFactor$    = "VARIABLE"

            CALL ChangeRunParameters(NumProbesPerDimension%,Np%,Nd%,Nt&,G,Alpha,Beta,DeltaT,Frep,PlaceInitialProbes$,InitialAcceleration$,RepositionFactor$,FunctionName$)

            Np% = NumProbesPerDimension%*Nd%

            REDIM XiMin(1 TO Nd%), XiMax(1 TO Nd%) : FOR i% = 1 TO Nd% : XiMin(i%) = -100## : XiMax(i%) = 100## : NEXT i%

            REDIM StartingXiMin(1 TO Nd%), StartingXiMax(1 TO Nd%) : FOR i% = 1 TO Nd% : StartingXiMin(i%) = XiMin(i%) : StartingXiMax(i%) = XiMax(i%) : NEXT i%

            IF PlaceInitialProbes$ = "2D GRID" THEN

                Np% = NumProbesPerDimension%^2 : REDIM R(1 TO Np%, 1 TO Nd%, 0 TO Np%) 'to create (Np/Nd) x (Np/Nd) grid

            END IF 'SPHERE

    CASE "F1" '(n-D)

            Nd%                   = 30
            NumProbesPerDimension% = 2
            Np%                   = NumProbesPerDimension%*Nd%

            Nt&      = 1000
            G        = 2##
            Alpha    = 2##
            Beta     = 2##
            DeltaT   = 1##
            Frep     = 0.5##

            PlaceInitialProbes$  = "UNIFORM ON-AXIS"
            InitialAcceleration$ = "ZERO"
            RepositionFactor$    = "VARIABLE"

            CALL ChangeRunParameters(NumProbesPerDimension%,Np%,Nd%,Nt&,G,Alpha,Beta,DeltaT,Frep,PlaceInitialProbes$,InitialAcceleration$,RepositionFactor$,FunctionName$)

            Np% = NumProbesPerDimension%*Nd%

            REDIM XiMin(1 TO Nd%), XiMax(1 TO Nd%) : FOR i% = 1 TO Nd% : XiMin(i%) = -100## : XiMax(i%) = 100## : NEXT i%

            REDIM StartingXiMin(1 TO Nd%), StartingXiMax(1 TO Nd%) : FOR i% = 1 TO Nd% : StartingXiMin(i%) = XiMin(i%) : StartingXiMax(i%) = XiMax(i%) : NEXT i%

            IF PlaceInitialProbes$ = "2D GRID" THEN

                Np% = NumProbesPerDimension%^2 : REDIM R(1 TO Np%, 1 TO Nd%, 0 TO Nt&) 'to create (Np/Nd) x (Np/Nd) grid

            END IF 'F1

    CASE "F2" '(n-D)

            Nd%                   = 30
            NumProbesPerDimension% = 2
            Np%                   = NumProbesPerDimension%*Nd%

            Nt&      = 1000
            G        = 2##
            Alpha    = 2##
            Beta     = 2##
            DeltaT   = 1##
            Frep     = 0.5##

            PlaceInitialProbes$  = "UNIFORM ON-AXIS"
            InitialAcceleration$ = "ZERO"
            RepositionFactor$    = "VARIABLE"

            CALL ChangeRunParameters(NumProbesPerDimension%,Np%,Nd%,Nt&,G,Alpha,Beta,DeltaT,Frep,PlaceInitialProbes$,InitialAcceleration$,RepositionFactor$,FunctionName$)

            Np% = NumProbesPerDimension%*Nd%

            REDIM XiMin(1 TO Nd%), XiMax(1 TO Nd%) : FOR i% = 1 TO Nd% : XiMin(i%) = -10## : XiMax(i%) = 10## : NEXT i%

            REDIM StartingXiMin(1 TO Nd%), StartingXiMax(1 TO Nd%) : FOR i% = 1 TO Nd% : StartingXiMin(i%) = XiMin(i%) : StartingXiMax(i%) = XiMax(i%) : NEXT i%

            IF PlaceInitialProbes$ = "2D GRID" THEN

                Np% = NumProbesPerDimension%^2 : REDIM R(1 TO Np%, 1 TO Nd%, 0 TO Nt&) 'to create (Np/Nd) x (Np/Nd) grid

            END IF

    CASE "F3" '(n-D)

            Nd%                   = 30
```

```
        NumProbesPerDimension% = 2
        Np%              = NumProbesPerDimension%*Nd%

        Nt&     = 1000
        G       = 2##
        Alpha   = 2##
        Beta    = 2##
        DeltaT  = 1##
        Frep    = 0.5##

        PlaceInitialProbes$  = "UNIFORM ON-AXIS"
        InitialAcceleration$ = "ZERO"
        RepositionFactor$    = "VARIABLE"

        CALL ChangeRunParameters(NumProbesPerDimension%,Np%,Nd%,Nt&,G,Alpha,Beta,DeltaT,Frep,PlaceInitialProbes$,InitialAcceleration$,RepositionFactor$,FunctionName$)

        Np% = NumProbesPerDimension%*Nd%

        REDIM XiMin(1 TO Nd%), XiMax(1 TO Nd%) : FOR i% = 1 TO Nd% : XiMin(i%) = -100## : XiMax(i%) = 100## : NEXT i%

        REDIM StartingXiMin(1 TO Nd%), StartingXiMax(1 TO Nd%) : FOR i% = 1 TO Nd% : StartingXiMin(i%) = XiMin(i%) : StartingXiMax(i%) = XiMax(i%) : NEXT i%

        IF PlaceInitialProbes$ = "2D GRID" THEN

            Np% = NumProbesPerDimension%^2 : REDIM R(1 TO Np%, 1 TO Nd%, 0 TO Nt&) 'to create (Np/Nd) x (Np/Nd) grid

        END IF

CASE "F4" '(n-D)

        Nd%              = 30
        NumProbesPerDimension% = 2
        Np%              = NumProbesPerDimension%*Nd%

        Nt&     = 1000
        G       = 2##
        Alpha   = 2##
        Beta    = 2##
        DeltaT  = 1##
        Frep    = 0.5##

        PlaceInitialProbes$  = "UNIFORM ON-AXIS"
        InitialAcceleration$ = "ZERO"
        RepositionFactor$    = "VARIABLE"

        CALL ChangeRunParameters(NumProbesPerDimension%,Np%,Nd%,Nt&,G,Alpha,Beta,DeltaT,Frep,PlaceInitialProbes$,InitialAcceleration$,RepositionFactor$,FunctionName$)

        Np% = NumProbesPerDimension%*Nd%

        REDIM XiMin(1 TO Nd%), XiMax(1 TO Nd%) : FOR i% = 1 TO Nd% : XiMin(i%) = -100## : XiMax(i%) = 100## : NEXT i%

        REDIM StartingXiMin(1 TO Nd%), StartingXiMax(1 TO Nd%) : FOR i% = 1 TO Nd% : StartingXiMin(i%) = XiMin(i%) : StartingXiMax(i%) = XiMax(i%) : NEXT i%

        IF PlaceInitialProbes$ = "2D GRID" THEN

            Np% = NumProbesPerDimension%^2 : REDIM R(1 TO Np%, 1 TO Nd%, 0 TO Nt&) 'to create (Np/Nd) x (Np/Nd) grid

        END IF

CASE "F5" '(n-D)

        Nd%              = 30
        NumProbesPerDimension% = 2
        Np%              = NumProbesPerDimension%*Nd%

        Nt&     = 1000
        G       = 2##
        Alpha   = 2##
        Beta    = 2##
        DeltaT  = 1##
        Frep    = 0.5##

        PlaceInitialProbes$  = "UNIFORM ON-AXIS"
        InitialAcceleration$ = "ZERO"
        RepositionFactor$    = "VARIABLE"

        CALL ChangeRunParameters(NumProbesPerDimension%,Np%,Nd%,Nt&,G,Alpha,Beta,DeltaT,Frep,PlaceInitialProbes$,InitialAcceleration$,RepositionFactor$,FunctionName$)

        Np% = NumProbesPerDimension%*Nd%

        REDIM XiMin(1 TO Nd%), XiMax(1 TO Nd%) : FOR i% = 1 TO Nd% : XiMin(i%) = -30## : XiMax(i%) = 30## : NEXT i%

        REDIM StartingXiMin(1 TO Nd%), StartingXiMax(1 TO Nd%) : FOR i% = 1 TO Nd% : StartingXiMin(i%) = XiMin(i%) : StartingXiMax(i%) = XiMax(i%) : NEXT i%

        IF PlaceInitialProbes$ = "2D GRID" THEN

            Np% = NumProbesPerDimension%^2 : REDIM R(1 TO Np%, 1 TO Nd%, 0 TO Nt&) 'to create (Np/Nd) x (Np/Nd) grid

        END IF

 CASE "F6" '(n-D) STEP

        Nd%              = 30
        NumProbesPerDimension% = 2 '20
        Np%              = NumProbesPerDimension%*Nd%

        Nt&     = 1000
        G       = 2##
        Alpha   = 2##
        Beta    = 2##
        DeltaT  = 1##
        Frep    = 0.5##

        PlaceInitialProbes$  = "UNIFORM ON-AXIS"
        InitialAcceleration$ = "ZERO"
        RepositionFactor$    = "VARIABLE" '"FIXED"

        CALL ChangeRunParameters(NumProbesPerDimension%,Np%,Nd%,Nt&,G,Alpha,Beta,DeltaT,Frep,PlaceInitialProbes$,InitialAcceleration$,RepositionFactor$,FunctionName$)

        Np% = NumProbesPerDimension%*Nd%

        REDIM XiMin(1 TO Nd%), XiMax(1 TO Nd%) : FOR i% = 1 TO Nd% : XiMin(i%) = -100## : XiMax(i%) = 100## : NEXT i%

        REDIM StartingXiMin(1 TO Nd%), StartingXiMax(1 TO Nd%) : FOR i% = 1 TO Nd% : StartingXiMin(i%) = XiMin(i%) : StartingXiMax(i%) = XiMax(i%) : NEXT i%
```

```
      IF PlaceInitialProbes$ = "2D GRID" THEN

           Np% = NumProbesPerDimension%^2 : REDIM R(1 TO Np%, 1 TO Nd%, 0 TO Nt&) 'to create (Np/Nd) x (Np/Nd) grid

      END IF

CASE "F7" '(n-D)

      Nd%                    = 30
      NumProbesPerDimension% = 2 '20
      Np%                    = NumProbesPerDimension%*Nd%

      Nt&       = 100          'BECAUSE THIS FUNCTION HAS A RANDOM COMPONENT !!
      G         = 2##
      Alpha     = 2##
      Beta      = 2##
      DeltaT    = 1##
      Frep      = 0.5##

      PlaceInitialProbes$  = "UNIFORM ON-AXIS"
      InitialAcceleration$ = "ZERO"
      RepositionFactor$    = "VARIABLE" '"FIXED"

      CALL ChangeRunParameters(NumProbesPerDimension%,Np%,Nd%,Nt&,G,Alpha,Beta,DeltaT,Frep,PlaceInitialProbes$,InitialAcceleration$,RepositionFactor$,FunctionName$)

      Np% = NumProbesPerDimension%*Nd%

      REDIM XiMin(1 TO Nd%), XiMax(1 TO Nd%) : FOR i% = 1 TO Nd% : XiMin(i%) = -1.28## : XiMax(i%) = 1.28## : NEXT i%

      REDIM StartingXiMin(1 TO Nd%), StartingXiMax(1 TO Nd%) : FOR i% = 1 TO Nd% : StartingXiMin(i%) = XiMin(i%) : StartingXiMax(i%) = XiMax(i%) : NEXT i%

      IF PlaceInitialProbes$ = "2D GRID" THEN

           Np% = NumProbesPerDimension%^2 : REDIM R(1 TO Np%, 1 TO Nd%, 0 TO Nt&) 'to create (Np/Nd) x (Np/Nd) grid

      END IF 'F7

CASE "F8" '(n-D)

      Nd%                    = 30
      NumProbesPerDimension% = 2 '4 '20
      Np%                    = NumProbesPerDimension%*Nd%

      Nt&       = 1000
      G         = 2##
      Alpha     = 2##
      Beta      = 2##
      DeltaT    = 1##
      Frep      = 0.5##

      PlaceInitialProbes$  = "UNIFORM ON-AXIS"
      InitialAcceleration$ = "ZERO"
      RepositionFactor$    = "VARIABLE" '"FIXED"

      CALL ChangeRunParameters(NumProbesPerDimension%,Np%,Nd%,Nt&,G,Alpha,Beta,DeltaT,Frep,PlaceInitialProbes$,InitialAcceleration$,RepositionFactor$,FunctionName$)

      Np% = NumProbesPerDimension%*Nd%

      REDIM XiMin(1 TO Nd%), XiMax(1 TO Nd%) : FOR i% = 1 TO Nd% : XiMin(i%) = -500## : XiMax(i%) = 500## : NEXT i%

      REDIM StartingXiMin(1 TO Nd%), StartingXiMax(1 TO Nd%) : FOR i% = 1 TO Nd% : StartingXiMin(i%) = XiMin(i%) : StartingXiMax(i%) = XiMax(i%) : NEXT i%

      IF PlaceInitialProbes$ = "2D GRID" THEN

           Np% = NumProbesPerDimension%^2 : REDIM R(1 TO Np%, 1 TO Nd%, 0 TO Nt&) 'to create (Np/Nd) x (Np/Nd) grid

      END IF 'F8

CASE "F9" '(n-D)

      Nd%                    = 30
      NumProbesPerDimension% = 2 '4 '20
      Np%                    = NumProbesPerDimension%*Nd%

      Nt&       = 1000
      G         = 2##
      Alpha     = 2##
      Beta      = 2##
      DeltaT    = 1##
      Frep      = 0.5##

      PlaceInitialProbes$  = "UNIFORM ON-AXIS"
      InitialAcceleration$ = "ZERO"
      RepositionFactor$    = "VARIABLE" '"FIXED"

      CALL ChangeRunParameters(NumProbesPerDimension%,Np%,Nd%,Nt&,G,Alpha,Beta,DeltaT,Frep,PlaceInitialProbes$,InitialAcceleration$,RepositionFactor$,FunctionName$)

      Np% = NumProbesPerDimension%*Nd%

      REDIM XiMin(1 TO Nd%), XiMax(1 TO Nd%) : FOR i% = 1 TO Nd% : XiMin(i%) = -5.12## : XiMax(i%) = 5.12## : NEXT i%

      REDIM StartingXiMin(1 TO Nd%), StartingXiMax(1 TO Nd%) : FOR i% = 1 TO Nd% : StartingXiMin(i%) = XiMin(i%) : StartingXiMax(i%) = XiMax(i%) : NEXT i%

      IF PlaceInitialProbes$ = "2D GRID" THEN

           Np% = NumProbesPerDimension%^2 : REDIM R(1 TO Np%, 1 TO Nd%, 0 TO Nt&) 'to create (Np/Nd) x (Np/Nd) grid

      END IF 'F9

CASE "F10" '(n-D) Ackley's Function

      Nd%                    = 30
      NumProbesPerDimension% = 2 '4 '20
      Np%                    = NumProbesPerDimension%*Nd%

      Nt&       = 1000
      G         = 2##
      Alpha     = 2##
      Beta      = 2##
      DeltaT    = 1##
      Frep      = 0.5##

      PlaceInitialProbes$  = "UNIFORM ON-AXIS"
```

```
       InitialAcceleration$ = "ZERO"
       RepositionFactor$   = "VARIABLE" '"FIXED"

       CALL ChangeRunParameters(NumProbesPerDimension%,Np%,Nd%,Nt&,G,Alpha,Beta,DeltaT,Frep,PlaceInitialProbes$,InitialAcceleration$,RepositionFactor$,FunctionName$)

       Np% = NumProbesPerDimension%*Nd%

       REDIM XiMin(1 TO Nd%), XiMax(1 TO Nd%) : FOR i% = 1 TO Nd% : XiMin(i%) = -32## : XiMax(i%) = 32## : NEXT i%

       REDIM StartingXiMin(1 TO Nd%), StartingXiMax(1 TO Nd%) : FOR i% = 1 TO Nd% : StartingXiMin(i%) = XiMin(i%) : StartingXiMax(i%) = XiMax(i%) : NEXT i%

       IF PlaceInitialProbes$ = "2D GRID" THEN

            Np% = NumProbesPerDimension%^2 : REDIM R(1 TO Np%, 1 TO Nd%, 0 TO Nt&) 'to create (Np/Nd) x (Np/Nd) grid

       END IF 'F10

CASE "F11" '(n-D)

       Nd%                    = 30
       NumProbesPerDimension% = 2 '4 '20
       Np%                    = NumProbesPerDimension%*Nd%

       Nt&      = 1000
       G        = 2##
       Alpha    = 2##
       Beta     = 2##
       DeltaT   = 1##
       Frep     = 0.5##

       PlaceInitialProbes$ = "UNIFORM ON-AXIS"
       InitialAcceleration$ = "ZERO"
       RepositionFactor$   = "VARIABLE" '"FIXED"

       CALL ChangeRunParameters(NumProbesPerDimension%,Np%,Nd%,Nt&,G,Alpha,Beta,DeltaT,Frep,PlaceInitialProbes$,InitialAcceleration$,RepositionFactor$,FunctionName$)

       Np% = NumProbesPerDimension%*Nd%

       REDIM XiMin(1 TO Nd%), XiMax(1 TO Nd%) : FOR i% = 1 TO Nd% : XiMin(i%) = -600## : XiMax(i%) = 600## : NEXT i%

       REDIM StartingXiMin(1 TO Nd%), StartingXiMax(1 TO Nd%) : FOR i% = 1 TO Nd% : StartingXiMin(i%) = XiMin(i%) : StartingXiMax(i%) = XiMax(i%) : NEXT i%

       IF PlaceInitialProbes$ = "2D GRID" THEN

            Np% = NumProbesPerDimension%^2 : REDIM R(1 TO Np%, 1 TO Nd%, 0 TO Nt&) 'to create (Np/Nd) x (Np/Nd) grid

       END IF 'F11

CASE "F12" '(n-D) Penalized #1

       Nd%                    = 30
       NumProbesPerDimension% = 2 '4 '20
       Np%                    = NumProbesPerDimension%*Nd%

       Nt&      = 1000
       G        = 2##
       Alpha    = 2##
       Beta     = 2##
       DeltaT   = 1##
       Frep     = 0.5##

       PlaceInitialProbes$ = "UNIFORM ON-AXIS"
       InitialAcceleration$ = "ZERO"
       RepositionFactor$   = "VARIABLE" '"FIXED"

       CALL ChangeRunParameters(NumProbesPerDimension%,Np%,Nd%,Nt&,G,Alpha,Beta,DeltaT,Frep,PlaceInitialProbes$,InitialAcceleration$,RepositionFactor$,FunctionName$)

       Np% = NumProbesPerDimension%*Nd%

       REDIM XiMin(1 TO Nd%), XiMax(1 TO Nd%) : FOR i% = 1 TO Nd% : XiMin(i%) = -50## : XiMax(i%) = 50## : NEXT i%

       REDIM StartingXiMin(1 TO Nd%), StartingXiMax(1 TO Nd%) : FOR i% = 1 TO Nd% : StartingXiMin(i%) = XiMin(i%) : StartingXiMax(i%) = XiMax(i%) : NEXT i%

       IF PlaceInitialProbes$ = "2D GRID" THEN

            Np% = NumProbesPerDimension%^2 : REDIM R(1 TO Np%, 1 TO Nd%, 0 TO Nt&) 'to create (Np/Nd) x (Np/Nd) grid

       END IF 'F12

CASE "F13" '(n-D) Penalized #2

       Nd%                    = 30
       NumProbesPerDimension% = 2 '4 '20
       Np%                    = NumProbesPerDimension%*Nd%

       Nt&      = 1000
       G        = 2##
       Alpha    = 2##
       Beta     = 2##
       DeltaT   = 1##
       Frep     = 0.5##

       PlaceInitialProbes$ = "UNIFORM ON-AXIS"
       InitialAcceleration$ = "ZERO"
       RepositionFactor$   = "VARIABLE" '"FIXED"

       CALL ChangeRunParameters(NumProbesPerDimension%,Np%,Nd%,Nt&,G,Alpha,Beta,DeltaT,Frep,PlaceInitialProbes$,InitialAcceleration$,RepositionFactor$,FunctionName$)

       Np% = NumProbesPerDimension%*Nd%

       REDIM XiMin(1 TO Nd%), XiMax(1 TO Nd%) : FOR i% = 1 TO Nd% : XiMin(i%) = -50## : XiMax(i%) = 50## : NEXT i%

       REDIM StartingXiMin(1 TO Nd%), StartingXiMax(1 TO Nd%) : FOR i% = 1 TO Nd% : StartingXiMin(i%) = XiMin(i%) : StartingXiMax(i%) = XiMax(i%) : NEXT i%

       IF PlaceInitialProbes$ = "2D GRID" THEN

            Np% = NumProbesPerDimension%^2 : REDIM R(1 TO Np%, 1 TO Nd%, 0 TO Nt&) 'to create (Np/Nd) x (Np/Nd) grid

       END IF 'F13

CASE "F14" '(2-D) Shekel's Foxholes

       Nd%                    = 2
       NumProbesPerDimension% = 4 '20
```

```
        Np%                      = NumProbesPerDimension%*Nd%

        Nt&     = 1000
        G       = 2##
        Alpha   = 2##
        Beta    = 2##
        DeltaT  = 1##
        Frep    = 0.5##

        PlaceInitialProbes$  = "UNIFORM ON-AXIS"
        InitialAcceleration$ = "ZERO"
        RepositionFactor$    = "VARIABLE" '"FIXED"

        CALL ChangeRunParameters(NumProbesPerDimension%,Np%,Nd%,Nt&,G,Alpha,Beta,DeltaT,Frep,PlaceInitialProbes$,InitialAcceleration$,RepositionFactor$,FunctionName$)

        Nd% = 2 'cannot change dimensionality of Shekel's Foxholes function!

        Np% = NumProbesPerDimension%*Nd%

        REDIM XiMin(1 TO Nd%), XiMax(1 TO Nd%) : FOR i% = 1 TO Nd% : XiMin(i%) = -65.536## : XiMax(i%) = 65.536## : NEXT i%

        REDIM StartingXiMin(1 TO Nd%), StartingXiMax(1 TO Nd%) : FOR i% = 1 TO Nd% : StartingXiMin(i%) = XiMin(i%) : StartingXiMax(i%) = XiMax(i%) : NEXT i%

        IF PlaceInitialProbes$ = "2D GRID" THEN

            Np% = NumProbesPerDimension%^2 : REDIM R(1 TO Np%, 1 TO Nd%, 0 TO Nt&) 'to create (Np/Nd) x (Np/Nd) grid

        END IF 'F14

CASE "F15" '(4-D) Kowalik's Function

        Nd%                      = 4
        NumProbesPerDimension% = 4 '20
        Np%                      = NumProbesPerDimension%*Nd%

        Nt&     = 1000
        G       = 2##
        Alpha   = 2##
        Beta    = 2##
        DeltaT  = 1##
        Frep    = 0.5##

        PlaceInitialProbes$  = "UNIFORM ON-AXIS"
        InitialAcceleration$ = "ZERO"
        RepositionFactor$    = "VARIABLE" '"FIXED"

        CALL ChangeRunParameters(NumProbesPerDimension%,Np%,Nd%,Nt&,G,Alpha,Beta,DeltaT,Frep,PlaceInitialProbes$,InitialAcceleration$,RepositionFactor$,FunctionName$)

        Nd% = 4 'cannot change dimensionality of Kowalik's Function!

        Np% = NumProbesPerDimension%*Nd%

        REDIM XiMin(1 TO Nd%), XiMax(1 TO Nd%) : FOR i% = 1 TO Nd% : XiMin(i%) = -5## : XiMax(i%) = 5## : NEXT i%

        REDIM StartingXiMin(1 TO Nd%), StartingXiMax(1 TO Nd%) : FOR i% = 1 TO Nd% : StartingXiMin(i%) = XiMin(i%) : StartingXiMax(i%) = XiMax(i%) : NEXT i%

CASE "F16" '(2-D) Camel Back

        Nd%                      = 2
        NumProbesPerDimension% = 4 '20
        Np%                      = NumProbesPerDimension%*Nd%

        Nt&     = 1000
        G       = 2##
        Alpha   = 2##
        Beta    = 2##
        DeltaT  = 1##
        Frep    = 0.5##

        PlaceInitialProbes$  = "UNIFORM ON-AXIS"
        InitialAcceleration$ = "ZERO"
        RepositionFactor$    = "VARIABLE" '"FIXED"

        CALL ChangeRunParameters(NumProbesPerDimension%,Np%,Nd%,Nt&,G,Alpha,Beta,DeltaT,Frep,PlaceInitialProbes$,InitialAcceleration$,RepositionFactor$,FunctionName$)

        Nd% = 2 'cannot change dimensionality of Camel Back function!

        Np% = NumProbesPerDimension%*Nd%

        REDIM XiMin(1 TO Nd%), XiMax(1 TO Nd%) : FOR i% = 1 TO Nd% : XiMin(i%) = -5## : XiMax(i%) = 5## : NEXT i%

        REDIM StartingXiMin(1 TO Nd%), StartingXiMax(1 TO Nd%) : FOR i% = 1 TO Nd% : StartingXiMin(i%) = XiMin(i%) : StartingXiMax(i%) = XiMax(i%) : NEXT i%

        IF PlaceInitialProbes$ = "2D GRID" THEN

            Np% = NumProbesPerDimension%^2 : REDIM R(1 TO Np%, 1 TO Nd%, 0 TO Nt&) 'to create (Np/Nd) x (Np/Nd) grid

        END IF 'F16

CASE "F17" '(2-D) Branin

        Nd%                      = 2
        NumProbesPerDimension% = 4 '20
        Np%                      = NumProbesPerDimension%*Nd%

        Nt&     = 1000
        G       = 2##
        Alpha   = 2##
        Beta    = 2##
        DeltaT  = 1##
        Frep    = 0.5##

        PlaceInitialProbes$  = "UNIFORM ON-AXIS"
        InitialAcceleration$ = "ZERO"
        RepositionFactor$    = "VARIABLE" '"FIXED"

        CALL ChangeRunParameters(NumProbesPerDimension%,Np%,Nd%,Nt&,G,Alpha,Beta,DeltaT,Frep,PlaceInitialProbes$,InitialAcceleration$,RepositionFactor$,FunctionName$)

        Nd% = 2 'cannot change dimensionality of Branin function!

        Np% = NumProbesPerDimension%*Nd%

        REDIM XiMin(1 TO Nd%), XiMax(1 TO Nd%) : XiMin(1) = -5## : XiMax(1) = 10## : XiMin(2) = 0## : XiMax(2) = 15##
```

```
        REDIM StartingXiMin(1 TO Nd%), StartingXiMax(1 TO Nd%) : FOR i% = 1 TO Nd% : StartingXiMin(i%) = XiMin(i%) : StartingXiMax(i%) = XiMax(i%) : NEXT i%

        IF PlaceInitialProbes$ = "2D GRID" THEN

            Np% = NumProbesPerDimension%^2 : REDIM R(1 TO Np%, 1 TO Nd%, 0 TO Nt&) 'to create (Np/Nd) x (Np/Nd) grid

        END IF 'F17

CASE "F18" '(2-D) Goldstein-Price

        Nd%                     = 2
        NumProbesPerDimension%  = 4 '20
        Np%                     = NumProbesPerDimension%*Nd%

        Nt&      = 1000
        G        = 2##
        Alpha    = 2##
        Beta     = 2##
        DeltaT   = 1##
        Frep     = 0.5##

        PlaceInitialProbes$   = "UNIFORM ON-AXIS"
        InitialAcceleration$  = "ZERO"
        RepositionFactor$     = "VARIABLE" '"FIXED"

        CALL ChangeRunParameters(NumProbesPerDimension%,Np%,Nd%,Nt&,G,Alpha,Beta,DeltaT,Frep,PlaceInitialProbes$,InitialAcceleration$,RepositionFactor$,FunctionName$)

        Nd% = 2 'cannot change dimensionality of Branin function!

        Np% = NumProbesPerDimension%*Nd%

        REDIM XiMin(1 TO Nd%), XiMax(1 TO Nd%) : XiMin(1) = -2## : XiMax(1) = 2## : XiMin(2) = -2## : XiMax(2) = 2##

        REDIM StartingXiMin(1 TO Nd%), StartingXiMax(1 TO Nd%) : FOR i% = 1 TO Nd% : StartingXiMin(i%) = XiMin(i%) : StartingXiMax(i%) = XiMax(i%) : NEXT i%

        IF PlaceInitialProbes$ = "2D GRID" THEN

            Np% = NumProbesPerDimension%^2 : REDIM R(1 TO Np%, 1 TO Nd%, 0 TO Nt&) 'to create (Np/Nd) x (Np/Nd) grid

        END IF 'F18

CASE "F19" '(3-D) Hartman's Family #1

        Nd%                     = 3
        NumProbesPerDimension%  = 4 '20
        Np%                     = NumProbesPerDimension%*Nd%

        Nt&      = 1000
        G        = 2##
        Alpha    = 2##
        Beta     = 2##
        DeltaT   = 1##
        Frep     = 0.5##

        PlaceInitialProbes$   = "UNIFORM ON-AXIS"
        InitialAcceleration$  = "ZERO"
        RepositionFactor$     = "VARIABLE" '"FIXED"

        CALL ChangeRunParameters(NumProbesPerDimension%,Np%,Nd%,Nt&,G,Alpha,Beta,DeltaT,Frep,PlaceInitialProbes$,InitialAcceleration$,RepositionFactor$,FunctionName$)

        Nd% = 3 'cannot change dimensionality of Hartman's Family!

        Np% = NumProbesPerDimension%*Nd%

        REDIM XiMin(1 TO Nd%), XiMax(1 TO Nd%) : FOR i% = 1 TO Nd% : XiMin(i%) = 0## : XiMax(i%) = 1## : NEXT i%

        REDIM StartingXiMin(1 TO Nd%), StartingXiMax(1 TO Nd%) : FOR i% = 1 TO Nd% : StartingXiMin(i%) = XiMin(i%) : StartingXiMax(i%) = XiMax(i%) : NEXT i%

CASE "F20" '(6-D) Hartman's Family #2

        Nd%                     = 6
        NumProbesPerDimension%  = 4 '20
        Np%                     = NumProbesPerDimension%*Nd%

        Nt&      = 1000
        G        = 2##
        Alpha    = 2##
        Beta     = 2##
        DeltaT   = 1##
        Frep     = 0.5##

        PlaceInitialProbes$   = "UNIFORM ON-AXIS"
        InitialAcceleration$  = "ZERO"
        RepositionFactor$     = "VARIABLE" '"FIXED"

        CALL ChangeRunParameters(NumProbesPerDimension%,Np%,Nd%,Nt&,G,Alpha,Beta,DeltaT,Frep,PlaceInitialProbes$,InitialAcceleration$,RepositionFactor$,FunctionName$)

        Nd% = 6 'cannot change dimensionality of Hartman's Family!

        Np% = NumProbesPerDimension%*Nd%

        REDIM XiMin(1 TO Nd%), XiMax(1 TO Nd%) : FOR i% = 1 TO Nd% : XiMin(i%) = 0## : XiMax(i%) = 1## : NEXT i%

        REDIM StartingXiMin(1 TO Nd%), StartingXiMax(1 TO Nd%) : FOR i% = 1 TO Nd% : StartingXiMin(i%) = XiMin(i%) : StartingXiMax(i%) = XiMax(i%) : NEXT i%

CASE "F21" '(4-D) Shekel's Family m=5

        Nd%                     = 4
        NumProbesPerDimension%  = 4 '20
        Np%                     = NumProbesPerDimension%*Nd%

        Nt&      = 1000
        G        = 2##
        Alpha    = 2##
        Beta     = 2##
        DeltaT   = 1##
        Frep     = 0.5##

        PlaceInitialProbes$   = "UNIFORM ON-AXIS"
        InitialAcceleration$  = "ZERO"
        RepositionFactor$     = "VARIABLE" '"FIXED"

        CALL ChangeRunParameters(NumProbesPerDimension%,Np%,Nd%,Nt&,G,Alpha,Beta,DeltaT,Frep,PlaceInitialProbes$,InitialAcceleration$,RepositionFactor$,FunctionName$)
```

```
        Nd% = 4 'cannot change dimensionality of Shekel's Family!

        Np% = NumProbesPerDimension%*Nd%

        REDIM XiMin(1 TO Nd%), XiMax(1 TO Nd%) : FOR i% = 1 TO Nd% : XiMin(i%) = 0## : XiMax(i%) = 10## : NEXT i%

        REDIM StartingXiMin(1 TO Nd%), StartingXiMax(1 TO Nd%) : FOR i% = 1 TO Nd% : StartingXiMin(i%) = XiMin(i%) : StartingXiMax(i%) = XiMax(i%) : NEXT i%

CASE "F22" '(4-D) Shekel's Family m=7

        Nd%                 = 4
        NumProbesPerDimension% = 4 '20
        Np%                 = NumProbesPerDimension%*Nd%

        Nt&    = 1000
        G      = 2##
        Alpha  = 2##
        Beta   = 2##
        DeltaT = 1##
        Frep   = 0.5##

        PlaceInitialProbes$  = "UNIFORM ON-AXIS"
        InitialAcceleration$ = "ZERO"
        RepositionFactor$    = "VARIABLE" '"FIXED"

        CALL ChangeRunParameters(NumProbesPerDimension%,Np%,Nd%,Nt&,G,Alpha,Beta,DeltaT,Frep,PlaceInitialProbes$,InitialAcceleration$,RepositionFactor$,FunctionName$)

        Nd% = 4 'cannot change dimensionality of Shekel's Family!

        Np% = NumProbesPerDimension%*Nd%

        REDIM XiMin(1 TO Nd%), XiMax(1 TO Nd%) : FOR i% = 1 TO Nd% : XiMin(i%) = 0## : XiMax(i%) = 10## : NEXT i%

        REDIM StartingXiMin(1 TO Nd%), StartingXiMax(1 TO Nd%) : FOR i% = 1 TO Nd% : StartingXiMin(i%) = XiMin(i%) : StartingXiMax(i%) = XiMax(i%) : NEXT i%

CASE "F23" '(4-D) Shekel's Family m=10

        Nd%                 = 4
        NumProbesPerDimension% = 4 '20
        Np%                 = NumProbesPerDimension%*Nd%

        Nt&    = 1000
        G      = 2##
        Alpha  = 2##
        Beta   = 2##
        DeltaT = 1##
        Frep   = 0.5##

        PlaceInitialProbes$  = "UNIFORM ON-AXIS"
        InitialAcceleration$ = "ZERO"
        RepositionFactor$    = "VARIABLE" '"FIXED"

        CALL ChangeRunParameters(NumProbesPerDimension%,Np%,Nd%,Nt&,G,Alpha,Beta,DeltaT,Frep,PlaceInitialProbes$,InitialAcceleration$,RepositionFactor$,FunctionName$)

        Nd% = 4 'cannot change dimensionality of Shekel's Family!

        Np% = NumProbesPerDimension%*Nd%

        REDIM XiMin(1 TO Nd%), XiMax(1 TO Nd%) : FOR i% = 1 TO Nd% : XiMin(i%) = 0## : XiMax(i%) = 10## : NEXT i%

        REDIM StartingXiMin(1 TO Nd%), StartingXiMax(1 TO Nd%) : FOR i% = 1 TO Nd% : StartingXiMin(i%) = XiMin(i%) : StartingXiMax(i%) = XiMax(i%) : NEXT i%

   CASE "PBM_1" '2-D

        Nd%                 = 2
        NumProbesPerDimension% = 2 '4 '20
        Np%                 = NumProbesPerDimension%*Nd%

        Nt&    = 100
        G      = 2##
        Alpha  = 2##
        Beta   = 2##
        DeltaT = 1##
        Frep   = 0.5##

        PlaceInitialProbes$  = "UNIFORM ON-AXIS"
        InitialAcceleration$ = "ZERO"
        RepositionFactor$    = "VARIABLE" '"FIXED"

        CALL ChangeRunParameters(NumProbesPerDimension%,Np%,Nd%,Nt&,G,Alpha,Beta,DeltaT,Frep,PlaceInitialProbes$,InitialAcceleration$,RepositionFactor$,FunctionName$)

        Nd% = 2 'cannot change dimensionality of PBM_1!

        Np% = NumProbesPerDimension%*Nd%

        REDIM XiMin(1 TO Nd%), XiMax(1 TO Nd%)

        XiMin(1) = 0.5## : XiMax(1) = 3## 'dipole length, L, in Wavelengths
        XiMin(2) = 0## : XiMax(2) = PI2 'polar angle, Theta, in Radians

        REDIM StartingXiMin(1 TO Nd%), StartingXiMax(1 TO Nd%) : FOR i% = 1 TO Nd% : StartingXiMin(i%) = XiMin(i%) : StartingXiMax(i%) = XiMax(i%) : NEXT i%

        NN% = FREEFILE : OPEN "INFILE.DAT" FOR OUTPUT AS #NN% : PRINT #NN%,"PBM1.NEC" : PRINT #NN%,"PBM1.OUT" : CLOSE #NN% 'NEC Input/Output Files

   CASE "PBM_2" '2-D

        AddNoiseToPBM2$ = "NO" '"YES" '"NO" '"YES"

        Nd%                 = 2
        NumProbesPerDimension% = 4 '20
        Np%                 = NumProbesPerDimension%*Nd%

        Nt&    = 100
        G      = 2##
        Alpha  = 2##
        Beta   = 2##
        DeltaT = 1##
        Frep   = 0.5##

        PlaceInitialProbes$  = "UNIFORM ON-AXIS"
        InitialAcceleration$ = "ZERO"
        RepositionFactor$    = "VARIABLE" '"FIXED"

        CALL ChangeRunParameters(NumProbesPerDimension%,Np%,Nd%,Nt&,G,Alpha,Beta,DeltaT,Frep,PlaceInitialProbes$,InitialAcceleration$,RepositionFactor$,FunctionName$)
```

```
        Nd% = 2 'cannot change dimensionality of PBM_2!

        Np% = NumProbesPerDimension%*Nd%

        REDIM XiMin(1 TO Nd%), XiMax(1 TO Nd%)

        XiMin(1) = 5## : XiMax(1) = 15## 'dipole separation, D, in Wavelengths
        XiMin(2) = 0## : XiMax(2) = Pi   'polar angle, Theta, in Radians

        REDIM StartingXiMin(1 TO Nd%), StartingXiMax(1 TO Nd%) : FOR i% = 1 TO Nd% : StartingXiMin(i%) = XiMin(i%) : StartingXiMax(i%) = XiMax(i%) : NEXT i%

        NN% = FREEFILE : OPEN "INFILE.DAT" FOR OUTPUT AS #NN% : PRINT #NN%,"PBM2.NEC" : PRINT #NN%,"PBM2.OUT" : CLOSE #NN%

    CASE "PBM_3" '2-D

        Nd%                    = 2
        NumProbesPerDimension% = 4 '20
        Np%                    = NumProbesPerDimension%*Nd%

        Nt&     = 100
        G       = 2##
        Alpha   = 2##
        Beta    = 2##
        DeltaT  = 1##
        Frep    = 0.5##

        PlaceInitialProbes$  = "UNIFORM ON-AXIS"
        InitialAcceleration$ = "ZERO"
        RepositionFactor$    = "VARIABLE" '"FIXED"

        CALL ChangeRunParameters(NumProbesPerDimension%,Np%,Nd%,Nt&,G,Alpha,Beta,DeltaT,Frep,PlaceInitialProbes$,InitialAcceleration$,RepositionFactor$,FunctionName$)

        Nd% = 2 'cannot change dimensionality of PBM_3!

        Np% = NumProbesPerDimension%*Nd%

        REDIM XiMin(1 TO Nd%), XiMax(1 TO Nd%)

        XiMin(1) = 0## : XiMax(1) = 4## 'Phase Parameter, Beta (0-4)
        XiMin(2) = 0## : XiMax(2) = Pi  'polar angle, Theta, in Radians

        REDIM StartingXiMin(1 TO Nd%), StartingXiMax(1 TO Nd%) : FOR i% = 1 TO Nd% : StartingXiMin(i%) = XiMin(i%) : StartingXiMax(i%) = XiMax(i%) : NEXT i%

        NN% = FREEFILE : OPEN "INFILE.DAT" FOR OUTPUT AS #NN% : PRINT #NN%,"PBM3.NEC" : PRINT #NN%,"PBM3.OUT" : CLOSE #NN%

    CASE "PBM_4" '2-D

        Nd%                    = 2
        NumProbesPerDimension% = 4 '6 '2 '4 '20
        Np%                    = NumProbesPerDimension%*Nd%

        Nt&     = 100
        G       = 2##
        Alpha   = 2##
        Beta    = 2##
        DeltaT  = 1##
        Frep    = 0.5##

        PlaceInitialProbes$  = "UNIFORM ON-AXIS"
        InitialAcceleration$ = "ZERO"
        RepositionFactor$    = "VARIABLE" '"FIXED"

        CALL ChangeRunParameters(NumProbesPerDimension%,Np%,Nd%,Nt&,G,Alpha,Beta,DeltaT,Frep,PlaceInitialProbes$,InitialAcceleration$,RepositionFactor$,FunctionName$)

        Nd% = 2 'cannot change dimensionality of PBM_4!

        Np% = NumProbesPerDimension%*Nd%

        REDIM XiMin(1 TO Nd%), XiMax(1 TO Nd%)

        XiMin(1) = 0.5##   : XiMax(1) = 1.5## 'ARM LENGTH (NOT Total Length), wavelengths (0.5-1.5)
        XiMin(2) = Pi/18## : XiMax(2) = Pi/2## 'Inner angle, Alpha, in Radians (Pi/18-Pi/2)

        REDIM StartingXiMin(1 TO Nd%), StartingXiMax(1 TO Nd%) : FOR i% = 1 TO Nd% : StartingXiMin(i%) = XiMin(i%) : StartingXiMax(i%) = XiMax(i%) : NEXT i%

        NN% = FREEFILE : OPEN "INFILE.DAT" FOR OUTPUT AS #NN% : PRINT #NN%,"PBM4.NEC" : PRINT #NN%,"PBM4.OUT" : CLOSE #NN%

    CASE "PBM_5"

        NumCollinearElements% = 6 '30 'EVEN or ODD: 6,7,10,13,16,24 used by PBM

        Nd%                    = NumCollinearElements% - 1
        NumProbesPerDimension% = 4 '20
        Np%                    = NumProbesPerDimension%*Nd%

        Nt&     = 100
        G       = 2##
        Alpha   = 2##
        Beta    = 2##
        DeltaT  = 1##
        Frep    = 0.5##

        PlaceInitialProbes$  = "UNIFORM ON-AXIS"
        InitialAcceleration$ = "ZERO"
        RepositionFactor$    = "VARIABLE" '"FIXED"

        CALL ChangeRunParameters(NumProbesPerDimension%,Np%,Nd%,Nt&,G,Alpha,Beta,DeltaT,Frep,PlaceInitialProbes$,InitialAcceleration$,RepositionFactor$,FunctionName$)

        Nd% = NumCollinearElements% - 1

        Np% = NumProbesPerDimension%*Nd%

        REDIM XiMin(1 TO Nd%), XiMax(1 TO Nd%) : FOR i% = 1 TO Nd% : XiMin(i%) = 0.5## : XiMax(i%) = 1.5## : NEXT i%

        REDIM StartingXiMin(1 TO Nd%), StartingXiMax(1 TO Nd%) : FOR i% = 1 TO Nd% : StartingXiMin(i%) = XiMin(i%) : StartingXiMax(i%) = XiMax(i%) : NEXT i%

        NN% = FREEFILE : OPEN "INFILE.DAT" FOR OUTPUT AS #NN% : PRINT #NN%,"PBM5.NEC" : PRINT #NN%,"PBM5.OUT" : CLOSE #NN%
' =======================================================================================
' NOTE - DON'T FORGET TO ADD NEW TEST FUNCTIONS TO FUNCTION ObjectiveFunction() ABOVE !!
' =======================================================================================

    END SELECT
```

```
    IF Nd% > 100 THEN Nt& = MIN(Nt&,200) 'to avoid array dimensioning problems

    DiagLength = 0## : FOR i% = 1 TO Nd% : DiagLength = DiagLength + (XiMax(i%)-XiMin(i%))^2 : NEXT i% : DiagLength = SQR(DiagLength) 'compute length of decision space principal
diagonal

END SUB 'GetFunctionRunParameters()

'--------------------------------

FUNCTION ParrottF4(R(),Nd%,p%,j&) 'Parrott F4 (1-D)

'MAXIMUM = 1 AT -0.0796875... WITH ZERO OFFSET (SEEMS TO WORK BEST WITH JUST 3 PROBES, BUT NOT ALLOWED IN THIS VERSION...)

'References:

'Beasley, D., D. R. Bull, and R. R. Martin, "A Sequential Niche Technique for Multimodal
'Function Optimization," Evol. Comp. (MIT Press), vol. 1, no. 2, 1993, pp. 101-125
'(online at http://citeseer.ist.psu.edu/beasley93sequential.html).

'Parrott, D., and X. Li, "Locating and Tracking Multiple Dynamic Optima by a Particle Swarm
'Model Using Speciation," IEEE Trans. Evol. Computation, vol. 10, no. 4, Aug. 2006, pp. 440-458.

LOCAL Z, x, offset AS EXT

    offset = 0##

    x = R(p%,1,j&)

    Z = EXP(-2##*LOG(2##)*((x-0.08##-offset)/0.854##)^2)*(SIN(5##*Pi*((x-offset)^0.75##-0.05##)))^6 'WARNING! This is a NATURAL LOG, NOT Log10!!!

    ParrottF4 = Z

END FUNCTION 'ParrottF4()

'----------------------------

FUNCTION SGO(R(),Nd%,p%,j&) 'SGO Function (2-D)

'MAXIMUM = -130.8323226... @ -(-2.8362075...,-2.8362075...) WITH ZERO OFFSET.

'Reference:

'Hsiao, Y., Chuang, C., Jiang, J., and Chien, C., "A Novel Optimization Algorithm: Space
'Gravitational Optimization," Proc. of 2005 IEEE International Conference on Systems, Man,
'and Cybernetics, 3, 2323-2328. (2005)

    LOCAL x1, x2, Z, t1, t2, SGOx1offset, SGOx2offset AS EXT

    SGOx1offset = 0## : SGOx2offset = 0##

'   SGOx1offset = 40## : SGOx2offset = 10##

    x1 = R(p%,1,j&) - SGOx1offset : x2 = R(p%,2,j&) - SGOx2offset

    t1 = x1^4 - 16##*x1^2 + 0.5##*x1 : t2 = x2^4 - 16##*x2^2 + 0.5##*x2

    Z = t1 + t2

    SGO = -Z

END FUNCTION 'SGO()

'------------------

FUNCTION GoldsteinPrice(R(),Nd%,p%,j&) 'Goldstein-Price Function (2-D)

'MAXIMUM = -3 @ (0,-1) WITH ZERO OFFSET.

'Reference:

'Cui, Z., Zeng, J., and Sun, G. (2006) 'A Fast Particle Swarm Optimization,' Int'l. J.
'Innovative Computing, Information and Control, vol. 2, no. 6, December, pp. 1365-1380.

    LOCAL Z, x1, x2, offset1, offset2, t1, t2 AS EXT

    offset1 = 0## : offset2 = 0##

'    offset1 = 20## : offset2 = -10##

    x1 = R(p%,1,j&)-offset1 : x2 = R(p%,2,j&)-offset2

    t1 = 1##+(x1+x2+1##)^2*(19##-14##*x1+3##*x1^2-14##*x2+6##*x1*x2+3##*x2^2)

    t2 = 30##+(2##*x1-3##*x2)^2*(18##-32##*x1+12##*x1^2+48##*x2-36##*x1*x2+27##*x2^2)

    Z = t1*t2

    GoldsteinPrice = -Z

END FUNCTION 'GoldsteinPrice()

'-----------

FUNCTION StepFunction(R(),Nd%,p%,j&) 'Step Function (n-D)

'MAXIMUM VALUE = 0 @ [Offset]^n.

'Reference:

'Yao, X., Liu, Y., and Lin, G., "Evolutionary Programming Made Faster,"
'IEEE Trans. Evolutionary Computation, Vol. 3, No. 2, 82-102, Jul. 1999.

    LOCAL Offset, Z AS EXT

    LOCAL i%

    Z = 0## : Offset = 0## '75.123## '0##

    FOR i% = 1 TO Nd%

        IF Nd% = 2 AND i% = 1 THEN Offset = 75 '75##
```

```
        IF Nd% = 2 AND i% = 2 THEN Offset = 35 '30 '35##

        Z = Z + INT((R(p%,i%,j&)-Offset) + 0.5##)^2

    NEXT i%

    StepFunction = -Z

END FUNCTION 'StepFunction()

'-----------

FUNCTION Schwefe1226(R(),Nd%,p%,j&) 'Schwefel Problem 2.26 (n-D)

'MAXIMUM = 12,569.5 @ [420.8687]^30 (30-D CASE).

'Reference:

'Yao, X., Liu, Y., and Lin, G., "Evolutionary Programming Made Faster,"
'IEEE Trans. Evolutionary Computation, Vol. 3, No. 2, 82-102, Jul. 1999.

    LOCAL Z, Xi AS EXT

    LOCAL i%

    Z = 0##

    FOR i% = 1 TO Nd%

        Xi = R(p%,i%,j&)

        Z = Z + Xi*SIN(SQR(ABS(Xi)))

    NEXT i%

    Schwefel226 = Z

END FUNCTION 'SCHWEFEL226()

'-----------

FUNCTION Colville(R(),Nd%,p%,j&) 'Colville Function (4-D)

'MAXIMUM = 0 @ (1,1,1,1) WITH ZERO OFFSET.

'Reference:

'Doo-Hyun, and Se-Young, O., "A New Mutation Rule for Evolutionary Programming Motivated from
'Backpropagation Learning," IEEE Trans. Evolutionary Computation, Vol. 4, No. 2, pp. 188-190,
'July 2000.

    LOCAL Z, x1, x2, x3, x4, offset AS EXT

    offset = 0## '7.123##

    x1 = R(p%,1,j&)-offset : x2 = R(p%,2,j&)-offset : x3 = R(p%,3,j&)-offset : x4 = R(p%,4,j&)-offset

    Z =  100##*(x2-x1^2)^2 + (1##-x1)^2  + _
         90##*(x4-x3^2)^2 + (1##-x3)^2  + _
         10.1##*((x2-1##)^2 + (x4-1##)^2) + _
         19.8##*(x2-1##)*(x4-1##)

    Colville = -Z

END FUNCTION 'Colville()

'-----------

FUNCTION Griewank(R(),Nd%,p%,j&) 'Griewank (n-D)

'Max of zero at (0,...,0)

'Reference: Yao, X., Liu, Y., and Lin, G., "Evolutionary Programming Made Faster,"
'IEEE Trans. Evolutionary Computation, Vol. 3, No. 2, 82-102, Jul. 1999.

    LOCAL Offset, Sum, Prod, Z, Xi AS EXT

    LOCAL i%

    Sum = 0## : Prod = 1##

    Offset = 75.123##

    FOR i% = 1 TO Nd%

        Xi = R(p%,i%,j&) - Offset

        Sum = Sum + Xi^2

        Prod = Prod*COS(Xi/SQR(i%))

    NEXT i%

    Z = Sum/4000## - Prod + 1##

    Griewank = -Z

END FUNCTION 'Griewank()

'-----------

FUNCTION Himmelblau(R(),Nd%,p%,j&) 'Himmelblau (2-D)

    LOCAL Z, x1, x2, offset AS EXT

    offset = 0##

    x1 = R(p%,1,j&)-offset : x2 = R(p%,2,j&)-offset

    Z = 200## - (x1^2 + x2 -11##)^2 - (x1+x2^2-7##)^2

    Himmelblau = Z
```

```
END FUNCTION 'Himmelblau()

'-----------

FUNCTION Rosenbrock(R(),Nd%,p%,j&) 'Rosenbrock (n-D)

'MAXIMUM = 0 @ [1,...,1]^n (n-D CASE).

'Reference: Yao, X., Liu, Y., and Lin, G., "Evolutionary Programming Made Faster,"
'IEEE Trans. Evolutionary Computation, Vol. 3, No. 2, 82-102, Jul. 1999.

    LOCAL Z, Xi, Xi1 AS EXT

    LOCAL i%

    Z = 0##

    FOR i% = 1 TO Nd%-1

        Xi  = R(p%,i%,j&) : Xi1 = R(p%,i%+1,j&)

        Z = Z + 100##*(Xi1-Xi^2)^2 + (Xi-1##)^2

    NEXT i%

    Rosenbrock = -Z

END FUNCTION 'ROSENBROCK()

'-----------

FUNCTION Sphere(R(),Nd%,p%,j&) 'Sphere (n-D)

'MAXIMUM = 0 @ [0,...,0]^n (n-D CASE).

'Reference: Yao, X., Liu, Y., and Lin, G., "Evolutionary Programming Made Faster,"
'IEEE Trans. Evolutionary Computation, Vol. 3, No. 2, 82-102, Jul. 1999.

    LOCAL Z, Xi, Xi1 AS EXT

    LOCAL i%

    Z = 0##

    FOR i% = 1 TO Nd%

        Xi  = R(p%,i%,j&)

        Z = Z + Xi^2

    NEXT i%

    Sphere = -Z

END FUNCTION 'SPHERE()

'-----------

FUNCTION F1(R(),Nd%,p%,j&) 'F1 (n-D)

'MAXIMUM = ZERO (n-D CASE).

'Reference:

    LOCAL Z, Xi AS EXT

    LOCAL i%

    Z = 0##

    FOR i% = 1 TO Nd%

        Xi = R(p%,i%,j&)

        Z = Z + Xi^2

    NEXT i%

    F1 = -Z

END FUNCTION 'F1

'-----------

FUNCTION F2(R(),Nd%,p%,j&) 'F2 (n-D)

'MAXIMUM = ZERO (n-D CASE).

'Reference:

    LOCAL Sum, prod, Z, Xi AS EXT

    LOCAL i%

    Z = 0## : Sum = 0## : Prod = 1##

    FOR i% = 1 TO Nd%

        Xi = R(p%,i%,j&)

        Sum  = Sum+ ABS(Xi)

        Prod = Prod*ABS(Xi)

    NEXT i%

    Z = Sum + Prod

    F2 = -Z
```

```
END FUNCTION 'F2

'-----------

FUNCTION F3(R(),Nd%,p%,j&) 'F3 (n-D)

'MAXIMUM = ZERO (n-D CASE).

'Reference:

    LOCAL Z, Xk, Sum AS EXT

    LOCAL i%, k%

    Z = 0##

    FOR i% = 1 TO Nd%

        Sum = 0##

        FOR k% = 1 TO i%

            Xk = R(p%,k%,j&)

            Sum = Sum + Xk

        NEXT k%

        Z = Z + Sum^2

    NEXT i%

    F3 = -Z

END FUNCTION 'F3

'-----------

FUNCTION F4(R(),Nd%,p%,j&) 'F4 (n-D)

'MAXIMUM = ZERO (n-D CASE).

'Reference:

    LOCAL Z, Xi, MaxXi AS EXT

    LOCAL i%

    MaxXi = -1E4200

    FOR i% = 1 TO Nd%

        Xi = R(p%,i%,j&)

        IF ABS(Xi) >= MaxXi THEN MaxXi = ABS(Xi)

    NEXT i%

    F4 = -MaxXi

END FUNCTION 'F4

'-----------

FUNCTION F5(R(),Nd%,p%,j&) 'F5 (n-D)

'MAXIMUM = ZERO (n-D CASE).

'Reference:

    LOCAL Z, Xi, XiPlus1 AS EXT

    LOCAL i%

    Z = 0##

    FOR i% = 1 TO Nd%-1

        Xi      = R(p%,i%,j&)

        XiPlus1 = R(p%,i%+1,j&)

        Z = Z + (100##*(XiPlus1-Xi^2)^2+(Xi-1##))^2

    NEXT i%

    F5 = -Z

END FUNCTION 'F5

'-----------

FUNCTION F6(R(),Nd%,p%,j&) 'F6

'MAXIMUM VALUE = 0 @ [Offset]^n.

'Reference:

'Yao, X., Liu, Y., and Lin, G., "Evolutionary Programming Made Faster,"
'IEEE Trans. Evolutionary Computation, Vol. 3, No. 2, 82-102, Jul. 1999.

    LOCAL Z AS EXT

    LOCAL i%

    Z = 0##

    FOR i% = 1 TO Nd%

        Z = Z + INT(R(p%,i%,j&) + 0.5##)^2

    NEXT i%
```

```
    F6 = -Z

END FUNCTION 'F6

'-----------

FUNCTION F7(R(),Nd%,p%,j&) 'F7

'MAXIMUM VALUE = 0 @ [Offset]^n.

'Reference:

'Yao, X., Liu, Y., and Lin, G., "Evolutionary Programming Made Faster,"
'IEEE Trans. Evolutionary Computation, Vol. 3, No. 2, 82-102, Jul. 1999.

    LOCAL Z, Xi AS EXT

    LOCAL i%

    Z = 0##

    FOR i% = 1 TO Nd%

        Xi = R(p%,i%,j&)

        Z = Z + i%*Xi^4

    NEXT i%

    F7 = -Z - RandomNum(0##,1##)

END FUNCTION 'F7

'-----------

FUNCTION F8(R(),Nd%,p%,j&) '(n-D) F8 [Schwefel Problem 2.26]

'MAXIMUM = 12,569.5 @ [420.8687]^30 (30-D CASE).

'Reference:

'Yao, X., Liu, Y., and Lin, G., "Evolutionary Programming Made Faster,"
'IEEE Trans. Evolutionary Computation, Vol. 3, No. 2, 82-102, Jul. 1999.

    LOCAL Z, Xi AS EXT

    LOCAL i%

    Z = 0##

    FOR i% = 1 TO Nd%

        Xi = R(p%,i%,j&)

        Z = Z - Xi*SIN(SQR(ABS(Xi)))

    NEXT i%

    F8 = -Z

END FUNCTION 'F8

'-----------

FUNCTION F9(R(),Nd%,p%,j&) '(n-D) F9 [Rastrigin]

'MAXIMUM = ZERO (n-D CASE).

'Reference:

'Yao, X., Liu, Y., and Lin, G., "Evolutionary Programming Made Faster,"
'IEEE Trans. Evolutionary Computation, Vol. 3, No. 2, 82-102, Jul. 1999.

    LOCAL Z, Xi AS EXT

    LOCAL i%

    Z = 0##

    FOR i% = 1 TO Nd%

        Xi = R(p%,i%,j&)

        Z = Z + (Xi^2 - 10##*COS(TwoPi*Xi) + 10##)^2

    NEXT i%

    F9 = -Z

END FUNCTION 'F9

'-----------

FUNCTION F10(R(),Nd%,p%,j&) '(n-D) F10 [Ackley's Function]

'MAXIMUM = ZERO (n-D CASE).

'Reference:

'Yao, X., Liu, Y., and Lin, G., "Evolutionary Programming Made Faster,"
'IEEE Trans. Evolutionary Computation, Vol. 3, No. 2, 82-102, Jul. 1999.

    LOCAL Z, Xi, Sum1, Sum2 AS EXT

    LOCAL i%

    Z = 0## : Sum1 = 0## : Sum2 = 0##

    FOR i% = 1 TO Nd%

        Xi   = R(p%,i%,j&)
```

```
        Sum1 = Sum1 + Xi^2

        Sum2 = Sum2 + COS(TwoPi*Xi)

    NEXT i%

    Z = -20##*EXP(-0.2##*SQR(Sum1/Nd%)) - EXP(Sum2/Nd%) + 20## + e

    F10 = -Z

END FUNCTION 'F10

'-----------

FUNCTION F11(R(),Nd%,p%,j&) '(n-D) F11

'MAXIMUM = ZERO (n-D CASE).

'Reference:

'Yao, X., Liu, Y., and Lin, G., "Evolutionary Programming Made Faster,"
'IEEE Trans. Evolutionary Computation, Vol. 3, No. 2, 82-102, Jul. 1999.

    LOCAL Z, Xi, Sum, Prod AS EXT

    LOCAL i%

    Z = 0## : Sum = 0## : Prod = 1##

    FOR i% = 1 TO Nd%

        Xi  = R(p%,i%,j&)

        Sum = Sum + (Xi-100##)^2

        Prod = Prod*COS((Xi-100##)/SQR(i%))

    NEXT i%

    Z = Sum/4000## - Prod + 1##

    F11 = -Z

END FUNCTION 'F11

'-----

FUNCTION u(Xi,a,k,m)

LOCAL Z AS EXT

    Z = 0##

    SELECT CASE Xi

        CASE > a  : Z = k*(Xi-a)^m

        CASE < -a : Z = k*(-Xi-a)^m

    END SELECT

    u = Z

END FUNCTION

'-----------

FUNCTION F12(R(),Nd%,p%,j&) '(n-D) F12, Penalized #1

'Ref: Yao(1999).  Max=0 @ (-1,-1,...,-1), -50=<Xi<50.

    LOCAL Offset, Sum1, Sum2, Z, X1, Y1, Xn, Yn, Xi, Yi, XiPlus1, YiPlus1 AS EXT

    LOCAL i%, m%, A$

    X1 = R(p%,1,j&)   : Y1 = 1## + (X1+1##)/4##

    Xn = R(p%,Nd%,j&) : Yn = 1## + (Xn+1##)/4##

    Sum1 = 0##

    FOR i% = 1 TO Nd%-1

        Xi      = R(p%,i%,j&)   : Yi      = 1## + (Xi+1##)/4##

        XiPlus1 = R(p%,i%+1,j&): YiPlus1 = 1## + (XiPlus1+1##)/4##

        Sum1 = Sum1 + (Yi-1##)^2*(1##+10##*(SIN(Pi*YiPlus1))^2)

    NEXT i%

    Sum1 = Sum1 + 10##*(SIN(Pi*Y1))^2 + (Yn-1##)^2

    Sum1 = Pi*Sum1/Nd%

    Sum2 = 0##

    FOR i% = 1 TO Nd%

        Xi = R(p%,i%,j&)

        Sum2 = Sum2 + u(Xi,10##,100##,4##)

    NEXT i%

    Z = Sum1 + Sum2

    F12 = -Z

END FUNCTION 'F12()

'------------------

FUNCTION F13(R(),Nd%,p%,j&) '(n-D) F13, Penalized #2
```

```
'Ref: Yao(1999).  Max=0 @ (1,1,...,1), -50=<Xi=<50.

    LOCAL Offset, Sum1, Sum2, Z, Xi, Xn, XiPlus1, X1 AS EXT

    LOCAL i%, m%, AS

    X1 = R(p%,1,j&) : Xn = R(p%,Nd%,j&)

    Sum1 = 0##

    FOR i% = 1 TO Nd%-1

        Xi  = R(p%,i%,j&) : XiPlus1 = R(p%,i%+1,j&)

        Sum1 = Sum1 + (Xi-1##)^2*(1##+(SIN(3##*Pi*XiPlus1))^2)

    NEXT i%

    Sum1 = Sum1 + (SIN(Pi*3##*X1))^2 +(Xn-1##)^2*(1##+(SIN(TwoPi*Xn))^2)

    Sum2 = 0##

    FOR i% = 1 TO Nd%

        Xi = R(p%,i%,j&)

        Sum2 = Sum2 + u(Xi,5##,100##,4##)

    NEXT i%

    Z = Sum1/10## + Sum2

    F13 = -Z

END FUNCTION 'F13()

'------------------

SUB FillArrayAij 'needed for function F14, Shekel's Foxholes

    Aij(1,1)=-32##  : Aij(1,2)=-16##  : Aij(1,3)=0##  : Aij(1,4)=16##  : Aij(1,5)=32##
    Aij(1,6)=-32##  : Aij(1,7)=-16##  : Aij(1,8)=0##  : Aij(1,9)=16##  : Aij(1,10)=32##
    Aij(1,11)=-32## : Aij(1,12)=-16## : Aij(1,13)=0## : Aij(1,14)=16## : Aij(1,15)=32##
    Aij(1,16)=-32## : Aij(1,17)=-16## : Aij(1,18)=0## : Aij(1,19)=16## : Aij(1,20)=32##
    Aij(1,21)=-32## : Aij(1,22)=-16## : Aij(1,23)=0## : Aij(1,24)=16## : Aij(1,25)=32##

    Aij(2,1)=-32##  : Aij(2,2)=-32##  : Aij(2,3)=-32## : Aij(2,4)=-32## : Aij(2,5)=-32##
    Aij(2,6)=-16##  : Aij(2,7)=-16##  : Aij(2,8)=-16## : Aij(2,9)=-16## : Aij(2,10)=-16##
    Aij(2,11)=0##   : Aij(2,12)=0##   : Aij(2,13)=0##  : Aij(2,14)=0##  : Aij(2,15)=0##
    Aij(2,16)=16##  : Aij(2,17)=16##  : Aij(2,18)=16## : Aij(2,19)=16## : Aij(2,20)=16##
    Aij(2,21)=32##  : Aij(2,22)=32##  : Aij(2,23)=32## : Aij(2,24)=32## : Aij(2,25)=32##

END SUB

'-----

FUNCTION F14(R(),Nd%,p%,j&) 'F14 (2-D) Shekel's Foxholes (INVERTED...)

    LOCAL Sum1, Sum2, Z, Xi AS EXT

    LOCAL i%, jj%

    Sum1 = 0##

    FOR jj% = 1 TO 25

        Sum2 = 0##

        FOR i% = 1 TO 2

            Xi = R(p%,i%,j&)

            Sum2 = Sum2 + (Xi-Aij(i%,jj%))^6

        NEXT i%

        Sum1 = Sum1 + 1##/(jj%+Sum2)

    NEXT j%

    Z = 1##/(0.002##+Sum1)

    F14 = -Z

END FUNCTION 'F14

'-----------

FUNCTION F16(R(),Nd%,p%,j&) 'F16 (2-D) 6-Hump Camel-Back

    LOCAL x1, x2, Z AS EXT

    x1 = R(p%,1,j&) : x2 = R(p%,2,j&)

    Z = 4##*x1^2 - 2.1##*x1^4 + x1^6/3## + x1*x2 - 4*x2^2 + 4*x2^4

    F16 = -Z

END FUNCTION 'F16

'-----------

FUNCTION F15(R(),Nd%,p%,j&) 'F15 (4-D) Kowalik's Function

'Global maximum = -0.0003075 @ (0.1928,0.1908,0.1231,0.1358)

    LOCAL x1, x2, x3, x4, Num, Denom, Z, Aj(), Bj() AS EXT

    LOCAL jj%

    REDIM Aj(1 TO 11), Bj(1 TO 11)

    Aj(1)  = 0.1957## : Bj(1)  = 1##/0.25##
```

```basic
    Aj(2)  = 0.1947## : Bj(2)  = 1##/0.50##
    Aj(3)  = 0.1735## : Bj(3)  = 1##/1.00##
    Aj(4)  = 0.1600## : Bj(4)  = 1##/2.00##
    Aj(5)  = 0.0844## : Bj(5)  = 1##/4.00##
    Aj(6)  = 0.0627## : Bj(6)  = 1##/6.00##
    Aj(7)  = 0.0456## : Bj(7)  = 1##/8.00##
    Aj(8)  = 0.0342## : Bj(8)  = 1##/10.00##
    Aj(9)  = 0.0323## : Bj(9)  = 1##/12.00##
    Aj(10) = 0.0235## : Bj(10) = 1##/14.00##
    Aj(11) = 0.0246## : Bj(11) = 1##/16.00##

    Z = 0##

    x1 = R(p%,1,j&) : x2 = R(p%,2,j&) : x3 = R(p%,3,j&) : x4 = R(p%,4,j&)

    FOR jj% = 1 TO 11

        Num  = x1*(Bj(jj%)^2+Bj(jj%)*x2)

        Denom = Bj(jj%)^2+Bj(jj%)*x3+x4

        Z = Z + (Aj(jj%)-Num/Denom)^2

    NEXT jj%

    F15 = -Z

END FUNCTION 'F15
'-----------

FUNCTION F17(R(),Nd%,p%,j&) 'F17, (2-D) Branin

'Global maximum = -0.398 @ (-3.142.12.275), (3.142,2.275), (9.425,2.425)

    LOCAL x1, x2, Z AS EXT

    x1 = R(p%,1,j&) : x2 = R(p%,2,j&)

    Z = (x2-5.1##*x1^2/(4##*Pi^2)+5##*x1/Pi-6##)^2 + 10##*(1##-1##/(8##*Pi))*COS(x1) + 10##

    F17 = -Z

END FUNCTION 'F17
'-----------

FUNCTION F18(R(),Nd%,p%,j&) 'Goldstein-Price 2-D Test Function

'Global maximum = -3 @ (0,-1)

    LOCAL Z, x1, x2, t1, t2 AS EXT

    x1 = R(p%,1,j&) : x2 = R(p%,2,j&)

    t1 = 1##+(x1+x2+1##)^2*(19##-14##*x1+3##*x1^2-14##*x2+6##*x1*x2+3##*x2^2)

    t2 = 30##+(2##*x1-3##*x2)^2*(18##-32##*x1+12##*x1^2+48##*x2-36##*x1*x2+27##*x2^2)

    Z = t1*t2

    F18 = -Z

END FUNCTION 'F18()
'-----------

FUNCTION F19(R(),Nd%,p%,j&) 'F19 (3-D) Hartman's Family #1

'Global maximum = 3.86 @ (0.114,0.556,0.852)

    LOCAL Xi, Z, Sum, Aji(), Cj(), Pji() AS EXT

    LOCAL i%, jj%, m%

    REDIM Aji(1 TO 4, 1 TO 3), Cj(1 TO 4), Pji(1 TO 4, 1 TO 3)

    Aji(1,1) = 3.0## : Aji(1,2) = 10## : Aji(1,3) = 30## : Cj(1) = 1.0##
    Aji(2,1) = 0.1## : Aji(2,2) = 10## : Aji(2,3) = 35## : Cj(2) = 1.2##
    Aji(3,1) = 3.0## : Aji(3,2) = 10## : Aji(3,3) = 30## : Cj(3) = 3.0##
    Aji(4,1) = 0.1## : Aji(4,2) = 10## : Aji(4,3) = 35## : Cj(4) = 3.2##

    Pji(1,1) = 0.36890## : Pji(1,2) = 0.1170## : Pji(1,3) = 0.2673##
    Pji(2,1) = 0.46990## : Pji(2,2) = 0.4387## : Pji(2,3) = 0.7470##
    Pji(3,1) = 0.10910## : Pji(3,2) = 0.8732## : Pji(3,3) = 0.5547##
    Pji(4,1) = 0.03815## : Pji(4,2) = 0.5743## : Pji(4,3) = 0.8828##

    Z = 0##

    FOR jj% = 1 TO 4

        Sum = 0##

        FOR i% = 1 TO 3

            Xi = R(p%,i%,j&)

            Sum = Sum + Aji(jj%,i%)*(Xi-Pji(jj%,i%))^2

        NEXT i%

        Z = Z + Cj(jj%)*EXP(-Sum)

    NEXT jj%

    F19 = Z

END FUNCTION 'F19
'-----------

FUNCTION F20(R(),Nd%,p%,j&) 'F20 (6-D) Hartman's Family #2

'Global maximum = 3.32 @ (0.201,0.150,0.477,0.275,0.311,0.657)
```

```
    LOCAL Xi, Z, Sum, Aji(), Cj(), Pji() AS EXT

    LOCAL i%, jj%, m%

    REDIM Aji(1 TO 4, 1 TO 6), Cj(1 TO 4), Pji(1 TO 4, 1 TO 6)

    Aji(1,1) = 10.0## : Aji(1,2) = 3.00## : Aji(1,3) = 17.0## : Cj(1) = 1.0##
    Aji(2,1) =  0.05## : Aji(2,2) = 10.0## : Aji(2,3) = 17.0## : Cj(2) = 1.2##
    Aji(3,1) =  3.00## : Aji(3,2) = 3.50## : Aji(3,3) =  1.70## : Cj(3) = 3.0##
    Aji(4,1) = 17.0## : Aji(4,2) = 8.00## : Aji(4,3) =  0.05## : Cj(4) = 3.2##

    Aji(1,4) =  3.5## : Aji(1,5) =  1.7## : Aji(1,6) =   8##
    Aji(2,4) =  0.1## : Aji(2,5) =    8## : Aji(2,6) = 14##
    Aji(3,4) =   10## : Aji(3,5) =  17## : Aji(3,6) =   8##
    Aji(4,4) =   10## : Aji(4,5) = 0.1## : Aji(4,6) = 14##

    Pji(1,1) = 0.13120## : Pji(1,2) = 0.1696## : Pji(1,3) = 0.5569##
    Pji(2,1) = 0.23290## : Pji(2,2) = 0.4135## : Pji(2,3) = 0.8307##
    Pji(3,1) = 0.23480## : Pji(3,2) = 0.1415## : Pji(3,3) = 0.3522##
    Pji(4,1) = 0.40470## : Pji(4,2) = 0.8828## : Pji(4,3) = 0.8732##

    Pji(1,4) = 0.0124## : Pji(1,5) = 0.8283## : Pji(1,6) = 0.5886##
    Pji(2,4) = 0.3736## : Pji(2,5) = 0.1004## : Pji(2,6) = 0.9991##
    Pji(3,4) = 0.2883## : Pji(3,5) = 0.3047## : Pji(3,6) = 0.6650##
    Pji(4,4) = 0.5743## : Pji(4,5) = 0.1091## : Pji(4,6) = 0.0381##

    Z = 0##

    FOR jj% = 1 TO 4

        Sum = 0##

        FOR i% = 1 TO 6

            Xi = R(p%,i%,j&)

            Sum = Sum + Aji(jj%,i%)*(Xi-Pji(jj%,i%))^2

        NEXT i%

        Z = Z + Cj(jj%)*EXP(-Sum)

    NEXT jj%

    F20 = Z

END FUNCTION 'F20

'-----------

FUNCTION F21(R(),Nd%,p%,j&) 'F21 (4-D) Shekel's Family m=5

'Global maximum = 10

    LOCAL Xi, Z, Sum, Aji(), Cj() AS EXT

    LOCAL i%, jj%, m%

    m% = 5 : REDIM Aji(1 TO m%, 1 TO 4), Cj(1 TO m%)

    Aji(1,1)  = 4## : Aji(1,2)  =  4## : Aji(1,3)  = 4## : Aji(1,4)  =  4## : Cj(1)  = 0.1##
    Aji(2,1)  = 1## : Aji(2,2)  =  1## : Aji(2,3)  = 1## : Aji(2,4)  =  1## : Cj(2)  = 0.2##
    Aji(3,1)  = 8## : Aji(3,2)  =  8## : Aji(3,3)  = 8## : Aji(3,4)  =  8## : Cj(3)  = 0.2##
    Aji(4,1)  = 6## : Aji(4,2)  =  6## : Aji(4,3)  = 6## : Aji(4,4)  =  6## : Cj(4)  = 0.4##
    Aji(5,1)  = 3## : Aji(5,2)  =  7## : Aji(5,3)  = 3## : Aji(5,4)  =  7## : Cj(5)  = 0.4##

    Z = 0##

    FOR jj% = 1 TO m%  'NOTE: Index jj% is used to avoid same variable name as j&

        Sum = 0##

        FOR i% = 1 TO 4 'Shekel's family is 4-D only

            Xi = R(p%,i%,j&)

            Sum = Sum + (Xi-Aji(jj%,i%))^2

        NEXT i%

        Z = Z + 1##/(Sum + Cj(jj%))

    NEXT jj%

    F21 = Z

END FUNCTION 'F21

'-----------

FUNCTION F22(R(),Nd%,p%,j&) 'F22 (4-D) Shekel's Family m=7

'Global maximum = 10

    LOCAL Xi, Z, Sum, Aji(), Cj() AS EXT

    LOCAL i%, jj%, m%

    m% = 7 : REDIM Aji(1 TO m%, 1 TO 4), Cj(1 TO m%)

    Aji(1,1)  = 4## : Aji(1,2)  =  4## : Aji(1,3)  = 4## : Aji(1,4)  =  4## : Cj(1)  = 0.1##
    Aji(2,1)  = 1## : Aji(2,2)  =  1## : Aji(2,3)  = 1## : Aji(2,4)  =  1## : Cj(2)  = 0.2##
    Aji(3,1)  = 8## : Aji(3,2)  =  8## : Aji(3,3)  = 8## : Aji(3,4)  =  8## : Cj(3)  = 0.2##
    Aji(4,1)  = 6## : Aji(4,2)  =  6## : Aji(4,3)  = 6## : Aji(4,4)  =  6## : Cj(4)  = 0.4##
    Aji(5,1)  = 3## : Aji(5,2)  =  7## : Aji(5,3)  = 3## : Aji(5,4)  =  7## : Cj(5)  = 0.4##
    Aji(6,1)  = 2## : Aji(6,2)  =  9## : Aji(6,3)  = 2## : Aji(6,4)  =  9## : Cj(6)  = 0.6##
    Aji(7,1)  = 5## : Aji(7,2)  =  5## : Aji(7,3)  = 3## : Aji(7,4)  =  3## : Cj(7)  = 0.3##

    Z = 0##

    FOR jj% = 1 TO m%  'NOTE: Index jj% is used to avoid same variable name as j&

        Sum = 0##
```

```
        FOR i% = 1 TO 4 'Shekel's family is 4-D only

            Xi = R(p%,i%,j&)

            Sum = Sum + (Xi-Aji(jj%,i%))^2

        NEXT i%

        Z = Z + 1##/(Sum + Cj(jj%))

    NEXT jj%

    F22 = Z

END FUNCTION 'F22

'-----------

FUNCTION F23(R(),Nd%,p%,j&) 'F23 (4-D) Shekel's Family m=10

'Global maximum = 10

    LOCAL Xi, Z, Sum, Aji(), Cj() AS EXT

    LOCAL i%, jj%, m%

    m% = 10 : REDIM Aji(1 TO m%, 1 TO 4), Cj(1 TO m%)

    Aji(1,1)  = 4## : Aji(1,2)  =    4## : Aji(1,3)  = 4## : Aji(1,4)  =   4## : Cj(1)  = 0.1##
    Aji(2,1)  = 1## : Aji(2,2)  =    1## : Aji(2,3)  = 1## : Aji(2,4)  =   1## : Cj(2)  = 0.2##
    Aji(3,1)  = 8## : Aji(3,2)  =    8## : Aji(3,3)  = 8## : Aji(3,4)  =   8## : Cj(3)  = 0.2##
    Aji(4,1)  = 6## : Aji(4,2)  =    6## : Aji(4,3)  = 6## : Aji(4,4)  =   6## : Cj(4)  = 0.4##
    Aji(5,1)  = 3## : Aji(5,2)  =    7## : Aji(5,3)  = 3## : Aji(5,4)  =   7## : Cj(5)  = 0.4##
    Aji(6,1)  = 2## : Aji(6,2)  =    9## : Aji(6,3)  = 2## : Aji(6,4)  =   9## : Cj(6)  = 0.6##
    Aji(7,1)  = 5## : Aji(7,2)  =    5## : Aji(7,3)  = 3## : Aji(7,4)  =   3## : Cj(7)  = 0.3##
    Aji(8,1)  = 8## : Aji(8,2)  =    1## : Aji(8,3)  = 8## : Aji(8,4)  =   1## : Cj(8)  = 0.7##
    Aji(9,1)  = 6## : Aji(9,2)  =    2## : Aji(9,3)  = 6## : Aji(9,4)  =   2## : Cj(9)  = 0.5##
    Aji(10,1) = 7## : Aji(10,2) = 3.6## : Aji(10,3) = 7## : Aji(10,4) = 3.6## : Cj(10) = 0.5##

    Z = 0##

    FOR jj% = 1 TO m%  'NOTE:  Index jj% is used to avoid same variable name as j&

        Sum = 0##

        FOR i% = 1 TO 4 'Shekel's family is 4-D only

            Xi = R(p%,i%,j&)

            Sum = Sum + (Xi-Aji(jj%,i%))^2

        NEXT i%

        Z = Z + 1##/(Sum + Cj(jj%))

    NEXT jj%

    F23 = Z

END FUNCTION 'F23

'=========================================================== END FUNCTION DEFINITIONS ==========================================================

SUB Plot2DbestProbeTrajectories(NumTrajectories%,M(),R(),XiMin(),XiMax(),Np%,Nd%,LastStep&,FunctionName$)

LOCAL TrajectoryNumber%, ProbeNumber%, StepNumber&, N%, M%, ProcID???

LOCAL MaximumFitness, MinimumFitness AS EXT

LOCAL BestProbeThisStep()

LOCAL BestFitnessThisStep(), TempFitness() AS EXT

LOCAL Annotation$, xCoord$, yCoord$, GnuPlotEXE$, PlotWithLines$

    Annotation$    = ""

    PlotWithLines$ = "YES" '"NO"

    NumTrajectories% = MIN(Np%,NumTrajectories%)

    GnuPlotEXE$ = "wgnuplot.exe"

'    ---------------- Get Min/Max Fitnesses ----------------

    MaximumFitness = M(1,0) : MinimumFitness = M(1,0)  'Note:  M(p%,j&)

    FOR StepNumber& = 0 TO LastStep&

        FOR ProbeNumber% = 1 TO Np%

            IF M(ProbeNumber%,StepNumber&) >= MaximumFitness THEN MaximumFitness = M(ProbeNumber%,StepNumber&)

            IF M(ProbeNumber%,StepNumber&) =< MinimumFitness THEN MinimumFitness = M(ProbeNumber%,StepNumber&)

        NEXT ProbeNumber%

    NEXT StepNumber%

'    ------------- Copy Fitness Array M() into TempFitness to Preserve M() ----------------

    REDIM TempFitness(1 TO Np%, 0 TO LastStep&)

    FOR StepNumber& = 0 TO LastStep&

        FOR ProbeNumber% = 1 TO Np%

            TempFitness(ProbeNumber%,StepNumber&) = M(ProbeNumber%,StepNumber&)

        NEXT ProbeNumber%

    NEXT StepNumber%
```

```
'      ----------- LOOP ON TRAJECTORIES -----------

    FOR TrajectoryNumber% = 1 TO NumTrajectories%

'          -------------- Get Trajectory Coordinate Data ----------------

        REDIM BestFitnessThisStep(0 TO LastStep&), BestProbeThisStep%(0 TO LastStep&)

        FOR StepNumber& = 0 TO LastStep&

            BestFitnessThisStep(StepNumber&) = TempFitness(1,StepNumber&)

            FOR ProbeNumber% = 1 TO Np%

                IF TempFitness(ProbeNumber%,StepNumber&) >= BestFitnessThisStep(StepNumber&) THEN

                    BestFitnessThisStep(StepNumber&) = TempFitness(ProbeNumber%,StepNumber&)

                    BestProbeThisStep%(StepNumber&)  = ProbeNumber%

                END IF

            NEXT ProbeNumber%

        NEXT StepNumber&

'      ----- Create Plot Data File -----

    N% = FREEFILE

    SELECT CASE TrajectoryNumber%

        CASE 1  : OPEN "t1"  FOR OUTPUT AS #N%
        CASE 2  : OPEN "t2"  FOR OUTPUT AS #N%
        CASE 3  : OPEN "t3"  FOR OUTPUT AS #N%
        CASE 4  : OPEN "t4"  FOR OUTPUT AS #N%
        CASE 5  : OPEN "t5"  FOR OUTPUT AS #N%
        CASE 6  : OPEN "t6"  FOR OUTPUT AS #N%
        CASE 7  : OPEN "t7"  FOR OUTPUT AS #N%
        CASE 8  : OPEN "t8"  FOR OUTPUT AS #N%
        CASE 9  : OPEN "t9"  FOR OUTPUT AS #N%
        CASE 10 : OPEN "t10" FOR OUTPUT AS #N%

    END SELECT

'      ----------- Write Plot File Data -----------

'     M% = freefile : open "BestProbebData" for output as #M% 'debug

'     print #M%, "  Step #  BestProbe#        x1              x2"

    FOR StepNumber& = 0 TO LastStep&

        PRINT #N%, USING$("######.######## ######.########",R(BestProbeThisStep%(StepNumber&),1,StepNumber&),R(BestProbeThisStep%(StepNumber&),2,StepNumber&))

'         PRINT #M%, USING$("#####      #####      ######.########
######.########",StepNumber&,BestProbeThisStep%(StepNumber&),R(BestProbeThisStep%(StepNumber&),1,StepNumber&),R(BestProbeThisStep%(StepNumber&),2,StepNumber&))

        TempFitness(BestProbeThisStep%(StepNumber&),StepNumber&) = MinimumFitness 'so that same max will not be found for next trajectory

    NEXT StepNumber&

    CLOSE #N%

'     Close #M%

    NEXT TrajectoryNumber%

'      ------------------------- Plot Trajectories -------------------------

    CALL CreateGNUplotINIfile(0.13##*ScreenWidth&,0.18##*ScreenHeight&,0.7##*ScreenHeight&,0.7##*ScreenHeight&)

    Annotation$ = ""

    N% = FREEFILE

    OPEN "cmd2d.gp" FOR OUTPUT AS #N%

        PRINT #N%, "set xrange ["+REMOVE$(STR$(XiMin(1)),ANY" ")+":"+REMOVE$(STR$(XiMax(1)),ANY" ")+"]"
        PRINT #N%, "set yrange ["+REMOVE$(STR$(XiMin(2)),ANY" ")+":"+REMOVE$(STR$(XiMax(2)),ANY" ")+"]"

        'PRINT #N%, "set label "    + Quote$ + Annotation$ + Quote$ + " at graph " + xCoord$ + "," + yCoord$
        PRINT #N%, "set grid xtics " + "10"
        PRINT #N%, "set grid ytics " + "10"
        PRINT #N%, "set grid mxtics"
        PRINT #N%, "set grid mytics"
        PRINT #N%, "show grid"
        PRINT #N%, "set title " + Quote$ + "2D "+ FunctionName$+" TRAJECTORIES OF PROBES WITH BEST\nFITNESSES (ORDERED BY FITNESS)" + "\n" + RunID$ + Quote$
        PRINT #N%, "set xlabel " + Quote$ + "x1\n\n"                      + Quote$
        PRINT #N%, "set ylabel " + Quote$ + "\nx2"                        + Quote$

        IF PlotWithLines$ = "YES" THEN

            SELECT CASE NumTrajectories%

                CASE 1  : PRINT #N%, "plot "+Quote$+"t1"+Quote$+" w 1 lw 3"
                CASE 2  : PRINT #N%, "plot "+Quote$+"t1"+Quote$+" w 1 lw 3,"+Quote$+"t2"+Quote$+" w 1"
                CASE 3  : PRINT #N%, "plot "+Quote$+"t1"+Quote$+" w 1 lw 3,"+Quote$+"t2"+Quote$+" w 1,"+Quote$+"t3"+Quote$+" w 1"
                CASE 4  : PRINT #N%, "plot "+Quote$+"t1"+Quote$+" w 1 lw 3,"+Quote$+"t2"+Quote$+" w 1,"+Quote$+"t3"+Quote$+" w 1,"+Quote$+"t4"+Quote$+" w 1"
                CASE 5  : PRINT #N%, "plot "+Quote$+"t1"+Quote$+" w 1 lw 3,"+Quote$+"t2"+Quote$+" w 1,"+Quote$+"t3"+Quote$+" w 1,"+Quote$+"t4"+Quote$+" w 1,"+Quote$+"t5"+Quote$+"
w 1"
                CASE 6  : PRINT #N%, "plot "+Quote$+"t1"+Quote$+" w 1 lw 3,"+Quote$+"t2"+Quote$+" w 1,"+Quote$+"t3"+Quote$+" w 1,"+Quote$+"t4"+Quote$+" w 1,"+Quote$+"t5"+Quote$+"
w 1,"+Quote$+"t6"+Quote$+" w 1"
                CASE 7  : PRINT #N%, "plot "+Quote$+"t1"+Quote$+" w 1 lw 3,"+Quote$+"t2"+Quote$+" w 1,"+Quote$+"t3"+Quote$+" w 1,"+Quote$+"t4"+Quote$+" w 1,"+Quote$+"t5"+Quote$+"
w 1,"+Quote$+"t6"+Quote$+" w 1,"+_
                                     Quote$+"t7"+Quote$+" w 1"
                CASE 8  : PRINT #N%, "plot "+Quote$+"t1"+Quote$+" w 1 lw 3,"+Quote$+"t2"+Quote$+" w 1,"+Quote$+"t3"+Quote$+" w 1,"+Quote$+"t4"+Quote$+" w 1,"+Quote$+"t5"+Quote$+"
w 1,"+Quote$+"t6"+Quote$+" w 1,"+_
                                     Quote$+"t7"+Quote$+" w 1,"     +Quote$+"t8"+Quote$+" w 1"
                CASE 9  : PRINT #N%, "plot "+Quote$+"t1"+Quote$+" w 1 lw 3,"+Quote$+"t2"+Quote$+" w 1,"+Quote$+"t3"+Quote$+" w 1,"+Quote$+"t4"+Quote$+" w 1,"+Quote$+"t5"+Quote$+"
w 1,"+Quote$+"t6"+Quote$+" w 1,"+_
                                     Quote$+"t7"+Quote$+" w 1,"     +Quote$+"t8"+Quote$+" w 1,"+Quote$+"t9"+Quote$+" w 1"
                CASE 10 : PRINT #N%, "plot "+Quote$+"t1"+Quote$+" w 1 lw 3,"+Quote$+"t2"+Quote$+" w 1,"+Quote$+"t3"+Quote$+" w 1,"+Quote$+"t4"+Quote$+" w 1,"+Quote$+"t5"+Quote$+"
w 1,"+Quote$+"t6"+Quote$+" w 1,"+_
```

```
                                        Quote$+"t7"+Quote$+" w 1,"    +Quote$+"t8"+Quote$+" w 1,"+Quote$+"t9"+Quote$+" w 1,"+Quote$+"t10"+Quote$+" w 1"

      END SELECT

    ELSE

      SELECT CASE NumTrajectories%

        CASE 1  : PRINT #N%, "plot "+Quote$+"t1"+Quote$+" lw 2"
        CASE 2  : PRINT #N%, "plot "+Quote$+"t1"+Quote$+" lw 2,"+Quote$+"t2"+Quote$
        CASE 3  : PRINT #N%, "plot "+Quote$+"t1"+Quote$+" lw 2,"+Quote$+"t2"+Quote$+" ,"+Quote$+"t3"+Quote$
        CASE 4  : PRINT #N%, "plot "+Quote$+"t1"+Quote$+" lw 2,"+Quote$+"t2"+Quote$+" ,"+Quote$+"t3"+Quote$+" ,"+Quote$+"t4"+Quote$
        CASE 5  : PRINT #N%, "plot "+Quote$+"t1"+Quote$+" lw 2,"+Quote$+"t2"+Quote$+" ,"+Quote$+"t3"+Quote$+" ,"+Quote$+"t4"+Quote$+" ,"+Quote$+"t5"+Quote$
        CASE 6  : PRINT #N%, "plot "+Quote$+"t1"+Quote$+" lw 2,"+Quote$+"t2"+Quote$+" ,"+Quote$+"t3"+Quote$+" ,"+Quote$+"t4"+Quote$+" ,"+Quote$+"t5"+Quote$+"
,"+Quote$+"t6"+Quote$
        CASE 7  : PRINT #N%, "plot "+Quote$+"t1"+Quote$+" lw 2,"+Quote$+"t2"+Quote$+" ,"+Quote$+"t3"+Quote$+" ,"+Quote$+"t4"+Quote$+" ,"+Quote$+"t5"+Quote$+"
,"+Quote$+"t6"+Quote$+" ,"+_
                                        Quote$+"t7"+Quote$
        CASE 8  : PRINT #N%, "plot "+Quote$+"t1"+Quote$+" lw 2,"+Quote$+"t2"+Quote$+" ,"+Quote$+"t3"+Quote$+" ,"+Quote$+"t4"+Quote$+" ,"+Quote$+"t5"+Quote$+"
,"+Quote$+"t6"+Quote$+" ,"+_
                                        Quote$+"t7"+Quote$+" ,"    +Quote$+"t8"+Quote$
        CASE 9  : PRINT #N%, "plot "+Quote$+"t1"+Quote$+" lw 2,"+Quote$+"t2"+Quote$+" ,"+Quote$+"t3"+Quote$+" ,"+Quote$+"t4"+Quote$+" ,"+Quote$+"t5"+Quote$+"
,"+Quote$+"t6"+Quote$+" ,"+_
                                        Quote$+"t7"+Quote$+" ,"    +Quote$+"t8"+Quote$+" ,"+Quote$+"t9"+Quote$
        CASE 10 : PRINT #N%, "plot "+Quote$+"t1"+Quote$+" lw 2,"+Quote$+"t2"+Quote$+" ,"+Quote$+"t3"+Quote$+" ,"+Quote$+"t4"+Quote$+" ,"+Quote$+"t5"+Quote$+"
,"+Quote$+"t6"+Quote$+" ,"+_
                                        Quote$+"t7"+Quote$+" ,"    +Quote$+"t8"+Quote$+" ,"+Quote$+"t9"+Quote$+" ,"+Quote$+"t10"+Quote$

      END SELECT

    END IF

  CLOSE #N%

  ProcID??? = SHELL(GnuPlotEXE$+" cmd2d.gp -") : CALL Delay(0.5##)

END SUB 'Plot2DbestProbeTrajectories()

'----

SUB Plot2DindividualProbeTrajectories(NumTrajectories%,M(),R(),XiMin(),XiMax(),Np%,Nd%,LastStep&,FunctionName$)

LOCAL ProbeNumber%, StepNumber&, N%, ProcID???

LOCAL Annotation$, xCoord$, yCoord$, GnuPlotEXE$, PlotWithLines$

  NumTrajectories% = MIN(Np%,NumTrajectories%)

  Annotation$   = ""

  PlotWithLines$ = "YES" '"NO"

  GnuPlotEXE$ = "wgnuplot.exe"

'  ------------- LOOP ON PROBES --------------

  FOR ProbeNumber% = 1 TO MIN(NumTrajectories%,Np%)

'  ----- Create Plot Data File -----

    N% = FREEFILE

    SELECT CASE ProbeNumber%

      CASE 1  : OPEN "p1"  FOR OUTPUT AS #N%
      CASE 2  : OPEN "p2"  FOR OUTPUT AS #N%
      CASE 3  : OPEN "p3"  FOR OUTPUT AS #N%
      CASE 4  : OPEN "p4"  FOR OUTPUT AS #N%
      CASE 5  : OPEN "p5"  FOR OUTPUT AS #N%
      CASE 6  : OPEN "p6"  FOR OUTPUT AS #N%
      CASE 7  : OPEN "p7"  FOR OUTPUT AS #N%
      CASE 8  : OPEN "p8"  FOR OUTPUT AS #N%
      CASE 9  : OPEN "p9"  FOR OUTPUT AS #N%
      CASE 10 : OPEN "p10" FOR OUTPUT AS #N%
      CASE 11 : OPEN "p11" FOR OUTPUT AS #N%
      CASE 12 : OPEN "p12" FOR OUTPUT AS #N%
      CASE 13 : OPEN "p13" FOR OUTPUT AS #N%
      CASE 14 : OPEN "p14" FOR OUTPUT AS #N%
      CASE 15 : OPEN "p15" FOR OUTPUT AS #N%
      CASE 16 : OPEN "p16" FOR OUTPUT AS #N%

    END SELECT

'  ----------- Write Plot File Data -----------

    FOR StepNumber& = 0 TO LastStep&

      PRINT #N%, USING$("######.######## #####.########",R(ProbeNumber%,1,StepNumber&),R(ProbeNumber%,2,StepNumber&))

    NEXT StepNumber%

    CLOSE #N%

  NEXT ProbeNumber%

'  --------------------------------------------- Plot Trajectories ---------------------------------------------

'usage:  CALL CreateGNUplotINIfile(PlotWindowULC_X%,PlotWindowULC_Y%,PlotWindowWidth%,PlotWindowHeight%)

  CALL CreateGNUplotINIfile(0.17##*ScreenWidth&,0.22##*ScreenHeight&,0.7##*ScreenHeight&,0.7##*ScreenHeight&)

  Annotation$ = ""

  N% = FREEFILE

  OPEN "cmd2d.gp" FOR OUTPUT AS #N%

    PRINT #N%, "set xrange ["+REMOVE$(STR$(XiMin(1)),ANY" ")+":"+REMOVE$(STR$(XiMax(1)),ANY" ")+"]"
    PRINT #N%, "set yrange ["+REMOVE$(STR$(XiMin(2)),ANY" ")+":"+REMOVE$(STR$(XiMax(2)),ANY" ")+"]"

    PRINT #N%, "set grid xtics " + "10"
    PRINT #N%, "set grid ytics " + "10"
    PRINT #N%, "set grid mxtics"
    PRINT #N%, "set grid mytics"
    PRINT #N%, "show grid"
    PRINT #N%, "set title " + Quote$ + "2D "+ FunctionName$+" INDIVIDUAL PROBE TRAJECTORIES\n(ORDERED BY PROBE #)" + "\n" + RunID$ + Quote$
```

```
        PRINT #N%, "set xlabel " + Quote$ + "x1\n\n"                              + Quote$
        PRINT #N%, "set ylabel " + Quote$ + "\nx2"                               + Quote$

        IF PlotWithLines$ = "YES" THEN

            SELECT CASE NumTrajectories%

                CASE 1  : PRINT #N%, "plot "+Quote$+"p1"  +Quote$+" w l lw l"
                CASE 2  : PRINT #N%, "plot "+Quote$+"p1"  +Quote$+" w l lw l,"+Quote$+"p2"+Quote$+" w l"
                CASE 3  : PRINT #N%, "plot "+Quote$+"p1"  +Quote$+" w l lw l,"+Quote$+"p2"+Quote$+" w l,"+Quote$+"p3"+Quote$+" w l"
                CASE 4  : PRINT #N%, "plot "+Quote$+"p1"  +Quote$+" w l lw l,"+Quote$+"p2"+Quote$+" w l,"+Quote$+"p3"+Quote$+" w l,"+Quote$+"p4"+Quote$+" w l"
                CASE 5  : PRINT #N%, "plot "+Quote$+"p1"  +Quote$+" w l lw l,"+Quote$+"p2"+Quote$+" w l,"+Quote$+"p3"+Quote$+" w l,"+Quote$+"p4"+Quote$+" w
l,"+Quote$+"p5"+Quote$+" w l"
                CASE 6  : PRINT #N%, "plot "+Quote$+"p1"  +Quote$+" w l lw l,"+Quote$+"p2"+Quote$+" w l,"+Quote$+"p3"+Quote$+" w l,"+Quote$+"p4"+Quote$+" w
l,"+Quote$+"p5"+Quote$+" w l,"+Quote$+"p6"+Quote$+" w l"
                CASE 7  : PRINT #N%, "plot "+Quote$+"p1"  +Quote$+" w l lw l,"+Quote$+"p2"+Quote$+" w l,"+Quote$+"p3"+Quote$+" w l,"+Quote$+"p4"+Quote$+" w
l,"+Quote$+"p5"+Quote$+" w l,"+Quote$+"p6"+Quote$+" w l,"+_
                                       Quote$+"p7"  +Quote$+" w l"
                CASE 8  : PRINT #N%, "plot "+Quote$+"p1"  +Quote$+" w l lw l,"+Quote$+"p2"+Quote$+" w l,"+Quote$+"p3"+Quote$+" w l,"+Quote$+"p4"+Quote$+" w
l,"+Quote$+"p5"+Quote$+" w l,"+Quote$+"p6"+Quote$+" w l,"+_
                                       Quote$+"p7"  +Quote$+" w l,"     +Quote$+"p8"+Quote$+" w l"
                CASE 9  : PRINT #N%, "plot "+Quote$+"p1"  +Quote$+" w l lw l,"+Quote$+"p2"+Quote$+" w l,"+Quote$+"p3"+Quote$+" w l,"+Quote$+"p4"+Quote$+" w
l,"+Quote$+"p5"+Quote$+" w l,"+Quote$+"p6"+Quote$+" w l,"+_
                                       Quote$+"p7"  +Quote$+" w l,"     +Quote$+"p8"+Quote$+" w l,"+Quote$+"p9"+Quote$+" w l"

                CASE 10 : PRINT #N%, "plot "+Quote$+"p1"  +Quote$+" w l lw l,"+Quote$+"p2"  +Quote$+" w l,"+Quote$+"p3"  +Quote$+" w l,"+Quote$+"p4"  +Quote$+" w
l,"+Quote$+"p5"+Quote$+" w l,"+Quote$+"p6"+Quote$+" w l,"+_
                                       Quote$+"p7"  +Quote$+" w l,"     +Quote$+"p8"  +Quote$+" w l,"+Quote$+"p9"  +Quote$+" w l,"+Quote$+"p10"+Quote$+" w l"
                CASE 11 : PRINT #N%, "plot "+Quote$+"p1"  +Quote$+" w l lw l,"+Quote$+"p2"  +Quote$+" w l,"+Quote$+"p3"  +Quote$+" w l,"+Quote$+"p4"  +Quote$+" w l,"+Quote$+"p5"
+Quote$+" w l,"+Quote$+"p6"  +Quote$+" w l,"+_
                                       Quote$+"p7"  +Quote$+" w l,"     +Quote$+"p8"  +Quote$+" w l,"+Quote$+"p9"  +Quote$+" w l,"+Quote$+"p10"+Quote$+" w
l,"+Quote$+"p11"+Quote$+" w l"
                CASE 12 : PRINT #N%, "plot "+Quote$+"p1"  +Quote$+" w l lw l,"+Quote$+"p2"  +Quote$+" w l,"+Quote$+"p3"  +Quote$+" w l,"+Quote$+"p4"  +Quote$+" w l,"+Quote$+"p5"
+Quote$+" w l,"+Quote$+"p6"  +Quote$+" w l,"+_
                                       Quote$+"p7"  +Quote$+" w l,"     +Quote$+"p8"  +Quote$+" w l,"+Quote$+"p9"  +Quote$+" w l,"+Quote$+"p10"+Quote$+" w
l,"+Quote$+"p11"+Quote$+" w l,"+Quote$+"p12"+Quote$+" w l"
                CASE 13 : PRINT #N%, "plot "+Quote$+"p1"  +Quote$+" w l lw l,"+Quote$+"p2"  +Quote$+" w l,"+Quote$+"p3"  +Quote$+" w l,"+Quote$+"p4"  +Quote$+" w l,"+Quote$+"p5"
+Quote$+" w l,"+Quote$+"p6"  +Quote$+" w l,"+_
                                       Quote$+"p7"  +Quote$+" w l,"     +Quote$+"p8"  +Quote$+" w l,"+Quote$+"p9"  +Quote$+" w l,"+Quote$+"p10"+Quote$+" w
l,"+Quote$+"p11"+Quote$+" w l,"+Quote$+"p12"+Quote$+" w l,"+_
                                       Quote$+"p13"  +Quote$+" w l"
                CASE 14 : PRINT #N%, "plot "+Quote$+"p1"  +Quote$+" w l lw l,"+Quote$+"p2"  +Quote$+" w l,"+Quote$+"p3"  +Quote$+" w l,"+Quote$+"p4"  +Quote$+" w l,"+Quote$+"p5"
+Quote$+" w l,"+Quote$+"p6"  +Quote$+" w l,"+_
                                       Quote$+"p7"  +Quote$+" w l,"     +Quote$+"p8"  +Quote$+" w l,"+Quote$+"p9"  +Quote$+" w l,"+Quote$+"p10"+Quote$+" w
l,"+Quote$+"p11"+Quote$+" w l,"+Quote$+"p12"+Quote$+" w l,"+_
                                       Quote$+"p13"+Quote$+" w l,"     Quote$+"p14"+Quote$+" w l"
                CASE 15 : PRINT #N%, "plot "+Quote$+"p1"  +Quote$+" w l lw l,"+Quote$+"p2"  +Quote$+" w l,"+Quote$+"p3"  +Quote$+" w l,"+Quote$+"p4"  +Quote$+" w l,"+Quote$+"p5"
+Quote$+" w l,"+Quote$+"p6"  +Quote$+" w l,"+_
                                       Quote$+"p7"  +Quote$+" w l,"     +Quote$+"p8"  +Quote$+" w l,"+Quote$+"p9"  +Quote$+" w l,"+Quote$+"p10"+Quote$+" w
l,"+Quote$+"p11"+Quote$+" w l,"+Quote$+"p12"+Quote$+" w l,"+_
                                       Quote$+"p13"+Quote$+" w l,"     Quote$+"p14"+Quote$+" w l,"+Quote$+"p15"+Quote$+" w l"
                CASE 16 : PRINT #N%, "plot "+Quote$+"p1"  +Quote$+" w l lw l,"+Quote$+"p2"  +Quote$+" w l,"+Quote$+"p3"  +Quote$+" w l,"+Quote$+"p4"  +Quote$+" w l,"+Quote$+"p5"
+Quote$+" w l,"+Quote$+"p6"+Quote$+" w l,"+Quote$+"p12"+Quote$+" w l,"+_
                                       Quote$+"p7"  +Quote$+" w l,"     +Quote$+"p8"  +Quote$+" w l,"+Quote$+"p9"  +Quote$+" w l,"+Quote$+"p10"+Quote$+" w
l,"+Quote$+"p11"+Quote$+" w l,"+Quote$+"p12"+Quote$+" w l,"+_
                                       Quote$+"p13"+Quote$+" w l,"     Quote$+"p14"+Quote$+" w l,"+Quote$+"p15"+Quote$+" w l,"+Quote$+"p16"+Quote$+" w l"
            END SELECT

        ELSE

            SELECT CASE NumTrajectories%

                CASE 1  : PRINT #N%, "plot "+Quote$+"p1"+Quote$+" lw l"
                CASE 2  : PRINT #N%, "plot "+Quote$+"p1"+Quote$+" lw l,"+Quote$+"p2"+Quote$
                CASE 3  : PRINT #N%, "plot "+Quote$+"p1"+Quote$+" lw l,"+Quote$+"p2"+Quote$+" ,"+Quote$+"p3"+Quote$
                CASE 4  : PRINT #N%, "plot "+Quote$+"p1"+Quote$+" lw l,"+Quote$+"p2"+Quote$+" ,"+Quote$+"p3"+Quote$+" ,"+Quote$+"p4"+Quote$
                CASE 5  : PRINT #N%, "plot "+Quote$+"p1"+Quote$+" lw l,"+Quote$+"p2"+Quote$+" ,"+Quote$+"p3"+Quote$+" ,"+Quote$+"p4"+Quote$+" ,"+Quote$+"p5"+Quote$
                CASE 6  : PRINT #N%, "plot "+Quote$+"p1"+Quote$+" lw l,"+Quote$+"p2"+Quote$+" ,"+Quote$+"p3"+Quote$+" ,"+Quote$+"p4"+Quote$+" ,"+Quote$+"p5"+Quote$+"
,"+Quote$+"p6"+Quote$
                CASE 7  : PRINT #N%, "plot "+Quote$+"p1"+Quote$+" lw l,"+Quote$+"p2"+Quote$+" ,"+Quote$+"p3"+Quote$+" ,"+Quote$+"p4"+Quote$+" ,"+Quote$+"p5"+Quote$+"
,"+Quote$+"p6"+Quote$+" ,"+_
                                       Quote$+"p7"+Quote$
                CASE 8  : PRINT #N%, "plot "+Quote$+"p1"+Quote$+" lw l,"+Quote$+"p2"+Quote$+" ,"+Quote$+"p3"+Quote$+" ,"+Quote$+"p4"+Quote$+" ,"+Quote$+"p5"+Quote$+"
,"+Quote$+"p6"+Quote$+" ,"+_
                                       Quote$+"p7"+Quote$+" ,"     +Quote$+"p8"+Quote$
                CASE 9  : PRINT #N%, "plot "+Quote$+"p1"+Quote$+" lw l,"+Quote$+"p2"+Quote$+" ,"+Quote$+"p3"+Quote$+" ,"+Quote$+"p4"+Quote$+" ,"+Quote$+"p5"+Quote$+"
,"+Quote$+"p6"+Quote$+" ,"+_
                                       Quote$+"p7"+Quote$+" ,"     +Quote$+"p8"+Quote$+" ,"+Quote$+"p9"+Quote$
                CASE 10 : PRINT #N%, "plot "+Quote$+"p1"+Quote$+" lw l,"+Quote$+"p2"  +Quote$+" ,"+Quote$+"p3"+Quote$+" ,"+Quote$+"p4"  +Quote$+" ,"+Quote$+"p5"+Quote$+"
,"+Quote$+"p6"+Quote$+" ,"+_
                                       Quote$+"p7"+Quote$+" ,"     +Quote$+"p8"  +Quote$+" ,"+Quote$+"p9"+Quote$+" ,"+Quote$+"p10"  +Quote$
                CASE 11 : PRINT #N%, "plot "+Quote$+"p1"+Quote$+" lw l,"+Quote$+"p2"  +Quote$+" ,"+Quote$+"p3"+Quote$+" ,"+Quote$+"p4"  +Quote$+" ,"+Quote$+"p5" +Quote$+"
,"+Quote$+"p6"+Quote$+" ,"+_
                                       Quote$+"p7"+Quote$+" ,"     +Quote$+"p8"  +Quote$+" ,"+Quote$+"p9"+Quote$+" ,"+Quote$+"p10"  +Quote$+" ,"+Quote$+"p11"+Quote$
                CASE 12 : PRINT #N%, "plot "+Quote$+"p1"+Quote$+" lw l,"+Quote$+"p2"  +Quote$+" ,"+Quote$+"p3"+Quote$+" ,"+Quote$+"p4"  +Quote$+" ,"+Quote$+"p5" +Quote$+"
,"+Quote$+"p6" +Quote$+" ,"+_
                                       Quote$+"p7"+Quote$+" ,"     +Quote$+"p8"  +Quote$+" ,"+Quote$+"p9"+Quote$+" ,"+Quote$+"p10"  +Quote$+" ,"+Quote$+"p11"+Quote$+"
,"+Quote$+"p12"+Quote$
                CASE 13 : PRINT #N%, "plot "+Quote$+"p1" +Quote$+" lw l,"+Quote$+"p2"  +Quote$+" ,"+Quote$+"p3"+Quote$+" ,"+Quote$+"p4"  +Quote$+" ,"+Quote$+"p5" +Quote$+" ,"
+Quote$+"p6" +Quote$+" ,"+_
                                       Quote$+"p7"+Quote$+" ,"     +Quote$+"p8"  +Quote$+" ,"+Quote$+"p9"+Quote$+" ,"+Quote$+"p10"  +Quote$+" ,"+Quote$+"p11"+Quote$+" ,"
+Quote$+"p12"+Quote$
                                       Quote$+"p13"+Quote$
                CASE 14 : PRINT #N%, "plot "+Quote$+"p1" +Quote$+" lw l,"+Quote$+"p2"  +Quote$+" ,"+Quote$+"p3"+Quote$+" ,"+Quote$+"p4"  +Quote$+" ,"+Quote$+"p5" +Quote$+" ,"
+Quote$+"p6" +Quote$+" ,"+_
                                       Quote$+"p7" +Quote$+" ,"     +Quote$+"p8"  +Quote$+" ,"+Quote$+"p9"+Quote$+" ,"+Quote$+"p10"  +Quote$+" ,"+Quote$+"p11"+Quote$+" ,"
+Quote$+"p12"+Quote$+" ,"+_
                                       Quote$+"p13"+Quote$+" ,"     +Quote$+"p14"+Quote$
                CASE 15 : PRINT #N%, "plot "+Quote$+"p1" +Quote$+" lw l,"+Quote$+"p2"  +Quote$+" ,"+Quote$+"p3" +Quote$+" ,"+Quote$+"p4" +Quote$+" ,"+Quote$+"p5" +Quote$+" ,"
+Quote$+"p6" +Quote$+" ,"+_
                                       Quote$+"p7" +Quote$+" ,"     +Quote$+"p8" +Quote$+" ,"+Quote$+"p9" +Quote$+" ,"+Quote$+"p10"+Quote$+" ,"+Quote$+"p11"+Quote$+" ,"
+Quote$+"p12"+Quote$+" ,"+_
```

```
                               Quote$+"p13"+Quote$+" ,"    +Quote$+"p14"+Quote$+" ,"+Quote$+"p15"+Quote$

             CASE 16 : PRINT #N%, "plot "+Quote$+"p1" +Quote$+" lw 1,"+Quote$+"p2" +Quote$+" ,"+Quote$+"p3" +Quote$+" ,"+Quote$+"p4" +Quote$+" ,"+Quote$+"p5" +Quote$+" ,"
+Quote$+"p6" +Quote$+" ,"+_
                               Quote$+"p7" +Quote$+" ,"    +Quote$+"p8" +Quote$+" ,"+Quote$+"p9" +Quote$+" ,"+Quote$+"p10"+Quote$+" ,"+Quote$+"p11"+Quote$+" ,"
+Quote$+"p12"+Quote$+" ,"+_
                               Quote$+"p13"+Quote$+" ,"    +Quote$+"p14"+Quote$+" ,"+Quote$+"p15"+Quote$+" ,"+Quote$+"p16"+Quote$
        END SELECT

     END IF

  CLOSE #N%

  ProcID??? = SHELL(GnuPlotEXE$+" cmd2d.gp -") : CALL Delay(0.5##)

END SUB 'Plot2DindividualProbeTrajectories()

'----

SUB Plot3DbestProbeTrajectories(NumTrajectories%,M(),R(),XiMin(),XiMax(),Np%,Nd%,LastStep&,FunctionName$) 'XYZZY

LOCAL TrajectoryNumber%, ProbeNumber%, StepNumber&, N%, M%, ProcID???

LOCAL MaximumFitness, MinimumFitness AS EXT

LOCAL BestProbeThisStep%()

LOCAL BestFitnessThisStep(), TempFitness() AS EXT

LOCAL Annotation$, xCoord$, yCoord$, zCoord$, GnuPlotEXE$, PlotWithLines$

  Annotation$      = ""
  PlotWithLines$   = "NO" '"YES" '"NO"
  NumTrajectories% = MIN(Np%,NumTrajectories%)
  GnuPlotEXE$ = "wgnuplot.exe"
'    --------------- Get Min/Max Fitnesses ----------------

  MaximumFitness = M(1,0) : MinimumFitness = M(1,0)  'Note:  M(p%,j&)

     FOR StepNumber& = 0 TO LastStep&

        FOR ProbeNumber% = 1 TO Np%

           IF M(ProbeNumber%,StepNumber&) >= MaximumFitness THEN MaximumFitness = M(ProbeNumber%,StepNumber&)

           IF M(ProbeNumber%,StepNumber&) =< MinimumFitness THEN MinimumFitness = M(ProbeNumber%,StepNumber&)

        NEXT ProbeNumber%

     NEXT StepNumber%
'    ------------- Copy Fitness Array M() into TempFitness to Preserve M() ----------------

     REDIM TempFitness(1 TO Np%, 0 TO LastStep&)

     FOR StepNumber& = 0 TO LastStep&

        FOR ProbeNumber% = 1 TO Np%

           TempFitness(ProbeNumber%,StepNumber&) = M(ProbeNumber%,StepNumber&)

        NEXT ProbeNumber%

     NEXT StepNumber%
'    ----------- LOOP ON TRAJECTORIES -----------

     FOR TrajectoryNumber% = 1 TO NumTrajectories%
'       --------------- Get Trajectory Coordinate Data ----------------

        REDIM BestFitnessThisStep(0 TO LastStep&), BestProbeThisStep%(0 TO LastStep&)

        FOR StepNumber& = 0 TO LastStep&

           BestFitnessThisStep(StepNumber&) = TempFitness(1,StepNumber&)

           FOR ProbeNumber% = 1 TO Np%

              IF TempFitness(ProbeNumber%,StepNumber&) >= BestFitnessThisStep(StepNumber&) THEN

                 BestFitnessThisStep(StepNumber&) = TempFitness(ProbeNumber%,StepNumber&)

                 BestProbeThisStep%(StepNumber&)  = ProbeNumber%

              END IF

           NEXT ProbeNumber%

        NEXT StepNumber&
'    ----- Create Plot Data File -----

     N% = FREEFILE

     SELECT CASE TrajectoryNumber%

        CASE 1  : OPEN "t1"  FOR OUTPUT AS #N%
        CASE 2  : OPEN "t2"  FOR OUTPUT AS #N%
        CASE 3  : OPEN "t3"  FOR OUTPUT AS #N%
        CASE 4  : OPEN "t4"  FOR OUTPUT AS #N%
        CASE 5  : OPEN "t5"  FOR OUTPUT AS #N%
        CASE 6  : OPEN "t6"  FOR OUTPUT AS #N%
        CASE 7  : OPEN "t7"  FOR OUTPUT AS #N%
        CASE 8  : OPEN "t8"  FOR OUTPUT AS #N%
        CASE 9  : OPEN "t9"  FOR OUTPUT AS #N%
        CASE 10 : OPEN "t10" FOR OUTPUT AS #N%

     END SELECT
```

```
'    ------------ Write Plot File Data ------------

    FOR StepNumber& = 0 TO LastStep&

        PRINT #N%, USING$("######.######## ######.########
######.########",R(BestProbeThisStep%(StepNumber&),1,StepNumber&),R(BestProbeThisStep%(StepNumber&),2,StepNumber&),R(BestProbeThisStep%(StepNumber&),3,StepNumber&))+CHR$(13)

        TempFitness(BestProbeThisStep%(StepNumber&),StepNumber&) = MinimumFitness 'so that same max will not be found for next trajectory

    NEXT StepNumber%

    CLOSE #N%

    NEXT TrajectoryNumber%

'   ------------------------- Plot Trajectories -------------------------

    'CALL CreateGNUplotINIfile(0.1##*ScreenWidth&,0.25##*ScreenHeight&,0.6##*ScreenHeight&,0.6##*ScreenHeight&)

    Annotation$ = ""

    N% = FREEFILE

    OPEN "cmd3d.gp" FOR OUTPUT AS #N%

    PRINT #N%, "set pm3d"
    PRINT #N%, "show pm3d"
    PRINT #N%, "set hidden3d"
    PRINT #N%, "set view 45, 45, 1, 1"

    PRINT #N%, "unset colorbox"

    PRINT #N%, "set xrange [" + REMOVE$(STR$(XiMin(1)),ANY"" ) + ":" + REMOVE$(STR$(XiMax(1)),ANY"" ) + "]"
    PRINT #N%, "set yrange [" + REMOVE$(STR$(XiMin(2)),ANY"" ) + ":" + REMOVE$(STR$(XiMax(2)),ANY"" ) + "]"
    PRINT #N%, "set zrange [" + REMOVE$(STR$(XiMin(3)),ANY"" ) + ":" + REMOVE$(STR$(XiMax(3)),ANY"" ) + "]"
'   PRINT #N%, "set label "   + Quote$ + Annotation$ + Quote$+" at graph "+xCoord$+","+yCoord$+","+zCoord$
'   PRINT #N%, "show label"
    PRINT #N%, "set grid xtics ytics ztics"
    PRINT #N%, "show grid"
    PRINT #N%, "set title "  + Quote$ + "3D " + FunctionName$ + " PROBE TRAJECTORIES" + "\n" + RunID$ + Quote$
    PRINT #N%, "set xlabel " + Quote$ + "x1"                          + Quote$
    PRINT #N%, "set ylabel " + Quote$ + "x2"                          + Quote$
    PRINT #N%, "set zlabel " + Quote$ + "x3"                          + Quote$

    IF PlotWithLines$ = "YES" THEN

        SELECT CASE NumTrajectories%

            CASE 1 : PRINT #N%, "splot "+Quote$+"t1"+Quote$+" w 1 lw 3"
            CASE 2 : PRINT #N%, "splot "+Quote$+"t1"+Quote$+" w 1 lw 3,"+Quote$+"t2"+Quote$+" w 1"
            CASE 3 : PRINT #N%, "splot "+Quote$+"t1"+Quote$+" w 1 lw 3,"+Quote$+"t2"+Quote$+" w 1,"+Quote$+"t3"+Quote$+" w 1"
            CASE 4 : PRINT #N%, "splot "+Quote$+"t1"+Quote$+" w 1 lw 3,"+Quote$+"t2"+Quote$+" w 1,"+Quote$+"t3"+Quote$+" w 1,"+Quote$+"t4"+Quote$+" w 1"
            CASE 5 : PRINT #N%, "splot "+Quote$+"t1"+Quote$+" w 1 lw 3,"+Quote$+"t2"+Quote$+" w 1,"+Quote$+"t3"+Quote$+" w 1,"+Quote$+"t4"+Quote$+" w 1,"+Quote$+"t5"+Quote$+" w 1"
            CASE 6 : PRINT #N%, "splot "+Quote$+"t1"+Quote$+" w 1 lw 3,"+Quote$+"t2"+Quote$+" w 1,"+Quote$+"t3"+Quote$+" w 1,"+Quote$+"t4"+Quote$+" w 1,"+Quote$+"t5"+Quote$+" w 1,"+Quote$+"t6"+Quote$+" w 1"
            CASE 7 : PRINT #N%, "splot "+Quote$+"t1"+Quote$+" w 1 lw 3,"+Quote$+"t2"+Quote$+" w 1,"+Quote$+"t3"+Quote$+" w 1,"+Quote$+"t4"+Quote$+" w 1,"+Quote$+"t5"+Quote$+" w 1,"+Quote$+"t6"+Quote$+" w 1,"+_
                                         Quote$+"t7"+Quote$+" w 1"
            CASE 8 : PRINT #N%, "splot "+Quote$+"t1"+Quote$+" w 1 lw 3,"+Quote$+"t2"+Quote$+" w 1,"+Quote$+"t3"+Quote$+" w 1,"+Quote$+"t4"+Quote$+" w 1,"+Quote$+"t5"+Quote$+" w 1,"+Quote$+"t6"+Quote$+" w 1,"+_
                                         Quote$+"t7"+Quote$+" w 1,"    +Quote$+"t8"+Quote$+" w 1"
            CASE 9 : PRINT #N%, "splot "+Quote$+"t1"+Quote$+" w 1 lw 3,"+Quote$+"t2"+Quote$+" w 1,"+Quote$+"t3"+Quote$+" w 1,"+Quote$+"t4"+Quote$+" w 1,"+Quote$+"t5"+Quote$+" w 1,"+Quote$+"t6"+Quote$+" w 1,"+_
                                         Quote$+"t7"+Quote$+" w 1,"    +Quote$+"t8"+Quote$+" w 1,"+Quote$+"t9"+Quote$+" w 1"
            CASE 10 : PRINT #N%, "splot "+Quote$+"t1"+Quote$+" w 1 lw 3,"+Quote$+"t2"+Quote$+" w 1,"+Quote$+"t3"+Quote$+" w 1,"+Quote$+"t4"+Quote$+" w 1,"+Quote$+"t5"+Quote$+" w 1,"+Quote$+"t6"+Quote$+" w 1,"+_
                                          Quote$+"t7"+Quote$+" w 1,"    +Quote$+"t8"+Quote$+" w 1,"+Quote$+"t9"+Quote$+" w 1,"+Quote$+"t10"+Quote$+" w 1"
        END SELECT

    ELSE

        SELECT CASE NumTrajectories%

            CASE 1 : PRINT #N%, "splot "+Quote$+"t1"+Quote$+" lw 2"
            CASE 2 : PRINT #N%, "splot "+Quote$+"t1"+Quote$+" lw 2,"+Quote$+"t2"+Quote$
            CASE 3 : PRINT #N%, "splot "+Quote$+"t1"+Quote$+" lw 2,"+Quote$+"t2"+Quote$+" ,"+Quote$+"t3"+Quote$
            CASE 4 : PRINT #N%, "splot "+Quote$+"t1"+Quote$+" lw 2,"+Quote$+"t2"+Quote$+" ,"+Quote$+"t3"+Quote$+" ,"+Quote$+"t4"+Quote$
            CASE 5 : PRINT #N%, "splot "+Quote$+"t1"+Quote$+" lw 2,"+Quote$+"t2"+Quote$+" ,"+Quote$+"t3"+Quote$+" ,"+Quote$+"t4"+Quote$+" ,"+Quote$+"t5"+Quote$
            CASE 6 : PRINT #N%, "splot "+Quote$+"t1"+Quote$+" lw 2,"+Quote$+"t2"+Quote$+" ,"+Quote$+"t3"+Quote$+" ,"+Quote$+"t4"+Quote$+" ,"+Quote$+"t5"+Quote$+" ,"+Quote$+"t6"+Quote$
            CASE 7 : PRINT #N%, "splot "+Quote$+"t1"+Quote$+" lw 2,"+Quote$+"t2"+Quote$+" ,"+Quote$+"t3"+Quote$+" ,"+Quote$+"t4"+Quote$+" ,"+Quote$+"t5"+Quote$+" ,"+Quote$+"t6"+Quote$+" ,"+_
                                         Quote$+"t7"+Quote$
            CASE 8 : PRINT #N%, "splot "+Quote$+"t1"+Quote$+" lw 2,"+Quote$+"t2"+Quote$+" ,"+Quote$+"t3"+Quote$+" ,"+Quote$+"t4"+Quote$+" ,"+Quote$+"t5"+Quote$+" ,"+Quote$+"t6"+Quote$+" ,"+_
                                         Quote$+"t7"+Quote$+" ,"     +Quote$+"t8"+Quote$
            CASE 9 : PRINT #N%, "splot "+Quote$+"t1"+Quote$+" lw 2,"+Quote$+"t2"+Quote$+" ,"+Quote$+"t3"+Quote$+" ,"+Quote$+"t4"+Quote$+" ,"+Quote$+"t5"+Quote$+" ,"+Quote$+"t6"+Quote$+" ,"+_
                                         Quote$+"t7"+Quote$+" ,"     +Quote$+"t8"+Quote$+" ,"+Quote$+"t9"+Quote$
            CASE 10 : PRINT #N%, "splot "+Quote$+"t1"+Quote$+" lw 2,"+Quote$+"t2"+Quote$+" ,"+Quote$+"t3"+Quote$+" ,"+Quote$+"t4"+Quote$+" ,"+Quote$+"t5"+Quote$+" ,"+Quote$+"t6"+Quote$+" ,"+_
                                          Quote$+"t7"+Quote$+" ,"    +Quote$+"t8"+Quote$+" ,"+Quote$+"t9"+Quote$+" ,"+Quote$+"t10"+Quote$
        END SELECT

    END IF

    CLOSE #N%

    ProcID??? = SHELL(GnuPlotEXE$+" cmd3d.gp -") : CALL Delay(0.5##)

END SUB 'Plot3DbestProbeTrajectories()

'----

FUNCTION HasFITNESSsaturated$(Nsteps&,j&,Np%,Nd%,M(),R(),DiagLength)

LOCAL A$

LOCAL k&, p%

LOCAL BestFitness, SumOfBestFitnesses, BestFitnessStepJ, FitnessSatTOL AS EXT
```

```
     A$ = "NO"

     FitnessSatTOL = 0.000001## 'tolerance for FITNESS saturation

     IF j& < Nsteps& + 10 THEN GOTO ExitHasFITNESSsaturated 'execute at least 10 steps after averaging interval before performing this check

     SumOfBestFitnesses = 0##

     FOR k& = j&-Nsteps&+1 TO j&

          BestFitness = M(k&,1)

          FOR p% = 1 TO Np%

               IF M(p%,k&) >= BestFitness THEN BestFitness = M(p%,k&)

          NEXT p%

          IF k& = j& THEN BestFitnessStepJ = BestFitness

          SumOfBestFitnesses = SumOfBestFitnesses + BestFitness

     NEXT k&

     IF ABS(SumOfBestFitnesses/Nsteps&-BestFitnessStepJ) =< FitnessSatTOL THEN A$ = "YES" 'saturation if (avg value - last value) are within TOL

ExitHasFITNESSsaturated:

     HasFITNESSsaturated$ = A$

END FUNCTION 'HasFITNESSsaturated$()

'-----------

FUNCTION HasDAVGsaturated$(Nsteps&,j&,Np%,Nd%,M(),R(),DiagLength)

LOCAL A$

LOCAL k&

LOCAL SumOfDavg, DavgStepJ AS EXT

LOCAL DavgSatTOL AS EXT

     A$ = "NO"

     DavgSatTOL = 0.0005## 'tolerance for DAVG saturation

     IF j& < Nsteps& + 10 THEN GOTO ExitHasDAVGsaturated 'execute at least 10 steps after averaging interval before performing this check

     DavgStepJ = DavgThisStep(j&,Np%,Nd%,M(),R(),DiagLength)

     SumOfDavg = 0##

     FOR k& = j&-Nsteps&+1 TO j& 'check this step and previous (Nsteps-1) steps

          SumOfDavg = SumOfDavg + DavgThisStep(k&,Np%,Nd%,M(),R(),DiagLength)

     NEXT k&

     IF ABS(SumOfDavg/Nsteps&-DavgStepJ) =< DavgSatTOL THEN A$ = "YES" 'saturation if (avg value - last value) are within TOL

ExitHasDAVGsaturated:

     HasDAVGsaturated$ = A$

END FUNCTION 'HasDAVGsaturated$()

'-----------

FUNCTION OscillationInDavg$(j&,Np%,Nd%,M(),R(),DiagLength)

LOCAL A$

LOCAL k&, NumSlopeChanges%

     A$ = "NO"

     NumSlopeChanges% = 0

     IF j& < 15 THEN GOTO ExitDavgOscillation 'wait at least 15 steps

     FOR k& = j&-10 TO j&-1 'check previous ten steps

          IF (DavgThisStep(k&,Np%,Nd%,M(),R(),DiagLength)-DavgThisStep(k&-1,Np%,Nd%,M(),R(),DiagLength))* _
          (DavgThisStep(k&+1,Np%,Nd%,M(),R(),DiagLength)-DavgThisStep(k&,Np%,Nd%,M(),R(),DiagLength)) < 0## THEN INCR NumSlopeChanges%

     NEXT j&

     IF NumSlopeChanges% >= 3 THEN A$ = "YES"

ExitDavgOscillation:

     OscillationInDavg$ = A$

END FUNCTION 'OscillationInDavg()

'------

FUNCTION DavgThisStep(j&,Np%,Nd%,M(),R(),DiagLength)

LOCAL BestFitness, TotalDistanceAllProbes, SumSQ AS EXT

LOCAL p%, k&, N%, i%, BestProbeNumber%, BestTimeStep&

'    ----------- Best Probe #, etc. -----------

     FOR k& = 0 TO j&

          BestFitness = M(1,k&)

          FOR p% = 1 TO Np%

               IF M(p%,k&) >= BestFitness THEN
```

```
                    BestFitness = M(p%,k&) : BestProbeNumber% = p% : BestTimeStep& = k&

                END IF

            NEXT p% 'probe #

        NEXT k& 'time step
'   --------- Average Distance to Best Probe -----------

        TotalDistanceAllProbes = 0##

        FOR p% = 1 TO Np%

            SumSQ = 0##

            FOR i% = 1 TO Nd%

                SumSQ = SumSQ + (R(BestProbeNumber%,i%,BestTimeStep&)-R(p%,i%,j&))^2 'do not exclude p%=BestProbeNumber%(j&) from sum because it adds zero

             NEXT i%

            TotalDistanceAllProbes = TotalDistanceAllProbes + SQR(SumSQ)

        NEXT p%

        DavgThisStep = TotalDistanceAllProbes/(DiagLength*(Np%-1)) 'but exclude best prove from average

END FUNCTION 'DavgThisStep()
'-----------

SUB PlotBestFitnessEvolution(Nd%,Np%,LastStep&,G,DeltaT,Alpha,Beta,Frep,M(),PlaceInitialProbes$,InitialAcceleration$,RepositionFactor$,FunctionName$,Gamma)

LOCAL BestFitness(), GlobalBestFitness AS EXT

LOCAL PlotAnnotation$, PlotTitle$

LOCAL p%, j&, N%

    REDIM BestFitness(0 TO LastStep&)

    CALL GetPlotAnnotation(PlotAnnotation$,Nd%,Np%,LastStep&,G,DeltaT,Alpha,Beta,Frep,M(),PlaceInitialProbes$,InitialAcceleration$,RepositionFactor$,FunctionName$,Gamma)

    GlobalBestFitness = M(1,0)

    FOR j& = 0 TO LastStep&

        BestFitness(j&) = M(1,j&)

        FOR p% = 1 TO Np%

            IF M(p%,j&) >= BestFitness(j&)   THEN BestFitness(j&)   = M(p%,j&)

            IF M(p%,j&) >= GlobalBestFitness THEN GlobalBestFitness = M(p%,j&)

        NEXT p% 'probe #

    NEXT j& 'time step

    N% = FREEFILE

    OPEN "Fitness" FOR OUTPUT AS #N%

        FOR j& = 0 TO LastStep&

            PRINT #N%, USING$("###### #######.#######",j&,BestFitness(j&))

        NEXT j&

    CLOSE #N%

    PlotAnnotation$ = PlotAnnotation$ + "Best Fitness = " + REMOVE$(STR$(ROUND(GlobalBestFitness,8)),ANY" ")

    PlotTitle$ = "Best Fitness vs Time Step\n" + "[" + REMOVE$(STR$(Np%),ANY" ") + " probes, "+REMOVE$(STR$(LastStep&),ANY" ")+" time steps]"

    CALL CreateGNUplotINIfile(0.1##*ScreenWidth&,0.1##*ScreenHeight&,0.6##*ScreenWidth&,0.6##*ScreenHeight&)

    CALL TwoDplot("Fitness","Best Fitness","0.7","0.7","Time Step\n\n.",".\n\nBest Fitness(X)", _
                  "","","","","","","","wgnuplot.exe"," with lines linewidth 2",PlotAnnotation$)

END SUB 'PlotBestFitnessEvolution()
'------

SUB PlotAverageDistance(Nd%,Np%,LastStep&,G,DeltaT,Alpha,Beta,Frep,M(),PlaceInitialProbes$,InitialAcceleration$,RepositionFactor$,FunctionName$,R(),DiagLength,Gamma)

LOCAL Davg(), BestFitness(), TotalDistanceAllProbes, SumSQ AS EXT

LOCAL PlotAnnotation$, PlotTitle$

LOCAL p%, j&, N%, i%, BestProbeNumber%(), BestTimeStep&()

    REDIM Davg(0 TO LastStep&), BestFitness(0 TO LastStep&), BestProbeNumber%(0 TO LastStep&), BestTimeStep&(0 TO LastStep&)

    CALL GetPlotAnnotation(PlotAnnotation$,Nd%,Np%,LastStep&,G,DeltaT,Alpha,Beta,Frep,M(),PlaceInitialProbes$,InitialAcceleration$,RepositionFactor$,FunctionName$,Gamma)
'   ----------- Best Probe #, etc. -----------

    FOR j& = 0 TO LastStep&

        BestFitness(j&) = M(1,j&)

        FOR p% = 1 TO Np%

            IF M(p%,j&) >= BestFitness(j&) THEN

                BestFitness(j&) = M(p%,j&) : BestProbeNumber%(j&) = p% : BestTimeStep&(j&) = j& 'only probe number is used at this time, but other data are computed for possible
future use.

            END IF
```

```
        NEXT p% 'probe #

    NEXT j& 'time step

    N% = FREEFILE

'  --------- Average Distance to Best Probe -----------

    FOR j& = 0 TO LastStep&

        TotalDistanceAllProbes = 0##

        FOR p% = 1 TO Np%

            SumSQ = 0##

            FOR i% = 1 TO Nd%

                SumSQ = SumSQ + (R(BestProbeNumber%(j&),i%,j&)-R(p%,i%,j&))^2 'do not exclude p%=BestProbeNumber%(j&) from sum because it adds zero

            NEXT i%

            TotalDistanceAllProbes = TotalDistanceAllProbes + SQR(SumSQ)

        NEXT p%

        Davg(j&) = TotalDistanceAllProbes/(DiagLength*(Np%-1)) 'but exclude best prove from average

    NEXT j&

'  ----------- Create Plot Data File -----------

    OPEN "Davg" FOR OUTPUT AS #N%

        FOR j& = 0 TO LastStep&

            PRINT #N%, USING$("###### #######.#######",j&,Davg(j&))

        NEXT j&

    CLOSE #N%

    PlotTitle$ = "Average Distance of " + REMOVE$(STR$(Np%-1),ANY" ") + " Probes to Best Probe\nNormalized to Size of Decision Space\n" + _
                 "[" + REMOVE$(STR$(Np%),ANY" ") + " probes, " + REMOVE$(STR$(LastStep&),ANY" ") + " time steps]"

    CALL CreateGNUplotINIfile(0.2##*ScreenWidth&,0.2##*ScreenHeight&,0.6##*ScreenWidth&,0.6##*ScreenHeight&)

    CALL TwoDplot("Davg",PlotTitle$,"0.7","0.9","Time Step\n\n.",".\n\n<D>/Ldiag", _
                  "","","","","","","wgnuplot.exe"," with lines linewidth 2",PlotAnnotation$)

END SUB 'PlotAverageDistance()

'------

SUB GetPlotAnnotation(PlotAnnotation$,Nd%,Np%,LastStep&,G,DeltaT,Alpha,Beta,Frep,M(),PlaceInitialProbes$,InitialAcceleration$,RepositionFactor$,FunctionName$,Gamma)

LOCAL A$

    A$ = "" : IF PlaceInitialProbes$ = "UNIFORM ON-AXIS" AND Nd% > 1 THEN A$ = " ("+REMOVE$(STR$(Np%/Nd%),ANY" ") + "/axis)"

    PlotAnnotation$ = RunID$ + "\n" + _
                      FunctionName$ + " Function" + " (" + FormatInteger$(Nd%) + "-D) \n"    + _
                      FormatInteger$(Np%) + " probes"      + A$ + "\n" + _
                      "G = " + FormatFP$(G,2)             + "\n" + _
                      "Alpha = "      + FormatFP$(Alpha,1)  + "\n" + _
                      "Beta = "       + FormatFP$(Beta,1)   + "\n" + _
                      "DelT = "       + FormatFP$(DeltaT,1) + "\n" + _
                      "Gamma = "      + FormatFP$(Gamma,3)  + "\n" + _
                      "Init Probes " + PlaceInitialProbes$  + "\n" + _
                      "Init Accel " + InitialAcceleration$ + "\n" + _
                      "Frep = "       + FormatFP$(Frep,3)    + " (" + RepositionFactor$ + ")\n"

END SUB

'------

SUB PlotBestProbeVsTimeStep(Nd%,Np%,LastStep&,G,DeltaT,Alpha,Beta,Frep,M(),PlaceInitialProbes$,InitialAcceleration$,RepositionFactor$,FunctionName$,Gamma)

LOCAL BestFitness AS EXT

LOCAL PlotAnnotation$, PlotTitle$

LOCAL p%, j&, N%, BestProbeNumber%()

    REDIM BestProbeNumber%(0 TO LastStep&)

    CALL GetPlotAnnotation(PlotAnnotation$,Nd%,Np%,LastStep&,G,DeltaT,Alpha,Beta,Frep,M(),PlaceInitialProbes$,InitialAcceleration$,RepositionFactor$,FunctionName$,Gamma)

    FOR j& = 0 TO LastStep&

        Bestfitness = M(1,j&)

        FOR p% = 1 TO Np%

            IF M(p%,j&) >= BestFitness THEN

                BestFitness = M(p%,j&) : BestProbeNumber%(j&) = p%

            END IF

        NEXT p% 'probe #

    NEXT j& 'time step

    N% = FREEFILE

    OPEN "Best Probe" FOR OUTPUT AS #N%

        FOR j& = 0 TO LastStep&

            PRINT #N%, USING$("###### ####",j&,BestProbeNumber%(j&))

        NEXT j&
```

```
    CLOSE #N%

    PlotTitle$ = "Best Probe Number vs Time Step\n" + "[" +REMOVE$(STR$(Np%),ANY" ") + " probes, " + REMOVE$(STR$(LastStep&),ANY" ") + " time steps]"

    CALL CreateGNUplotINIfile(0.15#*ScreenWidth&,0.15#*ScreenHeight&,0.6##*ScreenWidth&,0.6##*ScreenHeight&)

'USAGE: CALL TwoDplot(PlotFileName$,PlotTitle$,xCoord$,yCoord$,XaxisLabel$,YaxisLabel$,LogXaxis$,LogYaxis$,xMin$,xMax$,yMin$,yMax$,xTics$,yTics$,GnuPlotEXE$,LineType$,Annotation$)

    CALL TwoDplot("Best Probe",PlotTitle$,"0.7","0.7","Time Step\n\n.",".\n\nBest Probe #","","","","","0",NoSpaces$(Np%+1,0),"","","wgnuplot.exe"," pt 8 ps .5 lw 1",PlotAnnotation$) 'pt, pointtype; ps, pointsize; lw, linewidth

END SUB 'PlotBestProbeVsTimeStep()

'------

SUB DisplayBestFitness(Np%,Nd%,LastStep&,M(),R(),BestFitnessProbeNumber%,BestFitnessTimeStep&,FunctionName$)

LOCAL A$, B$, p%, i%, j&

LOCAL BestFitness AS EXT

    B$ = "" : IF Nd% > 1 THEN B$ = "s"

    BestFitness = M(1,0)

    FOR j& = 0 TO LastStep&

        FOR p% = 1 TO Np%

            IF M(p%,j&) >= BestFitness THEN

                BestFitness = M(p%,j&) : BestFitnessProbeNumber% = p% : BestFitnessTimeStep& = j&

            END IF

        NEXT p%

    NEXT j&

    A$ = FunctionName$ + CHR$(13) +_
        "Best Fitness = " + REMOVE$(STR$(ROUND(BestFitness,8)),ANY" ") + " returned by" + CHR$(13)         +_
        "Probe # "         + REMOVE$(STR$(BestFitnessProbeNumber%),ANY" ") +_
        " at Time Step " + REMOVE$(STR$(BestFitnessTimeStep&),ANY" ") + CHR$(13) + CHR$(13) + "P" + REMOVE$(STR$(BestFitnessProbeNumber%),ANY" ") + " coordinate" + B$ + ":" +
CHR$(13)

    FOR i% = 1 TO Nd% : A$ = A$ + STR$(i%)+"     "+REMOVE$(STR$(ROUND(R(BestFitnessProbeNumber%,i%,BestFitnessTimeStep&),8)),ANY" ")+CHR$(13) : NEXT i%

    MSGBOX(A$)

END SUB 'DisplayBestFitness()

'------

FUNCTION FormatInteger$(M%) : FormatInteger$ = REMOVE$(STR$(M%),ANY" ") : END FUNCTION

'------

FUNCTION FormatFP$(X,Ndigits%)

LOCAL A$

    IF X = 0## THEN

        A$ = "0." : GOTO ExitFormatFP

    END IF

    A$ = REMOVE$(STR$(ROUND(ABS(X),Ndigits%)),ANY" ")

    IF ABS(X) < 1## THEN

        IF X > 0## THEN

            A$ = "0" + A$

        ELSE

            A$ = "-0" + A$

        END IF

    ELSE

        IF X < 0## THEN A$ = "-" + A$

    END IF

ExitFormatFP:

    FormatFP$ = A$

END FUNCTION

'-----------

SUB InitialProbeDistribution(Np%,Nd%,Nt&,XiMin(),XiMax(),R(),PlaceInitialProbes$,Gamma)

LOCAL DeltaXi, DelX1, DelX2, Di AS EXT

LOCAL NumProbesPerAxis&, p%, i%, k%, NumX1points%, NumX2points%, x1pointNum%, x2pointNum%, A$

    SELECT CASE PlaceInitialProbes$

        CASE "UNIFORM ON-AXIS"

            IF Nd% > 1 THEN

                NumProbesPerAxis& = Np%\Nd% 'even #

            ELSE

                NumProbesPerAxis& = Np%
```

```
                END IF
                FOR i% = 1 TO Nd%
                    FOR p% = 1 TO Np%
                        R(p%,i%,0) = XiMin(i%) + Gamma*(XiMax(i%)-XiMin(i%))
                    NEXT Np%
                NEXT i%
                FOR i% = 1 TO Nd% 'place probes axis-by-axis (i% is axis [dimension] number)
                    DeltaXi = (XiMax(i%)-XiMin(i%))/(NumProbesPerAxis%-1)
                    FOR k% = 1 TO NumProbesPerAxis
                        p% = k% + NumProbesPerAxis%*(i%-1) 'probe #
                        R(p%,i%,0) = XiMin(i%) + (k%-1)*DeltaXi
                    NEXT k%
                NEXT i%
            CASE "UNIFORM ON-DIAGONAL"
                FOR p% = 1 TO Np%
                    FOR i% = 1 TO Nd%
                        DeltaXi = (XiMax(i%)-XiMin(i%))/(Np%-1)
                        R(p%,i%,0) = XiMin(i%) + (p%-1)*DeltaXi
                    NEXT i%
                NEXT p%
            CASE "2D GRID"
                NumProbesPerAxis% = SQR(Np%) : NumX1points% = NumProbesPerAxis% : NumX2points% = NumX1points% 'broken down for possible future use
                DelX1 = (XiMax(1)-XiMin(1))/(NumX1points%-1)
                DelX2 = (XiMax(2)-XiMin(2))/(NumX2points%-1)
                FOR x1pointNum% = 1 TO NumX1points%
                    FOR x2pointNum% = 1 TO NumX2points%
                        p% = NumX1points%*(x1pointNum%-1)+x2pointNum% 'probe #
                        R(p%,1,0) = XiMin(1) + DelX1*(x1pointNum%-1) 'x1 coord
                        R(p%,2,0) = XiMin(2) + DelX2*(x2pointNum%-1) 'x2 coord
                    NEXT x2pointNum%
                NEXT x1pointNum%
            CASE "RANDOM"
                FOR p% = 1 TO Np%
                    FOR i% = 1 TO Nd%
                        R(p%,i%,0) = XiMin(i%) + RandomNum(0##,1##)*(XiMax(i%)-XiMin(i%))
                    NEXT i%
                NEXT p%
        END SELECT
    END SUB 'InitialProbeDistribution()
'------
SUB InitialProbeAccelerations(Np%,Nd%,A(),InitialAcceleration$,MaxInitialRandomAcceleration,MaxInitialFixedAcceleration)
LOCAL p%, i%
LOCAL A$
LOCAL FixedInitialAcceleration AS EXT
    SELECT CASE InitialAcceleration$
        CASE "ZERO"
            FOR p% = 1 TO Np%
                FOR i% = 1 TO Nd%
                    A(p%,i%,0) = 0##
                NEXT i% 'coordinate #
            NEXT p% 'probe #
        CASE "FIXED"
            A$ = INPUTBOX$("Fixed Initial Probe"+CHR$(13)+"Acceleration? [0.001-"+REMOVE$(STR$(MaxInitialFixedAcceleration),ANY" ")+"]","PROBES' INITIAL ACCELERATION","0.5")
            FixedInitialAcceleration = VAL(A$)
            IF FixedInitialAcceleration < 0.001## OR FixedInitialAcceleration > MaxInitialFixedAcceleration THEN FixedInitialAcceleration = 0.5##
            FOR p% = 1 TO Np%
                FOR i% = 1 TO Nd%
                    A(p%,i%,0) = FixedInitialAcceleration
```

```basic
                NEXT i% 'coordinate

            NEXT p% 'probe #

        CASE "RANDOM"

            FOR p% = 1 TO Np%

                FOR i% = 1 TO Nd%

                    A(p%,i%,0) = RandomNum(0##,1##)*MaxInitialRandomAcceleration

                NEXT i% 'coordinate

            NEXT p% 'probe #

    END SELECT

END SUB 'InitialProbeAccelerations()

'------

SUB RetrieveErrantProbes(Np%,Nd%,j&,XiMin(),XiMax(),R(),M(),RepositionFactor$,Frep) 'note: M(), RepositionFcator$ passed but not used in thsi version

LOCAL p%, i%

    FOR p% = 1 TO Np%

        FOR i% = 1 TO Nd%

            IF R(p%,i%,j&) < XiMin(i%) THEN R(p%,i%,j&) = MAX(XiMin(i%) + Frep*(R(p%,i%,j&-1)-XiMin(i%)),XiMin(i%)) 'CHANGED 02-07-10

            IF R(p%,i%,j&) > XiMax(i%) THEN R(p%,i%,j&) = MIN(XiMax(i%) - Frep*(XiMax(i%)-R(p%,i%,j&-1)),XiMax(i%))

        NEXT i%

    NEXT p%

END SUB 'RetrieveErrantProbes()

'------

SUB ChangeRunParameters(NumProbesPerDimension%,Np%,Nd%,Nt&,G,Alpha,Beta,DeltaT,Frep,PlaceInitialProbes$,InitialAcceleration$,RepositionFactor$,FunctionName$)

LOCAL p%, DefaultValue$

    A$ = INPUTBOX$("# dimensions?","Change # Dimensions ("+FunctionName$+")",NoSpaces$(Nd%+0,0)) : Nd%    = VAL(A$) : IF Nd% < 1 OR Nd% > 500 THEN Nd% = 2

    IF Nd% > 1 THEN NumProbesPerDimension% = 2*((NumProbesPerDimension%+1)\2) 'require an even # probes on each axis to avoid overlapping at origin (in symmetrical spaces at least...)

    IF Nd% = 1 THEN NumProbesPerDimension% = MAX(NumProbesPerDimension%,3)    'at least 3 probes on x-axis for 1-D functions

    Np% = NumProbesPerDimension%*Nd%

    A$ = INPUTBOX$("# time steps?","Change # Steps ("+FunctionName$+")",NoSpaces$(Nt&+0,0)) : Nt&    = VAL(A$) : IF Nt& < 3                THEN Nt& = 50

    A$ = INPUTBOX$("Grav Const G?","Change G ("+FunctionName$+")",NoSpaces$(G,2))             : G      = VAL(A$) : IF G < -100##    OR G > 100## THEN G   = 2##

    A$ = INPUTBOX$("Alpha?","Change Alpha ("+FunctionName$+")",NoSpaces$(Alpha,2))            : Alpha  = VAL(A$) : IF Alpha < -50## OR Alpha > 50## THEN Alpha = 2##

    A$ = INPUTBOX$("Beta?","Change Beta ("+FunctionName$+")",NoSpaces$(Beta,2))               : Beta   = VAL(A$) : IF Beta  < -50## OR Beta  > 50## THEN Beta = 2##'

    A$ = INPUTBOX$("Delta T?","Change Delta-T ("+FunctionName$+")",NoSpaces$(DeltaT,2))       : DeltaT = VAL(A$) : IF DeltaT =< 0##                 THEN DeltaT = 1##

    A$ = INPUTBOX$("Frep [0-1]?","Change Frep ("+FunctionName$+")",NoSpaces$(Frep,3))         : Frep   = VAL(A$) : IF Frep < 0##    OR Frep > 1## THEN Frep = 0.5##

'   ------------ Initial Probe Distribution ------------

    SELECT CASE PlaceInitialProbes$
        CASE "UNIFORM ON-AXIS"     : DefaultValue$ = "1"
        CASE "UNIFORM ON-DIAGONAL" : DefaultValue$ = "2"
        CASE "2D GRID"             : DefaultValue$ = "3"
        CASE "RANDOM"              : DefaultValue$ = "4"
    END SELECT

    A$ = INPUTBOX$("Initial Probes?"+CHR$(13)+"1 - UNIFORM ON-AXIS"+CHR$(13)+"2 - UNIFORM ON-DIAGONAL"+CHR$(13)+"3 - 2D GRID"+CHR$(13)+"4 - RANDOM","Initial Probe Distribution ("+FunctionName$+")",DefaultValue$)

    IF VAL(A$) < 1 OR VAL(A$) > 4 THEN A$ = "1"

    SELECT CASE VAL(A$)
        CASE 1 : PlaceInitialProbes$ = "UNIFORM ON-AXIS"
        CASE 2 : PlaceInitialProbes$ = "UNIFORM ON-DIAGONAL"
        CASE 3 : PlaceInitialProbes$ = "2D GRID"
        CASE 4 : PlaceInitialProbes$ = "RANDOM"
    END SELECT

    IF Nd% = 1  AND PlaceInitialProbes$ = "UNIFORM ON-DIAGONAL" THEN PlaceInitialProbes$ = "UNIFORM ON-AXIS" 'cannot be diagonal in 1-D space

    IF Nd% <> 2 AND PlaceInitialProbes$ = "2D GRID" THEN PlaceInitialProbes$ = "UNIFORM ON-AXIS" '2D grid is available only in 2 dimensions!

'   ------------- Initial Acceleration -----------------

    SELECT CASE InitialAcceleration$
        CASE "ZERO"   : DefaultValue$ = "1"
        CASE "FIXED"  : DefaultValue$ = "2"
        CASE "RANDOM" : DefaultValue$ = "3"
    END SELECT

    A$ = INPUTBOX$("Initial Acceleration?"+CHR$(13)+"1 - ZERO"+CHR$(13)+"2 - FIXED"+CHR$(13)+"3 - RANDOM","Initial Acceleration ("+FunctionName$+")",DefaultValue$)

    IF VAL(A$) < 1 OR VAL(A$) > 3 THEN A$ = "1"

    SELECT CASE VAL(A$)
        CASE 1 : InitialAcceleration$ = "ZERO"
        CASE 2 : InitialAcceleration$ = "FIXED"
        CASE 3 : InitialAcceleration$ = "RANDOM"
    END SELECT

'   ----------- Reposition Factor ---------------

    SELECT CASE RepositionFactor$
```

```
        CASE "FIXED"    : DefaultValue$ = "1"
        CASE "VARIABLE" : DefaultValue$ = "2"
        CASE "RANDOM"   : DefaultValue$ = "3"
    END SELECT

    A$ = INPUTBOX$("Reposition Factor?"+CHR$(13)+"1 - FIXED"+CHR$(13)+"2 - VARIABLE"+CHR$(13)+"3 - RANDOM","Retrieve Probes ("+FunctionName$+")",DefaultValue$)

    IF VAL(A$) < 1 OR VAL(A$) > 3 THEN A$ = "1"

    SELECT CASE VAL(A$)
        CASE 1 : RepositionFactor$ = "FIXED"
        CASE 2 : RepositionFactor$ = "VARIABLE"
        CASE 3 : RepositionFactor$ = "RANDOM"
    END SELECT

END SUB 'ChangeRunParameters()

'------

FUNCTION NoSpaces$(X,NumDigits%) :  NoSpaces$ = REMOVE$(STR$(X,NumDigits%),ANY" ") : END FUNCTION

'-----------

FUNCTION TerminateNowForSaturation$(j&,Nd%,Np%,Nt&,G,DeltaT,Alpha,Beta,R(),A(),M())

LOCAL A$, i&, p%, NumStepsForAveraging&

LOCAL BestFitness, AvgFitness, FitnessTOL AS EXT 'terminate if avg fitness does not change over NumStepsForAveraging& time steps

    FitnessTOL = 0.00001## : NumStepsForAveraging& = 10

    A$ = "NO"

    IF j& >= NumStepsForAveraging+10 THEN 'wait until step 10 to start checking for fitness saturation

        AvgFitness = 0##

        FOR i& = j&-NumStepsForAveraging&+1 TO j& 'avg fitness over current step & previous NumStepsForAveraging&-1 steps

            BestFitness = M(1,i&)

            FOR p% = 1 TO Np%

                IF M(p%,i&) >= BestFitness THEN BestFitness = M(p%,i&)

            NEXT p%

            AvgFitness = AvgFitness + BestFitness

        NEXT i&

        AvgFitness = AvgFitness/NumStepsForAveraging&

        IF ABS(AvgFitness-BestFitness) < FitnessTOL THEN A$ = "YES" 'compare avg fitness to best fitness at this step

    END IF

    TerminateNowForSaturation$ = A$

END FUNCTION 'TerminateNowForSaturation$()

'-----------

FUNCTION MagVector(V(),N%) 'returns magnitude of Nx1 column vector V

LOCAL SumSq AS EXT

LOCAL i%

    SumSQ = 0## : FOR i% = 1 TO N% : SumSQ = SumSQ + V(i%)^2 : NEXT i% : MagVector = SQR(SumSQ)

END FUNCTION 'MagVector()

'---

FUNCTION UnitStep(X)

LOCAL Z AS EXT

    IF X < 0## THEN

        Z = 0##

    ELSE

        Z = 1##

    END IF

    UnitStep = Z

END FUNCTION 'UnitStep()

'---

SUB Plot1Dfunction(FunctionName$,XiMin(),XiMax(),R()) 'plots 1D function on-screen

LOCAL NumPoints%, i%, N%

LOCAL DeltaX, X AS EXT

    NumPoints% = 32001

    DeltaX = (XiMax(1)-XiMin(1))/(NumPoints%-1)

    N% = FREEFILE

    SELECT CASE FunctionName$

        CASE "ParrottF4" 'PARROTT F4 FUNCTION

            OPEN "ParrottF4" FOR OUTPUT AS #N%

                FOR i% = 1 TO NumPoints%
```

```basic
                R(1,1,0) = XiMin(1) + (i%-1)*DeltaX

                PRINT #N%, USING$("#.###### #.######",R(1,1,0),ParrotF4(R(),1,1,0))

            NEXT i%

        CLOSE #N%

        CALL CreateGNUplotINIfile(0.2##*ScreenWidth&,0.2##*ScreenHeight&,0.6##*ScreenWidth&,0.6##*ScreenHeight&)

        CALL TwoDplot("ParrottF4","Parrott F4 Function","0.7","0.7","X\n\n.".\n\nParrott F4(X)","","","0","1","0","1","","","wgnuplot.exe"," with lines linewidth 2","")

    END SELECT

END SUB

'------

SUB CLEANUP 'probe coordinate plot files

    IF DIR$("P1")  <> "" THEN KILL "P1"
    IF DIR$("P2")  <> "" THEN KILL "P2"
    IF DIR$("P3")  <> "" THEN KILL "P3"
    IF DIR$("P4")  <> "" THEN KILL "P4"
    IF DIR$("P5")  <> "" THEN KILL "P5"
    IF DIR$("P6")  <> "" THEN KILL "P6"
    IF DIR$("P7")  <> "" THEN KILL "P7"
    IF DIR$("P8")  <> "" THEN KILL "P8"
    IF DIR$("P9")  <> "" THEN KILL "P9"
    IF DIR$("P10") <> "" THEN KILL "P10"
    IF DIR$("P11") <> "" THEN KILL "P11"
    IF DIR$("P12") <> "" THEN KILL "P12"
    IF DIR$("P13") <> "" THEN KILL "P13"
    IF DIR$("P14") <> "" THEN KILL "P14"
    IF DIR$("P15") <> "" THEN KILL "P15"

END SUB

'------

SUB Plot2Dfunction(FunctionName$,XiMin(),XiMax(),R())

LOCAL A$

LOCAL NumPoints%, i%, k%, N%

LOCAL DelX1, DelX2, Z AS EXT

    SELECT CASE FunctionName$

        CASE "PBM_1","PBM_2","PBM_3","PBM_4","PBM_5" : NumPoints% = 25

        CASE ELSE : NumPoints% = 100

    END SELECT

    N% = FREEFILE : OPEN "TwoDplot.DAT" FOR OUTPUT AS #N%

    DelX1 = (XiMax(1)-XiMin(1))/(NumPoints%-1) : DelX2 = (XiMax(2)-XiMin(2))/(NumPoints%-1)

    FOR i% = 1 TO NumPoints%

        R(1,1,0) = XiMin(1) + (i%-1)*DelX1  'x1 value

        FOR k% = 1 TO NumPoints%

            R(1,2,0) = XiMin(2) + (k%-1)*DelX2  'x2 value

            Z = ObjectiveFunction(R(),2,1,0,FunctionName$)

            PRINT #N%, USING$("#####.###### ######.###### #######.#####^^^^^",R(1,1,0),R(1,2,0),Z)

        NEXT k%

        PRINT #N%, ""

    NEXT i%

    CLOSE #N%

    CALL CreateGNUplotINIfile(0.1##*ScreenWidth&,0.1##*ScreenHeight&,0.6##*ScreenWidth&,0.6##*ScreenHeight&)

    A$ = "" : IF INSTR(FunctionName$,"PBM_") > 0 THEN A$ = "Coarse "

    CALL ThreeDplot2("TwoDplot.DAT",A$+"Plot of "+FunctionName$+" Function","","0.6","0.6","1.2", _
                    "x1","x2","z=F(x1,x2)","","","wgnuplot.exe","","","","","")

END SUB

'------

    SUB TwoDplot3curves(NumCurves%,PlotFileName1$,PlotFileName2$,PlotFileName3$,PlotTitle$,Annotation$,xCoord$,yCoord$,XaxisLabel$,YaxisLabel$, _
                    LogXaxis$,LogYaxis$,xMin$,xMax$,yMin$,yMax$,xTics$,yTics$,GnuPlotEXES)

        LOCAL N%

        LOCAL LineSize$

        LineSize$ = "2"

        N% = FREEFILE

        OPEN "cmd2d.gp" FOR OUTPUT AS #N%

            IF LogXaxis$ = "YES" AND LogYaxis$ = "NO"  THEN PRINT #N%, "set logscale x"
            IF LogXaxis$ = "NO"  AND LogYaxis$ = "YES" THEN PRINT #N%, "set logscale y"
            IF LogXaxis$ = "YES" AND LogYaxis$ = "YES" THEN PRINT #N%, "set logscale xy"

            IF xMin$ <> "" AND xMax$ <> "" THEN  PRINT #N%, "set xrange ["+xMin$+":"+xMax$+"]"

            IF yMin$ <> "" AND yMax$ <> "" THEN PRINT #N%, "set yrange ["+yMin$+":"+yMax$+"]"

            PRINT #N%, "set label "+Quote$+AnnoTation$+Quote$+" at graph "+xCoord$+","+yCoord$
```

```
        PRINT #N%, "set grid xtics"
        PRINT #N%, "set grid ytics"
        PRINT #N%, "set xtics "+xTics$
        PRINT #N%, "set ytics "+yTics$
        PRINT #N%, "set grid mxtics"
        PRINT #N%, "set grid mytics"
        PRINT #N%, "set title " +Quote$+PlotTitle$+Quote$
        PRINT #N%, "set xlabel "+Quote$+XaxisLabel$+Quote$
        PRINT #N%, "set ylabel "+Quote$+YaxisLabel$+Quote$

        SELECT CASE NumCurves%

        CASE 1
        PRINT #N%, "plot " + Quote$ + PlotFileName1$ + Quote$ + " with lines linewidth " + LineSize$

        CASE 2
        PRINT #N%, "plot " + Quote$ + PlotFileName1$ + Quote$ + " with lines linewidth " + LineSize$+", " + _
                           Quote$ + PlotFileName2$ + Quote$ + " with lines linewidth " + LineSize$
        CASE 3
        PRINT #N%, "plot " + Quote$ + PlotFileName1$ + Quote$ + " with lines linewidth " + LineSize$+", " + _
                           Quote$ + PlotFileName2$ + Quote$ + " with lines linewidth " + LineSize$+", " + _
                           Quote$ + PlotFileName3$ + Quote$ + " with lines linewidth " + LineSize$
        END SELECT

    CLOSE #N%

    SHELL(GnuPlotEXE$+" cmd2d.gp -")

    CALL Delay(0.3)

    END SUB 'TwoDplot3Curves()

'---

FUNCTION Fibonacci&&(N%) 'RETURNS Nth FIBONACCI NUMBER

LOCAL i%, Fn&&, Fn1&&, Fn2&&

LOCAL A$

    IF N% > 91 OR N% < 0 THEN

        MSGBOX("ERROR!  Fibonacci argument"+STR$(N%)+" > 91.  Out of range or < 0...") : EXIT FUNCTION

    END IF

    SELECT CASE N%

        CASE 0: Fn&& = 1

        CASE ELSE

            Fn&& = 0 : Fn2&& = 1 : i% = 0

            FOR i% = 1 TO N%

                Fn&& = Fn1&& + Fn2&&

                Fn1&& = Fn2&&

                Fn2&& = Fn&&

            NEXT i% 'LOOP

    END SELECT

    Fibonacci&& = Fn&&

END FUNCTION 'Fibonacci&&()

'-----------

FUNCTION RandomNum(a,b) 'Returns random number X, a=< X < b.

    RandomNum = a + (b-a)*RND

END FUNCTION 'RandomNum()

'-----------

FUNCTION GaussianDeviate(Mu,Sigma) 'returns NORMAL (Gaussian) random deviate with mean Mu and standard deviation Sigma (variance = Sigma^2)

'Refs: (1) Press, W.H., Flannery, B.P., Teukolsky, S.A., and Vetterling, W.T., "Numerical Recipes: The Art of Scientific Computing,"
'           §7.2, Cambridge University Press, Cambridge, UK, 1986.
'       (2) Shinzato, T., "Box Muller Method," 2007, http://www.sp.dis.titech.ac.jp/~shinzato/boxmuller.pdf

LOCAL s, t, Z AS EXT

    s = RND : t = RND

    Z = Mu + Sigma*SQR(-2##*LOG(s))*COS(TwoPi*t)

    GaussianDeviate = Z

END FUNCTION 'GaussianDeviate()

'-----------

    SUB ContourPlot(PlotFileName$,PlotTitle$,Annotation$,xCoord$,yCoord$,zCoord$, _
                    XaxisLabel$,YaxisLabel$,ZaxisLabel$,zMin$,zMax$,GnuPlotEXE$,A$)
        LOCAL N%

        N% = FREEFILE

        OPEN "cmd3d.gp" FOR OUTPUT AS #N%

            PRINT #N%, "show surface"
            PRINT #N%, "set hidden3d"
            IF zMin$ <> "" AND zMax$ <> "" THEN  PRINT #N%, "set zrange ["+zMin$+":"+zMax$+"]"
            PRINT #N%, "set label "+Quote$+Annotation$+Quote$+" at graph "+xCoord$+","+yCoord$+","+zCoord$
            PRINT #N%, "show label"
            PRINT #N%, "set grid xtics ytics ztics"
            PRINT #N%, "show grid"
            PRINT #N%, "set title "+Quote$+PlotTitle$+Quote$
```

```
            PRINT #N%, "set xlabel "+Quote$+XaxisLabel$+Quote$
            PRINT #N%, "set ylabel "+Quote$+YaxisLabel$+Quote$
            PRINT #N%, "set zlabel "+Quote$+ZaxisLabel$+Quote$
            PRINT #N%, "splot "+Quote$+PlotFileName$+Quote$+A$  '" notitle with linespoints" 'A$'" notitle with lines"
        CLOSE #N%

        SHELL(GnuPlotEXE$+" cmd3d.gp -")

    END SUB 'ContourPlot()

'---

    SUB ThreeDplot(PlotFileName$,PlotTitle$,Annotation$,xCoord$,yCoord$,zCoord$, _
                   XaxisLabel$,YaxisLabel$,ZaxisLabel$,zMin$,zMax$,GnuPlotEXE$,A$)

        LOCAL N%, ProcessID???

        N% = FREEFILE

        OPEN "cmd3d.gp" FOR OUTPUT AS #N%

            PRINT #N%, "set pm3d"
            PRINT #N%, "show pm3d"
            IF zMin$ <> "" AND zMax$ <> "" THEN  PRINT #N%, "set zrange ["+zMin$+":"+zMax$+"]"
            PRINT #N%, "set label "+Quote$+AnnOtation$+Quote$+" at graph "+xCoord$+","+yCoord$+","+zCoord$
            PRINT #N%, "show label"
            PRINT #N%, "set grid xtics ytics ztics"
            PRINT #N%, "show grid"
            PRINT #N%, "set title "+Quote$+PlotTitle$+Quote$
            PRINT #N%, "set xlabel "+Quote$+XaxisLabel$+Quote$
            PRINT #N%, "set ylabel "+Quote$+YaxisLabel$+Quote$
            PRINT #N%, "set zlabel "+Quote$+ZaxisLabel$+Quote$
            PRINT #N%, "splot "+Quote$+PlotFileName$+Quote$+A$+" notitle"' with lines"
        CLOSE #N%

        SHELL(GnuPlotEXE$+" cmd3d.gp -") : CALL Delay(0.5##)

    END SUB 'ThreeDplot()

'---

    SUB ThreeDplot2(PlotFileName$,PlotTitle$,Annotation$,xCoord$,yCoord$,zCoord$, _
                    XaxisLabel$,YaxisLabel$,ZaxisLabel$,zMin$,zMax$,GnuPlotEXE$,A$,xStart$,xStop$,yStart$,yStop$)

        LOCAL N%

        N% = FREEFILE

        OPEN "cmd3d.gp" FOR OUTPUT AS #N%

            PRINT #N%, "set pm3d"
            PRINT #N%, "show pm3d"
            PRINT #N%, "set hidden3d"
            PRINT #N%, "set view 45, 45, 1, 1"

            IF zMin$ <> "" AND zMax$ <> "" THEN  PRINT #N%, "set zrange ["+zMin$+":"+zMax$+"]"

            PRINT #N%, "set xrange [" + xStart$ + ":" + xStop$ + "]"
            PRINT #N%, "set yrange [" + yStart$ + ":" + yStop$ + "]"

            PRINT #N%, "set label "  + Quote$ + AnnOtation$ + Quote$+" at graph "+xCoord$+","+yCoord$+","+zCoord$
            PRINT #N%, "show label"
            PRINT #N%, "set grid xtics ytics ztics"
            PRINT #N%, "show grid"
            PRINT #N%, "set title " + Quote$+PlotTitle$    + Quote$
            PRINT #N%, "set xlabel " + Quote$+XaxisLabel$  + Quote$
            PRINT #N%, "set ylabel " + Quote$+YaxisLabel$  + Quote$
            PRINT #N%, "set zlabel " + Quote$+ZaxisLabel$  + Quote$
            PRINT #N%, "splot "      + Quote$+PlotFileName$ + Quote$ + A$ + " notitle with lines"
        CLOSE #N%

        SHELL(GnuPlotEXE$+" cmd3d.gp -")

    END SUB 'ThreeDplot2()

'---

    SUB TwoDplot2Curves(PlotFileName1$,PlotFileName2$,PlotTitle$,Annotation$,xCoord$,yCoord$,XaxisLabel$,YaxisLabel$, _
                        LogXaxis$,LogYaxis$,xMin$,xMax$,yMin$,yMax$,xTics$,yTics$,GnuPlotEXE$,LineSize)

        LOCAL N%, ProcessID???

        N% = FREEFILE

        OPEN "cmd2d.gp" FOR OUTPUT AS #N%
            'print #N%, "set output "+Quote$+"test.plt"+Quote$ 'tried this 3/11/06, didn't work...

            IF LogXaxis$ = "YES" AND LogYaxis$ = "NO"  THEN PRINT #N%, "set logscale x"
            IF LogXaxis$ = "NO" AND LogYaxis$ = "YES" THEN PRINT #N%, "set logscale y"
            IF LogXaxis$ = "YES" AND LogYaxis$ = "YES" THEN PRINT #N%, "set logscale xy"

            IF xMin$ <> "" AND xMax$ <> "" THEN  PRINT #N%, "set xrange ["+xMin$+":"+xMax$+"]"

            IF yMin$ <> "" AND yMax$ <> "" THEN  PRINT #N%, "set yrange ["+yMin$+":"+yMax$+"]"

            PRINT #N%, "set label "+Quote$+AnnOtation$+Quote$+" at graph "+xCoord$+","+yCoord$
            PRINT #N%, "set grid xtics"
            PRINT #N%, "set grid ytics"
            PRINT #N%, "set xtics "+xTics$
            PRINT #N%, "set ytics "+yTics$
            PRINT #N%, "set grid mxtics"
            PRINT #N%, "set grid mytics"
            PRINT #N%, "set title "+Quote$+PlotTitle$+Quote$
            PRINT #N%, "set xlabel "+Quote$+XaxisLabel$+Quote$
            PRINT #N%, "set ylabel "+Quote$+YaxisLabel$+Quote$

            PRINT #N%, "plot "+Quote$+PlotFileName1$+Quote$+" with lines linewidth "+REMOVE$(STR$(LineSize),ANY" ")+","+ _
                              Quote$+PlotFileName2$+Quote$+" with points pointsize 0.05"+REMOVE$(STR$(LineSize),ANY" ")

        CLOSE #N%

        ProcessID??? = SHELL(GnuPlotEXE$+" cmd2d.gp -") : CALL Delay(0.5##)

    END SUB 'TwoDplot2Curves()
```

```
'----

    SUB Probe2Dplots(ProbePlotsFileList$,PlotTitle$,Annotation$,xCoord$,yCoord$,XaxisLabel$,YaxisLabel$, _
                    LogXaxis$,LogYaxis$,xMin$,xMax$,yMin$,yMax$,xTics$,yTics$,GnuPlotEXE$)

        LOCAL N%, ProcessID???

        N% = FREEFILE

        OPEN "cmd2d.gp" FOR OUTPUT AS #N%

            IF LogXaxis$ = "YES" AND LogYaxis$ = "NO"  THEN PRINT #N%, "set logscale x"
            IF LogXaxis$ = "NO"  AND LogYaxis$ = "YES" THEN PRINT #N%, "set logscale y"
            IF LogXaxis$ = "YES" AND LogYaxis$ = "YES" THEN PRINT #N%, "set logscale xy"

            IF xMin$ <> "" AND xMax$ <> "" THEN  PRINT #N%, "set xrange ["+xMin$+":"+xMax$+"]"

            IF yMin$ <> "" AND yMax$ <> "" THEN  PRINT #N%, "set yrange ["+yMin$+":"+yMax$+"]"

            PRINT #N%, "set label "+Quote$+Annotation$+Quote$+" at graph "+xCoord$+","+yCoord$
            PRINT #N%, "set grid xtics"
            PRINT #N%, "set grid ytics"
            PRINT #N%, "set xtics "+xTics$
            PRINT #N%, "set ytics "+yTics$
            PRINT #N%, "set grid mxtics"
            PRINT #N%, "set grid mytics"
            PRINT #N%, "set title "+Quote$+PlotTitle$+Quote$
            PRINT #N%, "set xlabel "+Quote$+XaxisLabel$+Quote$
            PRINT #N%, "set ylabel "+Quote$+YaxisLabel$+Quote$

            PRINT #N%, ProbePlotsFileList$

        CLOSE #N%

        ProcessID??? = SHELL(GnuPlotEXE$+" cmd2d.gp -") : CALL Delay(0.5##)

    END SUB 'Probe2Dplots()

'----

SUB Show2Dprobes(R(),Np%,Nt&,j&,XiMin(),XiMax,Frep,BestFitness,BestProbeNumber%,BestTimeStep&,FunctionName$,RepositionFactor$,Gamma)

    LOCAL N%, p%

    LOCAL A$, PlotFileName$, PlotTitle$, Symbols$

    LOCAL xMin$, xMax$, yMin$, yMax$

    LOCAL s1, s2, s3, s4 AS EXT

    PlotFileName$ = "REMOVE$(STR$(j&),ANY" ")+")"

    IF j& > 0 THEN 'PLOT PROBES AT THIS TIME STEP

        PlotTitle$ = "\nLOCATIONS OF "+REMOVE$(STR$(Np%),ANY" ") + " PROBES AT TIME STEP" + STR$(j&) + " / " + REMOVE$(STR$(Nt&),ANY" ") + "\n" + _
                    "Fitness = "+REMOVE$(STR$(ROUND(BestFitness,3)),ANY" ") + ", Probe #" + REMOVE$(STR$(BestProbeNumber%),ANY" ") + " at Step #" +
REMOVE$(STR$(BestTimeStep&),ANY" ") + _
                    " [Frep = "+REMOVE$(STR$(Frep),4),ANY" ") + " " + RepositionFactor$ + "]\n"

    ELSE 'PLOT INITIAL PROBE DISTRIBUTION

        PlotTitle$ = "\nLOCATIONS OF "+REMOVE$(STR$(Np%),ANY" ") + " INITIAL PROBES FOR " + FunctionName$ + " FUNCTION\n[gamma = "+STR$(ROUND(Gamma,3))+"]\n"

    END IF

    N% = FREEFILE : OPEN PlotFileName$ FOR OUTPUT AS #N%

        FOR p% = 1 TO Np% : PRINT #N%, USING$("######.#####    ######.#####",R(p%,1,j&),R(p%,2,j&)) : NEXT p%

    CLOSE #N%

    s1 = 1.1## : s2 = 1.1## : s3 = 1.1## : s4 = 1.1## 'expand plots axes by 10%

    IF XiMin(1) > 0## THEN s1 = 0.9##
    IF XiMax(1) < 0## THEN s2 = 0.9##
    IF XiMin(2) > 0## THEN s3 = 0.9##
    IF XiMax(2) < 0## THEN s4 = 0.9##

    xMin$ = REMOVE$(STR$(s1*XiMin(1),2),ANY" ")
    xMax$ = REMOVE$(STR$(s2*XiMax(1),2),ANY" ")
    yMin$ = REMOVE$(STR$(s3*XiMin(2),2),ANY" ")
    yMax$ = REMOVE$(STR$(s4*XiMax(2),2),ANY" ")

    CALL TwoDplot(PlotFileName$,PlotTitle$,"0.6","0.7","x1\n\n","\nx2","NO","NO",xMin$,xMax$,yMin$,yMax$,"5","5","wgnuplot.exe"," pointsize 1 linewidth 2","")

    KILL PlotFileName$ 'erase plot data file after probes have been displayed

END SUB 'ShowProbes()

'----

    SUB TwoDplot(PlotFileName$,PlotTitle$,xCoord$,yCoord$,XaxisLabel$,YaxisLabel$, _
                LogXaxis$,LogYaxis$,xMin$,xMax$,yMin$,yMax$,xTics$,yTics$,GnuPlotEXE$,LineType$,Annotation$)

        LOCAL N%, ProcessID???

        N% = FREEFILE

        OPEN "cmd2d.gp" FOR OUTPUT AS #N%

            IF LogXaxis$ = "YES" AND LogYaxis$ = "NO"  THEN PRINT #N%, "set logscale x"
            IF LogXaxis$ = "NO"  AND LogYaxis$ = "YES" THEN PRINT #N%, "set logscale y"
            IF LogXaxis$ = "YES" AND LogYaxis$ = "YES" THEN PRINT #N%, "set logscale xy"

            IF xMin$ <> "" AND xMax$ <> "" THEN  PRINT #N%, "set xrange ["+xMin$+":"+xMax$+"]"
            IF yMin$ <> "" AND yMax$ <> "" THEN  PRINT #N%, "set yrange ["+yMin$+":"+yMax$+"]"

            PRINT #N%, "set label "    + Quote$ + Annotation$ + Quote$ + " at graph " + xCoord$ + "," + yCoord$
            PRINT #N%, "set grid xtics " + XTics$
            PRINT #N%, "set grid ytics " + yTics$
            PRINT #N%, "set grid mxtics"
            PRINT #N%, "set grid mytics"
            PRINT #N%, "show grid"
```

```
        PRINT #N%, "set title " + Quote$+PlotTitle$+Quote$
        PRINT #N%, "set xlabel " + Quote$+XaxisLabel$+Quote$
        PRINT #N%, "set ylabel " + Quote$+YaxisLabel$+Quote$

        PRINT #N%, "plot "+Quote$+PlotFileName$+Quote$+" notitle"+LineType$

        CLOSE #N%

        ProcessID??? = SHELL(GnuPlotEXE$+" cmd2d.gp -") : CALL Delay(0.5##)

    END SUB 'TwoDplot()

'-----

    SUB CreateGNUplotINIfile(PlotWindowULC_X%,PlotWindowULC_Y%,PlotWindowWidth%,PlotWindowHeight%)

    LOCAL N%, WinPath$, A$, B$, WindowsDirectory$

    WinPath$ = UCASE$(ENVIRON$("Path"))'DIR$("C:\WINDOWS",23)
        DO
            B$ = A$

            A$ = EXTRACT$(WinPath$,";")

            WinPath$ = REMOVE$(WinPath$,A$+";")

            IF RIGHT$(A$,7) = "WINDOWS" OR A$ = B$ THEN EXIT LOOP

            IF RIGHT$(A$,5) = "WINNT"   OR A$ = B$ THEN EXIT LOOP

        LOOP

    WindowsDirectory$ = A$

    N% = FREEFILE

'   ---------- WGNUPLOT.INPUT FILE -----------
    OPEN WindowsDirectory$+"\wgnuplot.ini" FOR OUTPUT AS #N%

        PRINT #N%,"[WGNUPLOT]"
        PRINT #N%,"TextOrigin=0 0"
        PRINT #N%,"TextSize=640 150"
        PRINT #N%,"TextFont=Terminal,9"
        PRINT #N%,"GraphOrigin="+REMOVE$(STR$(PlotWindowULC_X%),ANY" ")+" "+REMOVE$(STR$(PlotWindowULC_Y%),ANY" ")
        PRINT #N%,"GraphSize=" +REMOVE$(STR$(PlotWindowWidth%),ANY" ")+" "+REMOVE$(STR$(PlotWindowHeight%),ANY" ")
        PRINT #N%,"GraphFont=Arial,10"
        PRINT #N%,"GraphColor=1"
        PRINT #N%,"GraphToTop=1"
        PRINT #N%,"GraphBackground=255 255 255"
        PRINT #N%,"Border=0 0 0 0"
        PRINT #N%,"Axis=192 192 192 2 2"
        PRINT #N%,"Line1=0 0 255 0 0"
        PRINT #N%,"Line2=0 255 0 0 1"
        PRINT #N%,"Line3=255 0 0 0 2"
        PRINT #N%,"Line4=255 0 255 0 3"
        PRINT #N%,"Line5=0 0 128 0 4"

    CLOSE #N%

    END SUB 'CreateGNUplotINIfile()

'------

    SUB Delay(NumSecs)

        LOCAL StartTime, StopTime AS EXT

        StartTime = TIMER

        DO UNTIL (StopTime-StartTime) >= NumSecs

            StopTime = TIMER

        LOOP

    END SUB 'Delay()

'-----

SUB MathematicalConstants
    EulerConst   = 0.5772156649015328606065512##
    Pi           = 3.14159265358979323846264338##
    Pi2          = Pi/2##
    Pi4          = Pi/4##
    TwoPi        = 2##*Pi
    FourPi       = 4##*Pi
    e            = 2.71828182845904523536028719##
    Root2        = 1.41421356237309550488##
END SUB

'-----

SUB AlphabetAndDigits
    Alphabet$    = "ABCDEFGHIJKLMNOPQRSTUVWXYZabcdefghijklmnopqrstuvwxyz"
    Digits$      = "0123456789"
    RunID$       = DATE$ + ", " + TIME$
END SUB

'------

SUB SpecialSymbols
    Quote$            = CHR$(34) 'Quotation mark '
    SpecialCharacters$ = "~`!@#$%^&*()_+"
END SUB

'-----

SUB EMconstants
    Mu0  = 4E-7##*Pi      'hy/meter
    Eps0 = 8.854##*1E-12  'fd/meter
    c    = 2.998E8##      'velocity of light, 1##/SQR(Mu0*Eps0)  'meters/sec
    eta0 = SQR(Mu0/Eps0) 'impedance of free space, ohms
END SUB
```

```
'------

SUB ConversionFactors
    Rad2Deg      = 180##/Pi
    Deg2Rad      = 1##/Rad2Deg
    Feet2Meters  = 0.3048##
    Meters2Feet  = 1##/Feet2Meters
    Inches2Meters = 0.0254##
    Meters2Inches = 1##/Inches2Meters
    Miles2Meters = 1609.344##
    Meters2Miles = 1##/Miles2Meters
    NautMi2Meters = 1852##
    Meters2NautMi = 1##/NautMi2Meters
END SUB

'------

SUB ShowConstants 'puts up msgbox showing all constants

LOCAL A$

A$ = _
"Mathematical Constants:"+CHR$(13)+_
"Euler const="+STR$(EulerConst)+CHR$(13)+_
"Pi="+STR$(Pi)+CHR$(13)+_
"Pi/2="+STR$(Pi2)+CHR$(13)+_
"Pi/4="+STR$(Pi4)+CHR$(13)+_
"2Pi="+STR$(TwoPi)+CHR$(13)+_
"4Pi="+STR$(FourPi)+CHR$(13)+_
"e="+STR$(e)+CHR$(13)+_
"Sqr2="+STR$(Root2)+CHR$(13)+CHR$(13)+_
"Alphabet, Digits & Special Characters:"+CHR$(13)+_
"Alphabet="+Alphabet$+CHR$(13)+_
"Digits="+Digits$+CHR$(13)+_
"quote="+Quote$+CHR$(13)+_
"Spec chars="+SpecialCharacters$+CHR$(13)+CHR$(13)+_
"E&M Constants:"+CHR$(13)+_
"Mu0="+STR$(Mu0)+CHR$(13)+_
"Eps0="+STR$(Eps0)+CHR$(13)+_
"c="+STR$(c)+CHR$(13)+_
"Eta0="+STR$(eta0)+CHR$(13)+CHR$(13)+_
"Conversion Factors:"+CHR$(13)+_
"Rad2Deg="+STR$(Rad2Deg)+CHR$(13)+_
"Deg2Rad="+STR$(Deg2Rad)+CHR$(13)+_
"Ft2meters="+STR$(Feet2Meters)+CHR$(13)+_
"Inches2Meters="+STR$(Inches2Meters)+CHR$(13)+_
"Meters2Inches="+STR$(Inches2Inches)+CHR$(13)+_
"Miles2Meters="+STR$(Miles2Meters)+CHR$(13)+_
"Meters2Miles="+STR$(Meters2Miles)+CHR$(13)+_
"NautMi2Meters="+STR$(NautMi2Meters)+CHR$(13)+_
"Meters2NautMi="+STR$(Meters2NautMi)+CHR$(13)+CHR$(13)

MSGBOX(A$)

END SUB

'------

SUB DisplayRmatrix(Np%,Nd%,Nt&,R())

LOCAL p%, i%, j&, A$

    A$ = "Position Vector Matrix R()"+CHR$(13)

    FOR p% = 1 TO Np%

        FOR i% = 1 TO Nd%

            FOR j& = 0 TO Nt&

                A$ = A$ + "R("+STR$(p%)+", "+STR$(i%)+", "+STR$(j&)+ ") ="+STR$(R(p%,i%,j&)) + CHR$(13)

            NEXT j&

        NEXT i%

    NEXT p%

    MSGBOX(A$)

END SUB

'------

SUB DisplayRmatrixThisTimeStep(Np%,Nd%,j&,R(),Gamma)

LOCAL p%, i%, A$, B$

    A$ = "Position Vector Matrix R() at step "+STR$(j&)+", Gamma ="+STR$(Gamma)+":"+CHR$(13)+CHR$(13)

    FOR p% = 1 TO Np%

        A$ = A$ + "Probe#"+REMOVE$(STR$(p%),ANY" ")+": "

        B$ = ""

        FOR i% = 1 TO Nd%

            B$ = B$ + "  " + USING$("####.##",R(p%,i%,j&))

        NEXT i%

        A$ = A$ + B$ + CHR$(13)

    NEXT p%

    MSGBOX(A$)

END SUB

'------
```

```
SUB DisplayAmatrix(Np%,Nd%,Nt&,A())

LOCAL p%, i%, j&, A$

    A$ = "Acceleration Vector Matrix A()"+CHR$(13)

    FOR p% = 1 TO Np%

        FOR i% = 1 TO Nd%

            FOR j& = 0 TO Nt&

                A$ = A$ + "A("+STR$(p%)+", "+STR$(i%)+", "+STR$(j&)+" ) ="+STR$(A(p%,i%,j&)) + CHR$(13)

            NEXT j&

        NEXT i%

    NEXT p%

    MSGBOX(A$)

END SUB
'------

SUB DisplayAmatrixThisTimeStep(Np%,Nd%,j&,A())

LOCAL p%, i%, A$

    A$ = "Acceleration matrix A() at step "+STR$(j&)+":"+CHR$(13)

    FOR p% = 1 TO Np%

        FOR i% = 1 TO Nd%

            A$ = A$ + "A("+STR$(p%)+", "+STR$(i%)+", "+STR$(j&)+" ) ="+STR$(A(p%,i%,j&)) + CHR$(13)

        NEXT i%

    NEXT p%

    MSGBOX(A$)

END SUB
'------

SUB DisplayMmatrix(Np%,Nt&,M())

LOCAL p%, j&, A$

    A$ = "Fitness Matrix M()"+CHR$(13)

    FOR p% = 1 TO Np%

        FOR j& = 0 TO Nt&

            A$ = A$ + "M("+STR$(p%)+", "+STR$(j&)+" ) ="+STR$(M(p%,j&)) + CHR$(13)

        NEXT j&

    NEXT p%

    MSGBOX(A$)

END SUB
'------

SUB DisplayMmatrixThisTimeStep(Np%,j&,M())

LOCAL p%, A$

    A$ = "Fitness matrix M() at step "+STR$(j&)+":"+CHR$(13)

    FOR p% = 1 TO Np%

        A$ = A$ + "M("+STR$(p%)+", "+STR$(j&)+" ) ="+STR$(M(p%,j&)) + CHR$(13)

    NEXT p%

    MSGBOX(A$)

END SUB
'------

SUB DisplayXiMinMax(Nd%,XiMin(),XiMax())

LOCAL i%, A$

    A$ = ""

    FOR i% = 1 TO Nd%

        A$ = A$ + "XiMin("+STR$(i%)+" ) = "+STR$(XiMin(i%))+"   XiMax("+STR$(i%)+" ) = "+STR$(XiMax(i%)) + CHR$(13)

    NEXT i%

    MSGBOX(A$)

END SUB
'------

SUB DisplayRunParameters2(FunctionName$,Nd%,Np%,Nt&,G,DeltaT,Alpha,Beta,Frep,PlaceInitialProbes$,InitialAcceleration$,RepositionFactor$)

LOCAL A$

    A$ = "Function = "+ FunctionName$+CHR$(13)+_
         "Nd = "+STR$(Nd%)+CHR$(13)+_
         "Np = "+STR$(Np%)+CHR$(13)+_
         "Nt = "+STR$(Nt&)+CHR$(13)+_
```

```
        "G    = "+STR$(G)+CHR$(13)+_
        "DeltaT  = "+STR$(DeltaT)+CHR$(13)+_
        "Alpha  = "+STR$(Alpha)+CHR$(13)+_
        "Beta   = "+STR$(Beta)+CHR$(13)+_
        "Frep   = "+STR$(Frep)+CHR$(13)+_
        "Init Probes: "+PlaceInitialProbes$+CHR$(13)+_
        "Init Accel:  "+InitialAcceleration$+CHR$(13)+_
        "Retrive Method: "+RepositionFactor$+CHR$(13)

    MSGBOX(A$)

END SUB

'------

SUB
Tabulate1DprobeCoordinates(Max1DprobesPlotted%,Nd%,Np%,LastStep&,G,DeltaT,Alpha,Beta,Frep,R(),M(),PlaceInitialProbes$,InitialAcceleration$,RepositionFactor$,FunctionName$,Gamma)

LOCAL N%, ProbeNum%, FileHeader$, A$, B$, C$, D$, E$, F$, H$, StepNum%, FieldNumber% 'kludgy, yes, but it accomplishes its purpose...

        CALL GetPlotAnnotation(FileHeader$,Nd%,Np%,LastStep&,G,DeltaT,Alpha,Beta,Frep,M(),PlaceInitialProbes$,InitialAcceleration$,RepositionFactor$,FunctionName$,Gamma)

        REPLACE "\n" WITH " , " IN FileHeader$

        FileHeader$ = LEFT$(FileHeader$,LEN(FileHeader$)-2)

        FileHeader$ = "PROBE COORDINATES" + CHR$(13) +_
                "-----------------" + CHR$(13) + FileHeader$

        N% = FREEFILE : OPEN "ProbeCoordinates.DAT" FOR OUTPUT AS #N%

            A$ = "   Step #    " : B$ = "   ------    " : C$ = ""

            FOR ProbeNum% = 1 TO Np% 'create out data file header

                SELECT CASE ProbeNum%
                    CASE   1 TO   9 : E$ = ""   : F$ = "           " : H$ = "        "
                    CASE  10 TO  99 : E$ = "-"  : F$ = "          " : H$ = "        "
                    CASE 100 TO 999 : E$ = "--" : F$ = "         " : H$ = "        "
                END SELECT

                A$ = A$ + "P" + NoSpaces$(ProbeNum%+0,0) + F$ 'note: adding zero to ProbeNum% necessary to convert to floating point...

                B$ = B$ + E$ + "--" + H$

                C$ = C$ + "######.###    "

'               C$ = C$ + "##.#######"

            NEXT ProbeNum%

            PRINT #N%, FileHeader$ + CHR$(13) : PRINT #N%, A$ : PRINT #N%, B$

            FOR StepNum% = 0 TO LastStep&

                D$ = USING$("######   ",StepNum&)

                FOR ProbeNum% = 1 TO Np% : D$ = D$ + USING$(C$,R(ProbeNum%,1,StepNum&)) : NEXT ProbeNum%

                PRINT #N%, D$

            NEXT StepNum&

        CLOSE #N%

END SUB 'Tabulate1DprobeCoordinates()

'------

SUB PlotIDprobePositions(Max1DprobesPlotted%,Nd%,Np%,LastStep&,G,DeltaT,Alpha,Beta,Frep,R(),M(),PlaceInitialProbes$,InitialAcceleration$,RepositionFactor$,FunctionName$,Gamma)
    'plots on-screen 1D function probe positions vs time step if Np =< 10

LOCAL ProcessID??, N%, n1%, n2%, n3%, n4%, n5%, n6%, n7%, n8%, n9%, n10%, n11%, n12%, n13%, n14%, n15%, ProbeNum%, StepNum&, A$

LOCAL PlotAnnotation$

    IF Np% > Max1DprobesPlotted% THEN EXIT SUB

    CALL CLEANUP 'delete old "Px" plot files, if any

    ProbeNum% = 0

    DO 'create output data files, probe-by-probe
        INCR ProbeNum% : n1% = FREEFILE : OPEN "P"+REMOVE$(STR$(ProbeNum%),ANY" ") FOR OUTPUT AS #n1% : IF ProbeNum% = Np% THEN EXIT LOOP
        INCR ProbeNum% : n2% = FREEFILE : OPEN "P"+REMOVE$(STR$(ProbeNum%),ANY" ") FOR OUTPUT AS #n2% : IF ProbeNum% = Np% THEN EXIT LOOP
        INCR ProbeNum% : n3% = FREEFILE : OPEN "P"+REMOVE$(STR$(ProbeNum%),ANY" ") FOR OUTPUT AS #n3% : IF ProbeNum% = Np% THEN EXIT LOOP
        INCR ProbeNum% : n4% = FREEFILE : OPEN "P"+REMOVE$(STR$(ProbeNum%),ANY" ") FOR OUTPUT AS #n4% : IF ProbeNum% = Np% THEN EXIT LOOP
        INCR ProbeNum% : n5% = FREEFILE : OPEN "P"+REMOVE$(STR$(ProbeNum%),ANY" ") FOR OUTPUT AS #n5% : IF ProbeNum% = Np% THEN EXIT LOOP
        INCR ProbeNum% : n6% = FREEFILE : OPEN "P"+REMOVE$(STR$(ProbeNum%),ANY" ") FOR OUTPUT AS #n6% : IF ProbeNum% = Np% THEN EXIT LOOP
        INCR ProbeNum% : n7% = FREEFILE : OPEN "P"+REMOVE$(STR$(ProbeNum%),ANY" ") FOR OUTPUT AS #n7% : IF ProbeNum% = Np% THEN EXIT LOOP
        INCR ProbeNum% : n8% = FREEFILE : OPEN "P"+REMOVE$(STR$(ProbeNum%),ANY" ") FOR OUTPUT AS #n8% : IF ProbeNum% = Np% THEN EXIT LOOP
        INCR ProbeNum% : n9% = FREEFILE : OPEN "P"+REMOVE$(STR$(ProbeNum%),ANY" ") FOR OUTPUT AS #n9% : IF ProbeNum% = Np% THEN EXIT LOOP
        INCR ProbeNum% : n10% = FREEFILE : OPEN "P"+REMOVE$(STR$(ProbeNum%),ANY" ") FOR OUTPUT AS #n10% : IF ProbeNum% = Np% THEN EXIT LOOP
        INCR ProbeNum% : n11% = FREEFILE : OPEN "P"+REMOVE$(STR$(ProbeNum%),ANY" ") FOR OUTPUT AS #n11% : IF ProbeNum% = Np% THEN EXIT LOOP
        INCR ProbeNum% : n12% = FREEFILE : OPEN "P"+REMOVE$(STR$(ProbeNum%),ANY" ") FOR OUTPUT AS #n12% : IF ProbeNum% = Np% THEN EXIT LOOP
        INCR ProbeNum% : n13% = FREEFILE : OPEN "P"+REMOVE$(STR$(ProbeNum%),ANY" ") FOR OUTPUT AS #n13% : IF ProbeNum% = Np% THEN EXIT LOOP
        INCR ProbeNum% : n14% = FREEFILE : OPEN "P"+REMOVE$(STR$(ProbeNum%),ANY" ") FOR OUTPUT AS #n14  : IF ProbeNum% = Np% THEN EXIT LOOP
        INCR ProbeNum% : n15% = FREEFILE : OPEN "P"+REMOVE$(STR$(ProbeNum%),ANY" ") FOR OUTPUT AS #n15% : IF ProbeNum% = Np% THEN EXIT LOOP
    LOOP

    ProbeNum% = 0

    DO 'output probe positions as a function of time step
        INCR ProbeNum% : FOR StepNum% = 0 TO LastStep& : PRINT #n1%,  USING$("######  ######.#######",StepNum&,R(ProbeNum%,1,StepNum&)) : NEXT StepNum% : IF ProbeNum% = Np% THEN
EXIT LOOP
        INCR ProbeNum% : FOR StepNum% = 0 TO LastStep& : PRINT #n2%,  USING$("######  ######.#######",StepNum&,R(ProbeNum%,1,StepNum&)) : NEXT StepNum% : IF ProbeNum% = Np% THEN
EXIT LOOP
        INCR ProbeNum% : FOR StepNum% = 0 TO LastStep& : PRINT #n3%,  USING$("######  ######.#######",StepNum&,R(ProbeNum%,1,StepNum&)) : NEXT StepNum% : IF ProbeNum% = Np% THEN
EXIT LOOP
        INCR ProbeNum% : FOR StepNum% = 0 TO LastStep& : PRINT #n4%,  USING$("######  ######.#######",StepNum&,R(ProbeNum%,1,StepNum&)) : NEXT StepNum% : IF ProbeNum% = Np% THEN
EXIT LOOP
        INCR ProbeNum% : FOR StepNum% = 0 TO LastStep& : PRINT #n5%,  USING$("######  ######.#######",StepNum&,R(ProbeNum%,1,StepNum&)) : NEXT StepNum% : IF ProbeNum% = Np% THEN
EXIT LOOP
```

```
          INCR ProbeNum% : FOR StepNum& = 0 TO LastStep& : PRINT #n6%,  USING$("###### #####.#######",StepNum&,R(ProbeNum%,1,StepNum&)) : NEXT StepNum& : IF ProbeNum% = Np% THEN
EXIT LOOP
          INCR ProbeNum% : FOR StepNum& = 0 TO LastStep& : PRINT #n7%,  USING$("###### #####.#######",StepNum&,R(ProbeNum%,1,StepNum&)) : NEXT StepNum& : IF ProbeNum% = Np% THEN
EXIT LOOP
          INCR ProbeNum% : FOR StepNum& = 0 TO LastStep& : PRINT #n8%,  USING$("###### #####.#######",StepNum&,R(ProbeNum%,1,StepNum&)) : NEXT StepNum& : IF ProbeNum% = Np% THEN
EXIT LOOP
          INCR ProbeNum% : FOR StepNum& = 0 TO LastStep& : PRINT #n9%,  USING$("###### #####.#######",StepNum&,R(ProbeNum%,1,StepNum&)) : NEXT StepNum& : IF ProbeNum% = Np% THEN
EXIT LOOP
          INCR ProbeNum% : FOR StepNum& = 0 TO LastStep& : PRINT #n10%, USING$("###### #####.#######",StepNum&,R(ProbeNum%,1,StepNum&)) : NEXT StepNum& : IF ProbeNum% = Np% THEN
EXIT LOOP
          INCR ProbeNum% : FOR StepNum& = 0 TO LastStep& : PRINT #n11, USING$("###### #####.#######",StepNum&,R(ProbeNum%,1,StepNum&)) : NEXT StepNum& : IF ProbeNum% = Np% THEN
EXIT LOOP
          INCR ProbeNum% : FOR StepNum& = 0 TO LastStep& : PRINT #n12%, USING$("###### #####.#######",StepNum&,R(ProbeNum%,1,StepNum&)) : NEXT StepNum& : IF ProbeNum% = Np% THEN
EXIT LOOP
          INCR ProbeNum% : FOR StepNum& = 0 TO LastStep& : PRINT #n13%, USING$("###### #####.#######",StepNum&,R(ProbeNum%,1,StepNum&)) : NEXT StepNum& : IF ProbeNum% = Np% THEN
EXIT LOOP
          INCR ProbeNum% : FOR StepNum& = 0 TO LastStep& : PRINT #n14%, USING$("###### #####.#######",StepNum&,R(ProbeNum%,1,StepNum&)) : NEXT StepNum& : IF ProbeNum% = Np% THEN
EXIT LOOP
          INCR ProbeNum% : FOR StepNum& = 0 TO LastStep& : PRINT #n15%, USING$("###### #####.#######",StepNum&,R(ProbeNum%,1,StepNum&)) : NEXT StepNum& : IF ProbeNum% = Np% THEN
EXIT LOOP
      LOOP

      ProbeNum% = 0

      DO 'close output data files
          INCR ProbeNum% : CLOSE #n1%  : IF ProbeNum% = Np% THEN EXIT LOOP
          INCR ProbeNum% : CLOSE #n2%  : IF ProbeNum% = Np% THEN EXIT LOOP
          INCR ProbeNum% : CLOSE #n3%  : IF ProbeNum% = Np% THEN EXIT LOOP
          INCR ProbeNum% : CLOSE #n4%  : IF ProbeNum% = Np% THEN EXIT LOOP
          INCR ProbeNum% : CLOSE #n5%  : IF ProbeNum% = Np% THEN EXIT LOOP
          INCR ProbeNum% : CLOSE #n6%  : IF ProbeNum% = Np% THEN EXIT LOOP
          INCR ProbeNum% : CLOSE #n7%  : IF ProbeNum% = Np% THEN EXIT LOOP
          INCR ProbeNum% : CLOSE #n8%  : IF ProbeNum% = Np% THEN EXIT LOOP
          INCR ProbeNum% : CLOSE #n9%  : IF ProbeNum% = Np% THEN EXIT LOOP
          INCR ProbeNum% : CLOSE #n10% : IF ProbeNum% = Np% THEN EXIT LOOP
          INCR ProbeNum% : CLOSE #n11% : IF ProbeNum% = Np% THEN EXIT LOOP
          INCR ProbeNum% : CLOSE #n12% : IF ProbeNum% = Np% THEN EXIT LOOP
          INCR ProbeNum% : CLOSE #n13% : IF ProbeNum% = Np% THEN EXIT LOOP
          INCR ProbeNum% : CLOSE #n14% : IF ProbeNum% = Np% THEN EXIT LOOP
          INCR ProbeNum% : CLOSE #n15% : IF ProbeNum% = Np% THEN EXIT LOOP
      LOOP

      ProbeNum% = 0 : A$ = ""

      DO 'create file string for plot command file
          INCR ProbeNum% : A$ = A$ + Quote$ + "P"+REMOVE$(STR$(ProbeNum%),ANY" ") + Quote$ + " w 1 lw 2, " : IF ProbeNum% = Np% THEN EXIT LOOP
          INCR ProbeNum% : A$ = A$ + Quote$ + "P"+REMOVE$(STR$(ProbeNum%),ANY" ") + Quote$ + " w 1 lw 2, " : IF ProbeNum% = Np% THEN EXIT LOOP
          INCR ProbeNum% : A$ = A$ + Quote$ + "P"+REMOVE$(STR$(ProbeNum%),ANY" ") + Quote$ + " w 1 lw 2, " : IF ProbeNum% = Np% THEN EXIT LOOP
          INCR ProbeNum% : A$ = A$ + Quote$ + "P"+REMOVE$(STR$(ProbeNum%),ANY" ") + Quote$ + " w 1 lw 2, " : IF ProbeNum% = Np% THEN EXIT LOOP
          INCR ProbeNum% : A$ = A$ + Quote$ + "P"+REMOVE$(STR$(ProbeNum%),ANY" ") + Quote$ + " w 1 lw 2, " : IF ProbeNum% = Np% THEN EXIT LOOP
          INCR ProbeNum% : A$ = A$ + Quote$ + "P"+REMOVE$(STR$(ProbeNum%),ANY" ") + Quote$ + " w 1 lw 2, " : IF ProbeNum% = Np% THEN EXIT LOOP
          INCR ProbeNum% : A$ = A$ + Quote$ + "P"+REMOVE$(STR$(ProbeNum%),ANY" ") + Quote$ + " w 1 lw 2, " : IF ProbeNum% = Np% THEN EXIT LOOP
          INCR ProbeNum% : A$ = A$ + Quote$ + "P"+REMOVE$(STR$(ProbeNum%),ANY" ") + Quote$ + " w 1 lw 2, " : IF ProbeNum% = Np% THEN EXIT LOOP
          INCR ProbeNum% : A$ = A$ + Quote$ + "P"+REMOVE$(STR$(ProbeNum%),ANY" ") + Quote$ + " w 1 lw 2, " : IF ProbeNum% = Np% THEN EXIT LOOP
          INCR ProbeNum% : A$ = A$ + Quote$ + "P"+REMOVE$(STR$(ProbeNum%),ANY" ") + Quote$ + " w 1 lw 2, " : IF ProbeNum% = Np% THEN EXIT LOOP
          INCR ProbeNum% : A$ = A$ + Quote$ + "P"+REMOVE$(STR$(ProbeNum%),ANY" ") + Quote$ + " w 1 lw 2, " : IF ProbeNum% = Np% THEN EXIT LOOP
          INCR ProbeNum% : A$ = A$ + Quote$ + "P"+REMOVE$(STR$(ProbeNum%),ANY" ") + Quote$ + " w 1 lw 2, " : IF ProbeNum% = Np% THEN EXIT LOOP
          INCR ProbeNum% : A$ = A$ + Quote$ + "P"+REMOVE$(STR$(ProbeNum%),ANY" ") + Quote$ + " w 1 lw 2, " : IF ProbeNum% = Np% THEN EXIT LOOP
          INCR ProbeNum% : A$ = A$ + Quote$ + "P"+REMOVE$(STR$(ProbeNum%),ANY" ") + Quote$ + " w 1 lw 2, " : IF ProbeNum% = Np% THEN EXIT LOOP
          INCR ProbeNum% : A$ = A$ + Quote$ + "P"+REMOVE$(STR$(ProbeNum%),ANY" ") + Quote$ + " w 1 lw 2, " : IF ProbeNum% = Np% THEN EXIT LOOP
      LOOP

      A$ = LEFT$(A$,LEN(A$)-2)

      CALL GetPlotAnnotation(PlotAnnotation$,Nd%,Np%,LastStep&,G,DeltaT,Alpha,Beta,Frep,M(),PlaceInitialProbes$,InitialAcceleration$,RepositionFactor$,FunctionName$,Gamma)

      N% = FREEFILE

      OPEN "cmd2d.gp" FOR OUTPUT AS #N%

          PRINT #N%, "set label "    + Quote$ + PlotAnnotation$ + Quote$ + " at graph 0.5,0.95"
          PRINT #N%, "set grid xtics"
          PRINT #N%, "set grid ytics"
          PRINT #N%, "set title " + Quote$ + "Evolution of "   + FunctionName$ + " Probe Positions"+ "\n" + RunID$ + Quote$
          PRINT #N%, "set xlabel " + Quote$ + "Time Step"       + Quote$
          PRINT #N%, "set ylabel " + Quote$ + "Probe Coordinate" + Quote$
          PRINT #N%, "plot "       + A$

      CLOSE #N%

      CALL CreateGNUplotINIfile(0.2##*ScreenWidth&,0.2##*ScreenHeight&,0.6##*ScreenWidth&,0.6##*ScreenHeight&) 'USAGE: CALL
CreateGNUplotINIfile(PlotWindowULC_X%,PlotWindowULC_Y%,PlotWindowWidth%,PlotWindowHeight%)

      ProcessID?? = SHELL("wgnuplot.exe"+" cmd2d.gp -") : CALL Delay(5##) 'before SUB Cleanup is called

END SUB

'------

SUB
DisplayRunParameters(FunctionName$,Nd%,Np%,Nt&,G,DeltaT,Alpha,Beta,Frep,R(),A(),M(),PlaceInitialProbes$,InitialAcceleration$,RepositionFactor$,RunCFO$,ShrinkDS$,CheckForEarlyTermi
nation$)

LOCAL A$, B$, YN&

      ShrinkDS$ = "NO"              : YN& = MSGBOX("Adaptively Shrink DS?",%MB_YESNO,"ADAPTIVE DS?")              : IF YN& = %IDYES THEN ShrinkDS$ = "YES"

      CheckForEarlyTermination$ = "NO" : YN& = MSGBOX("Check for Early Termination?",%MB_YESNO,"EARLY TERMINATION?") : IF YN& = %IDYES THEN CheckForEarlyTermination$ = "YES"

      B$ = "" : IF PlaceInitialProbes$ = "UNIFORM ON-AXIS" AND Nd% > 1 THEN B$ = "  ["+REMOVE$(STR$(Np%/Nd%),ANY" ") + "/axis]"

      RunCFO$ = "NO"

A$ = "RUN CFO WITH THE" + CHR$(13) +_
          "FOLLOWING PARAMETERS?"                           + CHR$(13) + CHR$(13) +_
          "Function "       + FunctionName$                 + " (" + REMOVE$(STR$(Nd%),ANY" ") + "-D)" + CHR$(13) +_
          "# time steps = " + REMOVE$(STR$(Nt&),ANY" ")     + CHR$(13) + _
          "Grav Const G = " + REMOVE$(STR$(G,2),ANY" ")     + CHR$(13) + _
          "Delta-T = "      + REMOVE$(STR$(DeltaT,3),ANY" ") + CHR$(13) + _
          "Exp Alpha = "    + REMOVE$(STR$(Alpha,3),ANY" ") + CHR$(13) + _
          "Exp Beta = "     + REMOVE$(STR$(Beta,3),ANY" ")  + CHR$(13) + _
          "Frep = "         + REMOVE$(STR$(Frep,4),ANY" ")  + " ("+RepositionFactor$ + ")" + CHR$(13) + _
          "Initial Probes: " + PlaceInitialProbes$          + CHR$(13) + _
          "Initial Accel: " + InitialAcceleration$          + CHR$(13) + _
```

```
         "Check for Early Termination? " + CheckForEarlyTermination$ + CHR$(13) + _
         "Shrink Decision Space? "        + ShrinkDS$ + CHR$(13) +CHR$(13)

'    lResult& = MSGBOX(txt$ [, [style&], title$])

     YN& = MSGBOX(A$,%MB_YESNO,"CONFIRM RUN")

     IF YN& = %IDYES THEN RunCFO$ = "YES"

END SUB

'------

SUB StatusWindow(FunctionName$,StatusWindowHandle???)

     GRAPHIC WINDOW "Run Progress, "+FunctionName$,0.08##*ScreenWidth&,0.08##*ScreenHeight&,0.25##*ScreenWidth&,0.17##*ScreenHeight& TO StatusWindowHandle???

     GRAPHIC ATTACH StatusWindowHandle???,0,REDRAW

     GRAPHIC FONT "Lucida Console",8,0 'Courier New",8,0 'Fixed width fonts

     GRAPHIC SET PIXEL (35,15) : GRAPHIC PRINT "  Initializing...     " : GRAPHIC REDRAW

END SUB

'------

SUB GetTestFunctionNumber(FunctionName$)

   LOCAL hDlg AS DWORD

   LOCAL N%, M%

   LOCAL FrameWidth&, FrameHeight&, BoxWidth&, BoxHeight&

   BoxWidth& = 276 : BoxHeight& = 300 : FrameHeight& = 80 : FrameHeight& = BoxHeight&-5

   DIALOG NEW 0, "CENTRAL FORCE OPTIMIZATION TEST FUNCTIONS",,, BoxWidth&, BoxHeight&, %WS_CAPTION OR %WS_SYSMENU, 0 TO hDlg

'----------------------------------------------------------------

   CONTROL ADD FRAME,  hDlg, %IDC_FRAME1, "Test Functions",     5, 2, FrameWidth&, FrameHeight&
   CONTROL ADD FRAME,  hDlg, %IDC_FRAME2, "GSO Test Functions", 95, 2, FrameWidth&, 255

   CONTROL ADD OPTION, hDlg, %IDC_Function_Number1, "Parrott F4",10,  14, 60, 10, %WS_GROUP OR %WS_TABSTOP
   CONTROL ADD OPTION, hDlg, %IDC_Function_Number2, "SGO"       ,  10,  24, 60, 10
   CONTROL ADD OPTION, hDlg, %IDC_Function_Number3, "Goldstein-Price", 10,  34, 60, 10
   CONTROL ADD OPTION, hDlg, %IDC_Function_Number4, "Step"      ,  10,  44, 60, 10
   CONTROL ADD OPTION, hDlg, %IDC_Function_Number5, "Schwefel 2.26", 10,  54, 60, 10
   CONTROL ADD OPTION, hDlg, %IDC_Function_Number6, "Colville",  10,  64, 60, 10
   CONTROL ADD OPTION, hDlg, %IDC_Function_Number7, "Griewank", 10,  74, 60, 10

   CONTROL ADD OPTION, hDlg, %IDC_Function_Number31, "PBM #1",    10,  84, 60, 10
   CONTROL ADD OPTION, hDlg, %IDC_Function_Number32, "PBM #2",    10,  94, 60, 10
   CONTROL ADD OPTION, hDlg, %IDC_Function_Number33, "PBM #3",    10, 104, 60, 10
   CONTROL ADD OPTION, hDlg, %IDC_Function_Number34, "PBM #4",    10, 114, 60, 10
   CONTROL ADD OPTION, hDlg, %IDC_Function_Number35, "PBM #5",    10, 124, 60, 10
   CONTROL ADD OPTION, hDlg, %IDC_Function_Number36, "Himmelblau",10, 134, 60, 10
   CONTROL ADD OPTION, hDlg, %IDC_Function_Number37, "Rosenbrock",10, 144, 60, 10
   CONTROL ADD OPTION, hDlg, %IDC_Function_Number38, "Sphere",    10, 154, 60, 10
   CONTROL ADD OPTION, hDlg, %IDC_Function_Number39, "Reserved",  10, 164, 60, 10
   CONTROL ADD OPTION, hDlg, %IDC_Function_Number40, "Reserved",  10, 174, 60, 10
   CONTROL ADD OPTION, hDlg, %IDC_Function_Number41, "Reserved",  10, 184, 60, 10
   CONTROL ADD OPTION, hDlg, %IDC_Function_Number42, "Reserved",  10, 194, 60, 10
   CONTROL ADD OPTION, hDlg, %IDC_Function_Number43, "Reserved",  10, 204, 60, 10
   CONTROL ADD OPTION, hDlg, %IDC_Function_Number44, "Reserved",  10, 214, 60, 10
   CONTROL ADD OPTION, hDlg, %IDC_Function_Number45, "Reserved",  10, 224, 60, 10
   CONTROL ADD OPTION, hDlg, %IDC_Function_Number46, "Reserved",  10, 234, 60, 10
   CONTROL ADD OPTION, hDlg, %IDC_Function_Number47, "Reserved",  10, 244, 60, 10
   CONTROL ADD OPTION, hDlg, %IDC_Function_Number48, "Reserved",  10, 254, 60, 10
   CONTROL ADD OPTION, hDlg, %IDC_Function_Number49, "Reserved",  10, 264, 60, 10
   CONTROL ADD OPTION, hDlg, %IDC_Function_Number50, "Reserved",  10, 274, 60, 10

' --------------------- Test Functions from GSO Paper --------------------
   CONTROL ADD OPTION, hDlg, %IDC_Function_Number8,  "f1", 120,  14, 40, 10
   CONTROL ADD OPTION, hDlg, %IDC_Function_Number9,  "f2", 120,  24, 40, 10
   CONTROL ADD OPTION, hDlg, %IDC_Function_Number10, "f3", 120,  34, 40, 10
   CONTROL ADD OPTION, hDlg, %IDC_Function_Number11, "f4", 120,  44, 40, 10
   CONTROL ADD OPTION, hDlg, %IDC_Function_Number12, "f5", 120,  54, 40, 10
   CONTROL ADD OPTION, hDlg, %IDC_Function_Number13, "f6", 120,  64, 40, 10
   CONTROL ADD OPTION, hDlg, %IDC_Function_Number14, "f7", 120,  74, 40, 10
   CONTROL ADD OPTION, hDlg, %IDC_Function_Number15, "f8", 120,  84, 40, 10
   CONTROL ADD OPTION, hDlg, %IDC_Function_Number16, "f9", 120,  94, 40, 10
   CONTROL ADD OPTION, hDlg, %IDC_Function_Number17, "f10", 120, 104, 40, 10
   CONTROL ADD OPTION, hDlg, %IDC_Function_Number18, "f11", 120, 114, 40, 10
   CONTROL ADD OPTION, hDlg, %IDC_Function_Number19, "f12", 120, 124, 40, 10
   CONTROL ADD OPTION, hDlg, %IDC_Function_Number20, "f13", 120, 134, 40, 10
   CONTROL ADD OPTION, hDlg, %IDC_Function_Number21, "f14", 120, 144, 40, 10
   CONTROL ADD OPTION, hDlg, %IDC_Function_Number22, "f15", 120, 154, 40, 10
   CONTROL ADD OPTION, hDlg, %IDC_Function_Number23, "f16", 120, 164, 40, 10
   CONTROL ADD OPTION, hDlg, %IDC_Function_Number24, "f17", 120, 174, 40, 10
   CONTROL ADD OPTION, hDlg, %IDC_Function_Number25, "f18", 120, 184, 40, 10
   CONTROL ADD OPTION, hDlg, %IDC_Function_Number26, "f19", 120, 194, 40, 10
   CONTROL ADD OPTION, hDlg, %IDC_Function_Number27, "f20", 120, 204, 40, 10
   CONTROL ADD OPTION, hDlg, %IDC_Function_Number28, "f21", 120, 214, 40, 10
   CONTROL ADD OPTION, hDlg, %IDC_Function_Number29, "f22", 120, 224, 40, 10
   CONTROL ADD OPTION, hDlg, %IDC_Function_Number30, "f23", 120, 234, 40, 10

   CONTROL SET OPTION  hDlg, %IDC_Function_Number1, %IDC_Function_Number1, %IDC_Function_Number3 'default to Parrott F4

'----------------------------------------------------------------

   CONTROL ADD BUTTON, hDlg, %IDOK, "&OK", 200, 0.45##*BoxHeight&, 50, 14

'----------------------------------------------------------------

   DIALOG SHOW MODAL hDlg CALL DlgProc

   CALL Delay(0.5##)

   IF FunctionNumber% < 1 OR FunctionNumber% > 38 THEN

      FunctionNumber% = 1 : MSGBOX("Error in function number...")
```

```
    END IF

    SELECT CASE FunctionNumber%

        CASE 1 : FunctionName$ = "ParrottF4"
        CASE 2 : FunctionName$ = "SGO"
        CASE 3 : FunctionName$ = "GP"
        CASE 4 : FunctionName$ = "STEP"
        CASE 5 : FunctionName$ = "SCHWEFEL_226"
        CASE 6 : FunctionName$ = "COLVILLE"
        CASE 7 : FunctionName$ = "GRIEWANK"
        CASE 8 : FunctionName$ = "F1"
        CASE 9 : FunctionName$ = "F2"
        CASE 10: FunctionName$ = "F3"
        CASE 11: FunctionName$ = "F4"
        CASE 12: FunctionName$ = "F5"
        CASE 13: FunctionName$ = "F6"
        CASE 14: FunctionName$ = "F7"
        CASE 15: FunctionName$ = "F8"
        CASE 16: FunctionName$ = "F9"
        CASE 17: FunctionName$ = "F10"
        CASE 18: FunctionName$ = "F11"
        CASE 19: FunctionName$ = "F12"
        CASE 20: FunctionName$ = "F13"
        CASE 21: FunctionName$ = "F14"
        CASE 22: FunctionName$ = "F15"
        CASE 23: FunctionName$ = "F16"
        CASE 24: FunctionName$ = "F17"
        CASE 25: FunctionName$ = "F18"
        CASE 26: FunctionName$ = "F19"
        CASE 27: FunctionName$ = "F20"
        CASE 28: FunctionName$ = "F21"
        CASE 29: FunctionName$ = "F22"
        CASE 30: FunctionName$ = "F23"
        CASE 31: FunctionName$ = "PBM_1"
        CASE 32: FunctionName$ = "PBM_2"
        CASE 33: FunctionName$ = "PBM_3"
        CASE 34: FunctionName$ = "PBM_4"
        CASE 35: FunctionName$ = "PBM_5"
        CASE 36: FunctionName$ = "HIMMELBLAU"
        CASE 37: FunctionName$ = "ROSENBROCK"
        CASE 38: FunctionName$ = "SPHERE"

    END SELECT

END SUB

'-----------

CALLBACK FUNCTION DlgProc() AS LONG

    '-------------------------------------------------------------------
    ' Callback procedure for the main dialog
    '-------------------------------------------------------------------
    LOCAL c, lRes AS LONG, sText AS STRING

    SELECT CASE AS LONG CBMSG

    CASE %WM_INITDIALOG' %WM_INITDIALOG is sent right before the dialog is shown.

    CASE %WM_COMMAND            ' <- a control is calling

        SELECT CASE AS LONG CBCTL  ' <- look at control's id

        CASE %IDOK                 ' <- OK button or Enter key was pressed

            IF CBCTLMSG = %BN_CLICKED THEN
                '-----------------------------------------
                ' Loop through the Function_Number controls
                ' to see which one is selected
                '-----------------------------------------
                FOR c = %IDC_Function_Number1 TO %IDC_Function_Number50

                    CONTROL GET CHECK CBHNDL, c TO lRes

                    IF lRes THEN EXIT FOR

                NEXT 'c holds the id for selected test function.

                FunctionNumber% = c-120

                DIALOG END CBHNDL

            END IF

        END SELECT

    END SELECT

END FUNCTION

'---------------------------- PBM ANTENNA BENCHMARK FUNCTIONS ----------------------------

'Reference for benchmarks PBM_1 through PBM_5:

'Pantoja, M F., Bretones, A. R., Martin, R. G., "Benchmark Antenna Problems for Evolutionary
'Optimization Algorithms," IEEE Trans. Antennas & Propagation, vol. 55, no. 4, April 2007,
'pp. 1111-1121

FUNCTION PBM_1(R(),Nd%,p%,j&) 'PBM Benchmark #1: Max D for Variable-Length CF Dipole

    LOCAL Z, LengthWaves, ThetaRadians AS EXT

    LOCAL N%, Nsegs%, FeedSegNum%

    LOCAL NumSegs$, FeedSeg$, HalfLength$, Radius$, ThetaDeg$, Lyne$, GainDB$

    LengthWaves = R(p%,1,j&)

    ThetaRadians = R(p%,2,j&)

    ThetaDeg$ = REMOVE$(STR$(ROUND(ThetaRadians*Rad2Deg,2)),ANY" ")

    IF TALLY(ThetaDeg$,".") = 0 THEN ThetaDeg$ = ThetaDeg$+"."
```

```basic
    Nsegs% = 2*(INT(100*LengthWaves)\2)+1 '100 segs per wavelength, must be an odd #, VOLTAGE SOURCE

    FeedSegNum% = Nsegs%\2 + 1 'center segment number, VOLTAGE SOURCE

    NumSegs$    = REMOVE$(STR$(Nsegs%),ANY" ")

    FeedSeg$    = REMOVE$(STR$(FeedSegNum%),ANY" ")

    HalfLength$ = REMOVE$(STR$(ROUND(LengthWaves/2##,6)),ANY" ")

    IF TALLY(HalfLength$,".") = 0 THEN HalfLength$ = HalfLength$+"."

    Radius$     = "0.00001" 'REMOVE$(STR$(ROUND(LengthWaves/1000##,6)),ANY" ")

    N% = FREEFILE

    OPEN "PBM1.NEC" FOR OUTPUT AS #N%

        PRINT #N%,"CM File: PBM1.NEC"
        PRINT #N%,"CM Run ID "+DATE$+" "+TIME$
        PRINT #N%,"CM Nd="+STR$(Nd%)+", p="STR$(p%)+", j="+STR$(j&)
        PRINT #N%,"CM R(p,1,j)="+STR$(R(p%,1,j&))+", R(p,2,j)="+STR$(R(p%,2,j&))
        PRINT #N%,"CE"
        PRINT #N%,"GW 1,"+NumSegs$+",0.,0.,-"+HalfLength$+",0.,0.,-"+HalfLength$+","+Radius$
        PRINT #N%,"GE"
        PRINT #N%,"EX 0,1,"+FeedSeg$+",0,1.,0." 'VOLTAGE SOURCE
        PRINT #N%,"FR 0,1,0,0,299.79564,0."
        PRINT #N%,"RP 0,1,1,1001,"+ThetaDeg$+",0.,0.,0.,1000." 'gain at 1000 wavelengths range
        PRINT #N%,"XQ"
        PRINT #N%,"EN"

    CLOSE #N%

'    - - ANGLES - -       - POWER GAINS -      - - - POLARIZATION - - -     - E(THETA) - - -     - E(PHI) - - -
'  THETA     PHI    VERT.   HOR.   TOTAL    AXIAL   TILT  SENSE   MAGNITUDE    PHASE     MAGNITUDE    PHASE
' DEGREES  DEGREES    DB     DB      DB     RATIO    DEG.               VOLTS/M   DEGREES    VOLTS/M   DEGREES
'  90.00     0.00    3.91 -999.99   3.91  0.00000    0.00 LINEAR   1.29504E-04    5.37   0.00000E+00   -5.24
'123456789x123456789x123456789x123456789x123456789x123456789x123456789x123456789x123456789x123456789x123456789x
'    10      20      30      40      50      60      70      80      90      100     110     120

    SHELL "n41_2k1.exe",0

    N% = FREEFILE

    OPEN "PBM1.OUT" FOR INPUT AS #N%

        WHILE NOT EOF(N%)

            LINE INPUT #N%, Lyne$

            IF INSTR(Lyne$,"DEGREES  DEGREES") > 0 THEN EXIT LOOP

        WEND 'position at next data line

        LINE INPUT #N%, Lyne$

    CLOSE #N%

    GainDB$ = REMOVE$(MID$(Lyne$,37,8),ANY" ")

    PBM_1 = 10^(VAL(GainDB$)/10##) 'Directivity

END FUNCTION 'PBM_1()

'----

FUNCTION PBM_2(R(),Nd%,p%,j&) 'PBM Benchmark #2: Max D for Variable-Separation Array of CF Dipoles

    LOCAL Z, DipoleSeparationWaves, ThetaRadians AS EXT

    LOCAL N%, i%

    LOCAL NumSegs$, FeedSeg$, Radius$, ThetaDeg$, Lyne$, GainDB$, Xcoord$, WireNum$

    DipoleSeparationWaves = R(p%,1,j&)

    ThetaRadians          = R(p%,2,j&)

    ThetaDeg$ = REMOVE$(STR$(ROUND(ThetaRadians*Rad2Deg,2)),ANY" ")

    IF TALLY(ThetaDeg$,".") = 0 THEN ThetaDeg$ = ThetaDeg$+"."

    NumSegs$ = "49"

    FeedSeg$ = "25"

    Radius$ = "0.00001"

    N% = FREEFILE

    OPEN "PBM2.NEC" FOR OUTPUT AS #N%

        PRINT #N%,"CM File: PBM2.NEC"
        PRINT #N%,"CM Run ID "+DATE$+" "+TIME$
        PRINT #N%,"CM Nd="+STR$(Nd%)+", p="STR$(p%)+", j="+STR$(j&)
        PRINT #N%,"CM R(p,1,j)="+STR$(R(p%,1,j&))+", R(p,2,j)="+STR$(R(p%,2,j&))
        PRINT #N%,"CE"

        FOR i% = -9 TO 9 STEP 2
            WireNum$ = REMOVE$(STR$((i%+11)\2),ANY" ")
            Xcoord$ = REMOVE$(STR$(i%*DipoleSeparationWaves/2##),ANY" ")
            PRINT #N%,"GW "+WireNum$+","+NumSegs$+","+Xcoord$+",0.,-0.25,"+Xcoord$+",0.,0.25,"+Radius$
        NEXT i%

        PRINT #N%,"GE"

        FOR i% = 1 TO 10
            PRINT #N%,"EX 0,"+REMOVE$(STR$(i%),ANY" ")+","+FeedSeg$+",0,1.,0." 'VOLTAGE SOURCE
        NEXT i%
        PRINT #N%,"FR 0,1,0,0,299.79564,0."
        PRINT #N%,"RP 0,1,1,1001,"+ThetaDeg$+",90.,0.,0.,1000." 'gain at 1000 wavelengths range
        PRINT #N%,"XQ"
        PRINT #N%,"EN"
```

```
    CLOSE #N%

'    - - ANGLES - -        - POWER GAINS -       - - - POLARIZATION - - -      - - - E(THETA) - - -       - - - E(PHI) - - -
'  THETA    PHI       VERT.   HOR.   TOTAL      AXIAL   TILT  SENSE   MAGNITUDE    PHASE     MAGNITUDE    PHASE
' DEGREES  DEGREES     DB      DB     DB         RATIO   DEG.          VOLTS/M    DEGREES      VOLTS/M    DEGREES
'   90.00    0.00     3.91  -999.99   3.91     0.00000   0.00  LINEAR  1.29504E-04    5.37   0.00000E+00    -5.24
'123456789x123456789x123456789x123456789x123456789x123456789x123456789x123456789x123456789x123456789x123456789x123456789x
'        10       20       30       40       50       60       70       80       90      100      110      120

    SHELL "n41_2k1.exe",0

    N% = FREEFILE

    OPEN "PBM2.OUT" FOR INPUT AS #N%

        WHILE NOT EOF(N%)

            LINE INPUT #N%, Lyne$

            IF INSTR(Lyne$,"DEGREES  DEGREES") > 0 THEN EXIT LOOP

        WEND 'position at next data line

        LINE INPUT #N%, Lyne$

    CLOSE #N%

    GainDB$ = REMOVE$(MID$(Lyne$,37,8),ANY" ")

    IF AddNoiseToPBM2$ = "YES" THEN

        Z = 10^(VAL(GainDB$)/10##) + GaussianDeviate(0##,0.4472##) 'Directivity with Gaussian noise (zero mean, 0.2 variance)

    ELSE

        Z = 10^(VAL(GainDB$)/10##) 'Directivity without noise

    END IF

    PBM_2 = Z

END FUNCTION 'PBM_2()

'----

FUNCTION PBM_3(R(),Nd%,p%,j&) 'PBM Benchmark #3: Max D for Circular Dipole Array

    LOCAL Beta, ThetaRadians, Alpha, ReV, ImV AS EXT

    LOCAL N%, i%

    LOCAL NumSegs$, FeedSeg$, Radius$, ThetaDeg$, Lyne$, GainDB$, Xcoord$, Ycoord$, WireNum$, ReEX$, ImEX$

    Beta         = R(p%,1,j&)

    ThetaRadians = R(p%,2,j&)

    ThetaDeg$ = REMOVE$(STR$(ROUND(ThetaRadians*Rad2Deg,2)),ANY" ")

    IF TALLY(ThetaDeg$,".") = 0 THEN ThetaDeg$ = ThetaDeg$+"."

    NumSegs$ = "49"

    FeedSeg$ = "25"

    Radius$ = "0.00001"

    N% = FREEFILE

    OPEN "PBM3.NEC" FOR OUTPUT AS #N%

        PRINT #N%,"CM File: PBM3.NEC"
        PRINT #N%,"CM Run ID "+DATE$+" "+TIME$
        PRINT #N%,"CM Nd="+STR$(Nd%)+", p="+STR$(p%)+", j="+STR$(j&)
        PRINT #N%,"CM R(p,1,j)="+STR$(R(p%,1,j&))+", R(p,2,j)="+STR$(R(p%,2,j&))
        PRINT #N%,"CE"

        FOR i% = 1 TO 8
            WireNum$ = REMOVE$(STR$(i%),ANY" ")

            SELECT CASE i%
                CASE 1 : Xcoord$ = "1"        : Ycoord$ = "0"
                CASE 2 : Xcoord$ = "0.70711"  : Ycoord$ = "0.70711"
                CASE 3 : Xcoord$ = "0"        : Ycoord$ = "1"
                CASE 4 : Xcoord$ = "-0.70711" : Ycoord$ = "0.70711"
                CASE 5 : Xcoord$ = "-1"       : Ycoord$ = "0"
                CASE 6 : Xcoord$ = "-0.70711" : Ycoord$ = "-0.70711"
                CASE 7 : Xcoord$ = "0"        : Ycoord$ = "-1"
                CASE 8 : Xcoord$ = "0.70711"  : Ycoord$ = "-0.70711"
            END SELECT

            PRINT #N%,"GW "+WireNum$+","+NumSegs$+","+Xcoord$+","+Ycoord$+",-0.25,"+Xcoord$+","+Ycoord$+",0.25,"+Radius$
        NEXT i%

        PRINT #N%,"GE"

        FOR i% = 1 TO 8
            Alpha = -COS(TwoPi*Beta*(i%-1))

            ReV   = COS(Alpha)
            ImV   = SIN(Alpha)

            ReEX$ = REMOVE$(STR$(ROUND(ReV,6)),ANY" ")
            ImEX$ = REMOVE$(STR$(ROUND(ImV,6)),ANY" ")

            IF TALLY(ReEX$,".") = 0 THEN ReEX$ = ReEX$+"."
            IF TALLY(ImEX$,".") = 0 THEN ImEX$ = ImEX$+"."

            PRINT #N%,"EX 0,"+REMOVE$(STR$(i%),ANY" ")+","+FeedSeg$+",0,"+ReEX$+","+ImEX$ 'VOLTAGE SOURCE
        NEXT i%

        PRINT #N%,"FR 0,1,0,0,299.79564,0."
        PRINT #N%,"RP 0,1,1,1001,"+ThetaDeg$+",0.,0.,0.,1000." 'gain at 1000 wavelengths range
```

```
        PRINT #N%,"XQ"
        PRINT #N%,"EN"

    CLOSE #N%

'     - - ANGLES - -       - POWER GAINS -      - - - POLARIZATION - - -    - - - E(THETA) - - -  - - - E(PHI) - - -
'  THETA      PHI     VERT.   HOR.   TOTAL    AXIAL   TILT  SENSE   MAGNITUDE    PHASE    MAGNITUDE    PHASE
' DEGREES   DEGREES    DB      DB      DB      RATIO   DEG.         VOLTS/M    DEGREES    VOLTS/M    DEGREES
'  90.00     0.00     3.91  -999.99   3.91   0.00000   0.00 LINEAR  1.29504E-04   5.37   0.00000E+00   -5.24
'123456789x123456789x123456789x123456789x123456789x123456789x123456789x123456789x123456789x123456789x123456789x
'     10       20      30      40      50      60       70      80      90      100      110      120

    SHELL "n41_2k1.exe",0

    N% = FREEFILE

    OPEN "PBM3.OUT" FOR INPUT AS #N%

        WHILE NOT EOF(N%)

            LINE INPUT #N%, Lyne$

            IF INSTR(Lyne$,"DEGREES  DEGREES") > 0 THEN EXIT LOOP

        WEND 'position at next data line

        LINE INPUT #N%, Lyne$

    CLOSE #N%

    GainDB$ = REMOVE$(MID$(Lyne$,37,8),ANY" ")

    PBM_3 = 10^(VAL(GainDB$)/10##) 'Directivity

END FUNCTION 'PBM_3()

'----

FUNCTION PBM_4(R(),Nd%,p%,j&) 'PBM Benchmark #4: Max D for Vee Dipole

    LOCAL TotalLengthWaves, AlphaRadians, ArmLength, Xlength, Zlength, Lfeed AS EXT

    LOCAL N%, i%, Nsegs%, FeedZcoord$

    LOCAL NumSegs$, Lyne$, GainDB$, Xcoord$, Zcoord$

    TotalLengthWaves = 2##*R(p%,1,j&)

    AlphaRadians     = R(p%,2,j&)

    Lfeed            = 0.01##

    FeedZcoord$      = REMOVE$(STR$(Lfeed),ANY" ")

    ArmLength = (TotalLengthWaves-2##*Lfeed)/2##

    Xlength = ROUND(ArmLength*COS(AlphaRadians),6)

    Xcoord$ = REMOVE$(STR$(Xlength),ANY" ") : IF TALLY(Xcoord$,".") = 0 THEN Xcoord$ = Xcoord$+"."

    Zlength = ROUND(ArmLength*SIN(AlphaRadians),6)

    Zcoord$ = REMOVE$(STR$(Zlength+Lfeed),ANY" ") : IF TALLY(Zcoord$,".") = 0 THEN Zcoord$ = Zcoord$+"."

    Nsegs% = 2*(INT(TotalLengthWaves*100)\2) 'even number, total # segs

    NumSegs$ = REMOVE$(STR$(Nsegs%\2),ANY" ") '# segs per arm

    N% = FREEFILE

    OPEN "PBM4.NEC" FOR OUTPUT AS #N%

        PRINT #N%,"CM File: PBM4.NEC"
        PRINT #N%,"CM Run ID "+DATE$+" "+TIME$
        PRINT #N%,"CM Nd="+STR$(Nd%)+", p="+STR$(p%)+", j="+STR$(j&)
        PRINT #N%,"CM R(p,1,j)="+STR$(R(p%,1,j&))+", R(p,2,j)="+STR$(R(p%,2,j&))
        PRINT #N%,"CE"

        PRINT #N%,"GW 1,5,0.,0.,-"+FeedZcoord$+",0.,0.,"+FeedZcoord$+",0.00001" 'feed wire, 1 segment, 0.01 wvln

        PRINT #N%,"GW 2,"+NumSegs$+",0.,0.,"+FeedZcoord$+","+Xcoord$+",0.,"+Zcoord$+",0.00001" 'upper arm

        PRINT #N%,"GW 3,"+NumSegs$+",0.,0.,-"+FeedZcoord$+","+Xcoord$+",0.,-"+Zcoord$+",0.00001" 'lower arm

        PRINT #N%,"GE"

        PRINT #N%,"EX 0,1,3,0,1.,0." 'VOLTAGE SOURCE

        PRINT #N%,"FR 0,1,0,0,299.79564,0."
        PRINT #N%,"RP 0,1,1,1001,90.,0.,0.,0.,1000." 'ENDFIRE gain at 1000 wavelengths range
        PRINT #N%,"XQ"
        PRINT #N%,"EN"

    CLOSE #N%

'     - - ANGLES - -       - POWER GAINS -      - - - POLARIZATION - - -    - - - E(THETA) - - -  - - - E(PHI) - - -
'  THETA      PHI     VERT.   HOR.   TOTAL    AXIAL   TILT  SENSE   MAGNITUDE    PHASE    MAGNITUDE    PHASE
' DEGREES   DEGREES    DB      DB      DB      RATIO   DEG.         VOLTS/M    DEGREES    VOLTS/M    DEGREES
'  90.00     0.00     3.91  -999.99   3.91   0.00000   0.00 LINEAR  1.29504E-04   5.37   0.00000E+00   -5.24
'123456789x123456789x123456789x123456789x123456789x123456789x123456789x123456789x123456789x123456789x123456789x
'     10       20      30      40      50      60       70      80      90      100      110      120

    SHELL "n41_2k1.exe",0

    N% = FREEFILE

    OPEN "PBM4.OUT" FOR INPUT AS #N%

        WHILE NOT EOF(N%)

            LINE INPUT #N%, Lyne$

            IF INSTR(Lyne$,"DEGREES  DEGREES") > 0 THEN EXIT LOOP
```

```
        WEND 'position at next data line

        LINE INPUT #N%, Lyne$

    CLOSE #N%

    GainDB$ = REMOVE$(MID$(Lyne$,37,8),ANY" ")

    PBM_4 = 10^(VAL(GainDB$)/10##) 'Directivity

END FUNCTION 'PBM_4()

'----

FUNCTION PBM_5(R(),Nd%,p%,j&) 'PBM Benchmark #5: N-element collinear array (Nd=N-1)

    LOCAL TotalLengthWaves, Di(), Ystart, Y1, Y2, SumDi AS EXT

    LOCAL N%, i%, q%

    LOCAL Lyne$, GainDB$

    REDIM Di(1 TO Nd%)

    FOR i% = 1 TO Nd%

        Di(i%) = R(p%,i%,j&) 'dipole separation, wavelengths

    NEXT i%

    TotalLengthWaves = 0##

    FOR i% = 1 TO Nd%

        TotalLengthwaves = TotalLengthWaves + Di(i%)

    NEXT i%

    TotalLengthWaves = TotalLengthWaves + 0.5## 'add half-wavelength of 1 meter at 299.8 MHz

    Ystart = -TotalLengthWaves/2##

    N% = FREEFILE

    OPEN "PBM5.NEC" FOR OUTPUT AS #N%

        PRINT #N%,"CM File: PBM5.NEC"
        PRINT #N%,"CM Run ID "+DATE$+" "+TIME$
        PRINT #N%,"CM Nd="+STR$(Nd%)+", p="STR$(p%)+", j="+STR$(j&)
        PRINT #N%,"CM R(p,1,j)="+STR$(R(p%,1,j&))+", R(p,2,j)="+STR$(R(p%,2,j&))
        PRINT #N%,"CE"

        FOR i% = 1 TO Nd%+1

            SumDi = 0##

            FOR q% = 1 TO i%-1

                SumDi = SumDi + Di(q%)

            NEXT q%

            Y1 = ROUND(Ystart + SumDi,6)

            Y2 = ROUND(Y1+0.5##,6) 'add one-half wavelength for other end of dipole

            PRINT #N%,"GW "+REMOVE$(STR$(i%),ANY" ")+",49,0.,"+REMOVE$(STR$(Y1),ANY" ")+",0.,0.,"+REMOVE$(STR$(Y2),ANY" ")+",0.,0.00001"

        NEXT i%

        PRINT #N%,"GE"

        FOR i% = 1 TO Nd%+1
            PRINT #N%,"EX 0,"+REMOVE$(STR$(i%),ANY" ")+",25,0,1.,0." 'VOLTAGE SOURCES
        NEXT i%

        PRINT #N%,"FR 0,1,0,0,299.79564,0."
        PRINT #N%,"RP 0,1,1,1001,90.,0.,0.,0.,1000." 'gain at 1000 wavelengths range
        PRINT #N%,"XQ"
        PRINT #N%,"EN"

    CLOSE #N%

'    - - ANGLES - -     - POWER GAINS -      - - - POLARIZATION - - -   - - - E(THETA) - - -    - - - E(PHI) - - -
'   THETA    PHI      VERT.   HOR.   TOTAL     AXIAL    TILT   SENSE   MAGNITUDE    PHASE    MAGNITUDE    PHASE
' DEGREES  DEGREES      DB     DB     DB       RATIO    DEG.            VOLTS/M    DEGREES     VOLTS/M    DEGREES
'   90.00    0.00      3.91 -999.99   3.91   0.00000    0.00 LINEAR   1.29504E-04    5.37   0.00000E+00    -5.24
'123456789x123456789x123456789x123456789x123456789x123456789x123456789x123456789x123456789x123456789x123456789x
'      10       20       30      40     50      60       70      80      90       100      110      120

    SHELL "n41_2k1.exe",0

    N% = FREEFILE

    OPEN "PBM5.OUT" FOR INPUT AS #N%

        WHILE NOT EOF(N%)

            LINE INPUT #N%, Lyne$

            IF INSTR(Lyne$,"DEGREES  DEGREES") > 0 THEN EXIT LOOP

        WEND 'position at next data line

        LINE INPUT #N%, Lyne$

    CLOSE #N%

    GainDB$ = REMOVE$(MID$(Lyne$,37,8),ANY" ")

    PBM_5 = 10^(VAL(GainDB$)/10##) 'Directivity

END FUNCTION 'PBM_5()
'*********************************************************** END PROGRAM 'CFO_02-07-2010.BAS'  *********************************************************************
```